\documentclass[showpacs,preprintnumbers,amsmath,amssymb]{revtex4}

\usepackage{epsfig}
\usepackage{graphicx}
\usepackage{dcolumn}
\usepackage{bm}
\usepackage{epstopdf}
\usepackage{subfigure}

\begin{document}

\title{The $SO(3)\times SO(3)\times U(1)$ Hubbard model on a square lattice 
in terms of $c$ and $\alpha\nu$ fermions and deconfined $\eta$-spinons and spinons}
\author{J. M. P. Carmelo} 
\affiliation{Centre of Physics, University of Minho, Campus Gualtar, P-4710-057 Braga, Portugal}
\affiliation{Beijing Computational Science Research Center, Beijing 100084, China}

\date{20 May 2011}


\begin{abstract}
In this paper a general description for the Hubbard model with nearest-neighbor transfer integral $t$ and on-site repulsion $U$
on a square lattice with $N_a^2\gg 1$ sites is introduced. It refers to three types of elementary objects whose occupancy configurations 
generate the state representations of the model extended global $SO(3)\times SO(3)\times U(1)$ symmetry recently
found in Ref. \cite{bipartite}. Such objects emerge from a suitable 
electron - rotated-electron unitary transformation. It is such that rotated-electron single and double occupancy are good quantum numbers for 
$U\neq 0$. The advantage of the description is that it accounts for the new found hidden
$U(1)$ symmetry in $SO(3)\times SO(3)\times U(1)=[SU(2)\times SU(2)\times U(1)]/Z_2^2$ beyond the well-known
$SO(4)=[SU(2)\times SU(2)]/Z_2$ model (partial) global symmetry. Specifically, the hidden
$U(1)$ symmetry state representations store full information on the
positions of the spins of the rotated-electron singly occupied sites relative to the remaining sites. Profiting from that complementary 
information, for the whole $U/4t>0$ interaction range independent spin state representations are 
naturally generated in terms of spin-$1/2$ spinon occupancy configurations in a spin
effective lattice. For all states such an effective lattice has as many sites as spinons. 
This allows the extension to intermediate $U/4t$ values of the usual large-$U/4t$ descriptions of the spin 
degrees of freedom of the electrons that singly occupy sites, now in terms of the spins of the singly-occupied sites rotated electrons. 
The operator description introduced in this paper brings about a more suitable scenario for handling the effects of hole doping.
Within it such effects are accounted for in terms of the residual interactions of the elementary objects whose occupancy configurations 
generate the state representations of the charge hidden $U(1)$ symmetry and
spin $SU(2)$ symmetry, respectively. This problem is investigated elsewhere. The most interesting physical
information revealed by the description refers to the model on the subspace generated by application of one- and two-electron
operators onto zero-magnetization ground states. (This is the square-lattice quantum liquid further studied
in Ref. \cite{companion}.) However, to access such an information one must start from the general
description introduced in this paper, which refers to the model in the full Hilbert space.  
\end{abstract}

\maketitle

\section{Introduction}

The Hubbard model on a bipartite lattice is the simplest realistic toy model for description of the electronic correlation effects in 
general many-electron problems with short-range interaction on such a lattice. The model involves two effective parameters: the in-plane nearest-neighbor
transfer integral $t$ and the effective on-site repulsion $U$. Despite that it is among the mostly studied models in condensed 
matter and ultra-cold atom physics, except for the one-dimensional (1D) bipartite lattice \cite{Lieb,Takahashi,Martins} there is no exact solution and few controlled approximations exist for finite $U/4t$ values. 

One of the few exact results for the model on a general bipartite lattice is that the ground state of the repulsive half-filled Hubbard model
is for finite number of lattice sites and spin density $m=0$ a spin singlet \cite{Lieb-89}. A method to achieve further useful information about the Hubbard
model on a bipartite lattice such as the square lattice \cite{companion}, honeycomb lattice \cite{Alejandro-10}, or cubic lattice \cite{cubic} involves the use of suitable 
systems of ultra-cold fermionic atoms. Indeed, one may expect very detailed experimental results over a wide range of parameters to be available \cite{Zoller}.
Preliminary results referring to the Mott-Hubbard insulating phase of the the Hubbard model on a cubic lattice are reported in Ref. \cite{cubic}.

There is another exact result concerning the Hubbard model on any bipartite lattice. It is that in addition to the spin $SU(2)$ symmetry the model has a second global $SU(2)$ symmetry \cite{Heilmann,Lieb-89}, called by some authors and in this paper $\eta$-spin symmetry \cite{Zhang}. A trivial result is that at $U=0$ the global symmetry of the Hubbard model on a bipartite lattice at vanishing chemical potential and magnetic field is $O(4)=SO(4)\times Z_2$. Here the factor $Z_2$ refers to the particle-hole transformation on a single spin under which the model Hamiltonian is not invariant for $U\neq 0$ and $SO(4)=[SU(2)\times SU(2)]/Z_2$ contains the two $SU(2)$ symmetries. Following work by Heilmann and Lieb \cite{Heilmann,Lieb-89}, Yang and Zhang considered the most natural possibility that the $SO(4)$ symmetry inherited from the $U=0$ Hamiltonian $O(4)=SO(4)\times Z_2$ symmetry was the model global symmetry for $U>0$ \cite{Zhang}.

However, a recent study of the problem by the author and collaborators reported in Ref. \cite{bipartite} reveals an exact extra hidden global $U(1)$ symmetry emerging for $U\neq 0$ in addition to $SO(4)$. It is related to the $U\neq 0$ local $SU(2)\times SU(2) \times U(1)$ gauge symmetry of the Hubbard model on a bipartite lattice with vanishing transfer integral \cite{U(1)-NL}. Such a local $SU(2)\times SU(2) \times U(1)$ gauge symmetry becomes for finite $U$ and $t$ a group of permissible unitary transformations. It is such that the corresponding local $U(1)$ canonical transformation is not the ordinary $U(1)$ gauge subgroup of electromagnetism. Instead it is a ``nonlinear" transformation \cite{U(1)-NL}. For $U\neq 0$ the related new found global symmetry of the model on any bipartite lattice is larger than $SO(4)$ and given by $[SU(2)\times SU(2)\times U(1)]/Z_2^2=[SO(4)\times U(1)]/Z_2=SO(3)\times SO(3)\times U(1)$. The factor $1/Z_2$ (and $1/Z_2^2$) in $SO(4)=[SU(2)\times SU(2)]/Z_2$ (and in $[SU(2)\times SU(2)\times U(1)]/Z_2^2$) imposes that $[S_{\eta}+S_s]$ is an integer number (and both $[S_s+S_c]$ and $[S_{\eta}+S_c]$ are integer numbers). Here $S_{\eta}$, $S_s$, and $S_c$ are the $\eta$-spin, the spin, and the eigenvalue of the generator of the new global $U(1)$ symmetry, respectively. The latter is found in Ref. \cite{bipartite} to be one half the number of rotated-electron singly occupied sites. This refers to any of the infinite electron - rotated-electron unitary transformations of Ref. \cite{Stein}, such that rotated-electron single and double occupancy are good quantum numbers for $U/4t\neq 0$. $2S_c$ is then the number of rotated-electron singly occupied sites. Within the present notation, $S^{x_3}_{\eta}= -[N_a^D-N]/2$ and $S^{x_3}_s= -[N_{\uparrow}-N_{\downarrow}]/2$ are the $\eta$-spin projection and spin projection, respectively, and $N_a^D\gg 1$ denotes the number of lattice sites. For the bipartite 1D and square lattices considered in this paper the labeling index $D$ in $N_a^D\equiv [N_a]^D$ reads $D=1$ and $D=2$, accounting for the $N_a\gg 1$ and $N_a^2=N_a\times N_a\gg 1$ lattice sites, respectively. The square and 1D lattices have spacing $a$ and length edge and chain length $L=N_a\,a$, respectively. 

An important point is that although addition of chemical-potential and magnetic-field operator terms to the Hubbard model on a bipartite lattice Hamiltonian lowers its symmetry, such terms commute with it. Therefore, the global symmetry being $SO(3)\times SO(3)\times U(1)$ implies that the set of independent rotated-electron occupancy configurations that generate the model energy and momentum eigenstates generate state representations of that global symmetry for all values of the electronic density $n$ and spin density $m$. It then follows that the total number of such independent representations must equal the Hilbert-space dimension, $4^{N_a^D}$. The results of Ref. \cite{bipartite} confirm that for the model on a bipartite lattice in its full Hilbert space the number of independent representations of the group $SO(3)\times SO(3)\times U(1)$ is indeed $4^{N_a^D}$. In contrast, the number of independent representations of the group $SO(4)$ is found to be smaller than the Hilbert-space dimension $4^{N_a^D}$. Studies of the Hubbard model on a bitartie lattice that rely on its transformation 
properties under symmetry operations \cite{Kampfer-05} should be extended to account for
the new hidden $U(1)$ symmetry.

The Hubbard model on the bipartite 1D lattice has an exact solution \cite{Lieb,Takahashi,Martins}, which can be shown to be fully 
consistent with its extended global symmetry. The exact solution of 1D integrable models can be reached by two different methods: 
The coordinate Bethe ansatz (BA) used by Bethe himself \cite{Bethe} and the more 
algebraic operator formulation usually called inverse scattering method \cite{ISM}. For the 1D Hubbard model such an algebraic formulation of the Bethe states refers to the transfer matrix of the classical coupled spin model, which is its ``covering" \cite{CM}. Indeed, within the inverse scattering method \cite{Martins,ISM} the central object to be diagonalized is the quantum transfer matrix rather than the underlying 1D Hubbard model. The transfer-matrix eigenvalues provide the spectrum of a set of conserved charges. The diagonalization of the charge degrees of freedom involves a transfer matrix associated with a charge monodromy matrix \cite{Martins}. Its off-diagonal entries are some of the creation and annihilation fields whose application onto a suitable vacuum generates the charge degrees of freedom of the energy eigenstates configurations. The solution of the spin degrees of freedom involves the 
diagonalization of the auxiliary transfer matrix associated with the spin monodromy 
matrix \cite{Martins}. Again, the off-diagonal entries of that matrix play the role of the creation and annihilation fields that generate the spin
degrees of freedom of the energy eigenstates configurations. 

The global symmetry of a solvable model is explicit in the algebraic operator formulation of its solution. This is why the solution of the 1D Hubbard model by the algebraic inverse scattering method \cite{Martins} was achieved only thirty years after that of the coordinate BA \cite{Lieb,Takahashi}. Indeed, it was expected that the charge and spin monodromy matrices of the former method had the same traditional 
Faddeev-Zamolodchikov ABCD form, found previously for the related 1D isotropic Heinsenberg model \cite{ISM}. Such an expectation was that consistent with the occurrence of a spin $SU(2)$ symmetry and a $\eta$-spin charge $SU(2)$ symmetry 
associated with a global $SO(4)=[SU(2)\times SU(2)]/Z_2$ symmetry \cite{Zhang}. If that was the whole global symmetry of the 1D Hubbard model, the charge and spin monodromy matrices would have the same ABCD form. 

Consistently with the recently found model extended global symmetry, all tentative schemes using charge and spin monodromy matrices of the same ABCD form failed to achieve the BA equations obtained by means of the coordinate BA \cite{Lieb,Takahashi}. As discussed in Appendix A, the problem was solved by Martins and Ramos, who used an appropriate representation of the charge and spin monodromy matrices, which allows for possible {\it hidden symmetries} \cite{Martins}. Indeed, the structure of the charge and spin monodromy matrices introduced by these authors is able to distinguish creation and annihilation fields as well as possible hidden symmetries. That for $U>0$ the 1D Hubbard model spin and charge degrees of freedom are associated with the $SU(2)$ and $U(2)=SU(2)\times U(1)$ symmetries, respectively, of $[SU(2)\times U(2)]/Z_2^2=SO(3)\times SO(3)\times U(1)$ rather than merely with two $SU(2)$ symmetries, is behind the different ABCD and ABCDF forms found in Ref. \cite{Martins} for the inverse-scattering method BA solution spin and charge monodromy matrices, respectively: Due to the extra $U(1)$ symmetry in $U(2)=SU(2)\times U(1)$, the latter matrix is larger than the former and involves more fields. 

Although the model on bipartite lattices other than the 1D lattice has no exact solution to be related to its global symmetry, that for $U>0$ such a symmetry is
$SO(3)\times SO(3)\times U(1)$ is expected to have important physical consequences. One of the goals of this paper is the introduction of a general quantum-object description for the Hubbard model on the bipartite square lattice to be used in Ref. \cite{companion} and elsewhere in the search of such physical consequences. To meet
that goal, the description introduced in this paper accounts for a set of independent rotated-electron occupancy configurations that generate the model state representations of the global $SO(3)\times SO(3)\times U(1)$ symmetry for all values of the electronic density $n$ and spin density $m$. For the 1D Hubbard model
such a complete set of states associated with the operator description introduced here are energy eigenstates and refer to the exact solution quantum numbers and set of conserving quantities. Due to the square-lattice Hubbard model lack of integrability and corresponding lack of an infinite number of conservation laws, for it only in the one- and two-electron subspace introduced in Section V are the above states energy eigenstates. Fortunately, such model in that subspace describes the important one- and two-electron physics. Furthermore, one of the motivations for our study focusing onto the Hubbard model on the square lattice follows from that model with a small three-dimensional (3D) uniaxial anisotropy perturbation being the simplest candidate toy model for describing the effects of electronics 
correlations in the high-$T_c$ superconductors \cite{ARPES-review,2D-MIT}.

The first step of our program involves an electron - rotated-electron unitary transformation of the type considered in Ref. \cite{Stein}.
A property specific to our transformation is that the
$U>0$ energy eigenstates can be generated from suitable chosen $U/4t\rightarrow\infty$ energy eigenstates upon application onto
the latter states of the corresponding electron - rotated-electron unitary operator. The related set of 
state representations of the group $SO(3)\times SO(3)\times U(1)$ that emerge from our
description are energy eigenstates yet are generated by exactly the same 
electron - rotated-electron unitary transformation from corresponding $U/4t\rightarrow\infty$ states. 
Their rotated electron occupancy configurations are simpler to describe in terms of three types 
of elementary objects directly related to the rotated electrons whose numbers and designations are: 
$M_s=2S_c$ spin-$1/2$ spinons, $M_{\eta}=[N_a^D-2S_c]$ 
$\eta$-spin-$1/2$ $\eta$-spinons, and $N_c=2S_c$ spin-less and $\eta$-spin-less 
charge $c$ fermions. The latter live on a lattice with $N_a^D =[N_c+N_c^h]$ sites identical
to the original lattice. Here $N_c^h= [N_a^D-2S_c]$ gives the number of
$c$ fermion holes. The relation of such objects to the rotated electrons is as follows.
The $M_s=2S_c$ spin-$1/2$ spinons describe the spin degrees of freedom 
of the $2S_c$ rotated electrons that singly occupy sites. The charge degrees
of freedom of such rotated electrons are described by the $N_c =2S_c$ $c$ fermions. The $M_{\eta}=[N_a^D-2S_c]$
$\eta$-spin-$1/2$ $\eta$-spinons describe the $\eta$-spin degrees
of freedom of the $[N_a^D-2S_c]$ sites doubly occupied and unoccupied by rotated electrons. 
Specifically, the $\eta$-spinons of $\eta$-spin projection $-1/2$ and $+1/2$ refer to the sites doubly
occupied and unoccupied, respectively, by rotated electrons. The remaining degrees of freedom
of such rotated-electron occupancy configurations are described by the $N_c^h= [N_a^D-2S_c]$ 
$c$ fermion holes. The expression of the rotated-electron
operators in terms of these three elementary objects is for $U>0$ identical to that of the electron
operators for large $U$ values in terms of the three objects obtained from
the exact transformation for separation of spin-$1/2$ fermions without constraints considered in Ref. \cite{Ostlund-06}.

Interestingly, $\eta$-spinon (and spinons) that are not invariant under the electron - rotated-electron unitary transformation
considered in this paper have $\eta$-spin $1/2$ (and spin $1/2$) but are confined within $\eta$-spin-neutral (and spin-neutral) 
$2\nu$-$\eta$-spinon (and $2\nu$-spinon) composite 
$\eta\nu$ fermions (and $s\nu$ fermions). Here $\nu=1,2,...$
is the number of confined $\eta$-spinon (and spinon) pairs. 
We emphasize that by ``confinement" is meant here that the spinons and $\eta$-spinons are bound
and anti-bound within such composite $s\nu$ and $\eta\nu$ fermions, respectively, alike for instance
protons and neutrons are bound within the nucleus. Whether the potential associated with such a
behavior is in some limit confining remains an interesting open question.
In turn, a well-defined number of deconfined spinons and deconfined $\eta$-spinons
are invariant under the electron - rotated-electron unitary transformation. Confined and deconfined spinons play 
an important role in the concept of deconfined quantum criticality \cite{Senthil-04}. The relation 
of the physical consequences of the model extended global symmetry
to such a concept is one of the interesting issues raised by our general
description to be investigated elsewhere.

In this paper often we discuss the use of our description for the model on a 1D lattice as well. 
Indeed our quantum-object description also applies to 1D.
However, due to the integrability of the 1D model associated with its
exact solution \cite{Lieb,Martins,Takahashi}, concerning some properties it leads to a 
different physics. Nonetheless there are also some common properties.
One of the procedures used to control the validity of the approximations used in 
the construction of our description profits from identifying
the common proprieties of the model on the square and 1D lattices. 
Some of these common properties are related to the model commuting with the generators of the group 
$SO(3)\times SO(3)\times U(1)$ both for the square and 1D bipartite lattices \cite{bipartite}.
As confirmed in Appendix A, for the 1D Hubbard model the states generated by the simple 
momentum occupancy configurations of the objects of our description are exact energy 
eigenstates for the whole Hilbert space and all finite values of the onsite repulsion $U$. For the Hubbard
model on the square lattice such occupancy configurations generate momentum eigenstates that 
are not in general energy eigenstates. Fortunately, they are so for that model in the one- and two-electron subspace defined in Section V,
which refers to the square-lattice quantum liquid introduced in that section and further studied in Ref. \cite{companion}.

The paper is organized as follows. A uniquely defined rotated-electron description 
for the Hubbrad model on the square lattice is the subject of Section II. In Section III the $c$ fermion, $\eta$-spin-$1/2$ $\eta$-spinon, and
spin-$1/2$ spinon and corresponding $c$, $\eta$-spin, and spin 
effective lattices are introduced. The vacua of the theory, 
the transformation laws under the electron - rotated-electron unitary
transformation of such objects, and the subspaces they refer to are issues also addressed
in that section. The composite $\alpha\nu$ bond particles and $\alpha\nu$ fermions, 
corresponding $\alpha\nu$ effective lattices, ground-state occupancy configurations,
and a complete set of momentum eigenstates
are the problems studied in Section IV. The one- and two-electron electron subspace
and the corresponding square-lattice quantum liquid are the subjects of Section V.
In Section VI the local $s1$ fermion operators and related $s1$ bond-particle operators
are expressed in terms of elementary spinon operators. Section VII contains the concluding remarks.
In Appendix A it is confirmed that the general operator description introduced in this 
paper for the Hubbard model on the square and 1D lattices is consistent with the exact 
solution of the 1D problem. For instance, it is confirmed that the discrete momentum values 
of the $c$ fermions and composite $\alpha\nu$ fermions are good quantum numbers. Furthermore, 
in that Appendix the relation of the creation and annihilation fields of the charge ABCDF algebra  
\cite{Martins} and more traditional spin ABCD Faddeev-Zamolodchikov algebra \cite{ISM} of the 
algebraic formulation of the 1D exact solution of Ref. \cite{Martins} to the $c$
and $\alpha\nu$ fermion operators is discussed. The consistency between the two corresponding 
operational representations is confirmed.   
Moreover, that the whole physics can be extracted from the Hubbard model in the lowest-weight state
(LWS) subspace spanned by the $\eta$-spin and spin LWSs is the subject addressed in Appendix B.
In Appendix C additional side information on the one- and two-electron subspace is provided.
Finally, Appendix D presents some technical results on the $s1$ bond-particle operators algebra.
  
\section{The model, a suitable rotated-electron description, 
and relation to the global $SO(3)\times SO(3)\times U(1)$ symmetry}

The Hubbard model on a square (or 1D) lattice with a very large number $N_a^D\gg 1$ of sites reads,
\begin{equation}
\hat{H} = t\,\hat{T} + U\,[N_a^D-\hat{Q}]/2 
\, ; \hspace{0.25cm}
\hat{T} = -\sum_{\langle\vec{r}_j\vec{r}_{j'}\rangle}\sum_{\sigma}[c_{\vec{r}_j,\sigma}^{\dag}\,c_{\vec{r}_{j'},\sigma}+h.c.] \, ;
\hspace{0.25cm}
{\hat{Q}} = \sum_{j=1}^{N_a^D}\sum_{\sigma =\uparrow
,\downarrow}\,\hat{n}_{\vec{r}_j,\sigma}\,(1- \hat{n}_{\vec{r}_j, -\sigma}) \, .
\label{H}
\end{equation}
Periodic boundary conditions and torus periodic boundary conditions are considered
for the 1D lattice for which $D=1$ and the square lattice for which
$D=2$, respectively. Moreover, in Eq. (\ref{H}) $t$ is the nearest-neighbor transfer integral, $\hat{T}$ is the kinetic-energy operator 
in units of $t$, and ${\hat{Q}}$ is the operator that counts the number of electron 
singly occupied sites. Hence the operator ${\hat{D}}=[{\hat{N}}-{\hat{Q}}]/2$
counts the number of electron doubly occupied sites.
Moreover, $\hat{n}_{{\vec{r}}_j,\sigma} = c_{\vec{r}_j,\sigma}^{\dag} c_{\vec{r}_j,\sigma}$
where $-\sigma=\uparrow$ (and $-\sigma=\downarrow$)
for $\sigma =\downarrow$ (and $\sigma =\uparrow$), ${\hat{N}} = \sum_{\sigma}
{\hat{N}}_{\sigma}$, and ${\hat{N}}_{\sigma}=\sum_{j=1}^{N_a^D}
\hat{n}_{{\vec{r}}_j,\sigma}$. 

The kinetic-energy operator $\hat{T}$ can be expressed in terms of the operators, 
\begin{eqnarray}
\hat{T}_0 & = & -\sum_{\langle\vec{r}_j\vec{r}_{j'}\rangle}\sum_{\sigma}[\hat{n}_{\vec{r}_{j},-\sigma}\,c_{\vec{r}_j,\sigma}^{\dag}\,
c_{\vec{r}_{j'},\sigma}\,\hat{n}_{\vec{r}_{j'},-\sigma} +
(1-\hat{n}_{\vec{r}_{j},-\sigma})\,c_{\vec{r}_j,\sigma}^{\dag}\,
c_{\vec{r}_{j'},\sigma}\,(1-\hat{n}_{\vec{r}_{j'},-\sigma})+h.c.] \, ,
\nonumber \\
\hat{T}_{+1} & = & -\sum_{\langle\vec{r}_j\vec{r}_{j'}\rangle}\sum_{\sigma}
\hat{n}_{\vec{r}_{j},-\sigma}\,c_{\vec{r}_j,\sigma}^{\dag}\,c_{\vec{r}_{j'},\sigma}\,(1-\hat{n}_{\vec{r}_{j'},-\sigma}) \, ,
\nonumber \\
\hat{T}_{-1} & = & -\sum_{\langle\vec{r}_j\vec{r}_{j'}\rangle}\sum_{\sigma}
(1-\hat{n}_{\vec{r}_{j},-\sigma})\,c_{\vec{r}_j,\sigma}^{\dag}\,
c_{\vec{r}_{j'},\sigma}\,\hat{n}_{\vec{r}_{j'},-\sigma} \, ,
\label{T-op}
\end{eqnarray}
as $\hat{T}= \hat{T}_0 + \hat{T}_{+1} + \hat{T}_{-1}$. These three kinetic operators play an 
important role in the physics. The operator $\hat{T}_0$ does not change electron double 
occupancy whereas the operators $\hat{T}_{+1}$ and $\hat{T}_{-1}$ change it by $+1$ 
and $-1$, respectively. 

We focus our attention onto ground states with 
hole concentration $x= [N_a^D-N]/N_a^D\geq 0$ and spin
density $m= [N_{\uparrow}-N_{\downarrow}]/N_a^D=0$ and their excited states. 
We are particularly interested in the
{\it LWS subspace} spanned by the LWSs of both the $\eta$-spin and
spin algebras. Such energy eigenstates refer to values
of $S_{\alpha}$ and $S^{x_3}_{\alpha}$ such that 
$S_{\alpha}=-S^{x_3}_{\alpha}$ for $\alpha =\eta,s$. 
The off-diagonal generators that generate the
non-LWSs from the LWSs commute with the operator $\hat{V}$ \cite{bipartite}.
Thus the whole physics can be extracted 
from the model (\ref{H}) in the LWS subspace,
as confirmed in Appendix B.

The studies of Ref. \cite{bipartite} consider unitary operators
$\hat{V}=\hat{V}(U/4t)$ and corresponding rotated-electron operators,
\begin{equation}
{\tilde{c}}_{\vec{r}_j,\sigma}^{\dag} =
{\hat{V}}^{\dag}\,c_{\vec{r}_j,\sigma}^{\dag}\,{\hat{V}}
\, ; \hspace{0.35cm}
{\tilde{c}}_{\vec{r}_j,\sigma} =
{\hat{V}}^{\dag}\,c_{\vec{r}_j,\sigma}\,{\hat{V}} 
\, ; \hspace{0.35cm}
{\tilde{n}}_{\vec{r}_j,\sigma} = {\tilde{c}}_{\vec{r}_j,\sigma}^{\dag}\,{\tilde{c}}_{\vec{r}_j,\sigma} \, .
\label{rotated-operators}
\end{equation}
Those are such that rotated-electron
single and double occupancy are good quantum
numbers for $U/4t>0$. The global $U(1)$ symmetry generator ${\tilde{S}}_c$
of eigenvalue $S_c$ reads \cite{bipartite},
\begin{equation}
{\tilde{S}}_c = {1\over 2}\sum_{j=1}^{N_a^D}\sum_{\sigma =\uparrow
,\downarrow}\,{\tilde{n}}_{\vec{r}_j,\sigma}\,(1- {\tilde{n}}_{\vec{r}_j,-\sigma}) \, .
\label{Or-ope}
\end{equation}
It follows that ${\tilde{S}}_c= {\hat{V}}^{\dag}\,{\hat{S}}_c\,{\hat{V}}$
where ${\hat{S}}_c= {1\over 2}{\hat{Q}}$ and the operator
${\hat{Q}}$ is given in Eq. (\ref{H}). As mentioned in the previous section,
$2S_c$ is the number of rotated-electron singly occupied sites. 
Most choices of the unitary operators $\hat{V}$ correspond 
to choices of $U/4t\rightarrow\infty$ sets
$\{\vert\Psi_{\infty}\rangle\}$ of $4^{N_a^D}$
energy eigenstates such that the states 
$\vert \Psi_{U/4t}\rangle={\hat{V}}^{\dag}\vert\Psi_{\infty}\rangle$
are not energy and momentum eigenstates for finite $U/4t$ values yet belong
to a subspace with fixed and well-defined values of $S_c$, $S_{\eta}$, 
and $S_s$. 

Let $\{\vert \Psi_{U/4t}\rangle\}$ be a
complete set of $4^{N_a^D}$ energy, momentum, $\eta$-spin,
$\eta$-spin projection, spin, and spin-projection eigenstates for
$U/4t>0$. In the limit $U/4t\rightarrow\infty$ such states correspond
to one of the many choices of sets $\{\vert\Psi_{\infty}\rangle\}$ of 
$4^{N_a^D}$ $U/4t$-infinite energy eigenstates. Both the sets of states
$\{\vert \Psi_{U/4t}\rangle\}$ and  $\{\vert \Psi_{\infty}\rangle\}$, respectively, are complete and the model Hilbert 
space is the same for all $U/4t>0$ values considered here. Hence it follows from
basic quantum mechanics Hilbert-space and operator properties that
for this choice there exists exactly one unitary operator ${\hat{V}}={\hat{V}}(U/4t)$
such that $\vert \Psi_{U/4t}\rangle ={\hat{V}}^{\dag}\vert\Psi_{\infty}\rangle$.
Here we consider such a unitary operator and corresponding generator
${\tilde{S}}_c= {\hat{V}}^{\dag}\,{\hat{S}}_c\,{\hat{V}}$ given in Eq.
(\ref{Or-ope}) and rotated-electron operators provided in Eq. (\ref{rotated-operators}).
The states $\vert \Psi_{U/4t}\rangle ={\hat{V}}^{\dag}\vert\Psi_{\infty}\rangle$
(one for each value of $U/4t>0$) that are 
generated from the same initial state $\vert\Psi_{\infty}\rangle$ 
belong to the same {\it $V$ tower}. 

A complete set $\{\vert \Phi_{U/4t}\rangle\}$ of related momentum eigenstates 
$\vert \Phi_{U/4t}\rangle ={\hat{V}}^{\dag}\vert\Phi_{\infty}\rangle$ 
is considered below in Section IV-E. Such states
are generated by occupancy configurations of the quantum objects of the general description
introduced in this paper. The unitary operator ${\hat{V}}^{\dag}$ appearing in the
general expression of such states is that which also appears in the general expression 
$\vert \Psi_{U/4t}\rangle ={\hat{V}}^{\dag}\vert\Psi_{\infty}\rangle$ of the above energy and momentum eigenstates.
However the states $\vert \Phi_{U/4t}\rangle$ are not in general energy eigenstates
of the Hubbard model on the square lattice. The interest of the states $\vert \Phi_{U/4t}\rangle ={\hat{V}}^{\dag}\vert\Phi_{\infty}\rangle$
is that those contained in the one- and two-electron subspace defined in Section V are both momentum 
and energy eigenstates of that model. Fortunately, the general description introduced in this paper 
is physically most useful and important for
the Hubbard model on the square lattice in that subspace, which refers to the square-lattice quantum liquid
further investigated in Ref. \cite{companion}.

Alike in Ref. \cite{bipartite}, we associate with any operator ${\hat{O}}$ an
operator ${\tilde{O}}={\hat{V}}^{\dag}\,{\hat{O}}\,{\hat{V}}$ that
has the same expression in terms of rotated-electron creation and 
annihilation operators as ${\hat{O}}$ in terms of electron creation and 
annihilation operators, respectively. Our convention is that marks placed 
over letters being a caret or a tilde denote operators. (An exception are the electron operators 
of Eq. (\ref{rotated-operators}), which we denote by $c_{\vec{r}_j,\sigma}^{\dag}$ and
$c_{\vec{r}_j,\sigma}$ rather than by ${\hat{c}}_{\vec{r}_j,\sigma}^{\dag}$ and
${\hat{c}}_{\vec{r}_j,\sigma}$, respectively.) Any operator ${\hat{O}}$ can then
be written in terms of rotated-electron creation and annihilation operators as,
\begin{eqnarray}
{\hat{O}} & = & {\hat{V}}\,{\tilde{O}}\,{\hat{V}}^{\dag}
= {\tilde{O}}+ [{\tilde{O}},{\hat{S}}\,] + {1\over
2}\,[[{\tilde{O}},{\hat{S}}\,],{\hat{S}}\,] + ... 
=  {\tilde{O}}+ [{\tilde{O}},{\tilde{S}}\,] + {1\over
2}\,[[{\tilde{O}},{\tilde{S}}\,],{\tilde{S}}\,] + ... \, ,
\nonumber \\
{\hat{S}} & = & -{t\over U}\,\left[\hat{T}_{+1} -\hat{T}_{-1}\right] 
+ {\cal{O}} (t^2/U^2) \, ; \hspace{0.25cm}
{\tilde{S}} = -{t\over U}\,\left[\tilde{T}_{+1} -\tilde{T}_{-1}\right] 
+ {\cal{O}} (t^2/U^2) \, .
\label{OOr}
\end{eqnarray}
The operator $\hat{S}$ appearing in this equation 
is related to the unitary operator as ${\hat{V}}^{\dag} = e^{{\hat{S}}}$ and
${\hat{V}} = e^{-{\hat{S}}}$. Although the general expression of ${\hat{S}}$ remains
unknown, an exact result is that it involves only the 
kinetic operators $\hat{T}_0$, $\hat{T}_{+1}$, and $\hat{T}_{-1}$
of Eq. (\ref{T-op}) and numerical $U/4t$ dependent coefficients \cite{companion,Stein}. 
For $U/4t>0$ that expression can be expanded in a series of $t/U$. The corresponding first-order
term has the universal form given in Eq. (\ref{OOr}). To arrive
to the expression of ${\hat{O}}$ in terms of the operator
${\tilde{S}}$ also given in Eq. (\ref{OOr}), the property 
that the operator ${\hat{V}}$ commutes with itself is used.
It implies that ${\hat{V}} = e^{-{\hat{S}}}={\tilde{V}} = e^{-{\tilde{S}}}$ and
${\hat{S}}={\tilde{S}}$. Hence both the operators
${\hat{V}}$ and ${\hat{S}}$ have the same expression in terms of
electron and rotated-electron creation and annihilation operators. This justifies why the expansion 
${\tilde{S}} = -(t/U)\,[\tilde{T}_{+1} -\tilde{T}_{-1}] + {\cal{O}} (t^2/U^2)$
given in that equation for the operator ${\tilde{S}}$ has the same form as that of ${\hat{S}}$.

The higher-order terms of the operator ${\tilde{S}}$ expression can be written as a product of 
operator factors. Their expressions involve the rotated
kinetic operators $\tilde{T}_0$, $\tilde{T}_{+1}$, and $\tilde{T}_{-1}$. 
The full expression of the operator $\hat{S}={\tilde{S}}$ can 
for $U/4t>0$ be written as ${\hat{S}}=-{\hat{S}}(\infty)-\delta {\hat{S}}=-{\tilde{S}}(\infty)-\delta {\tilde{S}}$.
Here ${\hat{S}}(\infty)={\tilde{S}}(\infty)$
corresponds to the operator $S(l)$ at $l=\infty$ defined in Eq. (61) 
of Ref. \cite{Stein}. Moreover, $\delta {\hat{S}}=\delta {\tilde{S}}$ has the general form
provided in Eq. (64) of that paper. The unitary operator ${\hat{V}} = e^{-{\hat{S}}}= e^{-{\tilde{S}}}$
considered here corresponds to exactly one choice of the coefficients $D^{(k)}(\bf{m})$ of 
that equation. Here the index $k=1,2,...$ refers to the number 
of rotated-electron doubly occupied sites. The problem of finding the explicit form 
of the operators ${\hat{V}} ={\tilde{V}}$ and ${\hat{S}}={\tilde{S}}$ is equivalent
to finding all coefficients $D^{(k)}(\bf{m})$ associated with the electron - rotated-electron
unitary transformation as defined in this paper.

For finite $U/4t$ values the Hamiltonian $\hat{H}$ of Eq. (\ref{H}) does not commute with 
the unitary operator ${\hat{V}} = e^{-{\hat{S}}}$. Hence when expressed in terms 
of the rotated-electron creation and annihilation operators of Eq. (\ref{rotated-operators}) it has an infinite 
number of terms. According to Eq. (\ref{OOr}) it reads, 
\begin{equation}
{\hat{H}} = {\hat{V}}\,{\tilde{H}}\,{\hat{V}}^{\dag}
= {\tilde{H}} + [{\tilde{H}},{\tilde{S}}\,] + {1\over
2}\,[[{\tilde{H}},{\tilde{S}}\,],{\tilde{S}}\,] + ... \, .
\label{HHr}
\end{equation}
The commutator $[{\tilde{H}},{\tilde{S}}\,]$ does not vanish
except for $U/4t\rightarrow\infty$ so that ${\hat{H}} \neq {\tilde{H}}$ for finite values of $U/4t$. 
Fortunately, for approximately $U/4t> 1$, 
out of the infinite terms on the right-hand-side of Eq. (\ref{HHr}) only the first few Hamiltonian terms play
an active role in the physics of the Hubbard model on the square lattice
in the one- and two-electron subspace \cite{companion}. 

Alike the operator ${\tilde{S}}$, the Hamiltonian expression in terms of rotated-electron operators
(\ref{HHr}) can be expanded in a series of $t/U$. 
Its terms generated up to fourth order in $t/U$ are within a unitary transformation the 
equivalent to the $t-J$ model with ring exchange and various correlated
hoppings \cite{HO-04}. Furthermore and in accordance to a general theorem proved in Ref. \cite{Taka}, at half filling the terms
of the Hamiltonian (\ref{HHr}) expansion in $t/U$ with odd powers in $t$ vanish due to 
the particle-hole symmetry and the resulting invariance of the spectrum under 
$t\rightarrow -t$. In turn, for finite hole concentration 
$x>0$ the expansion in powers of $t/U$ of the Hamiltonian (\ref{HHr}) 
involves terms with odd powers in $t$, absent at $x=0$. This is consistent
with the effects of increasing $U/4t$ being often different at $x=0$ and for $x>0$ \cite{poly}.

In Ref. \cite{bipartite} it is found that, in contrast
to the Hamiltonian, the three components of the momentum operator $\hat{{\vec{P}}}$,
three generators of the spin $SU(2)$ symmetry, and three generators
of the $\eta$-spin $SU(2)$ symmetry commute with the electron - rotated-electron 
unitary operator ${\hat{V}}={\tilde{V}}$. This also holds for the specific
choice of that operator associated with the rotated-electron description
introduced in this paper. Hence the above operators have
the same expression in terms of electron and rotated-electron 
creation and annihilation operators, so that the momentum operator reads,
\begin{equation}
\hat{{\vec{P}}}  = \sum_{\sigma=\uparrow ,\downarrow }\sum_{\vec{k}}\,\vec{k}\,
c_{\vec{k},\sigma }^{\dag }\,c_{\vec{k},\sigma } =
\sum_{\sigma=\uparrow ,\downarrow }\sum_{\vec{k}}\,\vec{k}\,
{\tilde{c}}_{\vec{k},\sigma }^{\dag }\,{\tilde{c}}_{\vec{k},\sigma } \, .
\label{P-invariant}
\end{equation}
Furthermore, the above-mentioned six generators are given by,
\begin{eqnarray}
{\hat{S}}_{\eta}^{x_3} & = & -{1\over 2}[N_a^D-\hat{N}] \, ;
\hspace{0.25cm}
{\hat{S}}_s^{x_3} = -{1\over 2}[{\hat{N}}_{\uparrow}- {\hat{N}}_{\downarrow}] \, ,
\nonumber \\
{\hat{S }}_{\eta}^{\dag} & = & \sum_{j=1}^{N_a^D}e^{i\vec{\pi}\cdot\vec{r}_j}\,c_{\vec{r}_j,\downarrow}^{\dag}\,
c_{\vec{r}_j,\uparrow}^{\dag} =\sum_{j=1}^{N_a^D}e^{i\vec{\pi}\cdot\vec{r}_j}\,{\tilde{c}}_{\vec{r}_j,\downarrow}^{\dag}\,
{\tilde{c}}_{\vec{r}_j,\uparrow}^{\dag} \, ;
\hspace{0.25cm}
{\hat{S}}_{\eta} = \sum_{j=1}^{N_a^D}e^{-i\vec{\pi}\cdot\vec{r}_j}\,c_{\vec{r}_j,\uparrow}\,c_{\vec{r}_j,\downarrow} 
=\sum_{j=1}^{N_a^D}e^{-i\vec{\pi}\cdot\vec{r}_j}\,{\tilde{c}}_{\vec{r}_j,\uparrow}\,{\tilde{c}}_{\vec{r}_j,\downarrow} \, ,
\nonumber \\
{\hat{S}}_s^{\dag} & = &
\sum_{j=1}^{N_a^D}\,c_{\vec{r}_j,\downarrow}^{\dag}\,c_{\vec{r}_j,\uparrow} =
\sum_{j=1}^{N_a^D}\,{\tilde{c}}_{\vec{r}_j,\downarrow}^{\dag}\,{\tilde{c}}_{\vec{r}_j,\uparrow} 
\, ; \hspace{0.25cm}
{\hat{S}}_s = \sum_{j=1}^{N_a^D}c_{\vec{r}_j,\uparrow}^{\dag}\,
c_{\vec{r}_j,\downarrow}=\sum_{j=1}^{N_a^D}{\tilde{c}}_{\vec{r}_j,\uparrow}^{\dag}\,
{\tilde{c}}_{j,\downarrow} \, .
\label{Scs}
\end{eqnarray}
Here for the model on the square (and 1D) lattice the vector $\vec{\pi}$ has 
Cartesian components $\vec{\pi}=[\pi,\pi]$ (and component $\pi$).

In contrast, except in the $U/4t\rightarrow\infty$ limit the generator ${\tilde{S}}_c$ of the charge independent 
$U(1)$ symmetry given in Eq. (\ref{Or-ope}) does not commute with
the unitary operator ${\hat{V}}$. This is behind the hidden character of such a symmetry.
On the contrary of the Hamiltonian, that generator has a complicated expression in terms
of electron creation and annihilation operators and a simple expression given in that equation
in terms of rotated-electron creation and annihilation operators. 
The operator of Eq. (\ref{Or-ope}) plus the six operators provided in Eq. (\ref{Scs}) 
are the seven generators of the group $[SO(4)\times U(1)]/Z_2=SO(3)\times SO(3)\times U(1)$ associated with
the global symmetry of the Hamiltonian (\ref{H}). 
 
\section{Three elementary quantum objects and 
corresponding $c$, $\eta$-spin, and spin effective lattices}

The rotated electrons as defined in this paper are not the ultimate objects whose simple occupancy configurations generate the 
state representations of the group $[SO(4)\times U(1)]/Z_2=SO(3)\times SO(3)\times U(1)$ associated with the global symmetry
of the model (\ref{H}). The rotated-electron occupancy configurations that generate such states are naturally expressed
in terms of those of three basic objects: $c$ fermions, spinons, and $\eta$-spinons. 

\subsection{Elementary quantum objects and their operators}

The local $c$ fermion annihilation operator $f_{\vec{r}_j,c}$ and corresponding creation operator 
$f_{\vec{r}_j,c}^{\dag}=(f_{\vec{r}_j,c})^{\dag}$ are constructed in terms of the rotated-electron 
operators of Eq. (\ref{rotated-operators}). The latter operator reads,
\begin{equation}
f_{\vec{r}_j,c}^{\dag} = (f_{\vec{r}_j,c})^{\dag} =
{\tilde{c}}_{\vec{r}_j,\uparrow}^{\dag}\,
(1-{\tilde{n}}_{\vec{r}_j,\downarrow})
+ e^{i\vec{\pi}\cdot\vec{r}_j}\,{\tilde{c}}_{\vec{r}_j,\uparrow}\,
{\tilde{n}}_{\vec{r}_j,\downarrow} 
\, ; \hspace{0.35cm}
f_{\vec{q}_j,c}^{\dag} =
{1\over {\sqrt{N_a^D}}}\sum_{j'=1}^{N_a^D}\,e^{+i\vec{q}_j\cdot\vec{r}_{j'}}\,
f_{\vec{r}_{j'},c}^{\dag} \, .
\label{fc+}
\end{equation}
In Eq. (\ref{fc+}) we have introduced as well the corresponding $c$ fermion momentum-dependent
operators and $e^{i\vec{\pi}\cdot\vec{r}_j}$ is $\pm 1$ depending on which
sub-lattice site $\vec{r}_j$ is on. (For the 1D lattice that phase factor can be written as $(-1)^j$.) 
The $c$ momentum band is studied in Ref. \cite{companion} and has the same
shape and momentum area as the electronic first-Brillouin zone.

Consistently with the $f_{\vec{r}_j,c}^{\dag}$ expression (\ref{fc+}), the three spinon local operators
$s^l_{\vec{r}_j}$ and the three $\eta$-spinon local operators $p^l_{\vec{r}_j}$
such that $l=\pm,x_3$, respectively, have for $U/4t>0$ the following expression in terms of 
rotated-electron operators,
\begin{equation}
s^l_{\vec{r}_j} = n_{\vec{r}_j,c}\,q^l_{\vec{r}_j} \, ; \hspace{0.50cm}
p^l_{\vec{r}_j} = (1-n_{\vec{r}_j,c})\,q^l_{\vec{r}_j} \, , 
\hspace{0.25cm} l =\pm,x_3 \, .
\label{sir-pir}
\end{equation}
Here $q^{\pm}_{\vec{r}_j}= q^{x_1}_{\vec{r}_j}\pm i\,q^{x_2}_{\vec{r}_j}$
where $x_1,x_2,x_3$ denotes the Cartesian coordinates of the operators $q^{x_i}_{\vec{r}_j}$
with $i=1,2,3$ and the the rotated quasi-spin operators read,
\begin{equation}
q^+_{\vec{r}_j} = ({\tilde{c}}_{\vec{r}_j,\uparrow}^{\dag}
- e^{i\vec{\pi}\cdot\vec{r}_j}\,{\tilde{c}}_{\vec{r}_j,\uparrow})\,
{\tilde{c}}_{\vec{r}_j,\downarrow} \, ; \hspace{0.25cm}
q^-_{\vec{r}_j} = (q^+_{\vec{r}_j})^{\dag} \, ;
\hspace{0.25cm}
q^{x_3}_{\vec{r}_j} = {1\over 2} - {\tilde{n}}_{\vec{r}_j,\downarrow} \, .
\label{rotated-quasi-spin}
\end{equation}
In addition, 
\begin{equation}
n_{\vec{r}_j,c} = f_{\vec{r}_j,c}^{\dag}\,f_{\vec{r}_j,c} \, ,
\label{n-r-c}
\end{equation}
is the $c$ fermion local density operator.

Since the electron - rotated-electron 
transformation generated by the operator $\hat{V}$
is unitary, the operators ${\tilde{c}}_{\vec{r}_j,\sigma}^{\dag}$ 
and ${\tilde{c}}_{\vec{r}_j,\sigma}$ of Eq. (\ref{rotated-operators})
have the same anticommutation relations as 
$c_{\vec{r}_j,\sigma}^{\dag}$ and $c_{\vec{r}_j,\sigma}$, respectively.
Straightforward manipulations based on Eqs.
(\ref{fc+})-(\ref{rotated-quasi-spin}) then lead
to the following algebra for the $c$ fermion operators,
\begin{equation}
\{f^{\dag}_{\vec{r}_j,c}\, ,f_{\vec{r}_{j'},c}\} = \delta_{j,j'} \, ;
\hspace{0.25cm}
\{f_{\vec{r}_j,c}^{\dag}\, ,f_{\vec{r}_{j'},c}^{\dag}\} =
\{f_{\vec{r}_j,c}\, ,f_{\vec{r}_{j'},c}\} = 0 \, ,
\label{albegra-cf}
\end{equation}
and $c$ fermion operators and rotated quasi-spin operators,
\begin{equation}
[f_{\vec{r}_j,c}^{\dag}\, ,q^l_{\vec{r}_{j'}}] =
[ f_{\vec{r}_j,c}\, ,q^l_{\vec{r}_{j'}}] = 0 \, .
\label{albegra-cf-s-h}
\end{equation}
In turn, the rotated quasi-spin operators $q^{x_1}_{\vec{r}_j}$,
$q^{x_2}_{\vec{r}_j}$, and $q^{x_3}_{\vec{r}_j}$ and the
related operators $q^{\pm}_{\vec{r}_j}= q^{x_1}_{\vec{r}_j}\pm i\,q^{x_2}_{\vec{r}_j}$
obey the following algebra,
\begin{equation}
[q^{x_p}_{\vec{r}_j}\, ,q^{x_{p'}}_{\vec{r}_{j'}}] =
i\,\delta_{j,j'}\sum_{p''} \epsilon_{pp'p''}\,q^{x_{p''}}_{\vec{r}_j} 
\, ; \hspace{0.15cm} p,p',p''=1,2,3 \, ,
\label{albegra-s-h}
\end{equation}
\begin{equation}
\{q^{+}_{\vec{r}_j},q^{-}_{\vec{r}_j}\} = 1 \, ,
\hspace{0.5cm}
\{q^{\pm}_{\vec{r}_j},q^{\pm}_{\vec{r}_j}\} = 0 \, ,
\label{albegra-qs-p-m}
\end{equation}
\begin{equation}
[q^{+}_{\vec{r}_j},q^{-}_{\vec{r}_{j'}}] = \delta_{j,j'}\,q^{x_3}_{\vec{r}_j}
\, ; \hspace{0.50cm}
[q^{\pm}_{\vec{r}_j},q^{\pm}_{\vec{r}_{j'}}]=0 \, .
\label{albegra-q-com}
\end{equation}
Hence the operators $q^{\pm}_{\vec{r}_j}$ anticommute 
on the same site and commute on different sites.
The same applies to three spinon operators $s^l_{\vec{r}_j}$ and three 
$\eta$-spinon operators $p^l_{\vec{r}_j}$, respectively, whose
expressions are given in Eq. (\ref{sir-pir}).

The relations provided in Eqs. (\ref{albegra-cf})-(\ref{albegra-q-com})
confirm that the $c$ fermions associated with the hidden global $U(1)$ symmetry
are $\eta$-spinless and spinless fermionic objects. They are consistent as well with 
the spinons and $\eta$-spinons being spin-$1/2$ and $\eta$-spin-$1/2$ objects, respectively, whose local operators
obey the usual corresponding $SU(2)$ algebras. 

On inverting the relations given in Eqs. (\ref{fc+}) and (\ref{rotated-quasi-spin}), the rotated-electron creation and/or 
annihilation operators of Eq. (\ref{rotated-operators}) are written in terms of the operators 
of the $c$ fermions and rotated quasi-spin operators. For the LWS subspace considered here this leads to,
\begin{equation}
{\tilde{c}}_{\vec{r}_j,\uparrow}^{\dag} =
f_{\vec{r}_j,c}^{\dag}\,\left({1\over 2} +
q^{x_3}_{\vec{r}_j}\right) + e^{i\vec{\pi}\cdot\vec{r}_j}\,
f_{\vec{r}_j,c}\,\left({1\over 2} - q^{x_3}_{\vec{r}_j}\right) \, ;
\hspace{0.25cm}
{\tilde{c}}_{\vec{r}_j,\downarrow}^{\dag} =
q^-_{\vec{r}_j}\,(f_{\vec{r}_j,c}^{\dag} -
e^{i\vec{\pi}\cdot\vec{r}_j}\,f_{\vec{r}_j,c}) \, .
\label{c-up-c-down}
\end{equation}

For $U/4t\rightarrow\infty$ the rotated electrons
become electrons. In that limit the $c$ fermion creation operators 
become the quasicharge annihilation operators of Ref. \cite{Ostlund-06} and
the spinon and $\eta$-spinon operators become the local spin and pseudospin operators, respectively,
of that reference. Consistently, in that limit Eqs. (\ref{fc+})-(\ref{c-up-c-down}) 
are equivalent to Eqs. (1)-(3) of Ref. \cite{Ostlund-06} with
the rotated-electron operators replaced by the
corresponding electron operators and the $c$ fermion
creation operator $f_{\vec{r}_j,c}^{\dag}$ replaced
by the quasicharge annihilation operator $\hat{c}_r$.

Since the transformation considered in Ref. \cite{Ostlund-06} 
does not introduce Hilbert-space constraints, suitable occupancy 
configurations of the objects associated with the local 
quasicharge, spin, and pseudospin operators generate a complete set 
of $U/4t\rightarrow\infty$ states $\{\vert\Psi_{\infty}\rangle\}$.
In Section IV-E a corresponding complete set of finite-$U/4t$ states of the
form $\vert \Phi_{U/4t}\rangle ={\hat{V}}^{\dag}\vert\Phi_{\infty}\rangle$
is constructed. Those are both state representations of the model global
$SO(3)\times SO(3)\times U(1)$ symmetry and momentum eigenstates.
In general the energy and momentum eigenstates
$\vert \Psi_{U/4t}\rangle ={\hat{V}}^{\dag}\vert\Psi_{\infty}\rangle$ of the
Hubbard model on the square lattice considered in Section II
are a superposition of a sub-set of states $\vert \Phi_{U/4t}\rangle$
with the same momentum eigenvalue.

\subsection{Interplay of the global symmetry
with the transformation laws under the 
operator $\hat{V}$: three basic effective lattices
and the theory vacua}

It follows from the results of Ref. \cite{bipartite} that
the $\eta$-spin $SU(2)$ and spin $SU(2)$ state representations correspond to 
the $\eta$-spin and spin degrees of freedom of independent rotated-electron
occupancy configurations of $[N_a^D-2S_c]$ sites and $2S_c$ sites,
respectively, of the original lattice. This applies to any choice of electron - rotated-electron unitary
transformation and thus applies as well to that associated with our description.
In turn, the state representations associated 
with the new-found hidden $U(1)$ symmetry of the global
$SO(3)\times SO(3)\times U(1)=[SU(2)\times SU(2)\times U(1)]/Z_2^2$ symmetry refer to the
relative positions in the original lattice of the sites involved in each of these two types of
configurations. In the present $N_a^D\gg 1$ limit it is useful to introduce the following numbers, 
\begin{equation}
N_{a_{\eta}}^D = N_a^D-2S_c 
\, ; \hspace{0.25cm}
N_{a_{s}}^D = 2S_c \, ,
\label{Na-eta-s}
\end{equation}
such that $N_a^D = N_{a_{\eta}}^D + N_{a_{s}}^D$.
Here the integer numbers $N_{a_{\eta}}^D$ and $N_{a_s}^D$ are the number of sites of
the original lattice singly occupied by rotated electrons and unoccupied plus doubly occupied
by rotated electrons, respectively. Below such numbers are found to play the role of number of sites of
a $\eta$-spin and spin effective lattice, respectively.
For the $D=2$ square lattice the number $N_a^2=N_a\times N_a$ 
is chosen so that the number $N_a$ of sites in a row or column  
is an integer. However, the designations $N_{a_{\eta}}^D$ and
$N_{a_s}^D$ do not imply that the corresponding numbers 
$N_{a_{\eta}}$ and $N_{a_{s}}$ are integers. In general they
are not integers. For finite values of $x$ and $(1-x)$ and $N_a^D\rightarrow\infty$
we use for the numbers $N_{a_{\eta}}$ and $N_{a_{s}}$ of sites in a row and column
of the $\eta$-spin and spin effective square lattices, respectively, 
the corresponding closest integer numbers.

Importantly, the numbers given in Eq. (\ref{Na-eta-s})
are fully controlled by the eigenvalue $S_c$ of the generator (\ref{Or-ope}) of the 
hidden global $U(1)$ symmetry. This
confirms the important role plaid by such a 
symmetry in the general description introduced in this paper.
The degrees of freedom of the rotated-electron occupancy configurations of each 
of the sets of $N_{a_{\eta}}^D=[N_a^D-2S_c]$ and  $N_{a_{s}}^D=2S_c$ sites of the original 
lattice that generate the $S_{\eta}$, $S_{\eta}^{x_3}$, $S_s$, $S_s^{x_3}$, $S_c$, and momentum eigenstates
studied in Section IV, which refer to state representations of the model global
$SO(3)\times SO(3)\times U(1)$ symmetry, naturally separate as follows: 

i) The occupancy configurations of the $c$ fermions associated with the 
operators $f_{\vec{r}_j,c}^{\dag}=(f_{\vec{r}_j,c})^{\dag}$ of Eq. (\ref{fc+}) where $j=1,...,N_a^D$
correspond to the state representations
of the hidden global $U(1)$ symmetry found in Ref. \cite{bipartite}. Such $c$
fermions live on the $c$ effective lattice. It is identical to the original lattice. Its occupancies are related to those of the 
rotated electrons: The number of $c$ fermion occupied and unoccupied sites 
is given by $N_c = N_{a_{s}}^D=2S_c$
and $N_c^h = N_{a_{\eta}}^D=[N_a^D-2S_c]$, respectively. Indeed,
the $c$ fermions occupy the sites singly occupied by the
rotated electrons. In turn, the rotated-electron doubly-occupied
and unoccupied sites are those unoccupied by the $c$ fermions. Hence the
$c$ fermion occupancy configurations describe the relative positions
in the original lattice of the $N_{a_{\eta}}^D=[N_a^D-2S_c]$ sites of the
{\it $\eta$-spin effective lattice} and $N_{a_{s}}^D=2S_c$ sites
of the {\it spin effective lattice}.

ii) The remaining degrees of freedom of rotated-electron
occupancies of the sets of $N_{a_{\eta}}^D=[N_a^D-2S_c]$ and  
$N_{a_{s}}^D=2S_c$ original-lattice sites correspond to the occupancy configurations 
associated with the $\eta$-spin $SU(2)$ symmetry
and spin $SU(2)$ symmetry state representations,
respectively. The occupancy configurations of the 
$N_{a_{\eta}}^D=[N_a^D-2S_c]$ sites of the
$\eta$-spin effective lattice and set of $N_{a_{s}}^D=2S_c$ sites
of the spin effective lattice are independent. The former configurations 
refer to the operators $p^l_{\vec{r}_j}$ of Eq. (\ref{sir-pir}), which
act only onto the $N_{a_{\eta}}^D=[N_a^D-2S_c]$ sites of
the $\eta$-spin effective lattice. The latter configurations
correspond to the operators 
$s^l_{\vec{r}_j}$ given in the same equation, which
act onto the $N_{a_{s}}^D=2S_c$ sites of
the spin effective lattice. This is assured by the operators $(1-n_{\vec{r}_j,c})$
and $n_{\vec{r}_j,c}$ in their expressions provided
in that equation, which play the role of projectors
onto the $\eta$-spin and spin effective lattice, respectively.

For $U/4t>0$ the degrees of freedom of each rotated-electron singly 
occupied site then separate into a spin-less $c$ fermion carrying the electronic
charge and a spin-down or spin-up spinon. Furthermore, the degrees of freedom of 
each rotated-electron doubly-occupied or unoccupied site separate 
into a $\eta$-spin-less {\it $c$ fermion hole} and a $\eta$-spin-down or 
$\eta$-spin-up $\eta$-spinon, respectively. The $\eta$-spin-down or 
$\eta$-spin-up $\eta$-spinon refers to the $\eta$-spin degrees of
freedom of a rotated-electron doubly-occupied
or unoccupied site, respectively, of the original lattice. 

Hence a key feature of our description is that for $U/4t>0$ its quantum objects correspond 
to rotated-electron configurations whose numbers of spin-down and
spin-up single occupied sites, double occupied sites, and unoccupied sites
are good quantum numbers. This is in contrast
to descriptions in terms of electron configurations, whose numbers of spin-down and
spin-up single occupied sites, double occupied sites, and unoccupied sites
are good quantum numbers only for $U/4t\gg 1$ \cite{2D-MIT,Feng,Fazekas,Xiao-Gang}.
  
The transformation laws under the 
electron - rotated-electron unitary transformation of the $\eta$-spinons (and spinons) as defined in this paper
play a major role in the description of the $\eta$-spin 
(and spin) $SU(2)$ state representations in terms of the occupancy configurations of the 
$N_{a_{\eta}}^D=[N_a^D-2S_c]$ sites of the $\eta$-spin
effective lattice (and set of $N_{a_{s}}^D=2S_c$ sites of the spin
effective lattice). Indeed, a well-defined number of
$\eta$-spinons (and spinons) remain invariant under that unitary 
transformation. Those are called deconfined $\pm 1/2$ $\eta$-spinons (and deconfined 
$\pm 1/2$ spinons). As further discussed below, they play the role of unoccupied 
sites of the $\eta$-spin (and spin) effective lattice. The 
values of the numbers $M^{de}_{\eta,\pm 1/2}$
of deconfined $\pm 1/2$ $\eta$-spinons and 
$M^{de}_{s,\pm 1/2}$ of deconfined $\pm 1/2$ spinons 
are fully controlled by the 
$\eta$-spin $S_{\eta}$ and $\eta$-spin projection $S_{\eta}^{x_3}=-{x\over 2}\,N_a^D$ 
and spin $S_{s}$ and spin projection $S_{s}^{x_3}=-{m\over 2}\,N_a^D$, respectively,
of the state under consideration as follows,
\begin{equation}
M^{de}_{\alpha} = [M^{de}_{\alpha,-1/2}+M^{de}_{\alpha,+1/2}]=2S_{\alpha} \, ;
\hspace{0.25cm}
M^{de}_{\alpha,\pm 1/2} = [S_{\alpha}\mp S_{\alpha}^{x_3}]
\, ; \hspace{0.25cm} \alpha = \eta \, , s \, .
\label{L-L}
\end{equation}
The invariance of such deconfined $\eta$-spinons (and deconfined spinons)
stems from the off diagonal generators of the $\eta$-spin
(and spin) algebra, which flip their $\eta$-spin (and spin), commuting with 
the unitary operator $\hat{V}$. This justifies why such generators
have for $U/4t>0$ the same expressions in terms of electron and
rotated-electron operators, as given in Eq. (\ref{Scs}).

It follows that the number of sites of the $\eta$-spin ($\alpha =\eta$) and spin
($\alpha =s$) effective lattice can be written as,
\begin{equation}
N_{a_{\alpha}}^D = [2S_{\alpha} + M^{co}_{\alpha}] =  
[M^{de}_{\alpha} + M^{co}_{\alpha}] = M_{\alpha} 
\, ; \hspace{0.25cm} \alpha = \eta \, , s \, .
\label{2S-2C}
\end{equation}
As justified below, here $M^{co}_{\alpha}$ is the confined $\eta$-spinon ($\alpha =\eta$) 
or confined spinon ($\alpha =s$) even number. The $\eta$-spin or spin degrees of freedom of the
occupancy configurations in a set of $2S_{\alpha}=M^{de}_{\alpha}$
sites out of the whole set of $N_{a_{\alpha}}^D$ sites have for $U/4t>0$ the same form in terms
of electrons and rotated electrons. In turn, for finite values of $U/4t$ and spin density $-(1-x)<m<(1-x)$
the corresponding $c$ fermion occupancy configurations that store the information
on the relative positions in the original lattice of the $N_{a_{s}}^D$ sites of the
spin effective lattice and $N_{a_{\eta}}^D=[N_a^D-N_{a_{s}}^D]$ sites of
the $\eta$-spin effective lattice are not invariant 
under the electron - rotated-electron unitary transformation.

Such an invariance of the $\eta$-spin degrees of freedom of the above occupancy configurations implies that in each
state representation there are exactly $2S_{\eta}=[M^{de}_{\eta,-1/2}+M^{de}_{\eta,+1/2}]=M^{de}_{\eta}$
sites such that $M^{de}_{\eta,-1/2}$ sites are doubly occupied and $M^{de}_{\eta,+1/2}$ sites are
unoccupied both by electrons and rotated electrons. Furthermore,
the invariance of the spin degrees of freedom of the sites singly occupied by
rotated electrons implies that there are
exactly $2S_{s}=[M^{de}_{s,-1/2}+M^{de}_{s,+1/2}]=M^{de}_{s}$ sites of the original lattice such that
$M^{de}_{s,-1/2}$ sites are singly occupied both for spin-down electrons and 
spin-down rotated electrons and $M^{de}_{s,+1/2}$ sites are singly occupied both for spin-up electrons and 
spin-up rotated electrons. The state representation is in general a superposition of such
occupancy configurations whose positions of the $[2S_{\eta}+2S_{s}]=[M^{de}_{\eta}+M^{de}_{s}]$ sites are different. 

In turn, out of the set of $[M^{co}_{\eta}+M^{co}_{s}]$ sites of the
original lattice left over, a set of $M^{co}_{\eta}/2$ sites are unoccupied
by rotated electrons, a set of $M^{co}_{\eta}/2$ sites are doubly occupied 
by rotated electrons, a set of $M^{co}_{s}/2$ sites are singly occupied by spin-up  
rotated electrons, and a set of $M^{co}_{s}/2$ sites are
singly occupied by spin-down rotated electrons.
However, in terms of electrons that set of $[M^{co}_{\eta}+M^{co}_{s}]$ 
sites of the original lattice has for finite $U/4t$ values
very involved occupancies. Indeed for electrons and
except for $U/4t\rightarrow\infty$ singly and doubly 
occupancy are not good quantum numbers and thus are
not conserved. The important point brought about by our description
is that due to the transformation laws under the electron - rotated-electron
unitary transformation this refers only to a sub-set of  $[M^{co}_{\eta}+M^{co}_{s}]$ sites out
of the $N_a^D$ sites of the original lattice. Indeed for the remaining
$[2S_{\eta}+2S_{s}]=[M^{de}_{\eta}+M^{de}_{s}]$ sites singly and doubly 
occupancy are good quantum numbers both for electrons and rotated electrons.

The site numbers $M^{co}_{\eta}\geq 0$ and $M^{co}_{s}\geq 0$ are good quantum
numbers given by,
\begin{equation}
M^{co}_{\eta} = [N_{a_{\eta}}^D - 2S_{\eta}] =
[N_a^D - 2S_c - 2S_{\eta}] \, ; \hspace{0.25cm}  
M^{co}_{s} = [N_{a_{s}}^D - 2S_{s}] = [2S_c - 2S_{s}] \, .
\label{C-C}
\end{equation}
Hence their values are fully determined by those of the eigenvalue $S_c $ of the hidden global $U(1)$ symmetry
generator and $\eta$-spin $S_{\eta}$ or spin $S_s$, respectively. This reveals that $M^{co}_{\eta}$ and $M^{co}_{s}$
are not independent quantum numbers. 

The physics behind the hidden $U(1)$ 
symmetry found in Ref. \cite{bipartite} includes that
brought about by the rotated-electron occupancy
configurations of the set of $[M^{co}_{\eta}+M^{co}_{s}]$ sites of Eq. (\ref{C-C}).   
The use of the corresponding model global 
$SO(3)\times SO(3)\times U(1)$ symmetry reveals that 
the numbers $M^{co}_{\alpha} = 0,2,4,...$ are always even integers.
Moreover, the application onto $S_{\alpha}=0$ states of the off-diagonal generators of the 
$\eta$-spin $(\alpha =\eta)$ or spin $(\alpha =s)$
algebra provided in Eq. (\ref{Scs}) gives zero. 
For such states $N_{a_{\alpha}}^D = M^{co}_{\alpha}$.
In turn, application of these generators onto $S_{\alpha}>0$ states flips the
$\eta$-spin ($\alpha=\eta$) or spin ($\alpha=s$)
of a deconfined $\eta$-spinon ($\alpha=\eta$) or deconfined spinon ($\alpha=s$)
but leaves invariant the rotated-electron occupancy 
configurations of the above considered set of $M^{co}_{\alpha}$ sites.  
It follows that such set of $M^{co}_{\eta}$ (and $M^{co}_{s}$) sites 
refers to $\eta$-spin-singlet (and spin-singlet) configurations involving
$M^{co}_{\eta}/2$ (and $M^{co}_{s}/2$) $-1/2$ $\eta$-spinons (and $-1/2$ spinons) and an equal number of $+1/2$ 
$\eta$-spinons (and $+1/2$ spinons). 

It follows from the above discussions that the total numbers of $\eta$-spinons ($\alpha =\eta$) and 
spinons ($\alpha =s$) read, 
\begin{eqnarray}
M_{\alpha} & = & N_{a_{\alpha}}^D = [M^{de}_{\alpha} + M^{co}_{\alpha}] = 
[M_{\alpha,-1/2} +M_{\alpha,+1/2}] \, ; \hspace{0.35cm}
M_{\alpha,\pm 1/2} = [M^{de}_{\alpha,\pm 1/2} + M^{co}_{\alpha}/2] 
\, ; \hspace{0.25cm} \alpha = \eta \, , s \, ,
\nonumber \\
M_{\eta} & = & N_{a_{\eta}}^D = [N_a^D - 2S_c] \, ; \hspace{0.35cm}
M_{s} = N_{a_{s}}^D  = 2S_c \, .
\label{M-L-C}
\end{eqnarray}

The $\eta$-spinon and spinon operator algebra refers to well-defined subspaces spanned by states
whose number of each of these basic objects is conserved and given in Eqs. (\ref{L-L}), (\ref{C-C}), and (\ref{M-L-C}). 
Hence in such subspaces the number $2S_c$ of 
rotated-electron singly occupied sites and the numbers $N_{a_{\eta}}^D$
and $N_{a_{s}}^D$ of sites of the $\eta$-spin and spin effective lattices,
respectively, are fixed. For hole concentrations $0\leq x<1$ and
maximum spin density $m=(1-x)$ (or $m=-(1-x)$) reached at some critical magnetic field
$H_c$ aligned parallel to the square-lattice plane for $D=2$ and 
pointing along the chain for $D=1$ the $c$ fermion operators are 
invariant under the electron - rotated-electron unitary 
transformation. There is a corresponding fully polarized  
vacuum $\vert 0_{\eta s}\rangle$ that remains
invariant under such a transformation. It reads,
\begin{equation}
\vert 0_{\eta s}\rangle = \vert 0_{\eta};N_{a_{\eta}}^D\rangle\times\vert 0_{s};N_{a_{s}}^D\rangle
\times\vert GS_c;2S_c\rangle \, .
\label{vacuum}
\end{equation}
Here the $\eta$-spin $SU(2)$ vacuum $\vert 0_{\eta};N_{a_{\eta}}^D\rangle$ 
associated with $N_{a_{\eta}}^D$ deconfined $+1/2$
$\eta$-spinons, the spin $SU(2)$ vacuum $\vert 0_{s};N_{a_{s}}^D\rangle$ 
with $N_{a_{s}}^D$ deconfined $+1/2$ spinons, and the $c$ $U(1)$
vacuum $\vert GS_c;2S_c\rangle$ with $N_c=2S_c$ $c$ fermions
remain invariant under the electron - rotated-electron unitary transformation. 
The latter vacuum may be expressed as $\prod_{{\vec{q}}}f^{\dag}_{{\vec{q}},c}\vert GS_c;0\rangle$.
Here the vacuum $\vert GS_c;0\rangle$ corresponds to the electron and
rotated-electron vacuum. Its form is that given in Eq. (\ref{vacuum}) with
$N_{a_{\eta}}^D=N_a^D$ and $N_{a_{s}}^D=2S_c=0$. 

For $U/4t>0$ only for a $m=(1-x)$ fully polarized state are the 
occupancy configurations of the state $\vert GS_c;2S_c\rangle$
and the corresponding $N_c=2S_c$ $c$ fermions 
invariant under the electron - rotated-electron unitary transformation. 
For the corresponding vacuum $\vert 0_{\eta};N_{a_{\eta}}^D\rangle$
(and $\vert 0_{s};N_{a_{s}}^D\rangle$), the $M_{\eta}=M^{de}_{\eta,+1/2}$ deconfined 
$+1/2$ $\eta$-spinons refer to $N_{a_{\eta}}^D=M^{de}_{\eta,+1/2}$ sites of
the original lattice unoccupied 
by rotated electrons (and the $M_{s}=M^{de}_{s,+1/2}$ 
deconfined $+1/2$ spinons to the spins of $N_{a_{s}}^D=M^{de}_{s,+1/2}$ 
spin-up rotated electrons that singly occupy sites of such
a lattice). At maximum spin density $m=(1-x)$ the $c$ fermions are the non-interacting spinless 
fermions that describe the charge degrees of freedom of the electrons of
the fully polarized ground state. At that spin density 
there are no electron doubly occupied sites and the quantum
problem is non-interacting for $U/4t>0$.

According to the analysis of Appendix B, full information about the quantum
problem described by the Hamiltonian (\ref{H})
can be achieved by defining it in the LWS subspace. Below in Section IV we then confirm that within the description
introduced in this paper, out of the 
$N_{a_{\alpha}}^D = M_{\alpha} = [M^{de}_{\alpha} + M^{co}_{\alpha}]=[2S_{\alpha} + M^{co}_{\alpha}]$
sites of the $\eta$-spin ($\alpha =\eta$) 
and spin ($\alpha =s$) effective lattice, the $2S_{\alpha}=M^{de}_{\alpha}$
sites whose occupancies $\eta$-spin ($\alpha =\eta$) and spin ($\alpha =s$) 
degrees of freedom are invariant under electron - rotated-electron
unitary transformation as defined in this paper play the role of unoccupied sites.
In turn, the remaining $M^{co}_{\alpha}$ sites play
the role of occupied sites. This is a natural consequence of
the $\eta$-spin $SU(2)$ vacuum $\vert 0_{\eta};N_{a_{\eta}}^D\rangle$ 
(and spin $SU(2)$ vacuum $\vert 0_{s};N_{a_{s}}^D\rangle$)
being for all $U/4t$ and $m$ values invariant under the electron - rotated-electron
unitary transformation. Indeed that vacuum is such that $N_{a_{\eta}}^D = 2S_{\eta}=M^{de}_{\eta} $
(and $N_{a_{s}}^D = 2S_{s}=M^{de}_{s} $), so that $M^{co}_{\eta}=0$ (and $M^{co}_{s}=0$.)

\subsection{Spacing and occupied and unoccupied sites of the $\eta$-spin and spin effective lattices}

Here we use the 1D problem to illustrate the independence of the $c$ effective lattice, $\eta$-spin effective lattice,
and spin effective lattice occupancy configurations. Indeed the 1D BA solution is constructed to inherently 
such configurations being independent. We then extend our analysis to the model on the square lattice.

\subsubsection{The $N$ rotated-electron amplitude}

For simplicity, in the following we denote the number of sites of the $c$,  $\eta$-spin, and spin effective lattices
of the 1D model by $N_a$, $N_{a_{\eta}}$, and $N_{a_{s}}$, respectively, rather than by our general notation 
$N_a^D$, $N_{a_{\eta}}^D$, and $N_{a_{s}}^D$ with $D=1$ for the 1D lattice.
(The  $c$ effective lattice is identical to the original lattice.)

As discussed above and confirmed below in Section IV-D and in Ref. \cite{bipartite},
the representations of the model global $SO(3)\times SO(3)\times U(1)=[SU(2)\times SU(2)\times U(1)]/Z_2^2$
symmetry are naturally expressed in terms of rotated-electron occupancy configurations rather than of electron configurations.
Hence a quantity of interest for our analysis is the amplitude associated with in a given
energy eigenstate $\vert \Psi_{U/4t}\rangle$, $N$ rotated electrons of spin projection $\sigma_1,\sigma_2,...,\sigma_N$,
rather than the corresponding $N$ electrons, being found at positions $x_1,x_2,...,x_N$, respectively,
\begin{equation}
f_{U/4t} (x_1,x_2,...,x_N) = \langle \Psi_{U/4t}\vert x_1,x_2,...,x_N \rangle =
\langle \Psi_{U/4t}\vert {\tilde{c}}_{x_1,\sigma_1}^{\dag} {\tilde{c}}_{x_2,\sigma_2}^{\dag}
...{\tilde{c}}_{x_N,\sigma_N}^{\dag}\vert 0\rangle \, .
\label{amplitude-1D-rot}
\end{equation}
Here the $N=0$ electron and rotated-electron vacuum $\vert 0\rangle$
refers to the vacuum $\vert 0_{\eta s}\rangle $ of Eq. (\ref{vacuum}) for $N_{a_{\eta}}=N_a$,
$N_{a_{s}}=0$, and $2S_c=0$ and $\vert \Psi_{U/4t}\rangle$ is an arbitrary energy
eigenstate. Obviously, the state 
${\tilde{c}}_{x_N,\sigma_N} {\tilde{c}}_{x_{N-1},\sigma_{N-1}}
...{\tilde{c}}_{x_1,\sigma_1}\vert \Psi_{U/4t}\rangle$ involved
in the amplitude (\ref{amplitude-1D-rot}) is not an energy eigenstate. Application
onto $\vert \Psi_{U/4t}\rangle$ of the product operator ${\tilde{c}}_{x_N,\sigma_N} {\tilde{c}}_{x_{N-1},\sigma_{N-1}}
...{\tilde{c}}_{x_1,\sigma_1}$ provides important information. As justified below,
before removing the rotated electrons and corresponding electrons
such an operator breaks the $M^{co}_{\eta}/2$ $\eta$-spin-neutral pairs involving $M^{co}_{\eta}/2$
confined $-1/2$ $\eta$-spinons and $M^{co}_{\eta}/2$ confined $+1/2$ $\eta$-spinons
as well as the $M^{co}_{s}/2$ spin-neutral pairs involving $M^{co}_{s}/2$
confined $-1/2$ spinons and $M^{co}_{s}/2$ confined $+1/2$ spinons. 

For the electron - rotated-electron unitary operator ${\hat{V}}^{\dag}$ as defined 
in Section II one can profit both from $\vert \Psi_{U/4t}\rangle ={\hat{V}}^{\dag}\vert\Psi_{\infty}\rangle$
being an energy eigenstate for $U/4t>0$ and the electronic vacuum $\vert 0\rangle$ 
associated with an empty original lattice being invariant under ${\hat{V}}^{\dag}$,
so that ${\hat{V}}^{\dag}\vert 0\rangle={\hat{V}}\vert 0\rangle=\vert 0\rangle$. Combining these two properties
with the unitarity of ${\hat{V}}^{\dag}$ one may express the amplitude (\ref{amplitude-1D-rot}) 
for the $N$ rotated electrons at finite $U/4t$ in terms of the corresponding amplitude 
of $N$ electrons for $U/4t\rightarrow\infty$,
\begin{eqnarray}
f_{U/4t} (x_1,x_2,...,x_N) & = & \langle\Psi_{\infty}\vert {\hat{V}}{\hat{V}}^{\dag}\,c_{x_1,\sigma_1}^{\dag}\,{\hat{V}}
{\hat{V}}^{\dag}\,c_{x_2,\sigma_2}^{\dag}\,{\hat{V}}...
{\hat{V}}^{\dag}\,c_{x_N,\sigma_N}^{\dag}\,{\hat{V}}\vert 0\rangle
\nonumber \\
& = & f_{\infty} (x_1,x_2,...,x_N) = \langle\Psi_{\infty}\vert c_{x_1,\sigma_1}^{\dag}c_{x_2,\sigma_2}^{\dag}...
c_{x_N,\sigma_N}^{\dag}\vert 0\rangle \, .
\label{amplitude-1D-rot-elec}
\end{eqnarray}
However we recall that in the $U/4t\rightarrow\infty$ limit electrons and rotated electrons become the same objects
so that the amplitude $f_{\infty} (x_1,x_2,...,x_N)$ also refers to $N$ rotated electrons.
Hence it is the amplitude $f_{U/4t} (x_1,x_2,...,x_N)$ of Eq. (\ref{amplitude-1D-rot}) 
for $U/4t\rightarrow\infty$. This reveals that for $U/4t>0$ and the rotated electrons
generated from the electrons by the electron - rotated-electron unitary operator ${\hat{V}}^{\dag}$
as defined in Section II the amplitude (\ref{amplitude-1D-rot}) is independent of $U/4t$.

Fortunately, the amplitude for $N$ electrons can be extracted from the exact BA solution.
For $U/4t$ finite it involves a permutation that is extremely complicated and useless for
practical calculations \cite{Ogata}. In turn, in the $U/4t\rightarrow\infty$ limit of Eq. (\ref{amplitude-1D-rot-elec}) 
it has a much simpler form first given in Eq. (2.23) of Ref. \cite{Woy} for energy eingenstates
that are LWSs of both the $\eta$-spin and spin algebras. It reads,
\begin{eqnarray}
f_{U/4t} (x_1,x_2,...,x_N) & = & f_{\infty} (x_1,x_2,...,x_N) 
\nonumber \\
& = & \phi_{U(1)}^{S_c} (x_1^s,x_2^s,...,x_{2S_c}^s)\times
\phi^{S_{\eta}}_{SU(2)} (x_1^d,x_2^d,...,x_{M^{co}_{\eta}/2}^d) \times
\phi^{S_{s}}_{SU(2)} (x_1^{s\downarrow},x_2^{s\downarrow},...,x_{M^{co}_{s}/2}^{s\downarrow}) \, ,
\nonumber \\
\phi_{U(1)}^{S_c} (x_1^s,x_2^s,...,x_{2S_c}^s) & = & (-1)^Q \left[\sum_{P}(-1)^P
e^{i\sum_{j=1}^{2S_c}k_{Pj}x^s_{Qj}}\right] = (-1)^Q\,\det\left[e^{i k_{Pj}x^s_{Qj}}\right] \, ,
\nonumber \\
\phi^{S_{\eta}}_{SU(2)} (x_1^d,x_2^d,...,x_{M^{co}_{\eta}/2}^d) & = &
e^{i\pi M^{co}_{\eta}/2}\,\phi_1 (x_1^d,x_2^d,...,x_{M^{co}_{\eta}/2}^d) \, ,
\nonumber \\
\phi^{S_{s}}_{SU(2)} (x_1^{s\downarrow},x_2^{s\downarrow},...,x_{M^{co}_{s}/2}^{s\downarrow}) & = &
\phi_2  (x_1^{s\downarrow},x_2^{s\downarrow},...,x_{M^{co}_{s}/2}^{s\downarrow}) \, .
\label{amplitude-1D}
\end{eqnarray}
Here we used the notations of our general description, which are those suitable to the model global 
$SO(3)\times SO(3)\times U(1)=[SU(2)\times SU(2)\times U(1)]/Z_2^2$ symmetry,
the permutation $Q$ is defined by the condition,
\begin{equation}
x_{Q1} \leq x_{Q2}\leq ... \leq x_{Q2S_c} \, ,
\label{Q}
\end{equation}
the summation $\sum_{P}$ is extended over all permutations $P$ of the BA real rapidity momenta $k_{Pj}$
and the corresponding determinant depends only on the spatial-coordinates of the $2S_c$ sites singly occupied by the rotated electrons of Eq. (\ref{Q}),
and not on their spins $\sigma_{Q1},\sigma_{Q2},...,\sigma_{Q2S_c}$. The functions $\phi_1 (x_1^d,x_2^d,...,x_{M^{co}_{\eta}/2}^d)$
and $\phi_2  (x_1^{s\downarrow},x_2^{s\downarrow},...,x_{M^{co}_{s}/2}^{s\downarrow})$ are obtained on taking
the $U/4t\rightarrow\infty$ limit of the expressions defined in Eq. (2.24) of Ref.
\cite{Woy} and in the paragraph after that equation. They are such that the corresponding functions
$\phi^{S_{\eta}}_{SU(2)} =e^{i\pi M^{co}_{\eta}/2}\,\phi_1$ and $\phi^{S_{s}}_{SU(2)} = \phi_2$ have
the following limiting behavior
\begin{equation}
\phi^{S_{\alpha}}_{SU(2)} = 1 \hspace{0.2cm}{\rm for}\hspace{0.2cm}M^{co}_{\alpha}/2=[N_{a_{\alpha}}/2-S_{\alpha}]
=[M_{\alpha}/2-S_{\alpha}] = 0 \, , 
\hspace{0.35cm} \alpha =\eta, s \, .
\label{Calpha-0}
\end{equation}

The BA solution refers either to a LWS or a HWS of both the $\eta$-spin and spin algebras. Expression
(\ref{amplitude-1D}) refers to a LWS for which
the sites occupied by the $N=[2S_c+M^{co}_{\eta}]=[2S_s +M^{co}_s+M^{co}_{\eta}]$ rotated electrons include the 
$N_{a_{s}}=2S_c=[2S_s +M^{co}_s]$ singly occupied sites whose spatial coordinates $x_1^s,x_2^s,...,x_{2S_c}^s$
are those in the argument of the function $\phi_{U(1)}^{S_c} (x_1^s,x_2^s,...,x_{2S_c}^s)$
plus the $M^{co}_{\eta}/2$ doubly occupied sites whose spatial coordinates $x_1^d,x_2^d,...,x_{M^{co}_{\eta}/2}^d$
are those in the argument of the function $\phi^{S_{\eta}}_{SU(2)} (x_1^d,x_2^d,...,x_{M^{co}_{\eta}/2}^d)$.
Moreover, the spatial coordinates of the $[N_{a_{\eta}}-M^{co}_{\eta}/2]$
rotated-electron unoccupied sites are those left
over by the above $[2S_c+M^{co}_{\eta}/2]=[2S_s +M^{co}_s+M^{co}_{\eta}/2]$ sites. In turn, the sites occupied by 
the $N_{\downarrow}=[M^{co}_s/2+M^{co}_{\eta}/2]$ spin-down rotated electrons include $M^{co}_s/2$ singly occupied sites 
whose spatial coordinates $x_1^{s\downarrow},x_2^{s\downarrow},...,x_{M^{co}_{s}/2}^{s\downarrow}$ are those in the argument of the function 
$\phi^{S_{s}}_{SU(2)} (x_1^{s\downarrow},x_2^{s\downarrow},...,x_{M^{co}_{s}/2}^{s\downarrow})$ plus $M^{co}_{\eta}/2$ doubly occupied sites whose
spatial coordinates are again those in the argument of the function $\phi^{S_{\eta}}_{SU(2)} (x_1^d,x_2^d,...,x_{M^{co}_{\eta}/2}^d)$.
On the other hand, the sites occupied by the $N_{\uparrow}=[2S_s+M^{co}_s/2+M^{co}_{\eta}/2]$ spin-up rotated 
electrons include $[2S_s+M^{co}_s/2]$ singly occupied sites plus $M^{co}_{\eta}/2$ doubly occupied sites.
While the spatial coordinates of the latter $M^{co}_{\eta}/2$ doubly occupied sites are the same
as those of the corresponding spin-down rotated electrons, the spatial coordinates of the $[2S_s+M^{co}_s/2]$
sites singly occupied by spin-up rotated electrons are those of the $N_{a_{s}}=2S_c=[2S_s +M^{co}_s]$ singly occupied sites 
in the argument of the function $\phi_{U(1)}^{S_c} (x_1^s,x_2^s,...,x_{2S_c}^s)$
except for the $M^{co}_s/2$ sites whose spatial coordinates $x_1^s,x_2^s,...,x_{M^{co}_{s}/2}^s$ are those of the singly occupied sites in the argument of the function 
$\phi^{S_{s}}_{SU(2)} (x_1^{s\downarrow},x_2^{s\downarrow},...,x_{M^{co}_{s}/2}^{s\downarrow})$, which refer 
only to spin-down rotated electrons.

Although the BA solution refers only to LWSs or HWSs, except for the $e^{i\pi M^{co}_{\eta}/2}$ phase factor in 
the $\phi^{S_{\eta}}_{SU(2)} (x_1^d,x_2^d,...,x_{M^{co}_{\eta}/2}^d)$ expression of Eq. (\ref{amplitude-1D}), 
which in the general case is replaced by $e^{i\pi M_{\eta,-1/2}}$, the amplitude $f_{U/4t} (x_1,x_2,...,x_N)$ expression given in that equation applies
to any energy eigenstate. Furthermore, in the general case the variables of the function 
$\phi^{S_{\eta}}_{SU(2)} (x_1^d,x_2^d,...,x_{M^{co}_{\eta}/2}^d)$ are as given in Eq. (\ref{amplitude-1D})
for $x>0$ but change to $\phi^{S_{\eta}}_{SU(2)} (x_1^h,x_2^h,...,x_{M^{co}_{\eta}/2}^h)$
for $x<0$. Here the spatial coordinates $x_1^h,x_2^h,...,x_{M^{co}_{\eta}/2}^h$ refer to
rotated-electron unoccupied sites associated with occupancy configurations whose
$\eta$-spin degrees of freedom are not invariant under the electron - rotated-electron
unitary transformation. Also the function
$\phi^{S_{s}}_{SU(2)} (x_1^{s\downarrow},x_2^{s\downarrow},...,x_{M^{co}_{s}/2}^{s\downarrow})$
remains the same for $m>0$ and is replaced by
$\phi^{S_{s}}_{SU(2)} (x_1^{s\uparrow},x_2^{s\uparrow},...,x_{M^{co}_{s}/2}^{s\uparrow})$
for $m<0$. The spatial coordinates $x_1^{s\uparrow},x_2^{s\uparrow},...,x_{M^{co}_{s}/2}^{s\uparrow}$ 
in the argument of the latter function refer to spin-up rotated-electron singly occupied sites 
associated with occupancy configurations whose spin degrees of freedom are not invariant 
under the electron - rotated-electron unitary transformation.

We emphasize that the $\eta$-spin function $\phi^{S_{\eta}}_{SU(2)}$ and spin function
$\phi^{S_{s}}_{SU(2)}$ do not depend on the spatial coordinates of the $M^{de}_{\eta,\pm 1/2}$ deconfined $\eta$-spinons 
and $M^{de}_{s,\pm 1/2}$ deconfined spinons, respectively. Indeed, for $x>0$ and $x<0$ the $\eta$-spin function 
$\phi^{S_{\eta}}_{SU(2)}$ only depends on the spatial coordinates of the $M^{co}_{\eta}/2$ confined 
$-1/2$ $\eta$-spinons and $M^{co}_{\eta}/2$ confined $+1/2$ $\eta$-spinons, respectively, whose $\eta$-spin-neutral
configurations are broken upon the application of the operator ${\tilde{c}}_{x_N,\sigma_N} {\tilde{c}}_{x_{N-1},\sigma_{N-1}}
...{\tilde{c}}_{x_1,\sigma_1}$ onto the energy eigenstate $\vert \Psi_{U/4t}\rangle$.
Similarly, for $m>0$ and $m<0$ the spin function 
$\phi^{S_{s}}_{SU(2)}$ only depends on the spatial coordinates of the $M^{co}_{s}/2$ confined 
$-1/2$ spinons and $M^{co}_{s}/2$ confined $+1/2$ spinons, respectively, whose spin-neutral
configurations are also broken upon the application of that operator onto the energy eigenstate under
consideration. 

Importantly and as given
in Eq. (\ref{amplitude-1D}) for the LWSs of both algebras, for all $4^{N_a}$ energy eigenstates the 
$N$ rotated-electron amplitude $f_{U/4t} (x_1,x_2,...,x_N)$ is a product of
three functions $\phi_{U(1)}^{S_c}$, $\phi^{S_{\eta}}_{SU(2)}$, and $\phi^{S_{s}}_{SU(2)}$ associated
with the $c$ hidden $U(1)$ symmetry, $\eta$-spin $SU(2)$ symmetry, and spin $SU(2)$ symmetry, respectively,
contained in the model global $SO(3)\times SO(3)\times U(1)=[SU(2)\times SU(2)\times U(1)]/Z_2^2$ symmetry.
Furthermore, for $U/4t>0$ that amplitude is independent
of $U/4t$. However, the energy of the corresponding occupancy
configurations strongly depends on $U/4t$. This follows from the Hamiltonian not commuting with
the electron - rotated-electron unitary operator ${\hat{V}}^{\dag}$. 
For $x>0$ (and $x<0$) excited states of physical interest have vanishing 
rotated-electron (and rotated-hole) double occupancy. Our discussion of the  
$N$ rotated-electron amplitude $f_{U/4t} (x_1,x_2,...,x_N)$ is limited to such states
yet, except for the above reported small changes its expression (\ref{amplitude-1D-rot}), applies 
to all $4^{N_a}$ energy eingenstates. In turn, at $x=0$ our study refers to both
states with vanishing and finite rotated-electron and rotated-hole double occupancy.

For the processes discussed below the large-$U/4t$ energy scale $U$ is for finite $U/4t$ values and $x\neq 0$ replaced by twice the 
absolute value of chemical potential, $2\vert\mu\vert$. This is because for $x\in (0,1)$ (and $x\in (-1,0)$) 
and finite $U/4t$ values the minimum energy for creation onto the ground state of a rotated-electron doubly occupied site
(and unoccupied site) is given by $2\vert\mu\vert$. In turn, at $x=0$ the chemical potential $\mu$ belongs to the range
$\mu \in (-\mu^0,\mu^0)$ whose energy width $2\mu^0\equiv \lim_{x\rightarrow 0}2\mu$  
equals the Mott-Hubbard gap. We use the convention that for $x\neq 0$ the chemical potential $\mu$
sign is that of the hole concentration $x\in (-1,1)$. Alike for 1D, we consider that for the square lattice
the magnetic field points in directions such that $H>0$ for spin 
density $m>0$ and $H<0$ for $m<0$. The minimum magnitude of the energy $\Delta_{D_{rot}}$ for creation of a number
$D_{rot}=M_{\eta,-1/2}=M^{de}_{\eta,-1/2}+M^{co}_{\eta}/2$ of rotated-electron doubly occupied sites
onto a $m=0$ and $x=0$ ground state and a $m=0$ and $x>0$ ground state is,
\begin{eqnarray}
{\rm min}\,\Delta_{D_{rot}} & = & \left[(\mu^0+\mu)M^{de}_{\eta,-1/2}+\mu^0\,M^{co}_{\eta}\right] \, ,
\hspace{0.2cm}{\rm at}\hspace{0.2cm}x=0\hspace{0.2cm}{\rm and}\hspace{0.2cm}\mu \in (-\mu^0,\mu^0) \, ,
\nonumber \\
& = & 2\mu\,[M^{de}_{\eta,-1/2}+M^{co}_{\eta}/2] \, ,\hspace{0.2cm}{\rm for}\hspace{0.2cm}x>0 \, ,
\label{min-D}
\end{eqnarray}
respectively. Similar expressions apply to the minimum magnitude of the energy $\Delta_{D_{rot}^h}$ for creation of a number
$D_{rot}^h=M_{\eta,+1/2}=M^{de}_{\eta,+1/2}+M^{co}_{\eta}/2$ of rotated-electron unoccupied sites
onto a $m=0$ and $x=0$ ground state and a $m=0$ and $x<0$ ground state, respectively, 
provided that $\mu$ is replaced by $-\mu$ and $M^{de}_{\eta,-1/2}$
by $M^{de}_{\eta,+1/2}$. Hence for $\mu \in (-\mu^0,\mu^0)$ the minimum energy for creation onto the $m=0$ and $x=0$ 
ground state of one rotated-electron doubly occupied site and rotated-electron unoccupied site is 
$(\mu^0+\mu)$ and $(\mu^0-\mu)$, respectively. For the chemical potential in the middle of the 
Mott-Hubbard gap so that $\mu=0$ at $x=0$ the minimum energy of either process is given by $\mu^0$.

For the model on the 1D (and square) lattice at $m=0$ the important $x=0$ energy scale $2\mu^0$ 
appearing in Eq. (\ref{min-D}), which is the Mott-Hubbard gap, has the following exact (and approximate) limiting behaviors,
\begin{equation}
2\mu^0 \approx {U\over \pi^2}\left({[8\pi]^2 t\over U}\right)^{D/2}e^{-2\pi \left({t\over U}\right)^{1/D}} 
\, , \hspace{0.25cm} U/4t\ll 1 \, ;
\hspace{0.5cm}
2\mu^0 \approx  [U - 4Dt] \, ,
\hspace{0.25cm}  U/4t\gg 1  \, ,
\hspace{0.25cm} D = 1,2\, .
\label{DMH}
\end{equation}
These half-filling results are consistent with the properties of the $x=0$ and $m=0$ absolute
ground state \cite{companion}. In turn, for $0<x<1$ and $m=0$ the chemical potential 
reads $\mu\approx U/2$ for $U/4t\rightarrow\infty$ and for finite $U/4t$ values is an increasing 
function of the hole concentration $x$ such that,
\begin{equation}
\mu^0\leq\mu (x)\leq\mu^1 \, ; \hspace{0.25cm}  0<x<1
\, , \hspace{0.15cm} m= 0 \, .
\label{mu-x}
\end{equation}
Here $\mu^1\equiv \lim_{x\rightarrow 1}\mu$. The related energy scale $2\mu^1$ reads,
\begin{equation}
2\mu^1 =  U + 4Dt \, ;
\hspace{0.25cm} D = 1,2 \, .
\label{mu-1}
\end{equation}
Expression (\ref{mu-1}) is exact both for the Hubbard
model on a 1D and square lattice. It can be explicitly derived for both lattices. It refers to the 
non-interacting limit of vanishing electronic density. For the model on the square lattice the
limiting behaviors reported in Eq. (\ref{DMH}) read $2\mu^0 \approx 64\,t\,e^{-2\pi\sqrt{t/U}}$ and $2\mu^0 \approx  U$
for $U/4t\ll 1$ and $U/4t\gg 1$, respectively. They are those of the related zero-temperature gap of Eq. (13) 
of Ref. \cite{Hubbard-T*-x=0}, which although showing up in the spin degrees of freedom
equals half the charge Mott-Hubbard gap \cite{companion}.

As discussed below in Section IV-C, $x>0$ (and $x<0$) ground states have vanishing rotated-electron
(and rotated-hole) double occupancy, the same applying for $x>0$ (and $x<0$) to the states of
excitation energy $\omega <2\vert\mu\vert$ that span the 
one- and two-electron subspace as defined in Section V. 
To understand the 1D amplitude expression (\ref{amplitude-1D}), we consider a configuration in which the first $N_{a_s}=2S_c=
[2S_s+M^{co}_s]=N_c$ sites in the original lattice are singly occupied and the remaining $N_{a_{\eta}}=[N_a-2S_c]=[2S_{\eta}+M^{co}_{\eta}]=N_c^h$ 
sites are unoccupied or doubly occupied by rotated electrons. In this configuration and considering processes of energy
lower than $2\vert\mu\vert$ the rotated electrons cannot 
move (except for the last one). Indeed either the Pauli principle or the onsite repulsion prevents it. Although 
in this configuration there is no direct interaction between the rotated electrons, through an 
intermediate state with energy $2\vert\mu\vert$ neighboring rotated electrons can see each other's spins, and 
rotated electrons with different spins can change position. 
We recall that for $U\rightarrow\infty$ the energy scale $2\vert\mu\vert$ reads $U$ for all values of $x$ and the usual 
$U/4t\gg 1$ discussion in terms of electrons is recovered \cite{Woy,Ogata}.
For instance, it follows from the above analysis that for $U/4t\gg 1$ the spins can move in the 
same way as do those in a Heisenberg chain. For $U/4t\gg 1$ the 
distribution of the spins will correspond to the eigenstates of the Heisenberg Hamiltonian. 

The situation with the rotated-electron unoccupied and doubly occupied sites is similar. Neighboring sites can observe each others 
occupancy through an intermediate state of relative energy whose absolute value is $2\vert\mu\vert$. Moreover, the same intermediate state 
makes it possible for a rotated-electron unoccupied and doubly occupied site to interchange position. (Thus for $U/4t\gg 1$ the 
distribution of the unoccupied and doubly occupied sites will be the same as the distribution 
of up and down spins in a Heisenberg chain.) 

The main point here is that neither the spinon distribution of the function 
$\phi^{S_{s}}_{SU(2)} (x_1^{s\sigma},x_2^{s\sigma},...,x_{M^{co}_{s}/2}^{s\sigma})$
where $\sigma =\downarrow$ and $\sigma =\uparrow$ for $m>0$ and $m<0$,
respectively, nor the $\eta$-spinon distribution associated with the relative distribution of the 
rotated-electron unoccupied and doubly occupied sites of the function
$\phi^{S_{\eta}}_{SU(2)} (x_1^d,x_2^d,...,x_{M^{co}_{\eta}/2}^d)$ for $x>0$
and $\phi^{S_{\eta}}_{SU(2)} (x_1^h,x_2^h,...,x_{M^{co}_{\eta}/2}^h)$ for $x<0$
do change if the chain of rotated-electron singly occupied sites is ``diluted'' by rotated-electron unoccupied and doubly occupied sites, 
making possible also direct propagation for the rotated electrons. Such a propagation is in turn described by 
independent occupancy configurations associated with the $c$ fermion 
distribution of the function $\phi_{U(1)}^{S_c} (x_1^s,x_2^s,...,x_{2S_c}^s)$. This important property is behind the independence of
the $c$ effective lattice, $\eta$-spin effective lattice, and spin effective lattice occupancy configurations.

Indeed, within our description the three degrees of freedom associated with the three types of rotated-electron occupancy configurations
behind the functions $\phi_{U(1)}^{S_c} (x_1^s,x_2^s,...,x_{2S_c}^s)$, $\phi^{S_{\eta}}_{SU(2)} (x_1^d,x_2^d,...,x_{M^{co}_{\eta}/2}^d)$,
and $\phi^{S_{s}}_{SU(2)} (x_1^{s\downarrow},x_2^{s\downarrow},...,x_{M^{co}_{s}/2}^{s\downarrow})$
in the amplitude general expression of Eq. (\ref{amplitude-1D}) are naturally described in terms of independent
$c$ fermion, $\eta$-spinon, and spinon occupancy configurations, respectively. The relationship of the operators of the $c$ fermions
to those of the rotated electrons is provided in Eq. (\ref{fc+}). That of the operators of the $\eta$-spinons and spinons
to the rotated-electron operators is given in Eqs. (\ref{sir-pir})-(\ref{rotated-quasi-spin}).

Except for $U/4t\gg 1$, the above independence of the three types of occupancy configurations 
is lost if one uses descriptions based on electrons rather than rotated electrons
as defined in Section II. Indeed and except for $U/4t\gg 1$, for finite $U/4t$ values the 
three independent degrees of freedom associated with the $c$ hidden $U(1)$ symmetry, $\eta$-spin $SU(2)$ symmetry, 
and spin $SU(2)$ symmetry, respectively, contained in the model global $SO(3)\times SO(3)\times U(1)=[SU(2)\times SU(2)\times U(1)]/Z_2^2$
symmetry, are difficult to separate in terms of electron occupancy configurations.

\subsubsection{Justification of the representation of the spin and $\eta$-spin effective lattices as square or 1D lattices}

Unlike in 1D, for the model on the square lattice the $N$ rotated-electron amplitudes cannot
be expressed in terms of simple permutations and site occupancy chain orders. Nonetheless,
due to the common interplay between the on-site electron correlations and the
global $SO(3)\times SO(3)\times U(1)=[SU(2)\times SU(2)\times U(1)]/Z_2^2$
symmetry, some of the physics is the same, particularly in the $N_a^D\rightarrow\infty$ limit. 
One may consider a configuration in which a original-lattice compact square domain 
of $N_{a_s}^2=2S_c=[2S_s+M^{co}_s]=N_c$ sites whose edge contains $N_{a_s}$ sites is singly occupied
by rotated electrons, and the complementary two-dimensional (2D) compact domain of $N_{a_{\eta}}^2=[N_a^2-2S_c]=
[2S_{\eta}+M^{co}_{\eta}]=N_c^h$ sites refers to rotated-electron unoccupied or doubly occupied sites. 
Obviously, in 2D there are many other shapes for compact domains of $N_{a_{\eta}}^2=[N_a^2-2S_c]=
[2S_{\eta}+M^{co}_{\eta}]=N_c^h$ sites. The square shape will be justified
below. In this configuration and considering processes of energy lower than $2\vert\mu\vert$, either the Pauli principle or the 
onsite repulsion prevents the rotated electrons to move (except for those on the lines referring
to the square edges separating the two 2D domains). 
Although in this configuration there is no direct interaction between the 
rotated electrons, through an intermediate state with energy $2\vert\mu\vert$ neighboring rotated electrons 
can see each other's spins, and rotated electrons with different spins can change position. 
Alike for 1D, the situation with the rotated-electron unoccupied and doubly occupied sites is similar
for a original-lattice compact square domain of $N_{a_{\eta}}^2=[N_a^2-2S_c]=[2S_{\eta}+M^{co}_{\eta}]=N_c^h$ 
sites. Unfortunately and unlike for 1D, for a finite system the corresponding spinon and $\eta$-spinon distributions may change if the 
the square-shape compact domain of rotated-electron singly occupied sites is ``diluted'' by rotated-electron unoccupied and doubly occupied sites.
Indeed, for a 2D system there is no order equivalent to the 1D uniquely defined chain order. 

However, within the $N_a^2\rightarrow\infty$ limit that the present description refers to the {\it dominant}
$c$ fermion occupancy configurations refer to an average uniform distribution of the $c$ effective lattice
occupied sites. Specifically, the $c$ fermion positions of such a uniform configuration correspond
to the average positions of the $c$ fermions in an energy eigenstate. Indeed, the $c$ fermion
momentum occupancy configurations of such a state are a superposition of all compatible real-space $c$ effective lattice
occupancy configurations. Fortunately, for the average configuration in which the rotated-electron singly occupied 
sites are uniformly ``diluted'' by rotated-electron unoccupied and doubly occupied sites the spinon and 
$\eta$-spinon distributions of the above original-lattice square-shape compact
domain of $N_{a_s}^2=2S_c=[2S_s+M^{co}_s]=N_c$ sites do not change. They may be described by corresponding occupancy configurations of an
effective square lattice with $N_{a_s}^2=2S_c=[2S_s+M^{co}_s]=N_c$ sites and edge length $L$
whose average spacing is for $x\neq 0$ larger than that of the original lattice. The same
arguments apply to the $\eta$-spinon distributions. 
This reveals that for the Hubbard model on the square lattice with
$N_a^2\rightarrow\infty$ sites the concepts of a spin effective square lattice and $\eta$-spin square effective 
lattice apply provided that $n=(1-x)$ and $x$ are finite, respectively.
 
That the above considered initial reference original-lattice compact domain of $N_{a_s}^2=2S_c=[2S_s+M^{co}_s]=N_c$ (and
$N_{a_{\eta}}^2=[N_a^2-2S_c]=[2S_{\eta}+M^{co}_{\eta}]=N_c^h$) sites has a square shape follows from only that compact domain shape transforming
into a corresponding uniform ``diluted'' domain fully contained in the original square lattice.
Any other initial reference compact-domain shape would transform into a uniform ``diluted'' domain that is
not fully contained in the original square lattice and thus is neither physically nor
mathematically acceptable. Here we are considering that the 2D crystal has a square 
shape whose edge is $L=a\,N_a$. If the square-lattice crystal has any other shape the compact domain
should have a similar shape with an area reduced by a factor of $2S_c/N_a^2$ for
the spin effective lattice (and of $[N_a^2-2S_c]/N_a^2$ for the $\eta$-spin effective lattice). 
Only then it transforms into a corresponding uniform ``diluted'' domain fully contained in the 
crystal. The final result is though the same, the obtained effective lattice being a square
lattice whose spacing is given by,
\begin{equation}
a_{\alpha} = {L\over N_{a_{\alpha}}} = {N_a\over N_{a_{\alpha}}}\, a
\, ; \hspace{0.25cm} \alpha = \eta \, , s \, ,
\label{a-alpha}
\end{equation}
where $N_{a_{\alpha}}=(N_{a_{\alpha}}^D)^{1/D}$.

This spinon and $\eta$-spinon distributions ``average invariance'' under ``site dilution'' 
emerging for the square lattice in the $N_a^D\rightarrow\infty$ limit is behind the 
description of such distributions in terms of the occupancy configurations of independent
$\eta$-spin and spin effective square lattices, respectively, and the direct propagation of $c$ fermions
in terms of occupancy configurations in an independent $c$ effective lattice. While for $0<x<1$ and
$0<m<n$ the former two lattices have both for 1D and 2D a number of sites smaller than $N_{a}^D$, the $c$
effective lattice is identical to the original lattice. 

Within the $N_a^D\gg 1$ limit of our description
the concept of a spin (and $\eta$-spin) effective lattice 
is well defined for finite values of the electronic density $n=(1-x)$
(and hole concentration $x$). The reasoning for the validity of the use of
the corresponding effective lattices occupancy configurations may be summarized 
by the two following statements:
\vspace{0.05cm}

1) The representation associated with the present description contains
full information about the relative positions of the sites of 
the $\eta$-spin and spin effective lattices in the 
original lattice. For each energy-eigenstate rotated-electron real-space occupancy configuration, 
that information is stored in the corresponding occupancy 
configurations of the $c$ fermions in 
their $c$ effective lattice. The latter lattice is identical to the original lattice. 
Such configurations correspond to the
state representations of the $U(1)$ symmetry in the subspaces
spanned by states with fixed 
values of $S_c$, $S_{\eta}$, and $S_s$. Indeed, the
sites of the $\eta$-spin (and spin) effective lattice have
in the original lattice the same real-space coordinates as
the sites of the $c$ effective lattice unoccupied (and
occupied) by $c$ fermions. 
\vspace{0.25cm}

2) Within the $N_a^D\gg 1$ limit that our description refers to, provided that
the electronic density $n=(1-x)$ (and hole concentration $x$) is finite,
the dominant $c$ effective lattice occupancy configurations of an energy eigenstate
of the Hubbard model on the square lattice
refer to a nearly uniform distribution of the $c$ fermions occupied sites
(and unoccupied sites). Hence due to the spinon and $\eta$~spinon distribution 
``average order'' emerging for the model on the square lattice
in the $N_a^D\rightarrow\infty$ limit, the spin and $\eta$-spin
effective lattices may be represented by square lattices. Moreover, the chain order 
invariance occurring for the 1D model both for the finite system and in that limit 
justifies why such effective lattices are 1D lattices. For both models
the corresponding spin effective lattice spacing $a_s$ and 
$\eta$-spin effective lattice spacing $a_{\eta}$ refers to the average spacing 
between the $c$ effective lattice occupied sites and between such a lattice 
unoccupied sites, respectively, given in Eq. (\ref{a-alpha}). Such spin and $\eta$-spin
effective lattices obey the physical requirement condition that in the  $x\rightarrow 0$ and
$x\rightarrow \pm 1$ limit, respectively, equal the original lattice. Note that in the $x\rightarrow 0$ (and
$x\rightarrow \pm 1$) limit one has that $N_{a_{s}}^D=N_a^D$ and the $\eta$-spin effective
lattice does not exist (and $N_{a_{\eta}}^D=N_a^D$ and the spin effective
lattice does not exist.)
\vspace{0.25cm}

The validity for $N_a^D\gg 1$ of the concept of a spin effective square lattice as constructed in this paper
is confirmed by the behavior of the expectation value $\delta d=\langle\Psi\vert \delta\hat{d} \vert\Psi\rangle$ 
of any energy eigenstate $\vert\Psi\rangle$. Here $\delta\hat{d}$ is the operator associated with the distance in real space of 
any of the $N_{a_{s}}^D$ sites of the spin effective lattice from the rotated-electron singly 
occupied site of the original lattice closest to it. The state $\vert\Psi\rangle$ belongs to a
subspace with fixed number of rotated-electron singly occupied sites.
The point is that $\delta d=\langle\Psi\vert \delta\hat{d} \vert\Psi\rangle$ vanishes in the $N_a^D\rightarrow\infty$ limit. 
We recall that the $N_{a_{s}}^D=2S_c$ sites of the spin effective lattice occupancies describe the spin
degrees of freedom of the $N_c=2S_c$ rotated-electron singly occupied sites of the
original lattice. The same applies to the $\eta$-spin effective lattice.

Consistently with the expression 
$a_{\alpha} = L/N_{a_{\alpha}} = [N_a/N_{a_{\alpha}}]\,a$
of Eq. (\ref{a-alpha}) where $\alpha = \eta , s$, the 
$\eta$-spin (and spin) effective lattice has both for 1D and 2D the 
same length and edge length $L$, respectively, as the original lattice. Furthermore, 
the requirement that for the 2D case when going 
through the whole crystal of square shape along the $ox_1$ or $ox_2$ directions 
a $\eta$-spinon (and spinon) passes an overall distance 
$L$ is met by an effective $\eta$-spin (and spin) square lattice.
Since the number of sites sum-rule $[N_{a_{\eta}}^D+N_{a_{s}}^D]=N_a^D$ holds,
the $\eta$-spin and spin effective lattices have in general a number of 
sites $N_{a_{\eta}}^D$ and $N_{a_{s}}^D$, respectively, smaller than that of the original lattice, $N_a^D$.
It follows that their lattice spacings (\ref{a-alpha}) are larger 
than that of the original lattice. 

The transformation laws under the electron - rotated-electron 
unitary transformation of the $\eta$-spin and spin degrees of freedom of the
rotated-electron occupancy configurations that generate the energy eigenstates
are in 1D behind the $\eta$-spin function $\phi^{S_{\eta}}_{SU(2)}$ and spin function
$\phi^{S_{s}}_{SU(2)}$ of the $N$ rotated-electron amplitude expression (\ref{amplitude-1D})
not depending on the spatial coordinates of the deconfined $\eta$-spinons 
of $\eta$-spin projections $-1/2$ and $+1/2$ and deconfined spinons
of spin projections $-1/2$ and $+1/2$, respectively. Both for the model on the 1D
and square lattice, it follows from such transformation laws that the $\eta$-spin and spin degrees of freedom of the
rotated-electron occupancy configurations that generate an energy eigenstate 
may be described in terms of $2S_{\eta}= M^{de}_{\eta}$ (and $2S_s =M^{de}_{s}$) ``unoccupied sites" 
and the $M^{co}_{\eta}$ (and $M^{co}_{s}$) ``occupied sites" of the $\eta$-spin 
(and spin) effective lattice. The corresponding $\eta$-spinon and spinon
occupancy configurations generate state representations of
the model global $SO(3)\times SO(3)\times U(1)$ symmetry. 
The $2S_{\eta}$ (and $2S_s$) ``unoccupied sites" of
such a $\eta$-spin (and spin) effective lattice correspond to $M^{de}_{\eta}=2S_{\eta}$ 
deconfined $\eta$-spinons 
(and $M^{de}_{s}=2S_s$ deconfined spinons)
that remain invariant under the electron - rotated-electron unitary 
transformation. In turn, the $M^{co}_{\eta}/2$ $+1/2$ $\eta$-spinons and $M^{co}_{\eta}/2$
$-1/2$ $\eta$-spinons (and $M^{co}_{s}/2$ $+1/2$ spinons and $M^{co}_{s}/2$
$-1/2$ spinons) that refer to the $M^{co}_{\eta}$  (and $M^{co}_{s}$) 
``occupied sites" of such a lattice do not remain invariant 
under that unitary transformation. 

For the Hubbard model on the square lattice the above concepts of $\eta$-spin and spin 
effective lattices ``occupied sites" and ``unoccupied sites" apply both to the energy
eigenstates and the related complete set of states introduced below in Section IV-E.

\subsubsection{The spin effective lattice for the $x>0$ and vanishing rotated-electron double occupancy subspace}

For $x>0$ the subspace of more physical interest is that of vanishing rotated-electron double occupancy.
For simplicity we consider here its LWS subspace, for which 
$M^{de}_{\eta} = M^{de}_{\eta,+1/2}=2S_{\eta}=x\,N_a^D$ in Eq. (\ref{L-L}) and $M^{co}_{\eta}=0$ in Eq. (\ref{2S-2C}) with
$\alpha =\eta$ in both these equations. Since $M^{co}_{\eta}=[N_{a_{\eta}}^D-2S_{\eta}]=[N_a^D-2S_c-2S_{\eta}]=0$
and thus $2S_c= (1-x)\,N_a^D$, for the model in that subspace the number of sites of the spin effective lattice $N_{a_{s}}^D =2S_c$ of
Eq. (\ref{Na-eta-s}) and its spacing given in Eq. (\ref{a-alpha}) for $\alpha =s$ simplify to,
\begin{equation}
N_{a_{s}}^D = M_s = (1-x)\,N_a^D
\, ; \hspace{0.35cm}
a_{s} = {a\over (1-x)^{1/D}} \, ,
\hspace{0.25cm} (1-x)> 1/N_a^D \, ,
\label{NNCC}
\end{equation}
respectively. Such a $M_{\eta,-1/2}=0$ subspace is spanned by ground states and their
excited states of energy below $2\mu$. 

For the 1D case, combination of the general expression (\ref{amplitude-1D}) with the boundary condition
$\phi^{S_{\eta}}_{SU(2)} = 1$ of Eq. (\ref{Calpha-0}) immediately leads for such a $x>0$ and $M_{\eta,-1/2}=0$ subspace to the
expression (2.14) of Ref. \cite{Ogata}, which refers to the $N$ electrons amplitude for
$U/4t\rightarrow\infty$. We recall that here it refers as well to the $N$ rotated-electrons amplitude for $U/4t>0$.

The sites of the $\eta$-spin lattice refer to those doubly occupied and unoccupied by rotated 
electrons in the original lattice. However, for the Hubbard model in the vanishing rotated-electron double occupancy 
subspace the concept of a $\eta$-spin lattice is useless. Indeed, for that subspace such a lattice either is 
empty ($x>0$) or does not exist ($x=0$). This is because the $\eta$-spin degrees of freedom
of the states that span that subspace are the same as those of
the $M^{co}_{\eta}=(N_a^D-2S_c-2S_{\eta})=0$ 
vacuum $\vert 0_{\eta};N_{a_{\eta}}^D\rangle$ of Eq. (\ref{vacuum}).
For states for which $S_c=N_a^D/2$ and thus $N_{a_{\eta}}^D=S_{\eta}=0$
the $\eta$-spin lattice does not exist and hence the spin effective lattice is identical to
the original lattice. In turn, for $S_c<N_a^D/2$ LWSs the $\eta$-spin degrees of freedom 
correspond to a single occupancy configuration of the $N_{a_{\eta}}^D=M^{de}_{\eta,+1/2}$ 
deconfined $+1/2$ $\eta$-spinons. 
Only for states and subspaces for which $N_{a_{\eta}}^D/N_a^D=[1-2S_c/N_a^D]$ is finite and the inequality $0<M^{co}_{\eta}<N_{a_{\eta}}^D$ holds
is the concept of a $\eta$-spin effective lattice useful. 

For $x>0$ the subspace resulting from the overlap of the $M_{\eta,-1/2}=0$ subspace 
with the one- and two-electron subspace defined below in Section V plays a major role
in the one- and two-electron physics. 

\section{Useful states generated by $c$ and $\alpha\nu$ fermion momentum occupancies}

The goal of this section is the introduction of a useful complete set of state representations of the
model global $SO(3)\times SO(3)\times U(1)$ symmetry. Such states are constructed to inherently 
within the $N_a^D\rightarrow\infty$ limit being momentum eigenstates. Besides a momentum eigenvalue,
such states have fixed values of $S_{\eta}$, $S_{\eta}^{x_3}$, $S_s$, $S_s^{x_3}$, and $S_c$.
The interest of such states is that for Hubbard model on the square lattice
in the one- and two-electron subspace introduced in Section V,
which refers to the square-lattice quantum liquid of Ref. \cite{companion}, they are both energy and momentum
eigenstates. Since they refer to a complete set of states, the general energy and momentum eigenstates 
$\vert \Psi_{U/4t}\rangle ={\hat{V}}^{\dag}\vert\Psi_{\infty}\rangle$ that span the whole Hilbert
space of that model are a superposition of a well-defined sub-set of such states with the same momentum eigenvalue and
the same $S_{\eta}$, $S_{\eta}^{x_3}$, $S_s$, $S_s^{x_3}$, and $S_c$ values and thus
$M^{co}_{\eta}=[N_a^D-2S_c-2S_{\eta}]$ and $M^{co}_{s}=[2S_c-2S_{s}]$ values.

As discussed in Appendix A, for the 1D Hubbard model, due to the occurrence of an infinite number
of conservation laws, such states are both energy and momentum eigenstates, so that the above superpositions refer to a single state. 
A preliminary version of the general rotated-electron operator description introduced here, which
lacked its relation to the model global $SO(3)\times SO(3)\times U(1)$ symmetry, was
presented in Ref. \cite{1D} for the particular case of the 1D Hubbard model. On
expressing the rotated-electron quantum numbers in terms of those of the exact
BA solution, a description in terms of $c$ fermions without internal structure and several
several branches of $\eta$-spin-singlet $2\nu$-$\eta$-spinon composite objects
called in this paper $\eta\nu$ fermions and several branches of spin-singlet $2\nu$-spinon composite objects
called in it $s\nu$ fermions emerges. Here $\nu=1,2,...$ is the number of $\eta$-spinon and spinon pairs.
In addition, the above states have a well-defined number of deconfined $\eta$-spinons and deconfined spinons. 
Inspired in the 1D exact solution, in the following we construct similar states for the model on the square lattice.

The complexity of the expressions of the $\eta\nu$ and $s\nu$ fermion operators in terms of $\eta$-spinon and 
spinon operators, respectively, increases upon increasing the number $2\nu$ of confined $\eta$-spinons and confined spinons. 
The expressions of the spin-neutral $2\nu$-spinon $s\nu$ fermion operators involve products of the spinon operators
$s^l_{\vec{r}_j}$. Similarly, those of the $\eta$-spin-neutral $2\nu$-$\eta$-spinon $\eta\nu$ 
fermion operators involve products of the $\eta$-spinon operators $p^l_{\vec{r}_j}$. The three spinon operators
$s^l_{\vec{r}_j}$ and three $\eta$-spinon operators $p^l_{\vec{r}_j}$ are expressed in terms of the rotated-electron operators
in Eqs. (\ref{sir-pir})-(\ref{rotated-quasi-spin}).The simplest general expressions of spin-neutral two-spinon $s1$ fermion
operators in terms of the spinon operators $s^l_{\vec{r}_j}$ are given in Section
VI both for the model on the 1D and square lattices.

\subsection{The composite $\alpha\nu$ fermions}

The global $SO(3)\times SO(3)\times U(1)$ symmetry of the model on any bipartite lattice implies 
that some features of the corresponding state representations are common to all such lattices. Our extension of the 
1D model $M^{co}_{s}=[2S_c-2S_{s}]$-site spin-neutral and $M^{co}_{\eta}=[N_a^D-2S_c-2S_{\eta}]$-site 
$\eta$-spin-neutral rotated-electron occupancy configurations in terms of those of composite 
$s\nu$ fermions and $\eta\nu$ fermions, respectively, to the model on the square lattice
accounts for the basic differences between the physics of the two models.

\subsubsection{The $M^{co}_{\eta}$-$\eta$-spinon and $M^{co}_{s}$-spinon configuration partitions}

Within the rotated-electron occupancy configurations that generate the exact energy eigenstates
$\vert \Psi_{U/4t}\rangle ={\hat{V}}^{\dag}\vert\Psi_{\infty}\rangle$ considered in Section II, there are $[M^{co}_{\eta}+M^{co}_{s}]$ 
sites out of the $N_a^D$ sites of the original lattice whose rotated-electron occupancy configurations are not
invariant under the electron - rotated-electron unitary transformation. The $\eta$-spin (and spin) degrees of freedom
of $M^{co}_{\eta}$ (and $M^{co}_{s}$) of such sites refer to $\eta$-spin-neutral (and spin-neutral) configurations involving $M^{co}_{\eta}/2$
$+1/2$ $\eta$-spinons (and $M^{co}_{s}/2$ $+1/2$ spinons) and an equal number $M^{co}_{\eta}/2$ of $-1/2$ $\eta$-spinons (and
$M^{co}_{s}/2$ of $-1/2$ spinons). 

This holds as well for the related momentum eigenstates $\vert \Phi_{U/4t}\rangle ={\hat{V}}^{\dag}\vert\Phi_{\infty}\rangle$ 
considered in the following. For the model on the square lattice, the latter states refer in general to a partition different
from that of the energy eigenstates $\vert \Psi_{U/4t}\rangle ={\hat{V}}^{\dag}\vert\Psi_{\infty}\rangle$ of 
the $\eta$-spin-neutral (and spin-neutral) configurations of such $M^{co}_{\eta}$ (and $M^{co}_{s}$) 
$\eta$-spinons (and spinons), in terms of smaller configurations. Specifically, for a given momentum eigenstate there is for each branch involving $\nu=1,2,...$ 
pairs of $\eta$-spinons (and spinons) a well-defined number $N_{\eta\nu}$ of $\eta$-spin-neutral $2\nu$-$\eta$-spinon composite $\eta\nu$ fermions 
(and $N_{s\nu}$ of spin-neutral $2\nu$-spinon composite $s\nu$ fermions). The set of such composite objects describes the 
$\eta$-spinon (and spinon) occupancy configurations of exactly $M^{co}_{\eta}$ (and $M^{co}_{s}$) sites of the original lattice. 
Hence the following two sum rules hold for all momentum eigenstates,
\begin{equation}
M^{co}_{\alpha} = [M_{\alpha} - 2S_{\alpha}] =
2\sum_{\nu =1}^{\infty}\nu\,N_{\alpha\nu} \, ; \hspace{0.35cm} \alpha = \eta \, , s \, .
\label{M-L-Sum}
\end{equation}
Such sum rules refer to subspaces spanned by states with fixed values 
of $S_c$, $S_{\eta}$, and $S_s$.

One $\eta$-spin neutral $2\nu$-$\eta$-spinon composite $\eta\nu$ fermion
describes the $\eta$-spin degrees of freedom of a $\eta$-spin-singlet 
occupancy configuration involving $\nu\leq M^{co}_{\eta}/2$ sites of the original lattice unoccupied by 
rotated electrons and an equal number of sites doubly occupied by rotated electrons.
The remaining degrees of freedom of such a rotated-electron occupancy 
configuration are described by $2\nu$ unoccupied sites of the $c$ effective lattice
whose spatial coordinates are those of the corresponding $2\nu$ sites of the
original lattice.

Similarly, one spin neutral $2\nu$-spinon composite $s\nu$ fermion
describes the spin degrees of freedom of a spin-singlet occupancy 
configuration involving $\nu\leq M^{co}_{s}/2$ sites of the original lattice singly occupied by 
spin-up rotated electrons and an equal number of sites singly occupied by spin-down rotated 
electrons. The remaining degrees of freedom of that rotated-electron occupancy 
configuration are described by $2\nu$ occupied sites of the $c$ effective lattice
whose spatial coordinates are those of the corresponding $2\nu$ sites of the
original lattice. 

For each $\eta\nu$ fermion branch (and $s\nu$ fermion branch), one may consider a $\eta\nu$ effective
lattice (and $s\nu$ effective lattice). It refers to occupancy configurations of $\eta$-spin-neutral (and spin-neutral) bonds of $2\nu$ confined 
$\eta$-spinons (and spinons). Hence each ``occupied site'' of such an effective lattice corresponds to $2\nu$
sites of the $\eta$-spin (and spin) effective lattice. In turn, the $M^{de}_{\eta}=2S_{\eta}$ (and $M^{de}_{s}=2S_{s}$)
$\eta$-spin (and spin) effective lattice sites referring to the deconfined $\eta$-spinons (and deconfined spinons) 
and some of such a lattice sites referring to $2\nu'$-$\eta$-spinon composite $\eta\nu'$ fermions (and 
$2\nu'$-spinon composite $s\nu'$ fermions) of $\nu'>\nu$ branches are found below to play the role of ``unoccupied sites'' 
of such a $\eta\nu$ effective lattice (and $s\nu$ effective lattice). The conjugate of the $\alpha\nu$ effective lattice site space
variables are the $\alpha\nu$ band discrete momentum values. Their number equals that of the
$\alpha\nu$ effective lattice sites. (For 1D, such a momentum values are good quantum numbers.)
As confirmed below in Section IV-D, the states generated by $c$ fermion and $\alpha\nu$ fermion occupancy 
configurations in the corresponding $c$ and $\alpha\nu$ momentum bands, respectively,
are state representations of the global $SO(3)\times SO(3)\times U(1)$ symmetry.

Within chromodynamics the quarks have color but all quark-composite physical 
particles are color-neutral \cite{Martinus}. Here the $\eta$-spinon (and spinons) that
are not invariant under the electron - rotated-electron unitary
transformation have $\eta$-spin $1/2$ (and spin $1/2$) but are
confined within $\eta$-spin-neutral (and spin-neutral) $2\nu$-$\eta$-spinon (and $2\nu$-spinon) 
composite $\eta\nu$ fermions (and $s\nu$ fermions).
The exact and detailed internal $2\nu$-spinon configuration and $2\nu$-$\eta$-spinon configuration of a composite 
$s\nu$ fermion and $\eta\nu$ fermion, respectively, is in general an involved
unsolved problem. (In 1D the exact BA solution takes implicitly into account such internal configurations.)
Fortunately, however, the problem simplifies for the model in the one- and two-electron subspace introduced in Section V
for which the only composite object that plays an active role is the two-spinon $s1$ fermion. Its internal
structure is an issue studied in Section VI. Such a two-spinon object is related to the resonating-valence-bond 
pictures considered long ago \cite{Fazekas,Pauling,Fradkin,Auerbach}
for spin-singlet occupancy configurations of ground states. 

In order to have control over the number of
state representations of the model global $SO(3)\times SO(3)\times U(1)$ symmetry, in the following
we consider all multi-$\eta$-spinon and multi-spinon composite objects. 
Full information on the $2\nu$-$\eta$-spinon
$(\alpha=\eta)$ or $2\nu$-spinon $(\alpha=s)$ configurations associated
with the internal degrees of freedom of the composite $\alpha\nu$ fermions
is not needed for the goals of this paper. Indeed, within the present $N_a^D\gg 1$ limit the problem of the internal degrees of freedom
of the composite $\alpha\nu$ fermions and $\alpha\nu$ bond particles 
separates from that of their positions in the corresponding effective lattices.  
The partial information on the internal degrees of freedom of the composite $\alpha\nu$ fermions
needed for our studies is accessed in the following by suitable use of their transformations 
laws under the electron - rotated-electron unitary transformation.
In turn, the deconfined $\eta$-spinons and deconfined spinons are invariant
under that transformation. Thus they are non-interacting deconfined objects that are not part of 
composite $\eta\nu$ fermions and composite $s\nu$ fermions, respectively.

Strong evidence that the extension from the 1D model to the Hubbard model on the square lattice
in the one- and two-electron subspace of the $c$ fermion and $s1$ fermion description that 
results from our general $c$ fermion and $\alpha\nu$ fermion description is
correct provided that the basic differences between the two models are accounted for
is given in Ref. \cite{companion}, concerning the half-filled spin spectrum. 
(States with finite occupancies of deconfined $\eta$-spinons or one $\eta 1$ fermion 
do not contribute to that gapless spectrum.)
The use of the description introduced here reveals that in terms of $s1$ fermion spinon
breaking and deconfined spinon processes the microscopic mechanisms that generate the coherent 
spectral-weight spin-wave spectrum are very simple. Specifically, the two-spinon $s1$ fermion description
is shown in Ref. \cite{companion} to render a non-perturbative many-electron problem studied in Ref. \cite{LCO-Hubbard-NuMi} by a complex
alternative method that involves summation of an infinite set of ladder diagrams
into a mere two-$s1$-fermion-hole spectrum, described by simple 
analytical expressions. Importantly, for $U/4t\approx 1.525$ and $t\approx 295$ meV the spin-wave spectrum of the 
parent compound La$_2$CuO$_4$ (LCO) \cite{LCO-neutr-scatt} is quantitatively described by the 
corresponding theoretical spectrum derived in Ref. \cite{companion} from simple spinon pair 
breaking $s1$ fermion processes. 
 
\subsubsection{Processes that conserve the number of sites of the $\eta$-spin and spin effective lattices}

The LWS vacuum (\ref{vacuum}) of the theory corresponds to $M_{\eta}=M^{de}_{\eta,+1/2}=N_{a_{\eta}}^D$ 
deconfined $+1/2$ $\eta$-spinons and $M_s =M^{de}_{s,+1/2}=N_{a_{s}}^D$ deconfined $+1/2$ spinons. 
(The corresponding HWS vacuum refers to $M_{\eta}=M^{de}_{\eta,-1/2}=N_{a_{\eta}}^D$
deconfined $-1/2$ $\eta$-spinons and $M_s =M^{de}_{s,-1/2}=N_{a_{s}}^D$ deconfined $-1/2$ spinons.)
Hence and as mentioned above, such objects play the role of ``unoccupied sites" of the $\eta$-spin 
and spin effective lattices, respectively. Consistently, the latter objects 
have vanishing energy. Relative to that vacuum their
$\eta$-spin and spin flip processes correspond 
to ``creation" processes of deconfined $-1/2$ $\eta$-spinons
and deconfined $-1/2$ spinons, respectively.

We start by considering processes that conserve the numbers  $M_{\eta}=N_{a_{\eta}}^D$ of
$\eta$-spinons and $M_s =N_{a_{s}}^D$ of spinons. Such processes also conserve
the number of $c$ fermions and refer to a subspace whose
LWS vacuum is that provided in Eq. (\ref{vacuum}).
In turn, creation (and annihilation) of a $c$ fermion involves annihilation
(and creation) of a $c$ fermion hole. The latter process involves
removal (and addition) of one site from (and to)
the $\eta$-spin effective lattice and addition 
(and removal) of one site to (and from) the spin effective lattice.

Within our LWS representation,
creation of a local $\eta\nu$ fermion involves the replacement of 
$2\nu=2,4,...$ deconfined $+1/2$ $\eta$-spinons
by a suitable $\eta$-spin-neutral configuration involving a number 
$\nu=1,2,...$ of $-1/2$ $\eta$-spinons and an equal number of $+1/2$ $\eta$-spinons.
The initial-state $2\nu=2,4,...$ deconfined $+1/2$ $\eta$-spinons
correspond to $2\nu=2,4,...$ sites of the $\eta$-spin effective lattice. 
In turn, creation of a local $s\nu$ fermion involves the 
replacement of $2\nu=2,4,...$ deconfined $+1/2$ spinons 
by a suitable spin-neutral configuration involving a number
$\nu=1,2,...$ of $-1/2$ spinons and an equal number of $+1/2$ spinons.
Again the initial-state $2\nu=2,4,...$ deconfined $+1/2$ spinons
refer to $2\nu=2,4,...$ sites of the spin effective lattice. 

Creation of local $\eta\nu$ (and $s\nu$) fermions always involves virtual processes 
where $2\nu=2,4,...$ deconfined $+1/2$ $\eta$-spinons (and $2\nu=2,4,...$
deconfined $+1/2$ spinons) are replaced by the
$\eta$-spin-singlet (and spin-singlet)
$2\nu$-site occupancy configurations of the local 
$\eta\nu$ fermions (and $s\nu$ fermions) in the
$\eta$-spin (and spin) effective lattice. For instance,
a given process for which two local $s1$ fermions of the initial state are replaced by
one local $s2$ fermion in the final state is
divided into two virtual processes. First,
two local $s1$ fermions are annihilated. This means that
the four sites of the spin effective lattice occupied
in the initial state by the local $s1$ fermions 
are under two spin-flip processes 
occupied in an intermediate virtual state
by four deconfined $+1/2$ spinons (annihilation
of two local $s1$ fermions). Second, one local $s2$
fermion is created on such four sites.
That involves two opposite
spin-flip processes and rearrangement of the
spinons associated with the creation of the 
local $s2$ fermion spin-neutral four-spinon occupancy
configurations in the spin effective lattice.

The $2\nu=2,4,...$ sites
of the spin effective lattice occupied by one
local $s\nu$ fermion correspond to the spin $SU(2)$
degrees of freedom of $\nu=1,2,...$ spin-up rotated-electron singly 
occupied sites and $\nu=1,2,...$ spin-down rotated-electron singly 
occupied sites. The corresponding hidden $U(1)$ symmetry degrees of freedom of such sites rotated-electron occupancies
are described by $2\nu=2,4,...$ occupied sites of the $c$
effective lattice. Therefore, these $2\nu=2,4,...$ rotated-electron singly 
occupied sites are described both by the local 
$s\nu$ fermion and $2\nu=2,4,...$ local $c$ fermions.

In turn, the $2\nu=2,4,...$ sites
of the $\eta$-spin effective lattice occupied by one
local $\eta\nu$ fermion correspond to the $\eta$-spin
$SU(2)$ degrees of freedom of $\nu=1,2,...$ rotated-electron doubly  
occupied sites and $\nu=1,2,...$ rotated-electron unoccupied 
sites. Their hidden $U(1)$ symmetry degrees of freedom
are described by $2\nu=2,4,...$ unoccupied sites of the $c$
effective lattice. As a result, the $\nu=1,2,...$ rotated-electron doubly  
occupied sites and $\nu=1,2,...$ rotated-electron unoccupied 
sites are described both by the local 
$\eta\nu$ fermion and $2\nu=2,4,...$ local $c$ fermion holes
($c$ effective lattice unoccupied sites.)

Note that any pair of local $\alpha\nu$ and $\alpha'\nu'$ fermions
always refer to two {\it different} sets of $2\nu=2,4,...$ and
$2\nu'=2,4,...$ sites, respectively, of the original lattice. 

\subsubsection{Processes that do not
conserve the number of sites of the $\eta$-spin and spin effective lattices}

Creation (and annihilation) of one local $c$ fermion is a process that
involves addition (and removal) of one site to 
(and from) the spin effective lattice and removal (and addition)
of one site from (and to) the $\eta$-spin effective lattice. Therefore,
the spin and $\eta$-spin effective lattices are exotic. Indeed the
number of their sites $N_{a_s}^D=2S_c$ and
$N_{a_{\eta}}^D=[N_a^D-2S_c]$, respectively, varies by $\pm 1$
and $\mp 1$ upon creation/annihilation of one
$c$ fermion. Such processes change the eigenvalue
$S_c$ of the generator (\ref{Or-ope}) of the hidden global $U(1)$
symmetry. A subspace with fixed $S_c$ value
and hence fixed $N_{a_{\eta}}^D=M_{\eta}=[N_a^D-2S_c]$
and $N_{a_s}^D=M_s=2S_c$ values is associated
with a well-defined vacuum $\vert 0_{\eta s}\rangle$ of form
given in Eq. (\ref{vacuum}). An excitation involving a change of
such values drives the system into a new subspace referring
to a different vacuum $\vert 0_{\eta s}\rangle$. 
In turn, $\eta$-spinon and spinon creation and annihilation
processes refer to excitations within the same
quantum-liquid subspace. It is associated with a uniquely defined
vacuum $\vert 0_{\eta s}\rangle$. 

It follows that from the point of view of the $\eta$-spin and spin
degrees of freedom, $c$ fermion creation and annihilation
processes correspond to a change of quantum system.
Indeed, the $\eta$-spin and spin lattices and corresponding
number of sites change along with the quantum-system
vacuum $\vert 0_{\eta s}\rangle$ of Eq. (\ref{vacuum}).
One can then say that for the $\eta$-spinon and spinon
representation there is a different quantum system for each
eigenvalue $S_c$ of the generator (\ref{Or-ope}) of the hidden global $U(1)$
symmetry. In turn, from the
point of view of the degrees of freedom associated with
the latter symmetry, the model (\ref{H}) corresponds
to a single quantum system. The local $c$ fermions live on a
lattice identical to the original lattice whose number of sites 
$N_a^D=[N_{a_s}^D+N_{a_{\eta}}^D]$ is fixed.

Creation (and annihilation) of one local $c$ fermion involves
a virtual process in which the $\eta$-spin
effective lattice removed (and added) site is occupied in the initial (and
final) state by a deconfined $+1/2$ $\eta$-spinon. 
Also the site of the spin effective lattice added 
(and removed) by such an elementary process is occupied in the final 
(and initial) state by a deconfined $+1/2$ spinon.

If, as occurs for one-electron addition (and removal),
creation (and annihilation) of a local 
$c$ fermion involves creation (and annihilation)
of a local $s1$ fermion, the overall process 
is divided into two virtual processes. 
For instance, complementarily to creation of the local $c$ fermion in 
its effective lattice, the virtual
processes occurring in the spin effective
lattice are the following. First, a site occupied by a deconfined 
$+1/2$ spinon is added to that lattice.
Second, a local $s1$ fermion is created. One of
the two initial-state deconfined $+1/2$ spinons involved in
the final-state two-site $s1$ bond configuration is that
located on the site added to the spin effective lattice.
(A corresponding momentum eigenstate involves the superposition
of many local configurations for which that site has different
positions.) 

On the other hand, if removal of a local $c$ fermion from its
effective lattice involves annihilation of a local $s1$
fermion, one has the following virtual
processes in the spin effective lattice. First, 
a local $s1$ fermion is annihilated. This
involves a rearrangement process.
It leads to the occupancy of its two sites
by two deconfined $+1/2$ spinons in the
intermediate virtual state. Second, the
site of the spin effective lattice occupied by  
one of these two deconfined spinons
is removed along with it.

Within the LWS representation
creation (and annihilation) of both one local $c$ 
fermion and one local $s1$ fermion corresponds
to creation (and annihilation) of a spin-down 
electron. In turn, the corresponding process of
creation (and annihilation) of a spin-up electron 
involves addition (and removal) of one 
local $c$ fermion to (and from) its effective lattice and addition
(and removal) of one site 
to (and from) the spin effective lattice.
In the final (and initial) state the latter site
is occupied by a deconfined $+1/2$ spinon.  
Creation and annihilation of local $c$ fermions
leads to addition and removal (and removal 
and addition) of sites in the spin (and $\eta$-spin)
effective lattice, respectively. Similarly, it is confirmed below
that creation of one local $\alpha\nu'$ fermion
gives rise to addition of $2(\nu'-\nu)$ sites to the $\alpha\nu$ effective
lattices of $\alpha\nu$ fermion branches such that $\nu<\nu'$. 

\subsection{The $\alpha\nu$ translation generators and corresponding $\alpha\nu$ band momenta}

For the $\alpha\nu$ effective lattice, one local $\alpha\nu$ fermion ``occupied site'' refers to $2\nu$ sites of the $\eta$-spin ($\alpha=\eta$)
or spin ($\alpha=s$) effective lattice. Alike for the corresponding latter lattice, the $2S_{\alpha}$ sites occupied by deconfined 
$\eta$-spinons ($\alpha=\eta$) or deconfined spinons ($\alpha=s$) are among those playing the role of the $\alpha\nu$ effective lattice
``unoccupied sites''. Unlike for the former lattice, it is found below that 
a number $2(\nu'-\nu)$ of sites of each $\alpha\nu'$ fermion with a number $\nu'>\nu$
of confined $\eta$-spinons ($\alpha=\eta$) or confined spinons ($\alpha=s$) play as well the role of 
$\alpha\nu$ effective lattice ``unoccupied sites''.

The conjugate variables of the $\alpha\nu$ effective lattice 
real-space coordinates are the discrete momentum values of the $\alpha\nu$ band.
As shown in Appendix A, for the 1D model such discrete momentum values
are the quantum numbers of the exact BA solution. For the Hubbard model
on the square lattice the $s1$ band momentum discrete
values of state representations belonging to the one- and two-electron subspace
introduced in Section V are good quantum numbers as well.

As reported in Ref. \cite{companion} for the $s1$ fermion operators,
the $\alpha\nu$ fermion operators can be generated from the operators of corresponding 
hard-core $\alpha\nu$ bond-particle operators as follows,
\begin{eqnarray}
f^{\dag}_{{\vec{r}}_{j},\alpha\nu} & = & e^{i\phi_{j,\alpha\nu}}\,
g^{\dag}_{{\vec{r}}_{j},\alpha\nu} \, ; \hspace{0.35cm}
\phi_{j,\alpha\nu} = \sum_{j'\neq j}f^{\dag}_{{\vec{r}}_{j'},\alpha\nu}
f_{{\vec{r}}_{j'},\alpha\nu}\,\phi_{j',j,\alpha\nu} \, ; \hspace{0.35cm}
\phi_{j',j,\alpha\nu} = \arctan \left({{x_{j'}}_2-{x_{j}}_2\over {x_{j'}}_1-{x_{j}}_1}\right) \, ,
\nonumber \\
f_{\vec{q}_j,\alpha\nu}^{\dag} & = & 
{1\over {\sqrt{N_{a_{\alpha\nu}}^D}}}\sum_{j'=1}^{N_{a_{\alpha\nu}}^D}\,e^{+i\vec{q}_j\cdot\vec{r}_{j'}}\,
f_{\vec{r}_{j'},\alpha\nu}^{\dag} \, ,
\hspace{0.15cm} (1-x)>0  \hspace{0.15cm}{\rm for}\hspace{0.15cm}
\alpha\nu = s1 \hspace{0.15cm}{\rm and}\hspace{0.15cm}
S_{\alpha}/N_a^D>0 \hspace{0.15cm}{\rm for}\hspace{0.15cm}
\alpha\nu\neq s1 \, .
\label{f-an-operators}
\end{eqnarray}
Here $\phi_{j,\alpha\nu}$ is the Jordan-Wigner phase \cite{companion,Wang} operator,
the indices $j'$ and $j$ refer to sites of the $\alpha\nu$ effective lattice,
and $f_{\vec{q}_j,\alpha\nu}^{\dag}$ are the corresponding momentum-dependent 
$\alpha\nu$ fermion operators. The number $N_{a_{\alpha\nu}}^D$ of discrete momentum values of the
$\alpha\nu$ momentum band equals that of sites of the $\alpha\nu$ effective lattice.
Its expression is derived below in Section IV-D. 

Alike in 1D and as illustrated in Section VI and Appendix D for the $s1$ bond-particle operators of the model
on the square lattice, the $\eta$-spin-neutral $2\nu$-$\eta$-spinon composite $\eta\nu$ 
bond-particle operators and spin-neutral $2\nu$-spinon composite $s\nu$ bond-particle operators
denoted in Eq. (\ref{f-an-operators}) by $g^{\dag}_{{\vec{r}}_{j},\alpha\nu}$ where $\alpha =\eta,s$
are constructed to inherently upon acting onto their $\alpha\nu$ effective lattice  
anticommuting on the same site and commuting on different sites.
Hence they are hard-core like and can be transformed onto fermionic operators, as given in that equation.
For $\nu>1$ the algebra behind their construction in terms of the elementary
$\eta$-spinon or spinon operators of Eqs. (\ref{sir-pir})-(\ref{rotated-quasi-spin})
is much more cumbersome than that of the two-spinon $s1$ bond particles studied in Section VI and Appendix D. 
Fortunately, the only property needed for the goals of this paper is that
upon acting onto their $\alpha\nu$ effective lattice they are hard-core like.

The expressions given in Eq. (\ref{f-an-operators}) apply to $\alpha\nu\neq s1$ branches provided that $S_{\alpha}/N_a^D>0$.
One can also handle the problem when $S_{\alpha}=0$ and $N_{\alpha\nu}/N_a^D\ll 1$ for a given $\alpha\nu\neq s1$ 
branch. Then provided that $N_{\alpha\nu'}=0$ for all remaining $\alpha\nu'$ branches with a number of $\eta$-spinon
or spinon pairs $\nu'>\nu$ one finds below that $N_{a_{\alpha\nu}}^D=N_{\alpha\nu}$. 
The momentum of the operators $f_{\vec{q}_j,\alpha\nu}^{\dag}$
is in that limiting case given by $\vec{q}_j \approx 0$. For the states that span the corresponding 
$S_{\alpha}=0$ and $N_{\alpha\nu}/N_a^D\ll 1$ subspace
all sites of the $\alpha\nu$ effective lattice are occupied and the $\alpha\nu$ momentum band is full. If $N_{\alpha\nu}$ is finite
one has $N_{a_{\alpha\nu}}^D= N_{\alpha\nu}$ discrete momentum values $\vec{q}_j\approx 0$ compactly 
distributed around zero momentum. Their Cartesian components momentum 
spacing is $2\pi/L$. A case of interest is when $S_{\alpha}=0$ and $N_{\alpha\nu}=1$. Then the $\alpha\nu$ effective lattice
has a single site and the corresponding $\alpha\nu$ band a single discrete momentum
value, $\vec{q} = 0$. In that case $\phi_{\alpha\nu}=\phi_{j,\alpha\nu}=0$. Hence $f^{\dag}_{{\vec{r}},\alpha\nu} 
= g^{\dag}_{{\vec{r}},\alpha\nu}$ and $f_{\vec{q},\alpha\nu}^{\dag}=f_{\vec{r},\alpha\nu}^{\dag}$
where $\vec{q} = 0$.

The operators $f_{\vec{q}_j,\alpha\nu}^{\dag}$ act onto subspaces with fixed values for the set of
numbers $S_{\alpha}$, $N_{\alpha\nu}$, and $\{N_{\alpha\nu'}\}$ for $\nu'>\nu$ branches. (Below it is shown that
this is equivalent to fixed values for the set of numbers $S_c$ and $\{N_{\alpha\nu'}\}$ for all $\nu'=1,2,...$ including $\nu$.)
Such subspaces are spanned by mutually neutral states. Those are states with fixed values for the numbers 
of $\alpha\nu$ fermions and $\alpha\nu$ fermion holes. Hence such states 
can be transformed into each other by $\alpha\nu$ band particle-hole processes. 
Creation of one $\alpha\nu$ fermion is a process that involves the transition between
two states belonging to different such subspaces.
It is a well-defined process whose generator is the product of two operators.
The first operator may add sites to or remove sites from the $\alpha\nu$ effective lattice. Alternatively,
it may introduce corresponding changes in the $\alpha\nu$ momentum band. The second operator
is the creation operator $f^{\dag}_{{\vec{r}},\alpha\nu}$ or $f^{\dag}_{{\vec{q}},\alpha\nu}$ appropriate to the excited-state subspace.

Provided that $(1-x)$ is finite for $s1$ fermions\cite{companion}
and $S_{\alpha}/N_a^D>0$ as $N_a^D\rightarrow\infty$ for $\alpha\nu\neq s1$ fermions, the phases 
$\phi_{j,\alpha\nu}$ given in Eq. (\ref{f-an-operators}) are 
associated with an effective vector potential \cite{Wang,Giu-Vigna},
\begin{eqnarray}
{\vec{A}}_{\alpha\nu} ({\vec{r}}_j) & = & \Phi_0\sum_{j'\neq j}
n_{\vec{r}_{j'},\alpha\nu}\,{{\vec{e}}_{x_3}\times ({\vec{r}}_{j'}-{\vec{r}}_{j})
\over ({\vec{r}}_{j'}-{\vec{r}}_{j})^2} \, ; \hspace{0.35cm} 
n_{\vec{r}_j,\alpha\nu} = f_{\vec{r}_j,\alpha\nu}^{\dag}\,f_{\vec{r}_j,\alpha\nu} \, ,
\nonumber \\
{\vec{B}}_{\alpha\nu} ({\vec{r}}_j) & = & {\vec{\nabla}}_{\vec{r}_j}\times {\vec{A}}_{\alpha\nu} ({\vec{r}}_j)
=  \Phi_0\sum_{j'\neq j}
n_{\vec{r}_{j'},\alpha\nu}\,\delta ({\vec{r}}_{j'}-{\vec{r}}_{j})\,{\vec{e}}_{x_3} \, .
\label{A-j-s1-3D}
\end{eqnarray}   
For the model on the square lattice the vector ${\vec{e}}_{x_3}$ 
appearing here is the unit vector perpendicular to the plane. (Often we use units such 
that the fictitious magnetic flux quantum is given by $\Phi_0=1$.)

The components of the microscopic momenta of the $\alpha\nu$ fermions are 
eigenvalues of the two (and one for 1D) $\alpha\nu$ translation generators 
${\hat{q}}_{\alpha\nu\,x_1}$ and ${\hat{q}}_{\alpha\nu\,x_2}$
in the presence of the fictitious magnetic field ${\vec{B}}_{\alpha\nu} ({\vec{r}}_j)$. That
seems to imply that for the model on the square lattice the components 
${q}_{x1}$ and ${q}_{x2}$ of the microscopic momenta $\vec{q}=[{q}_{x1},{q}_{x2}]$ refer to operators that do not commute.
However, in the subspaces where the operators $f_{\vec{q}_j,\alpha\nu}^{\dag}$ act onto
such are commuting operators.
Indeed, those subspaces are spanned by neutral states \cite{Giu-Vigna}. Since $[{\hat{q}}_{\alpha\nu\,x_1},{\hat{q}}_{\alpha\nu\,x_2}]=0$
in such subspaces, for the model on
the square lattice the $\alpha\nu$ fermions carry a microscopic momentum $\vec{q}=[{q}_{x1},{q}_{x2}]$
where the components ${q}_{x1}$ and ${q}_{x2}$ are well-defined simultaneously.

The momentum operator $\hat{{\vec{P}}}$ of Eq. (\ref{P-invariant})
commutes with the $\alpha\nu$ generators ${\hat{{\vec{q}}}}_{\alpha\nu}$ whose
eigenvalues are the $\alpha\nu$ fermion microscopic momenta $\vec{q}$. Consistently,
within our description it can be written as,
\begin{equation}
\hat{{\vec{P}}} = {\hat{{\vec{q}}}}_c 
+ \sum_{\nu =1}^{\infty}{\hat{{\vec{q}}}}_{s\nu} 
+ \sum_{\nu =1}^{\infty}{\hat{{\vec{q}}}}_{\eta\nu} 
+ \vec{\pi}\,{\hat{M}}_{\eta\, ,-1/2}  \, .
\label{P-c-alphanu}
\end{equation}
In this expression the $-1/2$ $\eta$-spinons momentum $\vec{\pi}$ results from the momentum
operator $\hat{{\vec{P}}}$ not commuting with the $\eta$-spin off-diagonal generators,
the operators ${\hat{M}}_{\eta\, ,\pm 1/2}$ and  ${\hat{M}}_{\eta}$ count the numbers 
$M_{\eta\, ,\pm 1/2}=[M^{de}_{\eta\,\pm1/2}+M^{co}_{\eta}/2]$ and $M_{\eta}=[M^{de}_{\eta}+M^{co}_{\eta}]$ of $\pm 1/2$ $\eta$-spinons
and $\eta$-spinons, respectively,
and the $c$ and $\alpha\nu$ translation generators read,
\begin{equation}
{\hat{{\vec{q}}}}_c  = \sum_{{\vec{q}}}{\vec{q}}\, \hat{N}_c ({\vec{q}})
\, ; \hspace{0.35cm}
{\hat{{\vec{q}}}}_{s\nu} = \sum_{{\vec{q}}}{\vec{q}}\, \hat{N}_{s\nu} ({\vec{q}})
\, ; \hspace{0.35cm}
{\hat{{\vec{q}}}}_{\eta\nu} = \sum_{{\vec{q}}}[\vec{\pi} -{\vec{q}}]\,\hat{N}_{\eta\nu} ({\vec{q}}) \, .
\label{m-generators}
\end{equation}
Here $\hat{N}_{c}({\vec{q}})$ and $\hat{N}_{\alpha\nu}({\vec{q}})$ are the momentum 
distribution-function operators,
\begin{equation}
\hat{N}_{c}({\vec{q}}) = f^{\dag}_{{\vec{q}},c}\,f_{{\vec{q}},c} \, ;
\hspace{0.35cm}
\hat{N}_{\alpha\nu}({\vec{q}}) = f^{\dag}_{{\vec{q}},\alpha\nu}\,f_{{\vec{q}},\alpha\nu} \, ,
\label{Nc-s1op}
\end{equation}
respectively. 

The $\eta$-spin flip process momentum appearing in the expressions given in the
above equations being $\vec{\pi}$ is consistent with the
$\eta$-spin-algebra off-diagonal generators ${\hat{S }}_{\eta}^{\dag}$ and ${\hat{S }}_{\eta}$ of Eq.
(\ref{Scs}) referring to that momentum. That
the $\eta\nu$ generators ${\hat{{\vec{q}}}}_{\eta\nu}$ of Eq. (\ref{m-generators})
involve $[\vec{\pi} -{\vec{q}}]$ instead of ${\vec{q}}$
is consistent as well with the anti-binding character of the $\eta\nu$ fermions reported below. 
For the 1D model the form of the momentum operator provided in Eq. (\ref{P-c-alphanu}) is that 
consistent with the exact solution, as discussed in Appendix A.

For both the Hubbard model on the square and 1D lattice the Hamiltonian $\hat{H}$ of 
Eq. (\ref{H}) and the momentum operator $\hat{{\vec{P}}}$ of Eqs. (\ref{P-invariant}) and
(\ref{P-c-alphanu}) obey within the $N_a^D\rightarrow\infty$ limit the commutation relations,
\begin{equation}
[\hat{H},\hat{{\vec{P}}}] = [\hat{H},{\hat{M}}_{\alpha\, ,\pm 1/2}]
= [\hat{H},{\hat{{\vec{q}}}}_c] = 0 \, ; \hspace{0.35cm} \alpha =\eta,s \, ,
\label{H-commutators}
\end{equation}
and
\begin{equation}
[\hat{{\vec{P}}},{\hat{M}}_{\alpha\, ,\pm 1/2}] = [\hat{{\vec{P}}},{\hat{{\vec{q}}}}_c]
= [\hat{{\vec{P}}},{\hat{{\vec{q}}}}_{\alpha\nu}] = 0 \, ; \hspace{0.35cm} \alpha =\eta,s
\, , \hspace{0.25cm} \nu =1,...,\infty \, .
\label{P-commutators}
\end{equation}
In turn, the set of commutators $[\hat{H},{\hat{{\vec{q}}}}_{\alpha\nu}]$
vanish for the 1D model whereas for the model on the square lattice one has in general
that $[\hat{H},{\hat{{\vec{q}}}}_{\alpha\nu}]\neq 0$. 

It follows from the momentum operator
expression (\ref{P-c-alphanu}) that the corresponding momentum eigenvalues $\vec{P}$
can be expressed as a sum of the filled $c$ and $\alpha\nu$ fermion microscopic momenta.        
For most subspaces the maximum value $\nu_{max}$ of the $\alpha\nu$ fermion $\eta$-spinon 
($\alpha =\eta$) or spinon ($\alpha =s$) pair number $\nu$ 
behaves as $\nu_{max}\rightarrow\infty$ in the $N_a^D\rightarrow\infty$ limit. 
In that case the corresponding set of $\alpha\nu$ translation generators 
${\hat{{\vec{q}}}}_{\alpha\nu}$ of Eq. (\ref{m-generators}) is infinite, $\nu =1,2,...,\infty$. 

We emphasize that as given in Eq. (\ref{H-commutators}), the commutator
$[\hat{H},{\hat{{\vec{q}}}}_c]$ vanishes. Indeed, for both the Hubbard model on the square and 1D lattice
the $c$ band momenta are good quantum numbers for the whole Hilbert space.
As further discussed below, the conservation of such momenta is related to both the
local separation of the electronic degrees of freedom due to the $U/4t\rightarrow\infty$ 
local $SU(2)\times SU(2) \times U(1)$ gauge symmetry and the unitarity of the operator $\hat{V}$.
For the square-lattice quantum liquid, whose construction performed in Section V
and Ref. \cite{companion} relies on the general description introduced in this paper,
both the $c$ band and $s1$ band discrete momentum values are good quantum numbers.
The shape of the $s1$ momentum band and of its
boundary as well as the form of the $s1$ and $c$ fermion energy dispersions
are problems addressed in Ref. \cite{companion} for the Hubbard model on the square lattice 
in the one- and two-electron subspace. 
      
Each subspace spanned by states with fixed values of $S_c$, $S_{\eta}$, 
and $S_s$ and hence also with fixed values of $M^{co}_{\eta}$ and $M^{co}_s$
can be divided into smaller subspaces spanned by states with fixed values for the
set of numbers $\{N_{\eta\nu}\}$ and $\{N_{s\nu}\}$, which must
obey the sum rules of Eq. (\ref{M-L-Sum}). The numbers $\{N_{\eta\nu}\}$ and $\{N_{s\nu}\}$
correspond to operators $\{{\hat{N}}_{\eta\nu}\}$ and $\{{\hat{N}}_{s\nu}\}$
that commute with both the momentum operator of Eqs. (\ref{P-invariant}) and
(\ref{P-c-alphanu}) and the seven generators  given in Eqs. (\ref{Or-ope}) and (\ref{Scs})
of the group $[SO(4)\times U(1)]/Z_2=SO(3)\times SO(3)\times U(1)$ associated with
the model global symmetry. In turn, the lack of
an infinite set of conservation laws implies that for 
the model on the square lattice the operators $\{{\hat{N}}_{\eta\nu}\}$ and $\{{\hat{N}}_{s\nu}\}$
do not commute in general with the Hamiltonian. Hence for that model the numbers $\{N_{\eta\nu}\}$ and $\{N_{s\nu}\}$ 
are not in general good quantum numbers. 

The same applies to the set of $\alpha\nu$ translation generators ${\hat{{\vec{q}}}}_{\alpha\nu}$ of Eq. (\ref{m-generators}) 
in the presence of the fictitious magnetic 
field ${\vec{B}}_{\alpha\nu}$ of Eq. (\ref{A-j-s1-3D}). As given in Eqs. (\ref{H-commutators})
and (\ref{P-commutators}), they 
commute with the momentum operator yet in general do not commute with the Hamiltonian of the 
model on the square lattice. For 1D both such $\alpha\nu$ translation generators
and the number operators of the set $\{{\hat{N}}_{\alpha\nu}\}$ commute with the Hamiltonian. Within the 
$N_a^D\rightarrow\infty$ limit there is an infinite number of such generators and numbers.
As discussed in Appendix A, they are associated with the set of infinite conservation laws
behind the model integrability \cite{Prosen}. For the Hubbard model on the square lattice in the 
one- and two-electron subspace most of such numbers vanish and a few of them become good quantum numbers. 

That the momentum operator commutes with the electron - rotated-electron unitary operator 
$\hat{V}$ implies that the momentum eigenvalues of the states 
$\vert \Phi_{U/4t}\rangle ={\hat{V}}^{\dag}\vert\Phi_{\infty}\rangle$ studied below in
Section IV-E are independent
of $U/4t$. They are the same as those of the corresponding states $\vert \Phi_{\infty}\rangle$.
For both the Hubbard model on the square and 1D lattice
it follows from Eqs. (\ref{P-c-alphanu}) and (\ref{m-generators}) that such momentum eigenvalues 
are given by,
\begin{equation}
\vec{P} = \sum_{{\vec{q}}}{\vec{q}}_j\, N_c ({\vec{q}})
+ \sum_{\nu =1}^{\infty}\sum_{{\vec{q}}}{\vec{q}}\, N_{s\nu} ({\vec{q}}) 
+ \sum_{\nu =1}^{\infty}\sum_{{\vec{q}}}[\vec{\pi} -{\vec{q}}]\,N_{\eta\nu} ({\vec{q}})
+ \vec{\pi}\,M_{\eta\, ,-1/2} \, .
\label{P-1-2-el-ss}
\end{equation}
The momentum distribution functions
$N_{c}({\vec{q}})$ and $N_{\alpha\nu}({\vec{q}})$ appearing here are  
such that application of the corresponding momentum distribution-function operators (\ref{Nc-s1op}) onto 
the momentum eigenstates $\vert \Phi_{U/4t}\rangle$ gives 
$\hat{N}_{c}({\vec{q}})\vert \Phi_{U/4t}\rangle=N_{c}({\vec{q}})\vert \Phi_{U/4t}\rangle$
and $\hat{N}_{\alpha\nu}({\vec{q}})\vert \Phi_{U/4t}\rangle=N_{\alpha\nu}({\vec{q}})\vert \Phi_{U/4t}\rangle$, respectively.
Indeed $N_{c}({\vec{q}})$ and $N_{\alpha\nu}({\vec{q}})$ are eigenvalues of such operators that read $1$ and 
$0$ for filled and unfilled momentum values, respectively. 
For the Hubbard model on the 1D lattice the validity of expression (\ref{P-1-2-el-ss}) is confirmed by 
the exact solution, the $c$ and $\alpha\nu$ fermion discrete momentum 
values being good quantum numbers. Hence as discussed in Appendix A, for 1D the momentum distribution-function operators
$\hat{N}_{c}({\vec{q}})$ and $\hat{N}_{\alpha\nu}({\vec{q}})$ of Eq. (\ref{Nc-s1op}) commute with
the Hamiltonian.

That the distributions $N_{c}({\vec{q}})=\langle\Phi_{U/4t}\vert f^{\dag}_{{\vec{q}},c}\,f_{{\vec{q}},c}\vert \Phi_{U/4t}\rangle$ and
$N_{\alpha\nu}({\vec{q}}) = \langle\Phi_{U/4t}\vert f^{\dag}_{{\vec{q}},\alpha\nu}\,f_{{\vec{q}},\alpha\nu}\vert \Phi_{U/4t}\rangle$
are given by $1$ or $0$ does not imply that the same applies to the rotated-electron momentum 
distribution $N ({\vec{k}})={1\over 2}\sum_{\sigma}N_{\sigma} ({\vec{k}})$ where
$N_{\sigma} ({\vec{k}})=\langle \Psi_{U/4t}\vert {\tilde{c}}_{\vec{k},\sigma}^{\dag}
{\tilde{c}}_{\vec{k},\sigma}\vert \Psi_{U/4t}\rangle =
\langle \Psi_{\infty}\vert c_{\vec{k},\sigma}^{\dag}
c_{\vec{k},\sigma}\vert \Psi_{\infty}\rangle$. For $U/4t>0$ such a rotated-electron momentum distribution 
is independent of $U/4t$, equaling the $U/t\rightarrow\infty$ electron momentum distribution. However, only
for the $x=0$ and $m=0$ ground state of both the model on the 1D and square lattice does the rotated-electron momentum 
distribution simply reads $N ({\vec{k}})=1$, alike $N_{c}({\vec{q}})$.
For the model on the 1D lattice the corresponding $x=0$, $m=0$, and $U/t\rightarrow\infty$ electron 
momentum distribution $N (k)=1$ is plotted in Fig. 3 (a) of Ref. \cite{88}.

In turn, for $x>0$ and $m=0$ ground states the rotated-electron momentum distribution is not merely 
given by $1$ or $0$, being a function of the rotated-electron momentum $\vec{k}$. 
It has contributions both from the $c$ fermion and spinon degrees of freedom. For the 1D model it
is studied in Ref. \cite{Ogata} in the $U/t\rightarrow\infty$ limit for electrons, on combining the exact BA ground-state 
wave function with numerical methods. (The distribution of that reference applies as well to rotated
electrons for $U/4t>0$.) That the energy spectrum depends on $U/4t$ is for the Hubbard model on the square lattice 
behind the $x>0$ and $m=0$ ground states belonging  in general to different $V$ towers for different values of $U/4t$. 
The exception is the $x=0$ and $m=0$ ground state, which belongs to the same $V$ tower
for all $U/4t>0$ values \cite{companion}.

In the following we provide further information on why for the model on the square
lattice the $c$ band momenta are conserved and the  momentum operator and
corresponding eigenvalues have the form given in Eqs. (\ref{P-c-alphanu}) and 
(\ref{P-1-2-el-ss}), respectively. In the $U/t\rightarrow\infty$ limit electron single and doubly occupancy 
become good quantum numbers and the electronic degrees of
freedom contributing to the momentum decouple into 
two main types: i) Those associated with electron hopping, which contribute to the kinetic energy; ii) Those associated with the  
$2\nu$-$\eta$-spinon composite $\eta\nu$ fermions and $2\nu$-spinon composite 
$s\nu$ fermion occupancy configurations, which do not contribute to the kinetic energy. 

The discrete momentum values associated with electron hopping and the finite kinetic energy are carried by 
quasicharge particles whose operator ${\mathcal{F}}^{\dag}_{{\vec{q}},c}$ is given by,
\begin{equation}
{\mathcal{F}}^{\dag}_{{\vec{q}},c}=
{1\over {\sqrt{N_a^D}}}\sum_{r}\,e^{+i\vec{q}\cdot r}\,\hat{c}_r 
\, ; \hspace{0.5cm} r \equiv\vec{r}_j \, .
\label{F+-cr}
\end{equation}
Here within the notation of Ref. \cite{Ostlund-06}, $r \equiv\vec{r}_j$ are the real-space coordinates
of the quasicharge particles of that reference. Such particles are the $c$ fermion holes in that limit. The
momentum values (ii) refer to the composite $\eta\nu$ and $s\nu$ fermions
whose spectrum is dispersionless for $U/t\rightarrow\infty$. As discussed below in Section IV-C, such objects
are anti-binding and binding neutral superpositions of $2\nu=2,4,...$ 
pseudospins (or $\eta$-spinons) and $2\nu=2,4,...$ spins
(our spinons), respectively. The corresponding local spin and pseudospin
operators have expressions in terms of electron operators similar to those
of Eqs. (\ref{sir-pir})-(\ref{rotated-quasi-spin}) of the corresponding spinon
and $\eta$-spinon operators, respectively, in terms of rotated-electron operators \cite{Ostlund-06}.
In the $U/4t\rightarrow\infty$ limit the $c$ fermion momentum is the part of the 
corresponding electronic momentum associated with the kinetic energy. Indeed
the electronic momentum has in addition contributions from the spin occupancy
configurations. In the $U/4t\rightarrow\infty$ limit such configurations are degenerate
and have vanishing energy. 

Note that in the $U/4t\rightarrow\infty$ limit the electronic
momentum is not a good quantum number. However, the part of it associated with
the kinetic energy alone, which results from the electronic motion of the singly
occupied sites relative to the remaining sites, is. The above separation of the electronic degrees of
freedom contributing to the momentum follows from in the $U/4t\rightarrow\infty$ limit
the Hubbard model on a bipartite lattice having a local $SU(2)\times SU(2) \times U(1)$ gauge symmetry,
stronger than the related global $[SU(2)\times SU(2)\times U(1)]/Z_2^2=[SO(4)\times U(1)]/Z_2=SO(3)\times SO(3)\times U(1)$
symmetry \cite{U(1)-NL}. Such a local gauge symmetry has indeed stronger consequences
in terms of the electronic degrees of freedom separation than a global symmetry alone.
(In that limit the finite-electron-double occupancy $\eta$-spin configurations are degenerate but
refer to infinite energy and thus are not accessible to the finite energy physics.)

Importantly, the finite-energy electronic degrees of freedom associated with the momentum values contributing to the
kinetic energy and not contributing to it are associated with the $c$ fermion local $U(1)$ gauge
symmetry and local spin $SU(2)$ gauge symmetry, respectively, of the model 
local $SU(2)\times SU(2) \times U(1)$ gauge symmetry. Such a local separation of the
electronic degrees of freedom associated with the momentum contributions
arising from the kinetic energy and spin configurations, respectively, which follows from two
independent local gauge symmetries, is behind the $c$ fermion momentum values
being good quantum numbers of the model on the square lattice in the $U/4t\rightarrow\infty$ limit. The point is that while 
the electronic momentum values have contributions both from the $c$ fermion and spin degrees 
of freedom so that the electrons are not true non-interacting spinless fermions, in the $U/4t\rightarrow\infty$ limit the $c$ fermions
are. The unitarity of the operator $\hat{V}$ then assures that such momentum values are
good quantum numbers of the model on the square lattice for $U/4t>0$ as well. 

This is why, as given in Eqs. (\ref{H-commutators}) and (\ref{P-commutators}),
for the whole interaction range $U/4t>0$ both
the Hamiltoninan and momentum operator commute with the $c$ translation generator
${\hat{{\vec{q}}}}_{c}$ of Eq. (\ref{m-generators}). And this applies both to the model on the
1D and square lattice. For $U/4t>0$ the distribution $N_c ({\vec{q}})$ is then an eigenvalue of the operator 
${\hat{N}}_c ({\vec{q}})$ of Eq. (\ref{Nc-s1op}). It reads $1$ and $0$ for filled and unfilled 
momentum values, respectively. Since for $U/4t>0$ the set of momentum eigenstates $\{\vert \Phi_{U/4t}\rangle\}$
introduced below in Section IV-E is complete, the corresponding energy
eigenstates $\vert \Psi_{U/4t}\rangle$ of the Hubbard model on the square
lattice can be expressed as superpositions of sub-sets of the former states with the same values for the
distribution $N_c ({\vec{q}})$.

Concerning the $s1$ fermion momentum values, it turns out that the $U/4t\rightarrow\infty$
local spin $SU(2)$ gauge symmetry also implies that they are good quantum numbers 
of the Hubbard model on the square lattice provided
that both the corresponding numbers $N_{s1}^h$ and $N_{s1}$ of unfilled and filled
discrete momentum values, respectively, are conserved. This condition
is not in general fulfilled for that model. Fortunately,
it is fulfilled for it in the one- and two-electron subspace as defined below
in Section V. The unitarity of the operator $\hat{V}$ then assures that in such
a subspace the $s1$ band momentum values are conserved for $U/4t>0$ as well.

The $U/4t>0$ model global $SO(3)\times SO(3)\times U(1)$ symmetry can be written as
$[SO(4)\times U(1)]/Z_2$. Both for the Hubbard model on the square 
and 1D lattices it is useful to rewrite the momentum eigenvalues $\vec{P}$ of Eq. (\ref{P-1-2-el-ss})
in terms of two contributions $\vec{P}_{U(1)}$ and ${\vec{P}}_{SO(4)}$ arising from the hidden $U(1)$ symmetry
and $SO(4)$ symmetry, respectively, state representation contributions. For both such models
the momentum contribution ${\vec{P}}_{SO(4)}=\vec{P} - \vec{P}_{U(1)} $ to the momentum eigenvalues,
 \begin{equation}
\vec{P} = \vec{P}_{U(1)} + {\vec{P}}_{SO(4)} \, ; \hspace{0.35cm}
\vec{P}_{U(1)} = \vec{P}_c = \sum_{{\vec{q}}}{\vec{q}}\, N_c ({\vec{q}})
 \, ; \hspace{0.35cm} {\vec{P}}_{SO(4)} = {\vec{P}}_{\eta} + {\vec{P}}_{s} \, ,
\label{P-1-2-el-ss-2D}
\end{equation}
is for $U/4t>0$ a good quantum number. This simply follows from $\vec{P}$ and $\vec{P}_{U(1)}$
being good quantum numbers as well. In expression (\ref{P-1-2-el-ss-2D}) 
$\vec{P}_{U(1)}$, ${\vec{P}}_{\eta}$, and ${\vec{P}}_{s}$ are the momentum contributions
arising from the $c$ fermion, $\eta$-spinon, and spinon occupancy configurations, respectively. The latter two contributions read
${\vec{P}}_{\eta}=\sum_{\nu}\sum_{{\vec{q}}}[\vec{\pi}-{\vec{q}}]\,N_{\eta\nu} ({\vec{q}})+\vec{\pi}\,M_{\eta\, ,-1/2}$
and ${\vec{P}}_{s}=\sum_{\nu}\sum_{{\vec{q}}}{\vec{q}}\, N_{s\nu} ({\vec{q}})$, respectively. 
While the $SO(4)$ momentum ${\vec{P}}_{SO(4)} = {\vec{P}}_{\eta} + {\vec{P}}_{s}$ given in Eq. (\ref{P-1-2-el-ss-2D}) 
is a good quantum number, the operators associated with ${\vec{P}}_{\eta}$ and ${\vec{P}}_{s}$ 
and the corresponding sets of translation generators ${\hat{{\vec{q}}}}_{\eta\nu}$ and ${\hat{{\vec{q}}}}_{s\nu}$,
respectively, of Eq. (\ref{m-generators}) labelled by the index $\nu =1,2,...$ do not commute in general 
with the Hamiltonian of the Hubbard model on the square lattice, unlike for 1D.

For $U/4t\rightarrow\infty$ all $\eta$-spin and spin configurations
are degenerate and do not contribute to the kinetic energy. Nevertheless
they lead to overall momentum contributions ${\vec{P}}_{\eta}$ and 
${\vec{P}}_{s}$, respectively. As mentioned above, an important point is that the 
$\alpha\nu$ fermion operators are defined in and act onto subspaces
spanned by mutually neutral states. For the Hubbard model on the square lattice
in these subspaces the $\alpha\nu$ translation generators $\hat{q}_{\alpha\nu\,x_1}$ and
$\hat{q}_{\alpha\nu\,x_2}$ commute. This implies that the $\alpha\nu$ translation 
generators ${\hat{{\vec{q}}}}_{\alpha\nu}$ of Eq. (\ref{m-generators})
in the presence of the fictitious magnetic field ${\vec{B}}_{\alpha\nu}$ 
of Eq. (\ref{A-j-s1-3D}) commute with the momentum operator both in the 
$U/4t\rightarrow\infty$ limit and for $U/4t$ finite, as given in Eq. (\ref{P-commutators}).
The momentum operator can then be written as in Eqs. (\ref{P-c-alphanu}) and (\ref{m-generators})
both for the Hubbard model on the 1D and square lattices.

The exact expressions of the numbers of $\alpha\nu$ band discrete momentum values
$N_{a_{\alpha\nu}}^D$ are found below in Section IV-D.
Thus for the model on the square lattice the momentum area $N_{a_{\alpha\nu}}^2\,[2\pi/L]^2$ and discrete momentum values number 
$N_{a_{\alpha\nu}}^2$ of the $\alpha\nu$ bands are known. 
In contrast, the shape of their boundary remains in general an open issue. The shape of the $c$ band is
the same as that of the first Brillouin zone. However, the discrete momentum values may
have small overall shifts under transitions between subspaces with different
$\sum_{\alpha\nu}N_{\alpha\nu}$ values. The studies of Ref. \cite{companion}
use several approximations to obtain useful information on the $c$ and $s1$ bands momentum values.
In that reference the corresponding $c$ and $s1$ fermion energy dispersions and velocities
are studied for the Hubbard model on the square lattice in the one- and two-electron subspace,
for which both the translation generators ${\hat{{\vec{q}}}}_{c}$ and ${\hat{{\vec{q}}}}_{s1}$ of Eq. (\ref{m-generators})
commute with that model Hamiltonian. 

\subsection{Ranges of the $c$ and $\alpha\nu$ fermion energies, their transformation laws,
and the ground-state occupancies}

Since the microscopic momenta of the $c$ fermions are good quantum numbers
both for the Hubbard model on the square and 1D lattices, one may define an energy dispersion $\epsilon_c (\vec{q})$ \cite{companion}.
For the 1D model also the $\alpha\nu$ energy dispersions $\epsilon_{\alpha\nu} (q)$ are
well defined. In turn, for the model on the square lattice an energy dispersion
$\epsilon_{\alpha\nu} (\vec{q})$ is for $\alpha\nu\neq s1$ branches well defined for the momentum
values for which such objects are invariant under the electron - rotated-electron
unitary transformation. For general $\alpha\nu$ fermion momentum 
values there is not in general a dispersion $\epsilon_{\alpha\nu} =\epsilon_{\alpha\nu} (\vec{q})$
defining a one-to-one correspondence between the energy $\epsilon_{\alpha\nu}$
and momentum $\vec{q}$. However, the range of the $\alpha\nu$ fermion
energy $\epsilon_{\alpha\nu}$ remains well defined. A $s1$ fermion energy dispersion $\epsilon_{s1} =\epsilon_{s1} (\vec{q})$ is 
well defined for the square-lattice quantum liquid \cite{companion}.

The quantum-object occupancy configurations of the ground state are found below. 
They are consistent with the energies for creation onto such states of  
our description quantum objects. For instance, the elementary 
energies $\epsilon_{s,-1/2} = 2\mu_B\,H$ and  $\epsilon_{\eta,-1/2} = 2\mu$ of Eq.
(\ref{energies}) of Appendix B correspond to creation onto a $m\geq 0$ and $x> 0$ 
ground state of a deconfined $-1/2$ spinon and a deconfined $-1/2$ $\eta$-spinon, respectively. 
Here $\mu_B$ is the Bohr magneton, $H$ the magnetic field, and $\mu$ the
chemical potential. The energy $\epsilon_{s,-1/2} = 2\mu_B\,H$ (and  $\epsilon_{\eta,-1/2} = 2\mu$)
refers to an elementary spin-flip (and $\eta$-spin-flip) process. It
transforms a deconfined $+1/2$ spinon (and
deconfined $+1/2$ $\eta$-spinon) into a deconfined $-1/2$ spinon 
(and deconfined $-1/2$ $\eta$-spinon). Such elementary energies control the range of
several physically important energy scales. Within our LWS
representation, a deconfined $+1/2$ spinon (and deconfined 
$+1/2$ $\eta$-spinon) has vanishing energy so that
$\epsilon_{s,+1/2} = 0$ (and $\epsilon_{\eta,+1/2} = 0$).
It follows that the energy of a pair of deconfined spinons (and deconfined $\eta$-spinons) 
with opposite projections is $2\mu_B\,H$ (and $2\mu$).
Indeed due to the invariance of such objects under the
electron - rotated-electron unitary transformation, 
they are not energy entangled and the total energy
is the sum of their individual energies.

In the following we confirm that ground states 
have no $\eta\nu$ fermions and no $s\nu$ fermions
with $\nu >1$ spinon pairs. The corresponding
energies $\epsilon_{\eta\nu}$ and $\epsilon_{s\nu}$, respectively,
considered below refer to creation onto the ground state of one of such objects. 
We start by providing a set of useful properties.
We emphasize that some of these properties are not
valid for descriptions generated by rotated-electron
operators associated with the general unitary operators ${\hat{V}}$ considered
in Ref. \cite{bipartite}. The following properties rather refer to
the specific operator description associated with the rotated-electron
operators ${\tilde{c}}_{\vec{r}_j,\sigma}^{\dag} =
{\hat{V}}^{\dag}\,c_{\vec{r}_j,\sigma}^{\dag}\,{\hat{V}}$ of Eq. (\ref{rotated-operators}) as
defined in Section II. 

Some of the following results are obtained from extension to the model on the square 
lattice of exact results extracted from the 1D model BA solution. However such an extension
accounts for the different physics of such models. Both for the model on the
square and 1D lattice, the range of the energy 
$\epsilon_{\alpha\nu}$ for addition onto the ground state of one $\alpha\nu$ fermion
derived below is that consistent with the interplay of the transformation laws of the $\alpha\nu$ 
fermions under the electron - rotated-electron unitary transformation 
with the model global $SO(3)\times SO(3)\times U(1)$ symmetry.

\subsubsection{The $\eta\nu$ fermion energy range}

Alike for 1D,  for the model on the square lattice one $\eta\nu$ fermion is
a $\eta$-spin-neutral anti-binding
configuration of a number $\nu=1,2,...$ of confined $-1/2$ $\eta$-spinons and 
an equal number of confined $+1/2$ $\eta$-spinons.
Symmetry implies that for $U/4t>0$ there is no energy overlap between the
$\epsilon_{\eta\nu}$ ranges corresponding to different $\nu=1,2,...$ branches.
Fermions belonging to neighboring 
$\eta\nu$ and $\eta\nu+1$ branches differ in the number of $\eta$-spinon pairs by one.
The requirement for the above lack of energy overlap 
is then that the energy bandwidth of the $\epsilon_{\eta\nu}$ range is smaller than or equal to $2\vert\mu\vert$. 
For all $x$ values, the energy scale $2\vert\mu\vert =[\epsilon_{\eta,-1/2} +\epsilon_{\eta,+1/2}]$
where $\mu =\mu^0$ at $x=0$
refers to the energy of a pair of deconfined $\eta$-spinons of opposite $\eta$-spin projection.
Such properties imply the following range for the energy $\epsilon_{\eta\nu}$,
\begin{equation}
2\nu\vert\mu\vert\leq \epsilon_{\eta\nu} < 2(\nu + i_{\eta\nu})\vert\mu\vert 
\, ; \hspace{0.25cm} 0\leq i_{\eta\nu} \leq 1\, ,
\label{energy-e-n}
\end{equation}
where $\mu =\mu^0$ at $x=0$.
Deconfined $\eta$-spinons are invariant under
the electron - rotated-electron unitary transformation. 
Thus they are non interacting and their energies are additive. 
Consistently, for all $x$ values $2\nu\vert\mu\vert$ is the energy of $\nu$ deconfined $-1/2$ $\eta$-spinons 
and $\nu$ deconfined $+1/2$ $\eta$-spinons. 
For instance, for $x>0$ the energy $2\nu\mu$ is as well that for creation of a
number $\nu=1,2,...$ of deconfined $-1/2$ $\eta$-spinons onto a $S_{\eta}=\nu$ ground
state with $2\nu$ deconfined $+1/2$ $\eta$-spinons.
Such a creation refers to $\nu$ $\eta$-spin-flip processes 
(transformation of $\nu$ rotated-electron unoccupied sites
into $\nu$ rotated-electron doubly occupied sites.)
For $m=0$ and $x>0$ the number $i_{\eta\nu}$ decreases continuously 
for increasing values of $U/4t$. It has the limiting
behaviors $i_{\eta\nu}\rightarrow 1$ for $U/4t\rightarrow 0$
and $i_{\eta\nu}\rightarrow 0$ for $U/4t\rightarrow\infty$.
Hence the $\epsilon_{\eta\nu}$ range vanishes for $U/4t\rightarrow\infty$. 
The latter behavior is associated with the full
degeneracy of the $\eta$-spin configurations reached
as $U/4t\rightarrow\infty$. In turn, at $m=0$ and $x=0$
the number $i_{\eta\nu}$ vanishes and $\epsilon_{\eta\nu}=2\nu\mu^0$
for the whole finite interaction range $U/4t>0$. As discussed below, this behavior follows
from the invariance under the electron - rotated-electron
unitary transformation of a $\eta\nu$ fermion created
onto a $x=0$ and $m=0$ ground state. 
  
\subsubsection{The $s\nu$ fermion energy range}

A $s\nu$ fermion is a spin-neutral binding
configuration of a number $\nu=1,2,...$ of confined $-1/2$ spinons and an equal number of confined $+1/2$ spinons.
Again, symmetry implies that for $U/4t>0$ there is no energy overlap
between the $\epsilon_{s\nu}$ ranges of different
$\nu=1,2,...$ branches. For $s\nu$ branches with a
number of spinon pairs $\nu >1$ such an energy range 
bandwidth is for the present binding
configurations and for the same reasoning as for the $\eta\nu$ fermion
smaller than or equal to $2\mu_B\,\vert H\vert$. For all $m$ values,
$2\mu_B\,\vert H\vert =[\epsilon_{s,-1/2} +\epsilon_{s,+1/2}]$ 
equals the energy of a pair of deconfined spinons of opposite spin projection.
Hence the range of the energy $\epsilon_{s\nu}$ for addition onto the 
ground state of one $s\nu$ fermion with $\nu >1$ spinon pairs is,
\begin{equation}
2(\nu - i_{s\nu})\mu_B\,\vert H\vert \leq \epsilon_{s\nu} \leq 2\nu\mu_B\,\vert H\vert 
\, ; \hspace{0.25cm} \nu > 1
\, , \hspace{0.15cm} 0\leq i_{s\nu} \leq 1 \, .
\label{energy-s-n}
\end{equation}
Deconfined spinons are invariant under
the electron - rotated-electron unitary transformation. Thus 
their energies are additive. It follows that for all $m$ values $2\nu\mu_B\,\vert H\vert$  is the energy of $\nu$ deconfined $-1/2$ spinons 
and $\nu$ deconfined $+1/2$ spinons. 
For example, for $m>0$ the energy $2\nu\mu_B\,H$ is that for creation of 
a number $\nu=1,2,...$ of deconfined $-1/2$ spinons onto a $S_{s}=\nu$ ground
state with $2\nu$ deconfined $+1/2$ spinons. Such a creation refers to
$\nu$ spin-flip processes. The number $i_{s\nu}$ decreases continuously 
for increasing values of $U/4t$. For any fixed $m$ value it has the limiting
behavior $i_{s\nu}\rightarrow 1$ for $U/4t\rightarrow 0$. Moreover, it is such 
that $i_{s\nu}<1$ and $i_{s\nu}\,2\mu_B\,H\rightarrow 0$ for $U/4t\rightarrow\infty$.
Hence the energy bandwidth of the $\epsilon_{s\nu}$ 
range vanishes for $U/4t\rightarrow\infty$.
Such a behavior is associated with the full
degeneracy of the spin configurations reached
for $U/4t\rightarrow\infty$. Note that at $m=0$ and thus $H=0$
one has that $\epsilon_{s\nu}=0$
for the whole finite interaction range $U/4t>0$. This behavior follows
from the invariance under the electron - rotated-electron
unitary transformation of a $s\nu$ fermion with $\nu>1$ spinon pairs
created onto a $m=0$ ground state. 

In turn, it is found below that for a $m=0$ and $x\geq 0$ ground state all sites of the 
$s1$ effective lattice are occupied. Hence the corresponding $s1$
momentum band is full. The range of the energy $-\epsilon_{s1}$ 
for removal from that state of one $s1$ fermion then is,
\begin{equation}
0 \leq -\epsilon_{s1} \leq {\rm max}\,\{W_{s1},\vert\Delta\vert\} \, .
\label{energy-s-in}
\end{equation}
Here $\vert\Delta\vert$ denotes the $s1$ fermion pairing energy per spinon considered in Ref. \cite{companion}.
For small finite hole concentrations $0<x\ll 1$ it vanishes both in the $U/4t\rightarrow 0$ 
and $U/4t\rightarrow \infty$ limits and goes through a maximum value at $U/4t=u_0\approx 1.3$.
At fixed $U/4t$ values it decreases for increasing $x$ and vanishes for $x>x_*$. Here
$x_*\in (0.23,0.28)$ for $U/4t\in (1.3,1.6)$ is a critical hole concentration below which the $s1$
fermion pairing refers to a spin short-range order \cite{companion}. 
For $U/4t>0$ and $x=0$ its magnitude $\vert\Delta\vert=\mu^0/2$ is larger than for $x\rightarrow 0$ and
the $s1$ fermion spinon pairing refers to an antiferromagnetic long-range order \cite{companion}.
On the other hand, at $m=0$ the energy scale $W_{s1}$ is the $s1$ fermion energy nodal bandwidth defined in Ref. \cite{companion}. 
Its maximum magnitude is reached at $U/4t=0$. For $U/4t>0$ it decreases monotonously for increasing values of $U/4t$, 
vanishing for $U/4t\rightarrow\infty$. That $W_{s1}\rightarrow 0$ for $U/4t\rightarrow\infty$ is associated with the full 
degeneracy of the spin configurations reached in that limit. In it the spectrum of the two-spinon composite
$s1$ fermions becomes dispersionless for the square-lattice quantum liquid of Ref. \cite{companion}. 

In the $U/4t\rightarrow\infty$ limit 
the $s1$ fermion occupancy configurations that generate the
spin degrees of freedom of spin-density
$m=0$ ground states considered below become for the 1D model
those of the spins of the spin-charge factorized wave function introduced 
by Woynarovich \cite{Woy}. It is associated with the $N$-electron amplitude $f_{\infty} (x_1,x_2,...,x_N)$ of Eq. (\ref{amplitude-1D-rot-elec}),
which was later rediscovered by Ogata and Shiba \cite{Ogata}. In turn, for the model
on the square lattice such configurations become in that limit and
within a mean-field approximation for the fictitious magnetic field 
${\vec{B}}_{s1}$ of Eq. (\ref{A-j-s1-3D}) for $\alpha\nu =s1$, those of a full lowest Landau level with 
$N_{s1}=N_{a_{s1}}^2=N/2$ one-$s1$-fermion degenerate states of the $2D$ quantum  
Hall effect \cite{companion}. Here $N_{a_{s1}}^2$ is the number of 
both sites of the $s1$ effective square lattice and $s1$ band discrete
momentum values. For finite $U/4t$ values and $x>0$ the degeneracy of the $N_{a_{s1}}^2=N/2$ 
one-$s1$ fermion states of the square-lattice quantum liquid studied in
Ref. \cite{companion} is removed by the emergence
of a finite-energy-bandwidth $s1$ fermion dispersion. However, the number of $s1$
band discrete momentum values remains being given by $N_{a_{s1}}^2=B_{s1}\,L^2/\Phi_0$.
In addition, the $s1$ effective lattice spacing remains reading $a_{s1}=l_{s1}/\sqrt{2\pi}$.
Here $l_{s1}\approx a/\sqrt{\pi(1-x)}$ is the fictitious-magnetic-field length. 

\subsubsection{The $c$ fermion energy range}
  
The energy $\epsilon_{c}$ for addition onto the ground state of 
one $c$ fermion of a given momentum and the energy $-\epsilon_{c}$ 
for removal from that state of such a $c$ fermion have the following ranges,
\begin{equation}
0 \leq \epsilon_{c}\leq W_c^h=[4Dt - W_c^p]   \, ; \hspace{0.5cm}
0 \leq -\epsilon_{c}\leq W_c^p  
\, , \hspace{0.25cm} D = 1,2 \, ,
\label{energy-c}
\end{equation}
respectively. Here $W_c^h=[4Dt-W_c^p]\in (0,4Dt)$ 
increases monotonously for increasing 
values of hole concentration $x\in (0,1)$.
The energy bandwidth $W_c^p$ depends little 
on $U/4t$. For $U/4t>0$ it has the following limiting behaviors,
\begin{equation}
W_c^p = 4Dt  \, ,
\hspace{0.25cm}  x = 0 \, ; \hspace{0.50cm}
W_c^p = 0  \, ,
\hspace{0.25cm}  x = 1 \, .
\label{W-c}
\end{equation}
The behaviors reported here for $\epsilon_{c}$ are justified in
Ref. \cite{companion}.

\subsubsection{Transformation laws of $\alpha\nu$ fermions and $c$ fermions under the electron - rotated-electron 
unitary transformation}

The minimum magnitude of the energy 
$\Delta_{D_{rot}}$ for creation of a number $D_{rot}=M_{\eta,-1/2}$ of rotated-electron doubly occupied sites
onto a $m=0$ and $x\geq 0$ ground state given in Eq. (\ref{min-D}) may be expressed in terms of both the 
numbers of deconfined $\eta$-spinons and $\eta\nu$ fermions as follows,
\begin{eqnarray}
{\rm min}\,\Delta_{D_{rot}} & = & (\mu^0+\mu)M^{de}_{\eta,-1/2}+\sum_{\nu=1}^{\infty}2\nu\mu^0\,N_{\eta\nu} 
\hspace{0.2cm}{\rm at}\hspace{0.2cm}x=0\hspace{0.2cm}{\rm and}\hspace{0.2cm}\mu \in (-\mu^0,\mu^0) \, ,
\nonumber \\
& = & 2\mu\,M^{de}_{\eta,-1/2}+\sum_{\nu=1}^{\infty}2\nu\mu\,N_{\eta\nu}
\hspace{0.2cm}{\rm for}\hspace{0.2cm}x>0 \, .
\label{min-D-N}
\end{eqnarray}
Here $\mu^0\equiv \lim_{x\rightarrow 0}\mu$ is the energy scale whose limiting behaviors are given in Eq. (\ref{DMH}).
Consistently, the following $\eta$-spinon and $\eta\nu$ fermion energy magnitudes hold,
\begin{eqnarray}
\epsilon_{\eta,\pm 1/2} & = &  (\mu^0\mp\mu) \, ; \hspace{0.35cm}
\epsilon_{\eta\nu} = 2\nu\mu^0\hspace{0.2cm}{\rm at}\hspace{0.2cm}x=0\hspace{0.2cm}{\rm and}\hspace{0.2cm}\mu \in (-\mu^0,\mu^0) \, ,
\nonumber \\
\epsilon_{\eta,-1/2} & = & 2\mu \, ; \hspace{0.35cm} \epsilon_{\eta,+1/2} = 0  \, ; \hspace{0.35cm} 
{\rm min}\,\epsilon_{\eta\nu} = 2\nu\mu\hspace{0.2cm}{\rm for}\hspace{0.2cm}x>0 \, ,
\nonumber \\
\epsilon_{\eta,-1/2} & = & 0 \, ; \hspace{0.35cm} \epsilon_{\eta,+1/2} = 2\vert\mu\vert \, ; \hspace{0.35cm} 
{\rm min}\,\epsilon_{\eta\nu} = 2\nu\vert\mu\vert\hspace{0.2cm}{\rm for}\hspace{0.2cm}x<0 \, .
\label{energy-eta}
\end{eqnarray}
The corresponding energy magnitudes concerning creation of deconfined spinons and 
$s\nu$ fermions with $\nu>1$ spinon pairs onto $x\geq 0$ ground states with arbitrary values of $m$ read, 
\begin{eqnarray}
\epsilon_{s,\pm 1/2} & = & \epsilon_{s\nu} = 0\hspace{0.2cm}{\rm at}\hspace{0.2cm}m=0\hspace{0.2cm}{\rm and}\hspace{0.2cm}\mu_B\,H=0 \, ,
\nonumber \\
\epsilon_{s,-1/2} & = & 2\mu_B\,H  \, ; \hspace{0.35cm} \epsilon_{s,+1/2} = 0  \, ; \hspace{0.35cm} 
{\rm max}\,\epsilon_{s\nu} = 2\nu\mu_B\,H \hspace{0.2cm}{\rm for}\hspace{0.2cm}\nu>1\hspace{0.2cm}{\rm and}\hspace{0.2cm}m>0 \, ,
\nonumber \\
\epsilon_{s,-1/2} & = & 0 \, ; \hspace{0.35cm} \epsilon_{s,+1/2} = 2\mu_B\,\vert H\vert \, ; \hspace{0.35cm} 
{\rm max}\,\epsilon_{s\nu} =  2\nu\mu_B\,\vert H\vert\hspace{0.2cm}{\rm for}\hspace{0.2cm}\nu>1\hspace{0.2cm}{\rm and}\hspace{0.2cm}m<0 \, .
\label{energy-s}
\end{eqnarray}
Hence for all hole concentrations $x$ and all spin densities $m$ the inequalities 
$\epsilon_{\eta\nu} \geq \nu [\epsilon_{\eta,-1/2} 
+\epsilon_{\eta,+1/2}] =2\nu\vert\mu\vert$ where $\mu=\mu^0$ at $x=0$ and $\epsilon_{s\nu} \leq \nu [\epsilon_{s,-1/2} 
+\epsilon_{s,+1/2}] = 2\nu\mu_B\,\vert H\vert$, respectively, hold. Furthermore, $[\epsilon_{\eta,-1/2} +\epsilon_{\eta,+1/2}]= 2\vert\mu\vert$
where again $\mu =\mu^0$ at $x=0$ and $[\epsilon_{s,-1/2} +\epsilon_{s,+1/2}]= 2\mu_B\,\vert H\vert$.

A related important property is that 
$\eta\nu$ fermions of any $\nu=1,2,...$ branch and $s\nu$ fermions with $\nu >1$ spinon pairs whose energy 
obeys the following relations,
\begin{eqnarray} 
\epsilon_{\eta\nu} & = & \nu [\epsilon_{\eta,-1/2} 
+\epsilon_{\eta,+1/2}] = 2\nu\vert\mu\vert \, , \hspace{0.25cm} \nu =1,2,...\hspace{0.2cm}{\rm for}\hspace{0.2cm}x\neq 0 
\nonumber \\
& = & \nu [\epsilon_{\eta,-1/2} 
+\epsilon_{\eta,+1/2}] = 2\nu\mu^0 \, , \hspace{0.25cm} \nu =1,2,...\hspace{0.2cm}{\rm at}\hspace{0.2cm}x=0
\nonumber \\ 
\epsilon_{s\nu} & = & \nu [\epsilon_{s,-1/2} 
+\epsilon_{s,+1/2}] = 2\nu\mu_B\,\vert H\vert 
\, , \hspace{0.25cm} \nu =2,3,... \, ,
\label{invariant-V}
\end{eqnarray}
remain invariant under the electron - rotated-electron
unitary transformation. Those are non-interacting objects such that
their energy is additive in the individual energies
of the corresponding $2\nu$ $\eta$-spinons
and $2\nu$ spinons, respectively. Therefore, for $U/4t>0$ they refer to the same occupancy configurations in terms
of both rotated electrons and electrons. This refers to the $\eta$-spin and spin degrees of freedom, respectively,
of such rotated-electron occupancy configurations. The corresponding rotated-electron degrees of
freedom associated with the $c$ fermion $U(1)$ symmetry are not in general invariant under
that transformation.

Another important property is that a $s\nu$ fermion with $\nu>1$ spinon pairs (and a $\eta\nu$ fermion) created onto an initial 
$S_{s}=0$ and $H=0$ (and $S_{\eta}=0$) ground state remains invariant under the electron - rotated-electron
unitary transformation. Indeed and as reported above, such an object has a uniquely defined energy $\epsilon_{s\nu}=0$
(and $\epsilon_{\eta\nu}=2\nu\mu^0$). Furthermore, its momentum is ${\vec{q}}={\vec{q}}_{s}=0$ (and ${\vec{q}}={\vec{q}}_{\eta}=0$). Hence it
obeys indeed to the condition (\ref{invariant-V}). In turn, $s\nu$ fermions with $\nu>1$ spinon pairs (and $\eta\nu$ fermions) created onto initial 
$S_{s}>0$ (and $S_{\eta}>0$) ground states may have energy $\epsilon_{s\nu}$ smaller than $2\nu\mu_B\,\vert H\vert$
(and energy $\epsilon_{\eta\nu}$ larger than $2\nu\vert\mu\vert$), so that they are not necessarily invariant
under the electron - rotated-electron unitary transformation.

To understand this property, we consider a set of $2\nu$ deconfined 
$\eta$-spinons  (and $2\nu$ deconfined spinons) including $\nu$ deconfined 
$+1/2$ $\eta$-spinons  (and $\nu$ deconfined $+1/2$ spinons) and $\nu$ deconfined 
$-1/2$ $\eta$-spinons  (and $\nu$ deconfined $-1/2$ spinons). Such a set of $2\nu$ deconfined 
$\eta$-spinons  (and $2\nu$ deconfined spinons) refers to a $S_{\eta} =\nu$ and
$S_{\eta}^{x_3} =0$ $\eta$-spin multiplet configuration of energy $2\nu\vert\mu\vert$ where $\mu=\mu^0$ at $x=0$
(and a $S_{s} =\nu$ and $S_{s}^{x_3} =0$ spin multiplet configuration of energy $2\nu\mu_B\,\vert H\vert$). 
In turn, the $2\nu$ $\eta$-spinons (and spinons) of a $\eta\nu$ fermion (and $s\nu$ fermion)  
correspond to a $S_{\eta} = S_{\eta}^{x_3} = 0$ $\eta$-spin singlet configuration of energy 
$\epsilon_{\eta\nu} \geq 2\nu\vert\mu\vert$ (and a $S_{s} =S_{s}^{x_3} =0$ spin singlet configuration of energy 
$\epsilon_{s\nu}\leq 2\nu\mu_B\,\vert H\vert$). The point is that only when the energy $\epsilon_{\eta\nu}$ 
(and $\epsilon_{s\nu}$) of a $\eta\nu$ fermion (and $s\nu$ fermion) $2\nu$-$\eta$-spinon 
$\eta$-spin singlet configuration (and $2\nu$-spinon spin singlet configuration) reads
$\epsilon_{\eta\nu} =2\nu\vert\mu\vert$ (and $\epsilon_{s\nu} =2\nu\mu_B\,\vert H\vert$), as given in Eq. (\ref{invariant-V}), is it
degenerated with that of the above $2\nu$-deconfined $\eta$-spinon $\eta$-spin multiplet configuration (and
$2\nu$-deconfined spinon spin multiplet configuration). Only when that occurs is such a $\eta\nu$ fermion (and $s\nu$ fermion)
invariant under the electron - rotated-electron unitary transformation. 

On the other hand, $\eta\nu$ fermions of any $\nu=1,2,...$ branch and $s\nu$ fermions with
$\nu >1$ spinon pairs whose energies obey the inequalities
$\epsilon_{\eta\nu} > 2\nu\vert\mu\vert$ and $\epsilon_{s\nu}< 2\nu\mu_B\,\vert H\vert$,
respectively, are not invariant under the electron - rotated-electron 
transformation. Furthermore, for finite $U/4t$ values 
$c$ fermions and $s1$ fermions are not in general invariant under that
transformation. 

For the 1D model the $\alpha\nu$ fermion energy $\epsilon_{\alpha\nu} (q)$ depends on
the $\alpha\nu$ fermion momentum $q$, which as discussed in Appendix A is a good quantum number.
For $U/4t>0$ and the $s\nu$ branches with $\nu>1$ spinon pairs and all $\eta\nu$ branches such momenta
belong to the range $q\in (-m\pi,+m\pi)$
and $q\in (-x\pi,+x\pi)$, respectively. Only at the limiting
momenta $q=q_{\eta}=\pm x\pi$ (and $q=q_{s}=\pm m\pi$)
is the invariance condition (\ref{invariant-V}) met
by the $\eta\nu$ fermion energy $\epsilon_{\eta\nu} (q)$ (and $s\nu$ fermion energy $\epsilon_{s\nu} (q)$
for $\nu>1$ branches). The $\eta\nu$ fermion energy dispersions $\epsilon_{\eta\nu} (q)$ are plotted for the $\eta 1$ (and
$\eta 2$) branches in Figs. 8 (a) and 9 (a) (and Figs. 8 (b) and 9 (b)) of Ref. \cite{1D-03} as a function of $q$ for several $U/4t$ values and
electronic density $n=1/2$ and $n=5/6$, respectively. (In that reference the $\eta\nu$ fermions
are called $c, \nu$ pseudoparticles.) The zero-energy level of Figs. 8 (a) and 9 (a) (and Figs. 8 (b) and 9 (b))
refers to the energy $2\vert\mu\vert$ (and $4\vert\mu\vert$) of the invariance condition (\ref{invariant-V}) for $\nu=1$
(and $\nu=2$).

For the Hubbard model on the square lattice the $\alpha\nu$ fermions whose energy obeys the invariance
condition (\ref{invariant-V}) have a well-defined momentum ${\vec{q}}_{\alpha}$, which can point in different
directions. At $x=0$ (and $m=0$) the momentum ${\vec{q}}_{\eta}$ (and ${\vec{q}}_{s}$) vanishes. This is alike for the above corresponding
momentum $q_{\eta}=\pm x\pi$ (and $q_{s}=\pm m\pi$) of the 1D model,
which vanishes at $x=0$ (and $m=0$). While at 1D the momentum $q_{\eta}=\pm x\pi$ 
(and $q_{s}=\pm m\pi$) can for $x>0$ (and $m>0$) have two values, for the square-lattice model 
the momentum ${\vec{q}}_{\eta}$ (and ${\vec{q}}_{s}$) can for $x>0$ (and $m>0$) point to several directions. 
For instance, for $x<x_*$ and $m=0$ where $x_*$ is the hole concentration considered in Ref.
\cite{companion} below which the ground-state $s1$ fermion spinon pairing is for $x>0$ associated with
a short-range spin order, a good approximation for the momentum ${\vec{q}}_{\eta}$ centered at $-\vec{\pi}=[-\pi,-\pi]$ is,
\begin{equation} 
{\vec{q}}_{\eta} = {\vec{q}}_{Fc}^{\,h} = q^h_{Fc}\,{\vec{e}}_{\phi_c} 
\, ; \hspace{0.25cm} q^h_{Fc} \approx 2\sqrt{x\,\pi} \, ; \hspace{0.25cm} \phi_{c} \in (0,2\pi) \, .
\label{q-Fc-h-m0}
\end{equation}
Here ${\vec{q}}_{Fc}^{\,h}$ is the $c$ fermion hole Fermi momentum considered in Ref. \cite{companion}
and ${\vec{e}}_{\phi_c}$ is a unit vector centered at $-\vec{\pi}$ of Cartesian components $[\cos \phi_{c},\sin \phi_{c}]$. 
As required, ${\vec{q}}_{\eta} \rightarrow 0$ as $x\rightarrow 0$.

\subsubsection{Ground state occupancies}

Consistently with the energy values given in Eqs. (\ref{energy-eta}) and (\ref{energy-s}),
one finds that both for the model on the 1D and 
square lattice in the subspace spanned by the $x>0$ and $m>0$ LWS 
ground states and their excited energy eigenstates of energy $\omega<{\rm min}\,\{2\mu,2\mu_B\,H\}$
the $\eta$-spinon and spinon numbers are given by $M_{\eta,-1/2}=M^{co}_{\eta}/2=M^{de}_{\eta,-1/2}=0$,
$M_{\eta,+1/2}=M^{de}_{\eta,+1/2} =2S_{\eta}=x\,N_a^D$ and $M_{s,-1/2}=M^{co}_{s}/2=N_{\downarrow}$,
$M_{s,+1/2}=M^{co}_{s}/2+M^{de}_{s ,+1/2} =N_{\uparrow}$, respectively, so that
$M^{de}_{s ,+1/2} = 2S_{s}=m\,N_a^D$. Hence for such energy eigenstates the numbers of
$\eta\nu$ fermions and $s\nu'$ fermions with $\nu'>1$ spinon pairs vanish and
the number of $c$ fermions is $N_c =2S_c = N$ and that of 
$s1$ fermions is conserved and reads $N_{s1}=N_{\downarrow}$.

Except that the $-1/2$ $\eta$-spinons and $-1/2$ spinons play the role
of the $+1/2$ $\eta$-spinons and $+1/2$ spinons, respectively, and
vice versa, similar results are reached for HWSs.
Comparison of the occupancies of the spin LWS
ground states ($m>0$) and spin HWS ground states ($m<0$) provides
useful information. From it and again consistently with the 
energy values of Eqs. (\ref{energy-eta}) and (\ref{energy-s}),
one finds that a $m=0$ ground state for which $N$ is even and $x\geq 0$ has
$M_{\eta,\pm1/2}$ and $N_c $ values as given above whereas $M_{s,\pm1/2}=M^{co}_{s}/2=N/2$ so that
$M^{de}_{s,\pm 1/2} = 0$ and $N_{s1}=N/2$. 

\subsection{The numbers of sites of the $\alpha\nu$ effective lattices and
discrete momentum values of the $\alpha\nu$ bands}

A local $\eta\nu$ (and $s\nu$) fermion refers to a well-defined superposition of $\eta$-spinon (and spinon) 
occupancy configurations involving $2\nu=2,4,...$ sites of the $\eta$-spin (and spin) effective lattice. 
For the model on the square lattice in the one- and two-electron subspace of most physical interest  
the only existing $\alpha\nu$ effective lattice is the $s1$ effective lattice. (In it the $s2$ and $\eta 1$
effective lattices either do not exist or have a single occupied site.) The number of sites, $N_{a_{s1}}^D$, 
of the $s1$ effective lattice plays a major role in the studies of that model one-electron and spin spectra \cite{companion}.
Alike all $\alpha\nu$ effective lattices, such an effective lattice is exotic, since the general expression for its 
number of sites depends on the values of $S_c$, $S_s$, and set of numbers $\{N_{s\nu}\}$. Hence the
$N_{a_{s1}}^D$ value is subspace dependent. 
In order to derive the expression of such a dependence one needs though to solve the general problem referring 
to all $\alpha\nu$ effective lattices. Indeed the values of the set $\{N_{a_{\alpha\nu}}^D\}$ of numbers of sites of
such lattices depend on each other. 

Our goal here is the thus derivation of the expression of the number of $\alpha\nu$ effective lattice sites
$N_{a_{\alpha\nu}}^D$ in terms of $S_c$, $S_s$, and set of numbers $\{N_{\alpha\nu'}\}$ where
$\nu'=1,2,...$. The real-space coordinate $\vec{r}_j$ of a local $\alpha\nu$ fermion
plays the role of its ``center of mass". The corresponding index $j=1,...,N_{a_{\alpha\nu}}^D$ enumerates the sites of
the $\alpha\nu$ effective lattice whose $N_{a_{\alpha\nu}}^D$ real-space coordinates 
have well-defined different values. Fortunately, within the present $N_a^D\gg 1$ limit,
(i) the internal structure of a local $\alpha\nu$ fermion 
and (ii) its position specified by the corresponding real-space coordinate $\vec{r}_j$ are separated problems. 
The present analysis refers only to the problem (ii). 
Concerning the internal structure of a local $\alpha\nu$ fermion, the only issue that matters for the present
analysis is that the $2\nu=2,4...$ sites of the $\eta$-spin 
(and spin) effective lattice occupied by a given local 
$\eta\nu$ (and $s\nu$) fermion correspond to
$2\nu=2,4,...$ sites of the original lattice that are not 
simultaneously occupied by any other composite fermions. 
The direct correspondence and relationship of the rotated-electron occupancy configurations to those
of the $c$ fermions, $\alpha\nu$ fermion, deconfined $\eta$-spinons,
and deconfined spinons has been constructed to inherently such a property holding.

The number of sites $N_{a_{\alpha\nu}}^D$ of the $\alpha\nu$ effective 
lattice is an integer number. However, for the model on the
square lattice for which $D=2$ the related number $N_{a_{\alpha\nu}}$ 
such that $N_{a_{\alpha\nu}}^2=N_{a_{\alpha\nu}}\times N_{a_{\alpha\nu}}$ is 
not in general an integer. Within the present $N_a^2\rightarrow\infty$
limit, we consider that it is the integer number closest to it.
This is why in that limit we use the notation $N_{a_{\alpha\nu}}^2$ 
for the $\alpha\nu$ effective square lattice number of sites.

The $\alpha\nu$ effective lattice and its $N_{a_{\alpha\nu}}^D$ sites
are well-defined concepts in a subspace in which the values of the set 
of numbers $N_c=2S_c$ and $\{N_{\alpha\nu'}\}$ with $\alpha=\eta, s$ and $\nu'=1,2,...$
remain fixed. (As confirmed below, this is equivalent to the $\eta$-spin $S_{\eta}$ ($\alpha=\eta$) or spin $S_{s}$ ($\alpha=s$)
and values of the set of numbers $\{N_{\alpha\nu'}\}$ with $\alpha=\eta, s$
and $\nu'=\nu,\nu +1,...$ remaining fixed.) One occupied site of the $\alpha\nu$ effective lattice corresponds
to $2\nu =2,4,...$ sites of the $\eta$-spin ($\alpha =\eta$)
or spin ($\alpha =s$) effective lattice. Hence the $N_{\alpha\nu}$ local fermions belonging to the same
$\alpha\nu$ branch occupy $N_{\alpha\nu}$ and $2\nu N_{\alpha\nu}$ sites
of the  $\alpha\nu$ effective lattice and  $\eta$-spin ($\alpha =\eta$)
or spin ($\alpha =s$) effective lattice, respectively.
Which sites of the $\eta$-spin (and spin) effective lattice
play the role of unoccupied sites of the $\eta\nu$ (and $s\nu$) effective lattice
is an issue fully determined by the number of $\eta$-spin-singlet (and spin-singlet)
configurations of each subspace with fixed values of $S_c$, $S_{\eta}$,
and $S_s$. Those are subspaces of the larger subspace that the 
$S_c>0$ vacuum of Eq. (\ref{vacuum}) refers to. The dimension of such
subspaces equals the product of the corresponding numbers of state representations of the 
$c$ fermion hidden $U(1)$ symmetry, $\eta$-spin $SU (2)$ symmetry ($\alpha =\eta$), and spin $SU (2)$
symmetry ($\alpha =s$), respectively. Those are the three symmetries contained in the model global 
$SO(3)\times SO(3)\times U(1)=[SU(2)\times SU(2)\times U(1)]/Z_2^2$ symmetry. 

The number of sites of the $\alpha\nu$ effective lattice, which equals that of $\alpha\nu$ band
discrete momentum values, has the general form,
\begin{equation}
N_{a_{\alpha\nu}}^D = [N_{\alpha\nu} + N^h_{\alpha\nu}] \, ,
\label{N*}
\end{equation}
where $N^h_{\alpha\nu}$ is the number of unoccupied sites whose expression we
derive in the following. To achieve such a goal, we follow the procedures of Ref. \cite{bipartite}
and divide the Hilbert space of the model (\ref{H})
in a set of subspaces spanned by the states 
with fixed values of $S_c$, $S_{\eta}$, and $S_s$ and
hence also of $N_c=2S_c$, $M_{\eta}=N_{a_{\eta}}^D=[N_a^D-2S_c]$,
and $M_{s}=N_{a_{s}}^D=2S_c$. We recall that 
for a subspace with fixed values of $S_c$, $S_{\eta}$,
and $S_s$, the number $M_{\eta}=N_{a_{\eta}}^D$
(and $M_{s}=N_{a_{s}}^D$) is both the total number of
$\eta$-spinons (and spinons) and the number of sites
of the $\eta$-spin (and spin) effective lattice.

According to the studies of Ref. \cite{bipartite}, the dimension of each 
such a subspace is,
\begin{equation}
d_r\cdot{\cal{N}}(S_{\eta} ,M_{\eta})\cdot{\cal{N}}(S_{s} ,M_{s}) \, .
\label{dimension}
\end{equation}
Here the dimension $d_r$ and the two $\alpha =\eta,s$ dimensions ${\cal{N}} (S_{\alpha},M_{\alpha})$ are given by,
\begin{equation}
d_r = {N_a^D\choose 2S_c} \, ; \hspace{0.35cm}
{\cal{N}} (S_{\alpha},M_{\alpha}) = (2S_{\alpha} +1)\left\{
{M_{\alpha}\choose M_{\alpha}/2-S_{\alpha}} - {M_{\alpha}\choose
M_{\alpha}/2-S_{\alpha}-1}\right\} \, , \hspace{0.15cm} \alpha=\eta,s \, .
\label{N-Sa-Ma}
\end{equation}
Those are the number of $c$ fermion $U(1)$ symmetry state representations and that of $\eta$-spin 
$SU(2)$ ($\alpha=\eta$) and spin $SU(2)$ ($\alpha =s$) symmetry state representations, respectively.

The dimension $d_r$ given in Eq. (\ref{N-Sa-Ma}) is here straightforwardly recovered 
as $d_r = {N_a^D\choose N_c}$. It equals the number 
of occupancy configurations of the $N_c=2S_c$ $c$ fermions 
in their $c$ band with $N_a^D$ discrete momentum values. 
On the other hand, the values of the numbers $N_{a_{\alpha\nu}}^D$ of discrete
momentum values of the $\alpha\nu$ band must exactly obey the following 
equality {\it for all} subspaces,
\begin{equation}
{{\cal{N}} (S_{\alpha},M_{\alpha})\over (2S_{\alpha} +1)} = \sum_{\{N_{\alpha\nu}\}}\, \prod_{\nu
=1}^{\infty}\,{N_{a_{\alpha\nu}}^D\choose N_{\alpha\nu}} \, , 
\hspace{0.25cm} \alpha=\eta,s \, .
\label{Ncs-cpb}
\end{equation}
Here ${N_{a_{\alpha\nu}}^D\choose N_{\alpha\nu}}$ 
is the number of occupancy configurations of the set of $N_{\alpha\nu}$ 
$\alpha\nu$ fermions in their $\alpha\nu$ band with $N_{a_{\alpha\nu}}^D$ 
discrete momentum values. The $\{N_{\alpha\nu}\}$ summation runs over 
{\it the whole} set of $N_{\alpha\nu}$ numbers that owing to
the conservation of the $\alpha =\eta,s$ numbers $M^{co}_{\eta}$ and $M^{co}_{s}$ of sites of the $\eta$-spin and
spin effective lattices, respectively, whose occupancy configurations are not invariant
under the electron - rotated-electron unitary transformation exactly obey the subspace sum rules
given in Eq. (\ref{M-L-Sum}). 

The general expression of the number $N_{a_{\alpha\nu}}^D$ of $\alpha\nu$ band 
discrete momentum values and thus of sites of the corresponding
$\alpha\nu$ effective lattice that obeys Eq. (\ref{Ncs-cpb}) for {\it all} subspaces 
is such that the $2S_{\eta}$ (and $2S_s$) sites of the 
$\eta$-spin (and spin) effective lattice occupied by $M^{de}_{\eta}=2S_{\eta}$
deconfined $\eta$-spinons (and $M^{de}_{s}=2S_{\eta}$ deconfined spinons) and 
$2(\nu'-\nu)$ sites out of the $2\nu'$ sites of that lattice occupied by each local $\eta\nu'$ (and $s\nu'$)
fermion such that $\nu'>\nu$ play the role of unoccupied sites of the 
$\eta\nu$ (and $s\nu$) effective lattice. Indeed the unique solution of the problem that assures the validity
of Eq. (\ref{Ncs-cpb}) with the $\{N_{\alpha\nu}\}$ summation running over 
the whole set of $N_{\alpha\nu}$ numbers that
obey the subspace sum-rule (\ref{M-L-Sum}) for {\it all} Hilbert-space subspaces
spanned by the states with fixed values of $S_c$, $S_{\eta}$, and $S_s$ refers to the following expression 
for the number of unoccupied sites of the $\alpha\nu$ effective lattice $N^h_{\alpha\nu}$
appearing in Eq. (\ref{N*}),
\begin{equation}
N^h_{\alpha\nu} = 
[M^{de}_{\alpha}+2\sum_{\nu'=\nu+1}^{\infty}(\nu'-\nu)N_{\alpha\nu'}] =
[N_{a_{\alpha}}^{D} - 
\sum_{\nu' =1}^{\infty}(\nu +\nu' - \vert \nu-\nu'\vert)N_{\alpha\nu'}] 
\, ; \hspace{0.25cm} \alpha = \eta, s \, ; \hspace{0.15cm} 
\nu =1,2,...,\infty \, ,
\label{N-h-an}
\end{equation}
where $M^{de}_{\alpha} =2S_{\alpha}$. Importantly, this expression is also that obtained from the BA exact solution for the 1D Hubbard
model. However, it is fully determined by state-representation dimension requirements of 
the global $SO(3)\times SO(3)\times U(1)$ symmetry
that apply to the Hubbard model on the square lattice as well. (Note that
the equivalence of the two expressions given in Eq. (\ref{N-h-an}) confirms that the $N_c=2S_c$ and $\{N_{\alpha\nu'}\}$ 
values with $\nu'=1,2,...$ remaining fixed is equivalent to the $\eta$-spin $S_{\eta}$ ($\alpha=\eta$) or 
spin $S_{s}$ ($\alpha=s$) and values of the set of numbers $\{N_{\alpha\nu'}\}$ with $\nu'=\nu,\nu +1,...$ 
remaining fixed as well. In both cases that implies that the $N_{a_{\alpha\nu}}^D$ value remains fixed.) 

The occupancies of the deconfined $\eta$-spinons and deconfined spinons give rise
to the usual factors $(2S_{\eta} +1)$ and $(2S_s +1)$, respectively, in the sub-space $SU(2)$ symmetry 
state-representation number expression ${\cal{N}} (S_{\eta},M_{\eta})$ and ${\cal{N}} (S_s,M_s)$ of 
Eqs. (\ref{N-Sa-Ma})  and (\ref{Ncs-cpb}). In Ref. \cite{bipartite} it is shown that the subspace-dimension summation,
\begin{equation}
{\cal{N}}_{tot} =  
\sum_{S_c=0}^{[N_a^D/2]}\,\sum_{S_{\eta}=0}^{[N_a^D/2-S_c]}\,
\sum_{S_s=0}^{S_c}{N_a^D\choose
2S_c} \prod_{\alpha =\eta,s}{[1+(-1)^{[2S_{\alpha}+2S_c]}]\over 2}
\,{\cal{N}}(S_{\alpha},M_{\alpha}) \, , 
\label{Ntot}
\end{equation}
gives indeed the dimension $4^{N_a^D}$ of the full Hilbert space.
In the present context this confirms that the set of momentum eigenstates $\vert \Phi_{U/4t}\rangle$
of our description whose general expressions are given below in Section IV-E is complete. 

From the use of the above equations one finds that the number 
of unoccupied sites of the $\alpha 1$ effective lattices reads,
\begin{equation}
N^h_{\alpha 1} = [N_{a_{\alpha}}^{D} - 2B_{\alpha}]
\, , \hspace{0.25cm}  \alpha = \eta \, , s \, ,
\label{Nh+Nh}
\end{equation}
where
\begin{equation}
B_{\alpha} = \sum_{\nu =1}^{\infty}N_{\alpha\nu} = 
{1\over 2}\,[N_{a_{\alpha}}^{D} - N^h_{\alpha 1}] \, , \hspace{0.15cm} \alpha = \eta \, , s \, .
\label{sum-rules}
\end{equation}
The number $N^h_{\alpha 1}$ given here equals that of $\alpha 1$ fermion holes in the $\alpha 1$ band. 
An important property discussed below in section IV-F is that for $U/4t>0$ and $x\geq 0$ the numbers
$B_{\eta}$ and $B_{s}$ are conserved for the Hubbard model on the square
lattice in subspaces such that $B_{\eta}/N_a^2\rightarrow 0$ and $[B_{s}-S_c+S_s]/N_a^2\rightarrow 0$
as $N_a^2\rightarrow\infty$. For the 1D model they are good quantum numbers for the
whole Hilbert space. (We could not prove that $B_{\eta}$ and $B_{s}$ are good quantum numbers 
for the Hubbard model on the square lattice in its full Hilbert space but that remains a possibility.)

Straightforward manipulations of the above equations lead to
the following general expressions for $S_{\eta}$, $S_s$, and $S_c$,
\begin{equation}
S_{\alpha} = {1\over 2} [N^h_{\alpha 1} - M^{co}_{\alpha} + 2B_{\alpha}]
\, , \hspace{0.25cm} \alpha = \eta \, , s 
\, ; \hspace{0.50cm}
S_c = {1\over 2}[N_a^D - N^h_{\eta 1} - 2B_{\eta}] =
{1\over 2} [N^h_{s1} + 2B_{s}] \, .
\label{S-S-S}
\end{equation}
The equality of the two $S_c$ expressions given Eq. (\ref{S-S-S}) implies that,
\begin{equation}
\sum_{\alpha =\eta,s} N^h_{\alpha 1} =
\left[N_a^D - \sum_{\alpha =\eta,s}\sum_{\nu =1}^{\infty}2N_{\alpha\nu}\right] =
[N_a^D - \sum_{\alpha =\eta,s}2B_{\alpha}] \, .
\label{sum-Nh+Nh}
\end{equation}
 
Provided that $N_{a_{\alpha\nu}}^D/N_a^D$ remains finite as $N_a^D\rightarrow\infty$, the related
$\alpha\nu$ effective lattices can for the 1D and square-lattice models be represented by 
1D and square lattices, respectively, of spacing,
\begin{equation}
a_{\alpha\nu} = {L\over N_{a_{\alpha\nu}}} = 
{N_a\over N_{a_{\alpha\nu}}}\, a
= {N_{a_{\alpha}}\over N_{a_{\alpha\nu}}}\, a_{\alpha}
\, ; \hspace{0.25cm} 
N_{a_{\alpha\nu}} \geq 1 \, .
\label{a-a-nu}
\end{equation}
Here $\nu = 1,2,...$ and $\alpha =\eta ,s$. 
The arguments behind the lattice geometry and the average distance $a_{\alpha\nu}$ 
between the sites of the $\alpha\nu$ effective lattice playing the role of lattice spacing
are similar to those used for the lattice geometry and spacing of the $\eta$-spin and spin effective lattices.
In turn, the corresponding $\alpha\nu$ bands whose number of
discrete momentum values is also given by $N_{a_{\alpha\nu}}^D$
are well defined even when $N_{a_{\alpha\nu}}^D$ is
given by a finite small number, $N_{a_{\alpha\nu}}^D=1,2,...$.

\subsection{A complete set of momentum eigenstates}

\subsubsection{Global symmetry state representations and corresponding momentum eigenstates}

We start by introducing the LWSs $\vert\Phi_{LWS;U/4t}\rangle ={\hat{V}}^{\dag}\vert\Phi_{LWS;\infty}\rangle$ 
associated with the general momentum eigenstates $\vert \Phi_{U/4t}\rangle = {\hat{V}}^{\dag}\vert \Phi_{\infty}\rangle$,
which refer to state representations of the model global $SO(3)\times SO(3)\times U(1)$ symmetry.
As justified in the following, the corresponding $U/4t\rightarrow\infty$ momentum eigenstates $\vert\Phi_{LWS;\infty}\rangle$ 
are of the form,
\begin{equation}
\vert \Phi_{LWS;\infty}\rangle = 
[\prod_{\alpha}\prod_{\nu}\prod_{{\vec{q}}\,'}{\mathcal{F}}^{\dag}_{{\vec{q}}\,',\alpha\nu}\vert 0_{\alpha};N_{a_{\alpha}^D}\rangle] 
[\prod_{{\vec{q}}}{\mathcal{F}}^{\dag}_{{\vec{q}},c}\vert GS_c;0\rangle] \, .
\label{LWS-full-el-infty}
\end{equation}
Here ${\mathcal{F}}^{\dag}_{{\vec{q}}\,',\alpha\nu}$ and ${\mathcal{F}}^{\dag}_{{\vec{q}},c}$ are creation operators
of a $U/4t\rightarrow\infty$ $\alpha\nu$ fermion of momentum ${\vec{q}}\,'$ and $c$ fermion of
momentum ${\vec{q}}$, respectively. The $\eta$-spin $SU(2)$ vacuum $\vert 0_{\eta};N_{a_{\eta}}^D\rangle$ 
containing $N_{a_{\eta}}^D=M^{de}_{\eta,+1/2}$ deconfined $+1/2$ $\eta$-spinons, the spin $SU(2)$ vacuum 
$\vert 0_{s};N_{a_{s}}^D\rangle$ containing $N_{a_{s}}^D=M^{de}_{a,+1/2}$ deconfined $+1/2$ spinons, and the $c$ fermion $U(1)$
vacuum $\vert GS_c;0\rangle$ appearing in this equation are those of the vacuum given in Eq. (\ref{vacuum}).
(In its expression, $\vert GS_c;2S_c\rangle=\prod_{{\vec{q}}}{\mathcal{F}}^{\dag}_{{\vec{q}},c}\vert GS_c;0\rangle$.)

We recall that for $U/4t\rightarrow\infty$ Eqs. (\ref{fc+})-(\ref{c-up-c-down}) are equivalent to Eqs. (1)-(3) of Ref. \cite{Ostlund-06} with
the $c$ fermion creation operator $f_{\vec{r}_j,c}^{\dag}$ replaced by the quasicharge annihilation operator $\hat{c}_r$.
Therefore, the operator ${\mathcal{F}}^{\dag}_{{\vec{q}},c}$ has the form given in Eq. (\ref{F+-cr}).
The holes of such quasicharge particles are for $U/4t\rightarrow\infty$ the spinless and $\eta$-spinless $c$ fermions. 
In that limit those describe the charge degrees of freedom
of the electrons of the singly occupied sites. Moreover, the spinons and $\eta$-spinons of the
$2\nu$-spinon operators ${\mathcal{F}}^{\dag}_{{\vec{q}}\,',s\nu}$ and
$2\nu$-$\eta$-spinon operators ${\mathcal{F}}^{\dag}_{{\vec{q}}\,',\eta\nu}$, respectively, are associated with the
local spin and pseudospin operators, respectively, defined in that reference.

Symmetry implies that for $U/4t>0$ the $N_{a_{\alpha\nu}}^D$ $\alpha\nu$ band
discrete momentum values of states belonging to the same $V$ tower are the same
and thus $U/4t$ independent. Importantly, the vacua of the $U/4t\rightarrow\infty$ states (\ref{LWS-full-el-infty}) are invariant 
under the electron - rotated-electron unitary transformation. It follows that for finite
values of $U/4t$ the corresponding $S_{\eta}$, $S_{\eta}^{x_3}$, $S_s$, $S_s^{x_3}$, $S_c$, and momentum eigenstates read,
\begin{eqnarray}
\vert \Phi_{LWS;U/4t}\rangle & = &
[\prod_{\alpha}\prod_{\nu}\prod_{{\vec{q}}\,'}f^{\dag}_{{\vec{q}}\,',\alpha\nu}\vert 0_{\alpha};N_{a_{\alpha}^D}\rangle] 
[\prod_{{\vec{q}}}f^{\dag}_{{\vec{q}},c}\vert GS_c;0\rangle] \, , 
\nonumber \\
f^{\dag}_{{\vec{q}}\,',\alpha\nu} & = &
{\hat{V}}^{\dag}\,{\mathcal{F}}^{\dag}_{{\vec{q}}\,',\alpha\nu}\,{\hat{V}} \, ; \hspace{0.25cm} f^{\dag}_{{\vec{q}},c} =
{\hat{V}}^{\dag}\,{\mathcal{F}}^{\dag}_{{\vec{q}},c}\,{\hat{V}} \, .
\label{LWS-full-el}
\end{eqnarray}
Unlike for the model on the square lattice, for the 1D model the states of form (\ref{LWS-full-el}) are energy eigenstates 
for the whole Hilbert space \cite{companion,1D}. In Appendix A it is confirmed that the discrete momentum
values of the $c$ and $\alpha\nu$ fermion operators appearing on the right-hand side of Eq. (\ref{LWS-full-el})
are indeed the good quantum numbers of the exact solution. Their occupancy configurations generate
the energy eigenstates $\vert \Psi_{LWS;U/4t}\rangle= \vert \Phi_{LWS;U/4t}\rangle$ of general form (\ref{LWS-full-el}). 

The general state representations of the global $SO(3)\times SO(3)\times U(1)$ symmetry
associated with the momentum eigenstates $\vert \Phi_{U/4t}\rangle$ are
generated by simple $c$ fermion and $\alpha\nu$ fermion occupancy configurations and deconfined 
$\eta$-spinon and deconfined spinon occupancies of the form,
\begin{eqnarray}
\vert \Phi_{U/4t}\rangle & = & 
\prod_{\alpha=\eta,s}\frac{({\hat{S}}^{\dag}_{\alpha})^{M^{de}_{\alpha,-1/2}}}{
\sqrt{{\cal{C}}_{\alpha}}}\vert \Phi_{LWS;U/4t}\rangle 
\nonumber \\
& = & 
[\prod_{\alpha}\frac{({\hat{S}}^{\dag}_{\alpha})^{M^{de}_{\alpha,-1/2}}}{
\sqrt{{\cal{C}}_{\alpha}}}\prod_{\nu}\prod_{{\vec{q}}\,'}f^{\dag}_{{\vec{q}}\,',\alpha\nu}\vert 0_{\alpha};N_{a_{\alpha}^D}\rangle] 
[\prod_{{\vec{q}}}f^{\dag}_{{\vec{q}},c}\vert GS_c;0\rangle] \, .
\label{non-LWS}
\end{eqnarray}
Here $\vert \Phi_{LWS;U/4t}\rangle$ are the states of Eq. (\ref{LWS-full-el})
and the normalization constant ${\cal{C}}_{\alpha}$ is given in Eq. 
(\ref{Calpha}) of Appendix B. 

The momentum values over which the products $\prod_{{\vec{q}}}$ and $\prod_{{\vec{q}}\,'}$, respectively,
of the expressions provided in Eqs. (\ref{LWS-full-el}) and (\ref{non-LWS})
run are those of the subspace of the state under consideration.
This justifies why the $c$ and $\alpha\nu$ bands have a set of
suitable discrete momentum values so that the corresponding generators on the vacua are simple products of
$f^{\dag}_{{\vec{q}},\alpha\nu}$ operators, as given in these equations.

The set of momentum eigenstates of form (\ref{non-LWS}) refers to an
orthogonal and normalized state basis. The number of such
states equals the dimension given in Eq. (\ref{Ntot}). Since in Ref. \cite{bipartite} it is confirmed that
such a dimension is $4^{N_a^D}$, the 
set of states $\{\vert \Phi_{U/4t}\rangle\}$ is complete. Consistently, for all values of the densities $n=(1-x)$ and $m$ the momentum eigenstates 
of general form (\ref{non-LWS}) correspond to the state representations of the 
$SO(3)\times SO(3)\times U(1)$ group counted in Section IV-D.

Also the related energy and momentum eigenstates $\vert \Psi_{U/4t}\rangle$ considered below
correspond to state representations of the $SO(3)\times SO(3)\times U(1)$ group. Their 
$\eta$-spin-neutral $\eta$-spinon and spin-neutral spinon occupancy configurations that
are not invariant under the electron - rotated-electron unitary transformation and involve 
$M^{co}_{\eta}$ and $M^{co}_{s}$ sites of the $\eta$-spin and spin effective lattices, respectively,
are in general different from those of the momentum eigenstates (\ref{non-LWS}). The latter 
are simpler, referring to $\eta\nu$ fermion and $s\nu$ fermion occupancy configurations, respectively.
     
\subsubsection{The general energy eigenstates}

The unitary operator $\hat{V}^{\dag}$ as defined in Section II has been constructed to 
inherently generating exact $U/4t>0$ energy and momentum eigenstates
$\vert \Psi_{U/4t}\rangle = {\hat{V}}^{\dag}\vert \Psi_{\infty}\rangle
=\prod_{\alpha}[({\hat{S}}^{\dag}_{\alpha})^{M^{de}_{\alpha,-1/2}}/\sqrt{{\cal{C}}_{\alpha}}]\vert \Psi_{LWS;U/4t}\rangle$. 
Within our description the corresponding set $\{\vert \Psi_{\infty}\rangle\}$
of $U/4t\rightarrow\infty$ energy and momentum eigenstates are suitably chosen according to the recipe 
reported in such a section. For the Hubbard model on the square lattice at finite $U/4t$ values
the energy eigenstates $\vert \Psi_{U/4t}\rangle$ are a suitable superposition of
$S_{\eta}$, $S_{\eta}^{x_3}$, $S_s$, $S_s^{x_3}$, $S_c$, ${\vec{P}}_{SO(4)}$, and momentum
eigenstates $\vert \Phi_{U/4t}\rangle$ (\ref{non-LWS}) of general form,
\begin{eqnarray}
\vert \Psi_{U/4t}\rangle & = & {1\over \sum_l \vert C_l\vert^2}\vert {\tilde{\Psi}}_{U/4t}\rangle
\, ; \hspace{0.35cm} C_l = \langle\Phi_{U/4t;l}\vert {\tilde{\Psi}}_{U/4t}\rangle \, ,
\nonumber \\
\vert {\tilde{\Psi}}_{U/4t}\rangle & = &  \sum_l C_l\,\vert \Phi_{U/4t;l}\rangle =
[\prod_{{\vec{q}}}f^{\dag}_{{\vec{q}},c}\vert GS_c;0\rangle]
[\prod_{\alpha}\frac{({\hat{S}}^{\dag}_{\alpha})^{M^{de}_{\alpha,-1/2}}}{
\sqrt{{\cal{C}}_{\alpha}}}\sum_l C_l\prod_{\nu (l)}\prod_{{\vec{q}}\,'(l)}
f^{\dag}_{{\vec{q}}\,',\alpha\nu}\vert 0_{\alpha};N_{a_{\alpha}^D}\rangle] \, .
\label{non-LWS-psi}
\end{eqnarray}
Here $\vert \Phi_{U/4t;l}\rangle$ denote momentum eigenstates of form
(\ref{non-LWS}) with different $\alpha\nu$ fermion momentum occupancy
configurations labelled by the index $l$ also appearing in the products $\prod_{\nu (l)}\prod_{{\vec{q}}\,'(l)}$.
All such states have the same $S_{\eta}$, $S_{\eta}^{x_3}$, $S_s$, $S_s^{x_3}$, and $S_c$
values and the same momentum eigenvalue and momentum ${\vec{P}}_{SO(4)}$. 
Their $c$ fermion momentum distribution function 
$N_{c}({\vec{q}})=\langle\Phi_{U/4t}\vert f^{\dag}_{{\vec{q}},c}\,f_{{\vec{q}},c}\vert \Phi_{U/4t}\rangle =
\langle\Psi_{U/4t}\vert f^{\dag}_{{\vec{q}},c}\,f_{{\vec{q}},c}\vert \Psi_{U/4t}\rangle$ is also the same,
consistently with the form of expression (\ref{non-LWS-psi}).

That for the Hubbard model on the square lattice the states $\vert \Phi_{U/4t;l}\rangle$ in the expansion (\ref{non-LWS-psi})
refer to different $\alpha\nu$ fermion momentum occupancy
configurations follows from that model Hamiltonian and the $\alpha\nu$ translation generators 
${\hat{{\vec{q}}}}_{\alpha\nu}$ of Eq. (\ref{m-generators}) in general not commuting.
Although the numbers $\{N_{\alpha\nu}\}$ are not in general good quantum numbers for the Hubbard 
model on the square lattice, the number $M^{co}_{\alpha}$ of confined $\eta$-spinons $(\alpha =\eta)$
and confined spinons $(\alpha =s)$ is. This means that rather being given by 
$M^{co}_{\alpha}=2\sum_{\nu=1}^{\infty}\nu \,N_{\alpha\nu}$, such a number results from a different
partition of the $M^{co}_{\alpha}/2$ neutral confined $\eta$-spinon $(\alpha =\eta)$
and confined spinon $(\alpha =s)$ pairs.

Consistently, the numbers $\{N^h_{\alpha\nu}\}$ given in Eq. (\ref{N-h-an}) are not in general good quantum numbers 
of the Hubbard model on the square lattice. (For the 1D model they are.) However, for a given energy eigenstate
$\vert \Psi_{U/4t}\rangle = \sum_l C_l\,\vert \Phi_{U/4t;l}\rangle$
of general form provided in Eq. (\ref{non-LWS-psi}) the values of the numbers of $\alpha\nu$ fermion holes
$N^h_{\alpha\nu}(l)$ of the corresponding momentum eigenstates $\vert \Phi_{U/4t;l}\rangle$ are
either even or odd integer numbers. Indeed the numbers,
\begin{equation}
P^h_{\alpha\nu} \equiv e^{i\pi\prod_{l} N^h_{\alpha\nu}(l)}=e^{i2\pi S_{\alpha}}=e^{i2\pi S_{c}}=e^{i\pi N} = \pm 1 
\, ; \hspace{0.25cm} \nu = 1, 2,... \, ; \hspace{0.15cm} \alpha = c, \eta \, ,
\label{Phs1}
\end{equation}
are good quantum numbers of both the Hubbard model on the square and 1D lattice.
For the energy eigenstate under consideration, the product $\prod_{l}$ appearing here runs over the same states 
as the sums $\sum_{l}$ in Eq. (\ref{non-LWS-psi}). If $N^h_{\alpha\nu}(l)$ is
an even or odd integer for all such states, then $\prod_{l} N^h_{\alpha\nu}(l)$ is as well
an even or odd integer, respectively. 

To arrive to the equality $e^{i\pi\prod_{l} N^h_{\alpha\nu}(l)}=e^{i2\pi S_{\alpha}}$ 
provided in Eq. (\ref{Phs1}) we use that $N^h_{\alpha\nu} = 
[2S_{\alpha}+2\sum_{\nu'=\nu+1}^{\infty}(\nu'-\nu)N_{\alpha\nu'}]$, as
given in Eq. (\ref{N-h-an}), where $2\sum_{\nu'=\nu+1}^{\infty}(\nu'-\nu)N_{\alpha\nu'}$
is always an even integer number. Furthermore, from Eq. (\ref{M-L-Sum}) one has that $2S_{\alpha}=[M_{\alpha}-M^{co}_{\alpha}]$
where $M^{co}_{\alpha}=2\sum_{\nu'=1}^{\infty}\nu' N_{\alpha\nu'}$ is again always an even integer number.
Moreover, $M_{\eta}=N_{a_{\eta}}^D=-2S_c+N_a^D$ where $N_a^D$ is an even integer
and $M_s =2S_c$, as given in Eq. (\ref{M-L-C}). This justifies why $e^{i2\pi S_{\alpha}}=e^{i2\pi S_{c}}$ and thus
$e^{i\pi\prod_{l} N^h_{\alpha\nu}(l)}=e^{i2\pi S_{c}}$. In turn, from Eq. (\ref{L-L}) one
has that $2S_{\alpha} = \pm 2S_{\alpha}^{x_3}
+2M^{de}_{\alpha,\pm1/2}$ where $2M^{de}_{\alpha,\pm1/2}$ is an even integer, $\pm 2S_{\eta}^{x_3} = \pm N \mp N_a^D$,
and $\pm 2S_{\eta}^{x_3} = \mp [N_{\uparrow}-N_{\downarrow}]$. Since the integer numbers
$N$ and $[N_{\uparrow}-N_{\downarrow}]$ have the same parity, one finally finds that
$e^{i\pi\prod_{l} N^h_{\alpha\nu}(l)}=e^{i\pi N}$.

The number $N^h_{\alpha\nu}$ plays an active role in the physics provided that the 
corresponding $\alpha\nu$ fermion branch has finite occupancy $N_{\alpha\nu}>0$ 
at least for one of the states contributing to the sum $\sum_{l}$ of Eq. (\ref{non-LWS-psi}).
For finite densities $n=(1-x)>0$ this always holds for the number $N^h_{s1}$ of $s1$ fermion holes, which
equals that of unoccupied sites of the $s1$ effective lattice. 

Alike the set of momentum eigenstates $\{\vert \Phi_{U/4t}\rangle\}$ of Eq.
(\ref{non-LWS}), the complete set of energy and momentum eigenstates
$\{\vert \Psi_{U/4t}\rangle\}$ of general form given in Eq. (\ref{non-LWS-psi}) refers
to state representations of the group $SO(3)\times SO(3)\times U(1)$.
The only difference is that the $M^{co}_{\eta}$-sites $\eta$-spin-neutral and $M^{co}_s$-sites 
spin-neutral $\eta$-spinon and spinon configurations, respectively, that are not
invariant under the electron - rotated-electron unitary transformation of the states 
$\{\vert \Psi_{U/4t}\rangle\}$ are linear superpositions of those of the
states $\{\vert \Phi_{U/4t}\rangle\}$.
In some limiting cases such as for the energy eigenstates that span the one- and two-electron 
subspace the momentum-eigenstate superposition
(\ref{non-LWS-psi}) reduces to a single state $\vert \Phi_{U/4t}\rangle$, so that $C_l=\delta_{l,l'}$. 
Here the index $l'$ refers to the $\alpha\nu$ fermion occupancy configurations of the
state $\vert \Phi_{U/4t}\rangle$ of form (\ref{non-LWS}) under consideration. 
(For 1D this applies to all states (\ref{non-LWS-psi}).)

\subsection{The good and quasi-good quantum numbers and selection rules}

That $S_{\eta}$, $S_{\eta}^{x_3}$, $S_s$, and $S_s^{x_3}$ are good quantum numbers for both the Hubbard model
on the 1D and square lattice implies that the numbers of deconfined $\eta$-spinons ($\alpha=\eta$) and 
deconfined spinons ($\alpha=s$) $M^{de}_{\alpha,\pm 1/2} = [S_{\alpha}\mp S_{\alpha}^{x_3}]$ of Eq. (\ref{L-L})
are good quantum numbers as well. The set of numbers $\{P^h_{\alpha\nu}\}$ of Eq. (\ref{Phs1})
and the $SO(4)$ momentum ${\vec{P}}_{SO(4)}$ are also conserved for the
model on both lattices. That the eigenvalue $S_c$ of the generator of the  
hidden $U(1)$ symmetry of the model global $SO(3)\times SO(3)\times U(1)$
symmetry is a good quantum number implies that the numbers
$N_{a_{s}}^D=N_c=M_s=2S_c$ and $N_{a_{\eta}}^D=N_c^h=M_{\eta}=[N_a^D-2S_c]$ are also good quantum
numbers. This is why the numbers $M^{co}_{\alpha}$ of Eqs. (\ref{2S-2C}), (\ref{C-C}), and
(\ref{M-L-Sum}), which can be expressed as $M^{co}_{\alpha} =[N_{a_{\alpha}}^D-2S_{\alpha}]$,
are good quantum numbers as well. The same then applies to the total numbers $M_{\alpha,\pm 1/2} = M^{de}_{\alpha,\pm 1/2} + M^{co}_{\alpha}/2$
of $\pm 1/2$ $\eta$-spinons ($\alpha=\eta$) and $\pm 1/2$ spinons ($\alpha=s$).

The lack of integrability of the Hubbard model on the square lattice is behind 
the $\alpha\nu$ translation generators ${\hat{{\vec{q}}}}_{\alpha\nu}$ of Eq. (\ref{m-generators})
not commuting with the Hamiltonian and the
set of numbers $\{N_{\alpha\nu}\}$ not being in general good quantum numbers. 
However, $B_{\eta}$ and $B_{s}$ are for $U/4t>0$ and $x\geq 0$ good quantum numbers 
in subspaces such that $B_{\eta}/N_a^2\rightarrow 0$ and $[B_{s}-S_c+S_s]/N_a^2\rightarrow 0$
as $N_a^2\rightarrow\infty$. For the model in such subspaces that is a necessary condition
for the $SO(4)$ momentum operator $\hat{{\vec{P}}}_{SO(4)}$ commuting with 
that model Hamiltonian. That as reported above that commutator vanishes confirms
that $B_{\eta}$ and $B_{s}$ are good quantum numbers in the above subspaces.
This is why they are called here quasi-good quantum numbers.
It follows from the relations (\ref{Nh+Nh}) that the $\eta 1$ fermion hole and $s 1$ fermion hole numbers
$N^h_{\eta 1}=[N_{a_{\eta}}^2 -2B_{\eta}]$ and $N^h_{s 1}=[N_{a_{s}}^2 -2B_{s}]$, respectively, are conserved as well. 
(It is possible that the numbers $B_{\eta}$ and $B_{s}$ and thus also the numbers $N^h_{\eta 1}$
and $N^h_{s 1}$ are conserved for the model on the square lattice in the whole Hilbert space,
yet we have not proved it.)

For some of the subspaces in which $B_{\eta}$ and $B_{s}$ have been shown to be good quantum numbers 
of the Hubbard model on the square lattice most of the $\alpha\nu$ fermion numbers 
$N_{\alpha\nu}$ vanish and the remaining are good quantum numbers. That happens
for subspaces spanned by states with fixed values for the conserving
numbers $M^{co}_{\eta}$, $M^{co}_{s}$, $B_{\eta}$, and $B_{s}$ provided that such values
are fulfilled by a unique choice for the corresponding values of the set of numbers $\{N_{\alpha\nu}\}$.
This occurs for the following subspaces:
\vspace{0.25cm}

A) Subspaces for which $[M^{co}_{\alpha}/2-B_{\alpha}]=0,1$ for $\alpha=\eta$ 
and/or $\alpha=s$. For such subspaces $N_{\alpha 1}= [2B_{\alpha}-M^{co}_{\alpha}/2]$ and
$N_{\alpha 2}=[M^{co}_{\alpha}/2-B_{\alpha}]$ are good quantum numbers and $N_{\alpha\nu}=0$ for $\nu >2$.  
Also $N_{a_{\alpha 1}}^2= [N_{a_{\alpha}}^2-M^{co}_{\alpha}/2]$ is conserved.
Furthermore, the $U/4t\rightarrow\infty$ local spin $SU(2)$ gauge symmetry for $\alpha =s$
or the $U/4t\rightarrow\infty$ local $\eta$-spin $SU(2)$ gauge symmetry for $\alpha =\eta$ also implies 
that the $\alpha 1$ band momenta are good quantum numbers 
for the Hubbard model on the square lattice in subspaces for which
both the corresponding numbers $N_{\alpha 1}^h$ and $N_{\alpha 1}$ of unfilled and filled
discrete momentum values, respectively, are conserved.
The unitarity of the operator $\hat{V}$ then assures that in such
subspaces the $\alpha 1$ band momentum values are conserved for $U/4t>0$ as well.
This is so for the present subspaces so that in them the Hamiltonian of the Hubbard model on the square lattice commutes with the $\alpha 1$ 
translation generators $\hat{q}_{\alpha 1\,x_1}$ and $\hat{q}_{\alpha 1\,x_2}$ 
in the presence of the corresponding fictitious magnetic field ${\vec{B}}_{\alpha 1}$.
\vspace{0.25cm}

B) A another type of subspace for which the $\alpha\nu$ fermion number $N_{\eta\nu}$ is
conserved is that for which $B_{\eta}=1$ and thus $M^{co}_{\eta}=2\nu$.
Then $N_{\eta\nu}=1$ and $N_{\eta\nu'}=0$ for $\nu'\neq \nu$. Such subspaces have no
no interest though for the problems studied in Sections V and VI.
\vspace{0.25cm}

For values of $[M^{co}_{\alpha}/2-B_{\alpha}]$ and $B_{\alpha}$ other than those of the subspaces (A) and (B)
there is no unique choice for the corresponding values of the set of numbers $\{N_{\alpha\nu}\}$ so 
that they are not good quantum numbers for the Hubbard model on the square lattice. Indeed,
states with the same $[M^{co}_{\alpha}/2-B_{\alpha}]$ and $B_{\alpha}$ values and different values for the
set of numbers $\{N_{\alpha\nu}\}$ may decay into each other. The general energy eigenstates (\ref{non-LWS-psi})
are superpositions of such states.

A particular case of a $[M^{co}_{\alpha}/2-B_{\alpha}]=0$ subspace refers to $M^{co}_{\eta}=2B_{\eta}=0$ 
and $M^{co}_s=2B_{s}=[2S_c-2S_s]$. For it $N_{a_{s1}}^2=[S_c+S_{s}]$, $N_{s1}=[S_c-S_s]$, and $N_{s1}^h=2S_{s}$.
The equality $N_{a_{\eta\nu}}^2=2S_{\eta}$ (and $N_{a_{s\nu}}^2=2S_{s}$) 
holds for all $\nu$ branches (and $\nu >1$ branches). For such a subspace $N_{\eta\nu}^h=2S_{\eta}$
(and $N_{s\nu}^h=2S_{s}$ for $\nu >1$) are good quantum numbers. 

For $S_s=0$, $x>0$, and $M_{\eta,-1/2}=0$ so that $B_{\eta}=M^{co}_{\eta}/2=0$ the Hubbard model on a square-lattice in a 
$[M^{co}_{s}/2-B_{s}]=0$ subspace of type (A) has in the $U/4t\rightarrow\infty$ limit an interesting physics. For 
such a subspace the $s1$ fermion occupancy configurations 
are in that limit and within a mean-field approximation for the fictitious magnetic field ${\vec{B}}_{s1}$ of Eq. (\ref{A-j-s1-3D})  
closely related to the physics of a full lowest Landau level of the 2D quantum Hall effect (QHE). 
From the use of Eqs. (\ref{N*}) and (\ref{N-h-an}) one finds $N_{a_{s1}}^2=N_{s1}=S_c$. That corresponds
to the case in which $\langle n_{\vec{r}_j,s1}\rangle\approx 1$. 
If in the mean field approximation one replaces the corresponding fictitious magnetic field by the 
average field created by all $s1$ fermions at a given position, one finds 
${\vec{B}}_{s1} ({\vec{r}}_j) \approx [\Phi_0/a_{s1}^2]\,{\vec{e}}_{x_3}$. The number of $s1$  
band discrete momentum values can then be written as,
\begin{equation}
N_{a_{s1}}^2 = {B_{s1}\,L^2\over \Phi_0} 
\, ; \hspace{0.35cm} \nu_{s1} = {N_{s1}\over N_{a_{s1}}^2} = 1 \, .
\label{N-B-Phi}
\end{equation}

The bandwidth of the $s1$ fermion energy $\epsilon_{s1}$ 
vanishes in the $U/4t\rightarrow\infty$ limit. Hence only in that
limit are the $N_{a_{s1}}^2$ one-$s1$-fermion states 
of Eq. (\ref{N-B-Phi}) degenerate. It follows that in such a limit $N_{a_{s1}}^2=B_{s1}\,L^2/\Phi_0$
plays the role of the number of degenerate states in each 
Landau level of the 2D QHE. In the subspace considered here the $s1$ fermion occupancies 
correspond to a full lowest Landau level with filling factor one, as given in Eq. (\ref{N-B-Phi}). 

Only for the $U/4t\rightarrow\infty$ limit there is
fully equivalence between the $s1$ fermion occupancy configurations
and, within a mean-field approximation for the fictitious magnetic field 
${\vec{B}}_{s1}$ of Eq. (\ref{A-j-s1-3D}), the 2D QHE with a $\nu_{s1}=1$ full lowest Landau level. 
In spite of the lack of state degeneracy emerging upon decreasing the value of $U/4t$,
there remains for finite $U/4t$ values some relation to the 2D QHE. The occurrence of QHE-type behavior 
for the $s1$ fermions in the square-lattice quantum liquid of Ref. \cite{companion} shows that a magnetic field is not 
essential to the 2D QHE physics. Indeed, the fictitious magnetic field ${\vec{B}}_{s1} ({\vec{r}}_j)$
arises from expressing the effects of the electronic correlations in terms of the 
$s1$ fermion interactions \cite{companion}.

The mean-field analysis associated with Eq. (\ref{N-B-Phi}) is consistent with 
in the $U/4t\rightarrow\infty$ limit the Hamiltonian of the Hubbard model on the square lattice 
in a subspace (A) commuting with the $s1$ translation
generator ${\hat{{\vec{q}}}}_{s1}$. Indeed, in that limit such an analysis refers to an effective
description where the Hamiltonian is the sum of a $c$ fermion kinetic-energy
term and a QHE like Hamiltonian for the $s1$ branch. The $s1$ translation
generators ${\hat{{\vec{q}}}}_{s1}$ then commute with all such Hamiltonian terms. 
This is consistent with in the $U/4t\rightarrow\infty$ limit
its eigenvalues being good quantum numbers for the model on the square 
lattice in a subspace (A). Since the electron - rotated-electron
transformation is unitary, such a commutation relation also holds for
$U/4t>0$. 

Last but nor least, the states that span the one- and two-electron subspace introduced in the following
are of type (A). This justifies why such states are energy eigenstates yet have the form (\ref{non-LWS}). 
Hence for $x>0$ and excitation energy below $2\mu$ for which
$M_{\eta,-1/2}=0$, the square-lattice quantum-liquid momentum values of the $c$, $s1$,
and $s2$ fermions are good quantum numbers \cite{companion}. For the half-filled Hubbard model
on the square lattice in the one- and two-electron subspace that holds as well for
the $\eta 1$ band momentum values. 

\section{The square-lattice quantum liquid: A two-component fluid
of charge $c$ fermions and spin-neutral two-spinon $s1$ fermions}

In this section we introduce a suitable one- and two-electron subspace. (For hole concentrations $x>0$ 
our studies refer to excitation energies below $2\mu$ for which $M_{\eta,-1/2}=0$ for that subspace.) 
For $x\geq 0$ the picture that emerges is that of a two-component quantum liquid
of charge $c$ fermions and spin neutral two-spinon $s1$ fermions.
It refers to the square-lattice quantum liquid further investigated in Ref. \cite{companion}. 

\subsection{The one- and two-electron subspace}

\subsubsection{General ${\cal{N}}$-electron subspaces}

We consider a $x\geq 0$ and $m=0$ ground state $\vert \Psi_{GS}\rangle$
whose $c$ and $s1$ fermion occupancies are those given in Section IV-C.
Application onto it of a ${\cal{N}}$-electron operator 
${\hat{O}}_{\cal{N}}$ generates a state,
\begin{equation}
{\hat{O}}_{{\cal{N}}}\vert \Psi_{GS}\rangle = \sum_j C_j \vert \Psi_{U/4t}(j)\rangle 
\, ; \hspace{0.5cm} C_j = \langle\Psi_{U/4t}(j)\vert{\hat{O}}_{{\cal{N}}}\vert \Psi_{GS}\rangle \, ,
\label{1-2-subspace}
\end{equation}
contained in a ${\cal{N}}$-electron subspace.
This is a subspace spanned by the set of energy eigenstates
$\{\vert \Psi_{U/4t}(j)\rangle\}$ such that $\sum_j \vert C_j\vert^2\approx 1$. It is thus associated
with a given ${\cal{N}}$-electron operator ${\hat{O}}_{{\cal{N}}}$. 
Generally, such operators can be written as a product of
\begin{equation}
{\cal{N}}=\sum_{l_{\eta},l_s=\pm 1} {\cal{N}}_{l_{\eta},l_s} \, ; \hspace{0.5cm} l = \pm 1 \,
, \label{d}
\end{equation}
electron creation and annihilation operators. Here
${\cal{N}}_{l_{\eta},l_s}$ is the number of electron creation and
annihilation operators in the operator ${\hat{O}}_{{\cal{N}}}$ expression for $l_{\eta}=-1$ and
$l_{\eta}=+1$, respectively, and with spin down and spin up for $l_s=-1$ and $l_s=+1$,
respectively.  

A general local ${\cal{N}}$-electron operator ${\hat{O}}_{{\cal{N}},j}$ refers to a product of ${\cal{N}}$
local electron creation and annihilation operators. For ${\cal{N}}>1$ such an
operator has a well defined local structure. It involves ${\cal{N}}_{-1,l_s}$ electron creation 
operators of spin projection $l_s/2$ and ${\cal{N}}_{+1,l_s}^l$ electron annihilation operators 
of spin projection $l_s/2$ whose real-space coordinates refer in general to a compact domain of 
neighboring lattice sites. Such a local
${\cal{N}}$-electron operator ${\hat{O}}_{{\cal{N}},j}$ may be labelled
by the real-space coordinate ${\vec{r}}_j$ of a corresponding central site.
A second type of ${\cal{N}}$-electron operator is denoted
by ${\hat{O}}_{{\cal{N}}} (\vec{k})$ and carries momentum $\vec{k}$. It is related to a local 
operator ${\hat{O}}_{{\cal{N}},j}$ by a Fourier transform. 

The general ${\cal{N}}$-electron operators 
${\hat{O}}_{{\cal{N}}}$ considered here belong to one of these two types and are such that
the ratio ${\cal{N}}/N_a^D$ vanishes in the thermodynamic limit. The operators
${\hat{O}}_{{\cal{N}}} (\vec{k})$ of physical interest correspond in general to operators
${\hat{O}}_{{\cal{N}},j}$ whose ${\cal{N}}$ elementary electronic operators create
or annihilate electrons in a compact domain of lattice sites.  
The more usual cases for the description of experimental studies correspond to the ${\cal{N}}=1$ one-electron and ${\cal{N}}=2$ 
two-electron operators. Therefore, in this section we are mostly interested in the corresponding one- and
two-electron subspace. 

Application onto a $x\geq 0$ and $m\geq 0$ ground state of a general
${\cal{N}}$-electron operator ${\hat{O}}_{{\cal{N}}}$ leads to electron
number deviations $\delta N=\delta N_{\uparrow}
+\delta N_{\downarrow}$ and $\delta N_{\uparrow}-\delta N_{\downarrow}$.
As a result of the expressions and relations of Sections III-B and IV-D,
such deviations may be expressed in terms of corresponding deviations in the
number of $c$ fermions, $\alpha\nu$ fermions, deconfined $\eta$-spinons,
and deconfined spinons as follows,
\begin{equation}
\delta N = \delta N_{c} + 2M^{de}_{\eta,-1/2} + M^{co}_{\eta} 
= \delta N_{c} + 2M^{de}_{\eta,-1/2} + 2\sum_{\nu =1}^{\infty}\nu\,N_{\eta\nu} \, ,
\label{DNu}
\end{equation}
and
\begin{equation}
\delta (N_{\downarrow} - N_{\uparrow}) = 2\delta N_{s1} -\delta N_{c} + 2M^{de}_{s,-1/2} +
2\sum_{\nu =2}^{\infty}\nu\,N_{s\nu} \, , 
\label{DNd}
\end{equation}
respectively. (Note that $M^{de}_{\eta,-1/2} =M^{co}_{\eta} =N_{\eta\nu}=M^{de}_{s,-1/2}=N_{s\nu}\vert_{\nu>1}=0$ 
for the initial ground state so that $\delta M^{de}_{\eta,-1/2}=M^{de}_{\eta,-1/2}$, 
$\delta M^{co}_{\eta} =M^{co}_{\eta}$, $\delta N_{\eta\nu}=N_{\eta\nu}$, 
$\delta M^{de}_{s,-1/2}=M^{de}_{s,-1/2}$, and $\delta N_{s\nu}\vert_{\nu>1}=N_{s\nu}\vert_{\nu>1}$.)
Hence only transitions to excited states associated with deviations obeying the sum
rules (\ref{DNu}) and (\ref{DNd}) are permitted. 
The electron number deviations (\ref{DNu}) and (\ref{DNd}) are associated with
sum rules obeyed by the numbers ${\cal{N}}_{l_{\eta},l_s}$ of Eq. (\ref{d}) specific to
the ${\cal{N}}$-electron operator ${\hat{O}}_{{\cal{N}}}$ under consideration, which read,
\begin{equation}
\delta N = \sum_{l_{\eta},l_s=\pm 1}(-l_{\eta})\,{\cal{N}}_{l_{\eta},l_s} \, ; \hspace{1cm} \delta
(N_{\downarrow}-N_{\uparrow}) = \sum_{l_{\eta},l_s=\pm
1}(l_{\eta}\,l_s)\,{\cal{N}}_{l_{\eta},l_s} \, . 
\label{NupdodllO}
\end{equation}

Furthermore, it is straightforward to show that useful exact selection rules hold for
excitations of well-defined initial ground states. For instance, the values of the numbers $M^{de}_{\eta} = [M^{de}_{\eta,-1/2} + M^{de}_{\eta,+1/2}] = 2S_{\eta}$
of deconfined $\eta$-spinons and $M^{de}_{\eta,\pm 1/2}$ of $\pm 1/2$ deconfined $\eta$-spinons generated by
application onto the $S_{\eta}=S_s=0$ ground state of a $\cal{N}$-electron operator
${\hat{O}}_{{\cal{N}}}$ are restricted to the following ranges,
\begin{eqnarray}
M^{de}_{\eta} & = & 2S_{\eta} = 0,1,...,{\cal{N}} \, ,
\nonumber \\
M^{de}_{\eta,\pm 1/2} & = & 0,1,...,\sum_{l_s=\pm 1}{\cal{N}}_{\pm 1,l_s} \, , 
\label{LsrcsL}
\end{eqnarray}
respectively. Here $S_{\eta}$ denotes the excited-state $\eta$-spin
and the numbers ${\cal{N}}_{l_{\eta},l_s}$ are those of Eq. (\ref{d})
specific to the $\cal{N}$-electron operator. 

In the case on an initial $S_s=0$ ground state with hole concentration $x>{\cal{N}}/N_a^D$ or $x= 0$ one finds that
the numbers $M^{de}_{s} = [M^{de}_{s,-1/2} + M^{de}_{s,+1/2}] = 2S_s$ and $M^{de}_{s,\pm 1/2}$ of deconfined spinons
generated by application onto that state of a $\cal{N}$-electron operator are restricted to the ranges,
\begin{eqnarray}
M^{de}_{s} & = & 0,1,...,{\cal{N}}\hspace{0.2cm}{\rm for}\hspace{0.2cm} x>{\cal{N}}/N_a^D \, ,
\nonumber \\
M^{de}_{s,\pm 1/2} & = & 0,1,...,\sum_{l_{\eta},l_s=\pm 1}\delta_{l_{\eta},\mp l_s}\,{\cal{N}}_{l_{\eta},l_s}
\hspace{0.2cm}{\rm for}\hspace{0.2cm} x>{\cal{N}}/N_a^D \, .
\nonumber \\
M^{de}_{s} & = & 0,1,...,,\sum_{l_s=\pm 1}{\cal{N}}_{+1,l_s}\hspace{0.2cm}{\rm at}\hspace{0.2cm}x=0  \, ,
\nonumber \\
M^{de}_{s,\pm 1/2} & = &  0,1,...,{\cal{N}}_{+1,\mp 1}\hspace{0.2cm}{\rm at}\hspace{0.2cm}x=0 \, . 
\label{srcsL}
\end{eqnarray}
The range restrictions of Eqs. (\ref{LsrcsL}) and (\ref{srcsL}) are exact for both the model on the square and 1D lattice,
as well as for any other bipartite lattice.

For $x>0$ we limit our study to the vanishing rotated-electron double occupancy subspace considered in Section III-C.
Consistently with the $\Delta_{D_{rot}}$ energy spectrum of Eqs. (\ref{min-D}) and (\ref{min-D-N}), 
this is accomplished merely by limiting the 
excitation energy to values below $2\mu$, so that the $M_{\eta,-1/2}=0$ 
constraint is automatically fulfilled: It follows from the form of such a spectrum that excited states with 
$\Delta_{D_{rot}}<2\mu$ have vanishing rotated-electron double occupancy. Indeed
creation of one rotated-electron doubly occupied site onto an initial $x>0$ and $m=0$ ground state
is a process of minimum energy $2\mu$. In turn, at $x=0$ we consider both states with vanishing and finite rotated-electron double occupancy.

Creation onto the $S_{\eta}= 0$, $\mu=0$, and $S_s=0$ ground state of one $\eta\nu$ fermion is a vanishing
momentum process whose finite energy is exactly given by $\epsilon_{\eta\nu} = 2\nu\mu^0$. That object then obeys 
the criterion of Eq. (\ref{invariant-V}) for invariance under the electron - rotated-electron unitary transformation. 
It follows that the $\eta$-spin degrees of freedom of such a $\eta\nu$ fermion exactly involve  
$\nu$ electron doubly occupied sites. Furthermore, creation onto an initial $x\geq 0$ and $S_s=0$ ground state of one $s\nu$ fermion
with a number $\nu >1$ of spinon pairs is a vanishing energy and momentum process.
Since vanishing spin $S_s=0$ refers to vanishing magnetic field $H=0$,
such an object obeys the criterion $\epsilon_{s\nu} = 2\nu\mu_B\,\vert H\vert = 0$ of
Eq. (\ref{invariant-V}). Thus it is invariant under the electron - rotated-electron unitary transformation. 
It follows that for $U/4t>0$ creation of such an object involves occupancy configurations whose 
spin degrees of freedom are similar in terms of both rotated-electron and electron occupancy configurations.
That reveals that such a $s\nu$ fermion describes
the spin degrees of freedom of a number $2\nu$ of electrons.

It then follows from the invariance under the electron - rotated-electron unitary transformation
of the $\eta$-spin and spin degrees of freedom of the above $\eta\nu$ fermion and $s\nu$ fermion,
respectively, that for $x>{\cal{N}}/N_a^D$ where ${\cal{N}}/N_a^D\rightarrow 0$ as $N_a^D\rightarrow\infty$
and excitation energy $\omega<2\mu$ and 
and any excitation energy at $x=0$ nearly the whole spectral weight generated by application onto the above ground states of 
$\cal{N}$-electron operators refers to a subspace spanned by excited states with numbers in the 
following range,
\begin{eqnarray}
M^{de}_{\eta} & = & 2S_{\eta} = M^{de}_{\eta,+1/2}= x\,N_a^D 
\hspace{0.2cm}{\rm for}\hspace{0.2cm} x>{\cal{N}}/N_a^D\hspace{0.2cm} {\rm and}\hspace{0.2cm}\omega<2\mu \, ,
\nonumber \\
M^{de}_{\eta,-1/2} & = & M^{co}_{\eta} =0
\hspace{0.2cm}{\rm for}\hspace{0.2cm} x>{\cal{N}}/N_a^D\hspace{0.2cm} {\rm and}\hspace{0.2cm}\omega<2\mu \, ,
\nonumber \\
M_{\eta} & = & M^{de}_{\eta} + M^{co}_{\eta} = 0,1,...,{\cal{N}}\hspace{0.2cm}{\rm at}\hspace{0.2cm}x=0 \, ,
\nonumber \\
M_{\eta,\pm 1/2} & = & M^{de}_{\eta,\pm 1/2} + M^{co}_{\eta}/2 = 0,1,...,\sum_{l_s=\pm 1}{\cal{N}}_{\pm 1,l_s}\hspace{0.2cm}{\rm at}\hspace{0.2cm}x=0 \, , 
\nonumber \\
M_{s} - 2B_s & = & M^{de}_{s} + M^{co}_{s} - 2B_s =
0,1,...,{\cal{N}}\hspace{0.2cm}{\rm for}\hspace{0.2cm}x>{\cal{N}}/N_a^D \, ,
\nonumber \\
& = & 0,1,...,\sum_{l_s=\pm 1}{\cal{N}}_{+1,l_s}\hspace{0.2cm}{\rm at}\hspace{0.2cm}x=0
\nonumber \\
M_{s,\pm 1/2} - B_s & = & M^{de}_{s,\pm 1/2} + M^{co}_{s}/2 
\nonumber \\
& = & 0,1,...,\sum_{l_{\eta},l_s=\pm 1}\delta_{l_{\eta},\mp l_s}\,{\cal{N}}_{l_{\eta},l_s} 
\hspace{0.2cm}{\rm for}\hspace{0.2cm}x>{\cal{N}}/N_a^D \, ,
\nonumber \\
& = &  0,1,...,{\cal{N}}_{+1,\mp 1}\hspace{0.2cm}{\rm at}\hspace{0.2cm}x=0 \, .
\label{srs0}
\end{eqnarray}
Here $M^{co}_{\alpha}=2\sum_{\nu=1}^{\infty}\nu \,N_{\alpha\nu}$ where
$\alpha =\eta, s$ and $B_s=\sum_{\nu=1}^{\infty}N_{s\nu}$,
as given in Eqs. (\ref{M-L-Sum}) and (\ref{sum-rules}), respectively. The quantities $M^{de}_{\eta}$,
$M^{de}_{\eta,\pm 1/2}$, $M^{co}_{\eta}$, $M^{de}_{s}$, $M^{de}_{s,\pm 1/2}$, and $M^{co}_s$ appearing in Eq. (\ref{srs0}) are good quantum numbers. Moreover,
according to the results of Section IV-F, provided that ${\cal{N}}/N_a^D
\rightarrow 0$ and $[B_s-S_c+S_s]/N_a^D\rightarrow 0$ as $N_a^D
\rightarrow\infty$, the number $B_s=\sum_{\nu}N_{s\nu}$ is a good quantum number for the model on the square
lattice, alike for 1D. 

The selection rules (\ref{LsrcsL}) and (\ref{srcsL}) are exact. In turn, for $x>0$ and $m=0$ initial ground states and excitation energy $\omega<2\mu$ 
(and an initial $x=0$ and $m=0$ ground state and any excitation energy) nearly the whole $\cal{N}$-electron 
spectral weight is generated by excited states whose numbers obey the approximate selection rules given
in Eq. (\ref{srs0}). Indeed excited states with numbers $[M^{co} - 2B_s]>{\cal{N}}$ for $x>0$ (and $M_{\eta}>{\cal{N}}$ 
and $[M_{s} - 2B_s]>\sum_{l_s=\pm 1}{\cal{N}}_{+1,l_s}$ at
$x=0$) generate a very small amount yet non vanishing $\cal{N}$-electron spectral weight.

Why in spite of the invariance under the electron - rotated-electron unitary transformation
of the $s\nu$ fermions with a number $\nu >1$ of spinon pairs created onto an
initial $x\geq 0$ and $m=0$ ground state (and that of the $\eta\nu$ fermions created onto an
initial $x=0$ and $m=0$ ground state) are the selection rules provided in Eq. (\ref{srs0}) not exact? The reason
is that while the spin degrees of freedom of the $2\nu$-electron occupancy configurations involved in a $s\nu$ fermion
are exactly described by that object, their hidden $U(1)$ symmetry degrees of freedom are not invariant
under the electron - rotated-electron unitary transformation. The same applies to the 
$\eta$-spin degrees of freedom of the $2\nu$-electron occupancy configurations exactly described by
a $\eta\nu$ fermion created onto a $x=0$ ground state. Their corresponding hidden $U(1)$ 
symmetry degrees of freedom are not in general invariant under that transformation.
The spin degrees of freedom of the $2\nu$-electron occupancy configurations involved in a $s\nu$ fermion
are for $U/4t>0$ exactly the same as those of the corresponding $2\nu$-rotated-electron occupancy configurations.
In turn, the occupancy configurations of the $2\nu$ $c$ fermions that describe 
the hidden $U(1)$ symmetry degrees of freedom of the $2\nu$ rotated electrons under consideration are slightly 
different from those of the corresponding $2\nu$ electrons. The same applies to the $\eta$-spin
degrees of freedom and hidden $U(1)$ symmetry degrees of freedom 
of the $2\nu$ electrons and corresponding $2\nu$ rotated electrons involved
in a $\eta\nu$ fermion created onto a $x=0$ ground state. The former and the latter are
and are not invariant under that transformation. Hence that the selection rules of Eq. (\ref{srs0}) are 
not exact, yet are a very good approximation, stems from the lack of invariance under the electron - rotated-electron unitary 
transformation of the degrees of freedom associated with the hidden $U(1)$ 
symmetry of the Hubbard model. This applies both to the model on the 1D and square lattice.

The $\cal{N}$-electron spectral weight generated by excited states of initial $x>0$ and $m=0$ ground states 
of excitation energy $\omega<2\mu$ and numbers $[M_{s} - 2B_s]>{\cal{N}}$ is extremely small.
The same applies to excited states of initial $x=0$ and $m=0$ ground states of numbers $M_{\eta}>{\cal{N}}$ 
and $[M_{s} - 2B_s]>\sum_{l_s=\pm 1}{\cal{N}}_{+1,l_s}$. Therefore, in this paper we define the 
${\cal{N}}$-electron subspace as that spanned by an initial $x\geq 0$ and $m=0$ ground state plus the set of excited states whose numbers obey
the approximate selection rules given in Eq. (\ref{srs0}). Note that the latter set of excited states depends on the
specific $\cal{N}$-electron operator under consideration. (For hole concentrations $x>0$ this definition refers
to excitation energy $\omega<2\mu$.)

\subsubsection{The one- and two-electron subspace}

The concept of a ${\cal{N}}$-electron subspace as defined above refers to a specific operator. In contrast,
rather than referring to a specific ${\cal{N}}$-electron operator, the one- and two subspace as defined here is 
the set of ${\cal{N}}=1$ and ${\cal{N}}=2$ subspaces associated with the one-electron operator and 
all simple two-electron operators, respectively. Besides the ${\cal{N}}=1$ one-electron operator ${\hat{O}}_1 (\vec{k}) = c_{\vec{k},\sigma}$ (measured in the
angle-resolved photoelectron spectroscopy), this includes a set of ${\cal{N}}=2$ 
operators ${\hat{O}}_{{\cal{N}}} (\vec{k})$ such as the spin-projection $\sigma$ density operator
${\hat{O}}_{2}^{\sigma sd} (\vec{k}) = [1/\sqrt{N_a^D}]\,\sum_{\vec{k}'}c^{\dagger}_{\vec{k}+\vec{k}',\sigma}
c_{\vec{k}',\sigma}$, the transverse spin-density operator ${\hat{O}}_{2}^{sdw} (\vec{k}) = 
[1/\sqrt{N_a^D}]\,\sum_{\vec{k}'}c^{\dagger}_{\vec{k}+\vec{k}',\uparrow} c_{\vec{k}',\downarrow}$, 
and the charge density operator (measured in density-density electron energy loss spectroscopy and inelastic X-ray scattering).
The latter operator is written in terms of the above spin-up and spin-down density operators. Moreover, the set
of ${\cal{N}}=2$ operators includes several superconductivity operators whose pairing symmetries
are in general different at 1D and for the square lattice. The local operators ${\hat{O}}_{{\cal{N}},j}$ corresponding 
to the operators ${\hat{O}}_{{\cal{N}}} (\vec{k})$ whose explicit expression is provided above read
${\hat{O}}_{1,j} = c_{{\vec{r}}_j,\sigma}$, ${\hat{O}}_{2,j}^{\sigma sd} =
c^{\dagger}_{{\vec{r}}_j,\sigma} c_{{\vec{r}}_j,\sigma}$, and ${\hat{O}}_{2,j}^{sdw} =
c^{\dagger}_{{\vec{r}}_j,\uparrow} c_{{\vec{r}}_j,\downarrow}$, respectively. 

In the case of excitations of $x>0$ and $m=0$ ground states, in the remaining of this paper we are mostly interested in the subspace obtained from the
overlap of the one- and two-electron subspace with the vanishing rotated-electron double occupancy subspace considered in Section III-C.
Such a subspace is the one- and two-electron subspace for excitation energy $\omega<2\mu$. 
For finite hole concentrations this it is the subspace of interest for the one- and two-electron physics. 
In turn, concerning the excitations of a $x=0$ and $m=0$ ground state we consider the whole one- and two-electron 
subspace, which refers both to the spin lower-Hubbard band physics and one-electron and charge upper-Hubbard band physics.

As discussed above concerning the general $\cal{N}$-electron spectral weight, for the model on the 1D and square 
lattices there is for $x>0$ an extremely small amount of one- and two-electron
spectral weight that for excitation energy $\omega<2\mu$ is generated by states 
that do not obey the approximate selection rules of Eq. (\ref{srs0}) for ${\cal{N}}=1,2$. Nearly all
such very small amount of spectral weight refers to $N_{s3}=1$ excited states. (States with $N_{s4}=1$ or $N_{s2}=2$ 
generate nearly no spectral weight.) That very small weight is neglected 
within our definition of the one- and two-electron subspace, which
refers to excitation energies below $2\mu$. In turn, concerning the one- and
two-electron excitations of a $x=0$ and $m=0$ ground state the very small amount of spectral
weight generated by states that do not obey the approximate selection rules of Eq. (\ref{srs0}) 
refers to $N_{s3}=1$ and/or $N_{\eta 2}=1$ or $N_{\eta 1}=2$ excited states. 
That very small weight is also neglected within our definition of the one- and two-electron subspace.
(Both the $N_{\eta 2}=1$ and $N_{\eta 1}=2$ excited states have energy much larger than the upper-Hubbard band
$M^{de}_{\eta,-1/2}=1,2$ or $N_{\eta 1}=1$ excited states that obey the selection rules of Eq. (\ref{srs0}) for ${\cal{N}}=1,2$.)

Initial $x> 0$ and $m=0$ ground states and their excited states of energy $\omega<2\mu$ that span 
the one- and two-electron subspace considered in this paper 
have no $-1/2$ $\eta$-spinons, no $\eta\nu$ fermions, and no $s\nu'$ fermions with $\nu'>2$ spinon pairs, so that 
$N_{\eta\nu}=0$ and $N_{s\nu'}=0$ for $\nu'>2$. In turn, initial $x= 0$ and $m=0$ ground states and their excited 
states that span the one- and two-electron subspace as defined here have no $\eta\nu$ fermions with $\nu>1$ 
$\eta$-spinon pairs and no $s\nu'$ fermions with $\nu'>2$ spinon pairs, so that $N_{\eta\nu}=0$ for $\nu>1$ 
and $N_{s\nu'}=0$ for $\nu'>2$. Thus, consistently with the approximate selection rules of Eq. (\ref{srs0}), the values of the 
object numbers of such ground states and their excited states that span the subspace defined here are restricted 
to the following ranges,
\begin{eqnarray}
M^{de}_{\eta} & = & 2S_{\eta} = M^{de}_{\eta,+1/2}= x\,N_a^D 
\hspace{0.2cm}{\rm for}\hspace{0.2cm} x>0\hspace{0.2cm} {\rm and}\hspace{0.2cm}\omega<2\mu \, ,
\nonumber \\
M^{de}_{\eta,-1/2} & = & M^{co}_{\eta} =0
\hspace{0.2cm}{\rm for}\hspace{0.2cm} x>0\hspace{0.2cm}{\rm and}\hspace{0.2cm}\omega<2\mu \, ,
\nonumber \\
M_{\eta} & = & 0,...,{\cal{N}}\hspace{0.2cm}{\rm for}\hspace{0.2cm}{\cal{N}}=1,2\hspace{0.2cm}{\rm at}\hspace{0.2cm}x=0 \, ,
\nonumber \\
M_{\eta,\pm 1/2} & = & 0,...,\sum_{l_s=\pm 1}{\cal{N}}_{\pm 1,l_s}
\hspace{0.2cm}{\rm for}\hspace{0.2cm}\sum_{l_s=\pm 1}{\cal{N}}_{\pm 1,l_s}=0,1,2\hspace{0.2cm}{\rm at}\hspace{0.2cm}x=0 \, , 
\nonumber \\
M_{s} - 2B_s & = & 0,...,{\cal{N}}\hspace{0.2cm}{\rm for}\hspace{0.2cm}{\cal{N}}=1,2\hspace{0.2cm}{\rm and}\hspace{0.2cm}x>0 \, ,
\nonumber \\
& = & 0,...,\sum_{l_s=\pm 1}{\cal{N}}_{+1,l_s}\hspace{0.2cm}{\rm for}\hspace{0.2cm}\sum_{l_s=\pm 1}{\cal{N}}_{+1,l_s}=0,1,2\hspace{0.2cm}{\rm at}\hspace{0.2cm}x=0
\nonumber \\
M_{s,\pm 1/2} - B_s & = & 0,...,\sum_{l_{\eta},l_s=\pm 1}\delta_{l_{\eta},\mp l_s}\,{\cal{N}}_{l_{\eta},l_s}
\hspace{0.2cm}{\rm for}\hspace{0.2cm}\sum_{l_{\eta},l_s=\pm 1}\delta_{l_{\eta},\mp l_s}\,{\cal{N}}_{l_{\eta},l_s}=0,1,2\hspace{0.2cm}{\rm for}\hspace{0.2cm}x>0 \, ,
\nonumber \\
& = &  0,...,{\cal{N}}_{+1,\mp 1}\hspace{0.2cm}{\rm for}\hspace{0.2cm}{\cal{N}}_{+1,\mp 1}=0,1,2\hspace{0.2cm}{\rm at}\hspace{0.2cm}x=0 \, .
\label{srs-0-ss}
\end{eqnarray}
We emphasize that the maximum values of the numbers $[\sum_{l_s}{\cal{N}}_{\pm 1,l_s}]$, $[\sum_{l_s}{\cal{N}}_{+1,l_s}]$,
$[\sum_{l_{\eta},l_s}\delta_{l_{\eta},\mp l_s}\,{\cal{N}}_{l_{\eta},l_s}]$, and ${\cal{N}}_{+1,\mp 1}$ appearing 
in this equation must be consistent with the inequality requirement ${\cal{N}}=\sum_{l_{\eta},l_s}{\cal{N}}_{l_{\eta},l_s}\leq 2$. 
Furthermore, the hole concentrations in the inequality $x>0$ and equality $x=0$ also appearing here
refer to the initial ground states and in the inequality $x>0$ we have neglected $1/N_a^D$ and $2/N_a^D$
corrections. We recall that the numbers ${\cal{N}}=1,2$ and ${\cal{N}}_{l_{\eta},l_s}$ correspond
to a specific ${\cal{N}}$-electron operator ${\hat{O}}_{{\cal{N}}}$ whose application onto a ground
state $\vert \Psi_{GS}\rangle$ generates ${\cal{N}}$-electron excited states,
as given in Eq. (\ref{1-2-subspace}). The subspace defined here refers though to the whole set of such subspaces
associated with the one-electron operator and the set of simple two-electron operators mentioned above. 

Fortunately, the subspace spanned by states whose numbers have values in the ranges given in Eq. (\ref{srs-0-ss}) is a $[M^{co}_{s}/2-B_{s}]=N_{s2}=0,1$  
and $[M^{co}_{\eta}/2-B_{\eta}]=0$ subspace (A), as defined in Section IV-F. For initial $x>0$ and $m=0$ (and $x=0$ and $m=0$) ground states 
it is such that $M^{co}_{s}=[2N_{s1}+4N_{s2}]$, $B_{s}=[N_{s1}+N_{s2}]$, and $M^{co}_{\eta}=B_{\eta}=0$ (and 
$M^{co}_{s}=[2N_{s1}+4N_{s2}]$, $B_{s}=[N_{s1}+N_{s2}]$, and $M^{co}_{\eta}/2=B_{\eta}=N_{\eta 1}=0,1$.) 
Hence the following $c$ and $s1$ fermion numbers are conserved both for the subspaces with initial $x=0$ and $x>0$ ground states,
\begin{equation}
N_{a_{c}}^D = [N_{c} + N^h_{c}] = 
N_{a}^D  \, ;
\hspace{0.25cm}
N_{c} = 2S_c \, ;
\hspace{0.25cm}
N^h_{c} = N_{a}^D - 2S_c \, ,
\label{Nac-Nhc}
\end{equation}
and
\begin{equation}
N_{a_{s1}}^D = [N_{s1} + N^h_{s1}] = 
[S_c + S_s] \, ;
\hspace{0.25cm}
N_{s1} = [S_c - S_s -2N_{s2}]  \, ;
\hspace{0.25cm}
N^h_{s 1} = [2S_s +2N_{s2}] =0,1,2 \, ,
\label{Nas1-Nhs1}
\end{equation}
respectively. The hidden $U(1)$ symmetry generator eigenvalue $S_c$ appearing here and the $\eta$-spin
$S_{\eta}$ and spin $S_s$ have in the present subspace the following values,
\begin{eqnarray}
S_c & = & {1\over 2}(1-x)\,N_a^D = N/2\hspace{0.2cm}{\rm for}\hspace{0.2cm}x>0\hspace{0.2cm}{\rm and}\hspace{0.2cm}\omega<2\mu 
\nonumber \\
& = & N/2 - M^{de}_{\eta,-1/2} - N_{\eta 1}\hspace{0.2cm}{\rm for}\hspace{0.2cm}{\rm the}
\hspace{0.2cm}{\rm initial}\hspace{0.2cm}x=0\hspace{0.2cm}{\rm GS}\hspace{0.2cm}{\rm excitations} \, ,
\nonumber \\
S_{\eta} & = & {1\over 2}x\,N_a^D\hspace{0.2cm}{\rm for}\hspace{0.2cm} x>0\hspace{0.2cm}{\rm and}\hspace{0.2cm}\omega<2\mu 
\nonumber \\
& = & {1\over 2}x\,N_a^D + M^{de}_{\eta,-1/2}=0,{1\over 2},1\hspace{0.2cm}{\rm for}\hspace{0.2cm}{\rm the}\hspace{0.2cm}{\rm initial}\hspace{0.2cm}x=0\hspace{0.2cm}{\rm GS}\hspace{0.2cm}{\rm excitations} \, ,
\nonumber \\
S_s & = & S_c -N_{s1}-2N_{s2} =0,{1\over 2},1\hspace{0.2cm}{\rm for}\hspace{0.2cm}x\geq 0 \, .
\label{NN-SSS}
\end{eqnarray}

Furthermore, for the subspace considered here the conserving number of Eq. (\ref{Phs1})
associated with the $\alpha\nu=s1$ branch simplifies to,
\begin{equation}
P^h_{s1} \equiv e^{i\pi N^h_{s1}}=e^{i2\pi S_{s}}=e^{i2\pi S_{c}}=e^{i\pi N} = \pm 1 \, .
\label{Ns1h-general}
\end{equation}
Combination of this exact relation between $N^h_{s1}$ and the number of electrons $N$
with the expressions and values of Eqs. (\ref{srs-0-ss}) and (\ref{Nas1-Nhs1}) 
reveals that the one- and two-electron subspace considered here is
spanned by states whose deviation $\delta N_c^h$ in the number of $c$ band holes
and number $N_{s1}^h$ of $s1$ band holes may only have the following values, 
\begin{eqnarray}
\delta N_c^h & = & -2\delta S_c = - \delta N = 0, \mp 1, \mp 2 \, ,
\nonumber \\
N_{s1}^h & = & 2S_s+2N_{s2} = 0, 1, 2 \, .
\label{deltaNcs1}
\end{eqnarray}
(The initial $x\geq 0$ and $m=0$ ground states have zero holes in the $s1$ band so that $\delta N_{s1}^h=N_{s1}^h$ 
for their excited states.) As discussed below in Section VI, for $N_{s1}^h=0$ ground states and their charge excited states 
all $M_s=2S_c$ spinons are confined within the two-spinon bonds of the $N_{s1}=M_s/2$ $s1$ fermions. 

The number $P^h_{s1}=\pm 1$ of Eq. (\ref{Phs1}) is associated with an
important exact selection rule. One of its consequences is that one- and two-electron excitations of
$x\geq 0$ and $m=0$ ground states contain no states with an even
and odd number $N^h_{s1}$ of $s1$ fermion holes in the $s1$ momentum band, respectively.
For such initial ground states we consider that $N$ is an even integer
number. Since, as given in Eq. (\ref{Phs1}), $e^{i\pi\prod_{l} N^h_{s1}(l)}=e^{i\pi N}$, 
for one-electron excited states for which the deviation in the
value of $N$ reads $\delta N=\pm 1$ the number of $s1$ fermion holes $N^h_{s1}$ 
must be always an odd integer. On the other hand, for both $\delta N=0$ and $\delta N=\pm 2$
two-electron excited states $N^h_{s1}$ must be
always an even integer. Such exact selection rules play an important role in the one- and two-electron
spectra of the square-lattice quantum liquid further studied in Ref. \cite{companion}.

That quantum liquid refers to the Hamiltonian 
(\ref{H}) in the one- and two-electron subspace as defined here for initial $x\geq 0$ and $m=0$ ground 
states. For initial $x>0$ and $m=0$ ground states and excitation energy below $2\mu$
the one- and two-electron subspace as defined here
is spanned by the states of Table 4 of Ref. \cite{companion}. Those are generated by creation or annihilation 
of $\vert\delta N_c^h\vert=0,1,2$ holes in the
$c$ momentum band and $N_{s1}^h=0,1,2$ holes in the $s1$ band 
plus small-momentum and low-energy particle-hole processes in the $c$ band. 
The charge excitations of such initial ground states
consist of a single particle-hole process in the $c$ band of arbitrary momentum and 
energy compatible with its momentum and energy bandwidths, 
plus small-momentum and low-energy $c$ fermion particle-hole processes.
Such charge excitations correspond to state
representations of the global $U(1)$ symmetry and
refer to the type of states denoted by ``charge" in the Table 4 of Ref. \cite{companion}.
The one-electron spin-doublet excitations correspond 
to the four types of states denoted by ``$\pm1\sigma$el." in
that table where $+1$ and $-1$ denotes creation and
annihilation, respectively, and $\sigma =\uparrow ,\downarrow$.
The spin-singlet and spin-triplet excitations refer to the four types of states 
denoted by ``singl.spin" and ``tripl.spin".
The two-electron excitations whose electrons are in a spin-singlet
configuration and those whose two created or annihilated electrons are 
in a spin-triplet configuration correspond to the five types of 
states ``$\pm 2\uparrow\downarrow$el." and ``$\pm 2\sigma$el."   
where $+2$ and $-2$ denotes creation and
annihilation, respectively, of two electrons.

\subsection{Confirmation that for 1D nearly the whole one-electron spectral weight is generated 
by processes obeying the ranges of Eqs. (\ref{srs-0-ss}) and (\ref{deltaNcs1})}

The transformation laws under the electron - rotated-electron unitary transformation
of deconfined spinons, deconfined $\eta$-spinons, and $\alpha\nu$ fermions of
several branches were used above to show that nearly the whole one- and two-electron spectral weight  
of the excitations of the Hubbard model on the 1D and square
lattices is generated by processes obeying the ranges (\ref{srs-0-ss})
and (\ref{deltaNcs1}).

Upon expressing one- or two-electron operators ${\hat{O}}_{\cal{N}}$ in terms of rotated-electron 
creation and annihilation operators, they have in general an infinite number of terms, as given
on the right-hand side of the first equation of (\ref{OOr}). Such rotated-electron operator terms, which
generate the excitations ${\hat{O}}_{\cal{N}}\vert \Psi_{GS}\rangle$ of Eq. (\ref{1-2-subspace}), may be expressed
in terms of the $c$ fermion operators, spinon operators, and $\eta$-spinon
operators given in Eqs. (\ref{fc+})-(\ref{rotated-quasi-spin}).
This is done on using the operator relations provided in Eq. (\ref{c-up-c-down}).
Concerning the contributions to the general operator expression given in Eq. (\ref{OOr}), which contains
commutators involving the operator ${\tilde{S}} = -(t/U)\,[\tilde{T}_{+1} -\tilde{T}_{-1}] 
+ {\cal{O}} (t^2/U^2)$, to fulfill such a task one takes into account that independently of their form, the additional 
higher-order operator terms ${\cal{O}} (t^2/U^2)$ are products of the kinetic operators 
$\tilde{T}_0$, $\tilde{T}_{+1}$, and $\tilde{T}_{-1}$ of Eq. (\ref{T-op}).

From such an analysis, one finds that the elementary processes associated with
the one- and two-electron subspace number value ranges of Eq. (\ref{srs-0-ss}) are fully generated by 
the leading-order operator ${\tilde{O}}$, which in our case is a one- or two-electron
operator ${\hat{O}}_{\cal{N}}$. In turn, the processes generated by 
the operator terms containing commutators involving the operator ${\tilde{S}}$
refer to excitations whose number value ranges are different
from those provided in that equation. This confirms that
such processes generate very little one- and two-electron
spectral weight, consistently with the exact number restrictions of Eqs. (\ref{LsrcsL}) and (\ref{srcsL}) and
the approximate number restrictions of Eq. (\ref{srs-0-ss}). 

For the Hubbard model on the 1D lattice the spectral-weight
distributions can be explicitly calculated by the pseudofermion
dynamical theory associated with the model exact solution \cite{V,TTF}, 
exact diagonalization of small chains \cite{1EL-1D}, and other methods. 
The relative one-electron spectral weight generated by different types of 
microscopic processes is studied in Ref. \cite{1EL-1D}.
The results of that reference confirm the dominance of
the processes associated with the number value ranges provided in
Eq. (\ref{srs-0-ss}). They refer specifically to operators ${\hat{O}}={\hat{O}}_{\cal{N}}={\hat{O}}_1$ and 
${\tilde{O}}={\tilde{O}}_{\cal{N}}={\tilde{O}}_1$ 
that are electron and rotated-electron, respectively, creation or
annihilation operators. Such studies confirm that the operator ${\tilde{O}}$
generates {\it all} processes associated with the number value ranges
of Eq. (\ref{srs-0-ss}). In addition, it also generates some of the non-dominant processes. That
is confirmed by the weights given in Table 1 of Ref.
\cite{1EL-1D}, which correspond to the dominant processes
associated with only these ranges. The small missing
weight refers to excitations whose number value ranges
are not those of Eq. (\ref{srs-0-ss}) but whose
weight is also generated by the operator ${\tilde{O}}$.
Indeed, that table refers to $U/4t\rightarrow\infty$
so that ${\hat{O}}={\tilde{O}}$ and the operator terms of 
the ${\hat{O}}$ expression provided in
Eq. (\ref{OOr}) containing commutators involving the 
operator ${\tilde{S}}$ vanish. 

For finite values of $U/4t$ all dominant processes associated 
with the number value ranges of Eq. (\ref{srs-0-ss})
are also generated by the operator ${\tilde{O}}$. In turn,
the small spectral weight associated with excitations whose
number value ranges are different from those are generated both
by that operator and the operator terms of the ${\hat{O}}$ expression of
Eq. (\ref{OOr}) containing commutators involving the 
operator ${\tilde{S}}$. For the model on the 1D lattice the small 
one-electron spectral weight generated by the non-dominant 
processes is largest at half filling and $U/4t\approx 1$. 

The particle-hole symmetry of the $x=0$ and $\mu=0$ ground state implies that
the relative spectral-weight contributions from different types of one-electron addition
excitations given in Fig. 2 of Ref. \cite{1EL-1D} for the 1D model at half filling  
leads to similar corresponding relative weights for   
half-filling one-electron removal. Analysis of that figure 
confirms that for the corresponding one-electron removal
spectrum the dominant processes associated 
with the number value ranges of Eq. (\ref{srs-0-ss})
refer to the states called holon - 1 $s1$ hole states in figure 1 of Ref. \cite{1EL-1D}. 
Within our notation such upper-Hubbard band states have one
deconfined $+1/2$ $\eta$-spinon and  one
deconfined $-1/2$ $\eta$-spinon for one-electron removal and addition, respectively.
Their minimum relative weight of about $0.95$ is reached
at $U/4t\approx 1$. For other hole concentrations $x>0$ 
and values of $U/4t$ the relative weight of one-electron states 
associated with the number value ranges of Eq. (\ref{srs-0-ss}) 
is always larger than $0.95$, as confirmed from analysis of 
Figs. 1 and 2 and the data provided in Table 1 of that reference.

For the Hubbard model on the square lattice the
explicit derivation of one- and two-electron spectral
weights is a more involved problem. The number value ranges
of Eq. (\ref{srs-0-ss}) also apply, implying similar results for the relative spectral weights
of one- and two-electron excitations. We emphasize that this is consistent with
the exact range restrictions of Eqs. (\ref{LsrcsL}) and (\ref{srcsL}), the exact selection rule associated with 
the conservation of the number $P^h_{s1}$ of Eq. (\ref{Ns1h-general}),
and the transformation laws under the electron - rotated-electron unitary transformation
of deconfined spinons, deconfined $\eta$-spinons, and $\alpha\nu$ fermions.
Such transformation laws are behind the approximate number value ranges
of Eq. (\ref{srs-0-ss}) being valid both for the
Hubbard model on the 1D and square lattices. 

\subsection{The spin and $s1$ effective lattices
for the one- and two-electron subspace}

According to the restrictions and numbers values of
Eqs. (\ref{srs-0-ss}) and (\ref{deltaNcs1}), the states
that span the one- and two-electron subspace may involve none 
or one $s2$ fermion. As confirmed in the studies of Ref. \cite{companion},
it is convenient to express the one- and two-electron excitation spectrum relative
to initial $x\geq 0$ and $m=0$ ground states in terms of the deviations in the numbers 
of $c$ effective lattice unoccupied sites and $s1$ effective lattice unoccupied sites. Those 
are given explicitly in Eq. (\ref{deltaNcs1}) and equal the corresponding deviations in the 
numbers of $c$ band fermion holes and $s1$ band fermion holes, respectively.
Note that for $x>0$ and $\omega<2\mu$ states the $s1$ fermion
related numbers provided in Eq. (\ref{Nas1-Nhs1}) can be written as 
$N_{a_{s1}}^D=[N/2+S_s]$, $N_{s1}=[N/2-S_s -2N_{s2}]$, and 
$N_{s1}^h=[2S_s +2N_{s2}]=0,1,2$
where $S_s=0$ for $N_{s2}=1$ and $S_s=0,1/2,1$ for $N_{s2}=0$.
In turn, for excited states of the $x=0$ and $m=0$ ground state they read
$N_{a_{s1}}^D=[N/2-M^{de}_{\eta,-1/2}-N_{\eta 1}+S_s]$,
$N_{s1}=[N/2-M^{de}_{\eta,-1/2}-N_{\eta 1}-S_s -2N_{s2}]$, and 
$N_{s1}^h=[2S_s +2N_{s2}]=0,1,2$ where $S_s$ and $N_{s2}$
may have the same values as above and $M^{de}_{\eta,-1/2}=0$ for $N_{\eta 1}=1$ and
$M^{de}_{\eta,-1/2}=0,1,2$  for $N_{\eta 1}=0$. Provided that $n=(1-x)$ is
finite, the corrections to $N_{a_{s1}}^D\approx N/2$ and
$N_{s1}\approx N/2$ are in both cases of the order of $1/N_a^D$, whereas 
$N_{s1}^h=[2S_s +2N_{s2}]=0,1,2$ has the same expression and allowed values.

As discussed above, for $N_{s2}=1$ spin-singlet excited energy eigenstates the single $s2$ fermion has vanishing energy 
and momentum. Consistently with Eq. (\ref{invariant-V}),
for vanishing magnetic field $H=0$ it is invariant under the electron - rotated-electron
unitary transformation. The same applies to the single $\eta 1$ fermion of $N_{\eta 1}=1$ 
$\eta$-spin-singlet excited states of the $x=0$, $\mu=0$, and $m=0$ ground state. 
Therefore, the only effect of creation
and annihilation of such two objects is in the numbers of sites and occupied sites of the $s1$ effective lattice.
Their creation can then be merely accounted for by small changes in the occupancies of the 
discrete momentum values of the $s1$ band. Hence and consistently with the additional information
provided in Appendix C, the only 
composite object whose internal occupancy configurations are 
important for the physics of the Hamiltonian (\ref{H}) in the one- and two-electron subspace 
is the spin-neutral two-spinon $s1$ fermion and related spin-singlet two-spinon $s1$ bond particle \cite{companion}.

It turns out that for the Hubbard model in the one- and two-electron 
subspace and alike for the $s2$ fermion and/or the $\eta 1$ fermion,
the presence of deconfined spinons is felt through the numbers of occupied and unoccupied sites of
the $s1$ effective lattice. For excited states of $x\geq 0$ and
$m=0$ ground states the number of deconfined $\eta$-spinons equals that of the unoccupied 
sites of the $c$ effective lattice. For excited states of the $x= 0$, $\mu =0$, and
$m=0$ ground state the presence of deconfined $\eta$-spinons is 
felt in addition through the above numbers of sites and occupied sites of
the $s1$ effective lattice. Those may be rewritten as
$N_{a_{s1}}^D=[N_a^D/2-M^{de}_{\eta}/2-N_{\eta 1}+S_s]$ and
$N_{s1}=[N_a^D/2-M^{de}_{\eta}/2-N_{\eta 1}-S_s -2N_{s2}]$, respectively, where
$M^{de}_{\eta}=2S_{\eta}=0,1,2$. Therefore, when acting onto the one- and two-electron subspace
as defined in this paper, the Hubbard model refers to a two-component quantum liquid that
can be described only in terms of $c$ fermions and $s1$ fermions. 
For excited states of $x>0$ and $m=0$ ground states this analysis applies to excitation energies $\omega<2\mu$.
For those of the $x=0$, $\mu=0$, and $m=0$ ground state the $\eta$-spin degrees of freedom
are in addition behind the finite energy $\Delta_{D_{rot}}$ given in Eq. (\ref{min-D-N}) and
the related energy $\Delta_{D_{rot}}^h$. Those are associated with rotated-electron doubly occupied sites 
and rotated-electron unoccupied sites, respectively, of excited states with finite occupancy $M^{de}_{\eta}=1,2$ of 
deconfined $\eta$-spinons or $N_{\eta 1}=1$ of a single $\eta 1$ fermion. However, the magnitude of that energy is fixed for each branch
of excitations of that ground state. Hence for it and its excited states the square-lattice quantum liquid may 
again be described solely in terms of $c$ fermions and $s1$ fermions.

For excited states of $x>0$ and $m=0$ ground states belonging to the one- and two-electron 
subspace the spin effective lattice has a number of sites given by $N_{a_{s}}^D =(1-x)\,N_a^D$.
For those of the $x=0$, $\mu=0$, and $m=0$ ground state it reads
$N_{a_{s}}^D =[N_a^D-M^{de}_{\eta}-2N_{\eta 1}]$ so that
$N_{a_{s}}^D$ may have the values $N_a^D$, $[N_a^D-1]$, and $[N_a^D-2]$.
For $x>0$ its value is smaller than that of the original lattice. Within the
$N_a^D\gg 1$ limit one may neglect corrections of the order $1/N_a^D$ 
so that for both types of excited states the lattice spacing $a_s$ is that provided 
in Eq. (\ref{a-alpha}) for $\alpha = s$. For the model on the square lattice it reads $a_{s} \approx a/\sqrt{1-x}$, 
as given in Eq. (\ref{NNCC}). Both it and its general expression given in Eq. (\ref{a-alpha}) 
are such that the area $L^2=[a_s\times N_{a_{s}}]^2=[a\times N_a]^2$ of the system is preserved. 
As discussed in Section III-C, the concept of a spin effective lattice is valid only within the $N_a^D\gg 1$ limit 
that our description refers to. For the model on the square lattice, the $s1$ fermion spinon occupancy 
configurations considered in Section VI are expected to be a good approximation 
provided that the ratio $N_{a_s}^2/N_a^2$ and thus the electronic density $n=(1-x)$
remain finite as $N_a^2\rightarrow\infty$. This is met for the hole-concentration
range $x\in (0,x_*)$ considered in the studies of Ref. \cite{companion}.

Within the present $N_a^2\gg 1$ limit there is for the  
one- and two-electron subspace of the model on the square lattice
commensurability between the real-space distributions of the $N_{a_{s1}}^2\approx N_{s1}$
sites of the $s1$ effective lattice and the $N_{a_{s}}^2\approx 2N_{s1}$ 
sites of the spin effective lattice. For $(1-x)\geq 1/N_a^2$ and $N_a^2
\gg 1$ the spin effective lattice has $N_{a_s}^2\approx (1-x)\,N_a^2$ sites and from the use of 
the expression given in Eq. (\ref{Nas1-Nhs1}) for the number of $s1$ effective
lattice sites $N_{a_{s1}}^2$ and Eq. (\ref{a-a-nu}) for the
corresponding spacing $a_{s1}$ we find,
\begin{equation}
a_{s1} = a_s\,\sqrt{{2\over1+{2S_s\over (1-x)N_a^2}}}\approx \sqrt{2}\,a_s\, \left(1-{2S_s\over 2(1-x)}{1\over N_a^2}\right)
\approx \sqrt{2}\,a_s \, , \hspace{0.25cm} S_s =0, {1\over 2}, 1 \, .
\label{a-a-s1-sube}
\end{equation}

The general description introduced in this paper refers to a very large
number of sites $N_a^D\gg 1$. Although very large, we assume that $N_a^D$ is finite and
only in the end of any calculation take the $N_a^D\rightarrow\infty$ limit. For $N_a^D\gg 1$
very large but finite the $m=0$ ground state spin effective lattice is full and both at $x=0$ 
and for $x>0$ such a state is a spin-singlet state. (For $m=0$ and $x=0$ this agrees with
the exact theorem of Ref. \cite{Lieb-89}.) 
For $N^h_{s1}=0$ states such as such $x\geq 0$ and $m=0$ ground states
and their charge excited states the spin effective square lattice has two
well-defined sub-lattices, which we call sub-lattice 1 and 2, respectively. 
(For the $N^h_{s1}=1,2$ states of the present subspace the spin effective lattice has 
two bipartite lattices as well, with the one or two extra sites accounted for by suitable boundary conditions.)
The two spin effective sub-lattices have spacing $a_{s1} \approx \sqrt{2}\,a_s$. 
The fundamental translation vectors of the sub-lattices 1 and 2 read,
\begin{equation}
{\vec{a}}_{s1} = {a_{s1}\over\sqrt{2}}({\vec{e}}_{x_1}+{\vec{e}}_{x_2})
\, , \hspace{0.25cm} 
{\vec{b}}_{s1} = -{a_{s1}\over\sqrt{2}}({\vec{e}}_{x_1}-{\vec{e}}_{x_2}) \, ,
\label{a-b-s1}
\end{equation}
respectively. Here ${\vec{e}}_{x_1}$ and ${\vec{e}}_{x_2}$ are unit vectors pointing in the direction associated with
the Cartesian coordinates $x_1$ and $x_2$, respectively. As further discussed in Section VI, 
the vectors given in this equation are the fundamental translation vectors 
of the $s1$ effective square lattice.

In the case of $x\geq 0$, $m=0$, and $N^h_{s1}=0$ ground states 
whose $s1$ momentum band is full and all $N_{a_{s1}}^2=N_{a_{s1}}\times N_{a_{s1}}$ sites of the $s1$ effective 
square lattice are occupied we consider that the square root $N_{a_s}$ of the number $N_{a_s}^2=N_{a_{s}}\times N_{a_{s}}$ 
of sites of the corresponding spin effective square lattice is an integer. Although the square root $N_{a_{s1}}$ of
the number $N_{a_{s1}}^2=N_{a_{s1}}\times N_{a_{s1}}$ of sites of the $s1$ effective lattice is not in general
an integer number, within the present $N_a^2\gg 1$ limit we consider that it is the closest 
integer to it.

\subsection{The square-lattice quantum liquid of $c$ and $s1$ fermions}

It follows from the the results reported in the previous sections and from the complementary technical analysis 
of Appendix C that when acting onto the one- and two-electron subspace as defined in this paper, the Hubbard 
model on a 1D or square lattice refers to a two-component quantum liquid described
in terms of two types of objects on the corresponding effective lattices
and momentum bands: The charge $c$ fermions and spin-neutral two-spinon $s1$ fermions.
The one- and two-electron subspace can be divided into smaller subspaces
that conserve $S_c$ and $S_s$. When expressed in terms of $c$ and $s1$ fermion operators, the
Hubbard model on a square lattice in the one- and two-electron
subspace is the square-lattice quantum liquid further studied in Ref. \cite{companion}.

Appendix C provides further information about why the square-lattice quantum liquid corresponding to the Hubbard model 
on the square lattice in the one- and two-electron subspace may be described only by $c$ and $s1$ fermions 
on their $c$ and $s1$ effective lattices, respectively. 
Furthermore, the specific form that the momentum 
eigenstates of Eq. (\ref{non-LWS}) have in the one- and two-electron subspace is provided in Eqs.
(\ref{non-LWS-c-s1}) and (\ref{LWS-full-el-c-s1}) of Appendix C. Consistently with the results of Section IV-F 
concerning the subspaces of type (A), for Hubbard model on the square lattice in that subspace such states are energy 
eingenstates. However, we recall that the spin and $s1$ effective lattices whose occupancy 
configurations generate the spin degrees of freedom of such states refer to an approximation valid
only within the $N_a^D\rightarrow\infty$ limit of our description. Hence although the
states of Eqs. (\ref{non-LWS-c-s1}) and (\ref{LWS-full-el-c-s1}) of Appendix C are
exact energy eigenstates, our description refers to an approximate representation of such states. 
Since for $x>0$ and excitation energy $\omega <2\mu$ they span all subspaces of the one- and two-electron
subspace that conserve $S_c$ and $S_s$, they span the
latter subspace as well. We recall that states with a single $s2$ fermion have
also the general form provided Eqs. (\ref{non-LWS-c-s1}) and (\ref{LWS-full-el-c-s1}) of Appendix C. 
As discussed in Section V-C and that Appendix, the presence of that vanishing-energy,
vanishing-momentum, and spin-neutral four-spinon object is accounted for
the values of the numbers $N_{s1} =[S_c - S_s - 2N_{s2}]$ and
$N^h_{s 1}=[2S_s +2N_{s2}]=0,1,2$ of Eq. (\ref{Nas1-Nhs1}).

The quantum-liquid $c$ fermions are $\eta$-spinless and spinless
objects without internal degrees of freedom and 
structure whose effective lattice 
is identical to the original lattice. For the complete set of $U/4t>0$ 
energy eigenstates that span the full Hilbert space, the occupied sites (and unoccupied
sites) of the $c$ effective lattice correspond to those
singly occupied (and doubly occupied plus unoccupied)
by the rotated electrons. The corresponding $c$ band
has the same shape and momentum area as the
first Brillouin zone \cite{companion}.

In contrast, the quantum-liquid composite spin-neutral two-spinon $s1$ fermions 
have internal structure. Thus the spinon occupancy configurations that describe
such objects are for the one- and two-electron subspace 
a more complex problem discussed below in Section VI. It is simplified by the
property that the number of unoccupied sites of the $s1$ effective lattice 
is in that subspace limited to the values $N_{s1}^h=0,1,2$. 

The energy eigenstates $\vert \Psi_{U/4t}\rangle$ of general form given in Eq. (\ref{non-LWS-c-s1}) of Appendix C
that for $x>0$ and excitation energy $\omega <2\mu$ span the one- and two-electron subspace have
numbers $N_{s2}=N_{a_{s2}}^D=0,1$ and $N_{s1}\approx N_{a_{s1}}^D$ such that
$N_{s1}^h=[N_{a_{s1}}^D-N_{s1}]=0,1,2$. As mentioned in Section IV-F, for the model on the square lattice
the spacing $a_{s1} \approx \sqrt{2}\,a_s=\sqrt{2/(1-x)}\,a$ of Eq. (\ref{a-a-s1-sube})
is directly related to a fictitious magnetic-field length $l_{s1}$ associated with the 
field of Eq. (\ref{A-j-s1-3D}) for the particular case of the $\alpha\nu =s1$ branch. Indeed, in that subspace one has that
$\langle n_{\vec{r}_j,s1}\rangle\approx 1$ and such a fictitious magnetic field reads
${\vec{B}}_{s1} ({\vec{r}}_j) \approx \Phi_0\sum_{j'\neq j}\delta ({\vec{r}}_{j'}-{\vec{r}}_{j})\,{\vec{e}}_{x_3}$.
It acting on one $s1$ fermion differs from zero only at the positions
of other $s1$ fermions. In the mean-field approximation one replaces it
by the average field created by all $s1$ fermions at position $\vec{r}_j$. This gives,
${\vec{B}}_{s1} ({\vec{r}}_j) \approx \Phi_0\,n_{s1} (\vec{r}_j)\,{\vec{e}}_{x_3}
\approx \Phi_0\,[N_{a_{s1}}^2/L^2]\,{\vec{e}}_{x_3}=[\Phi_0/a_{s1}^2]\,{\vec{e}}_{x_3}$. 
One then finds that the number $N_{a_{s1}}^2$
of the $s1$ band discrete momentum values equals $[B_{s1}\,L^2]/\Phi_0$. In addition, the
$s1$ effective lattice spacing $a_{s1}$ is expressed in terms to the fictitious 
magnetic-field length $l_{s1}\approx a/\sqrt{\pi(1-x)}$ as $a_{s1}^2=2\pi\,l_{s1}^2$.
This is consistent with each $s1$ fermion having a flux
tube of one flux quantum on average attached to it. 

As further discussed in Ref. \cite{companion}, for the present one- and two-electron subspace of the model on the square lattice
the $s1$ fermion problem is then related to the Chern-Simons theory \cite{Giu-Vigna}.
Indeed the number of flux quanta being one is consistent with the
$s1$ fermion and $s1$ bond-particle wave functions obeying Fermi and Bose statistics, respectively. 
Hence the composite $s1$ fermion consists of two spinons in a spin-singlet configuration plus an infinitely thin flux tube attached 
to it. Thus, each $s1$ fermion appears to carry a fictitious magnetic solenoid
with it as it moves around in the $s1$ effective lattice.

That the square-lattice quantum liquid is constructed to inherently 
the $c$ and $s1$ fermion discrete momentum values 
being good quantum numbers is behind the suitability of the present description in terms
of occupancy configurations of the $c$ and $s1$ effective lattices and corresponding
$c$ and $s1$ band discrete momentum values. The latter $c$ and $s1$ values are the conjugate of the real-space 
coordinates of the $c$ and $s1$ effective lattice, respectively. 
Are the approximations used in the construction of the $s1$ effective lattice inconsistent
with the $s1$ band discrete momentum values being good quantum numbers?
The answer is no. Indeed, such approximations concern the relative positions of 
the $j=1,...,N_{a_{s1}}^2$ sites of the $s1$ effective lattice \cite{companion}.
Those control the shape of the $s1$ momentum band boundary. They do not
affect the $s1$ band discrete momentum values being good quantum numbers.
At $x=0$ the spin effective lattice is identical to the original square lattice and
the $s1$ effective lattice is one of its two sub-lattices. Consistently, at $x=0$ and $m=0$
the boundary of the $s1$ momentum band is accurately known. Then
the $s1$ band coincides with an antiferromagnetic reduced Brillouin 
zone of momentum area $2\pi^2$ such that $\vert q_{x_1}\vert+\vert q_{x_2}\vert\leq\pi$ \cite{companion}.
In turn, it is known that for $x>0$ and $m=0$ the the $s1$ band boundary encloses a 
smaller momentum area $(1-x)2\pi^2$ yet its precise shape remains an open issue.
The related problems of the $c$ and $s1$ momentum bands and corresponding energy dispersions and velocities
are studied in Ref. \cite{companion}.     
     
\section{The $s1$ fermion operators and the related $s1$ bond-particle operators algebra}

In this section we call {\it configuration states} the spinon occupancy configurations in the spin effective lattice 
associated with the corresponding occupancy of the $N_{s1}$ local $s1$ fermions over the $N_{a_{s1}}^D$ 
sites of the $s1$ effective lattice. A local $s1$ fermion operator is generated from the corresponding
$s1$ bond particle operator $g^{\dag}_{{\vec{r}}_{j},s1}$ by the transformation
$f^{\dag}_{{\vec{r}}_{j},s1} = e^{i\phi_{j,s1}}\,g^{\dag}_{{\vec{r}}_{j},s1}$, as given in
Eq. (\ref{f-an-operators}) for the $\alpha\nu =s1$ branch. Here we consider the most general
scenario according to which the $s1$ bond particle
is a superposition of all independent spin effective lattice two-site bonds
centered at its real-space coordinate ${\vec{r}}_{j}$. This includes two-site bonds
of all lengths, the only restriction being that they must be centered at the $s1$
bond particle real-space coordinate ${\vec{r}}_{j}$.
Within our general scheme, each two-site bond is associated with a coefficient whose exact magnitude remains
unknown. The exact 1D BA solution implicitly accounts for the magnitudes of such coefficients
whereas for the model on the square lattice their exact magnitudes and bond and $x$ dependences
remain open problems. For instance, some such coefficients might vanish. Depending on the values of
the two-site-bond coefficients, our general formalism may describe both spin systems with
different types of long-range or short-range orders and disordered spin systems. The spin degrees
of freedom of the 1D model is an example of a spin disordered system.

A $x\geq 0$ and $m=0$ ground state and its charge excited states 
belonging to the one- and two-electron subspace are $N_{s1}^h=0$ 
energy eigenstates whose spin
degrees of freedom are described by 
the $N^h_{s1}=0$ configuration state studied in the following.
Within the use of suitable boundary conditions for the unoccupied site or two unoccupied sites 
of the $N^h_{s1}=1,2$ configuration states, similar results are obtained for such states.
For $N_{s1}^h>0$ excited states, each configuration state refers to well-defined positions of the 
$N^h_{s1}$ unoccupied sites. 

The $s1$ fermion operators $f_{\vec{q}_j,s1}^{\dag}$ in the expression
of the generators of the energy eigenstates whose general form is given in Eq. (\ref{non-LWS-c-s1})
of Appendix C are a superposition of the local $s1$ fermion operators 
provided in Eq. (\ref{f-an-operators}) for the $\alpha\nu =s1$ branch. The spin degrees of freedom of 
such states are generated by superpositions of the configuration states. 
While the states of Eq. (\ref{non-LWS-c-s1}) of Appendix C refer to 
$x>0$ and excitation energy $\omega <2\mu$, the same applies to the
excited states of the $x=0$, $\mu=0$, and $m=0$ ground state that span the one- and two-electron subspace.

\subsection{Independent two-site bonds}

We consider torus periodic boundary conditions for the spin effective square lattice 
with $N_{a_s}\times N_{a_s}$ sites of the Hubbard model on the square lattice, alike 
for the original lattice. That implies periodic boundary conditions for the $N_{a_s}$
rows and $N_{a_s}$ columns. Periodic boundary conditions are used for the one-chain 
spin effective lattice of the 1D model.
The spin effective lattice has in the present case two sub-lattices, which
we have named above sub-lattices 1 and 2. For the model on the square lattice their fundamental 
translation vectors are given in Eq. (\ref{a-b-s1}).
The real-space coordinates of the sites of each of such sub-lattices correspond to 
a possible choice of those of the $s1$ effective lattice. Indeed, there is for the $N^h_{s1}=0$ configuration
state a gauge ``symmetry" between the representations in terms of the occupancies of the
two alternative choices of real-space coordinates of
the $s1$ effective lattice. In Appendix B of Ref. \cite{companion} it is confirmed that they refer
to two alternative and equivalent representations of the $N^h_{s1}=0$ configuration
state. The possibility of the use of the real-space coordinates of either of the two corresponding choices of $s1$ effective 
lattices to label the $N^h_{s1}=0$ configuration state
is associated with the occurrence of a gauge structure \cite{Xiao-Gang}.
The real-space coordinates ${\vec{r}}_{j}$ of such two 
sub-lattices have $j=1,...,N_{a_{s1}}^D$ sites, 
the fundamental translation vectors being those given in 
Eq. (\ref{a-b-s1}). The real-space coordinates ${\vec{r}}_{j}$ of the local $s1$ fermions and
corresponding $s1$ bond particles are chosen to correspond to those of one of these two
sub-lattices. Throughout the remaining of this section the sub-lattices called sub-lattice 1 and 
sub-lattice 2 in Section V-C are the sub-lattice of the spin effective lattice of a $N_{s1}^h=0$ configuration state whose real-space
coordinates are and are not the same as those of the $s1$ effective lattice, respectively.  

A two-site bond connects two sites of the spin effective
lattice whose spinons have opposite spin projection and correspond to
a two-site spin-singlet configuration defined below.
Each $s1$ bond particle is a suitable superposition
of a well-defined set of two-site bonds. Such a spin-singlet
two-spinon $s1$ bond particle is related to the
resonating-valence-bond pictures for spin-singlet 
occupancy configurations of ground states studied in Refs. \cite{Fazekas,Pauling}.
The advantage of our rotated-electron description
is that the $s1$ bond particle is well defined for all values of $U/4t>0$ 
and not only for $U/4t\gg 1$. This is consistent with its two spinons 
referring to spins of the sites singly occupied by rotated electrons rather than electrons. 
Indeed most schemes used previously for the Hubbard model on the square lattice and related
models involving singly-occupied-site spins refer in general to 
large values of $U/4t\gg 1$ \cite{2D-MIT,Feng,Fazekas,Xiao-Gang}.
Here the $s1$ bond particles have been constructed to inherently involving
spinon occupancy configurations of sites of the spin effective lattice, which for $U/4t>0$ refer
only to the sites of the original lattice singly occupied by 
rotated electrons.

For the $N^h_{s1}=0$ configuration state
studied here the above bond superposition includes $2D=2,4$ families of 
two-site bonds, each family having $N_{s1}/2D$ different types of
such bonds: $N_{s1}/2D$ is the largest number of independent 
two-site bonds with the same bond center that exist for the 
above-considered boundary conditions. (Above and
in the remaining of this section we denote often
the number of family two-site bonds by $2D=2,4$ where
two and four is the number of such families for
the model on the 1D and square lattice, respectively.)
Two-site bond independence means here that for 
a given bond center all two-site bonds involve 
different pairs of sites and each site belongs to only one pair. 

The set of independent two-site bonds with the same bond center belong to the same two-site bond family. 
Each two-site bond of a given family has some local $s1$ fermion weight, which in some cases may 
vanish. For a $s1$ bond particle of real-space coordinate ${\vec{r}}_{j}$ 
there are $2D=2,4$ families of two-site bonds. The two-site bonds of each family
are centered at one of the $2D=2,4$ points of real-space coordinate
${\vec{r}}_{j}+{\vec{r}_{d,l}}^{\,0}$. Here the indices $d=1,2$ 
for $D=2$, $d=1$ for $D=1$, and $l=\pm 1$
uniquely define the two-site bond family and ${\vec{r}_{d,l}}^{\,0}$ is the
primary link vector. It connects the spin effective lattice site of real-space coordinate
${\vec{r}}_{j}$ to the center of the four (and two for 1D) two-site bonds
of real-space coordinate ${\vec{r}}_{j}+{\vec{r}_{d,l}}^{\,0}$ 
involving that site and its spin effective lattice nearest-neighboring sites. 
Note that the former site and the latter four sites
(and two sites for 1D) belong to sub-lattice 1 and
2, respectively. On choosing one of the 
two sub-lattices of the spin effective lattice to be
sub-lattice 1 and thus playing the role
of $s1$ effective lattice and representing the states
in terms of the occupancy configurations of the latter
lattice we say that there is a change of gauge structure \cite{Xiao-Gang}.
On accounting for all $N_{s1}/2D$ independent two-site bonds for each 
of the $2D$ two-site bond centers needed to describe a $s1$ bond 
particle of real-space coordinate ${\vec{r}}_{j}$ we consider
the most general situation. The unknown exact configuration refers
to some choice of the two-site bond local $s1$ fermion weights considered below,
some of which may vanish.
  
A $s1$ bond particle of real-space coordinate ${\vec{r}}_{j}$
involves $N_{a_s}^D/2=N_{s1}$ two-site bonds. This is consistent with each family
having $N_{a_s}^D/4D=N_{s1}/2D$ two-site bonds of different type. The two-site bond type
is labeled by an index $g=0,1,...,[N_{s1}/2D-1]$
uniquely defined below. Each two-site bond of a $s1$ bond particle of real-space coordinate
${\vec{r}}_{j}$ involves two sites of coordinates 
$\vec{r}-\vec{r}_{d,l}^{\,g}$ and $\vec{r}+
\vec{r}_{d,l}^{\,g}$ where ${\vec{r}}_{j}=\vec{r}-\vec{r}_{d,l}^{\,0}$. Hence
the two-site bond center $\vec{r}\equiv\vec{r}_{j}+{\vec{r}_{d,l}}^{\,0}$ is 
the middle point located half-way between the two sites.
(The link vector $\vec{r}_{d,l}^{\,g}$ is defined below.)
The real-space coordinates ${\vec{r}}_{j}=\vec{r}-\vec{r}_{d,l}^{\,g}$
and $\vec{r}+\vec{r}_{d,l}^{\,g}$ belong to the sub-lattice 1
and sub-lattice 2 of the spin effective lattice, respectively.
For each family there are $N_{a_s}^D/4D=N_{s1}/2D$ link vectors
$\vec{r}_{d,l}^{\,g}$, which read,
\begin{equation}
\vec{r}_{d,l}^{\,g} = {\vec{r}_{d,l}}^{\,0}
+ {\vec{T}}_{d,l}^{\,g} \, ; \hspace{0.50cm}
{\vec{r}_{d,l}}^{\,0} = l\,{a_s\over 2}\,{\vec{e}}_{x_d}
\, ; \hspace{0.50cm} g=0,1,...,[N_{s1}/2D-1] \, ,
\label{r-r0-T}
\end{equation}
where the index $d$ has values $d=1,2$, the index $l$ has values $l = \pm 1$, and ${\vec{T}}_{d,l}^{\,g}$ is a $T$ vector. 
It has Cartesian components ${\vec{T}}_{d,l}^{\,g}=[T_{d,l,1}^{\,g},T_{d,l,2}^{\,g}]$ 
for the square lattice and reads $\vec{T}_{1,l}^{\,g}=[T_{1,l,1}^{\,g}]$
for 1D. There 
are $N_{s1}/2D$ $T$ vectors ${\vec{T}}_{d,l}^{\,g}$,
one for each choice of the following Cartesian components,
\begin{eqnarray}
T_{d,l,i}^{\,g} & = & l\,a_s\,N_{i} 
\, ; \hspace{0.50cm} i = 1, 2 \, ,
\nonumber \\
N_d & = & 0,1,...,N_{a_s}/4 -1  
\, ; \hspace{0.15cm}
N_{\bar{d}} = -N_{a_s}/4 +1,...,-1,0,1,...,N_{a_s}/4 \, .
\label{xd-xd}
\end{eqnarray}
Here $d = 1,2$, $\bar{1} = 2$, $\bar{2} = 1$,
$ l = \pm 1$ and $N_d$ and $N_{\bar{d}}$ are consecutive integer 
numbers. The expressions provided in Eq. (\ref{r-r0-T}) apply to the
1D lattice provided that only the index value $d=1$
is considered. The single 1D component of the $T$ 
vectors is given by $T_{1,l,1}^{\,g} = l\,a_s\,N_1$ where
$N_1 = 0,1,...,N_{a_s}/4 -1$. 

The two-site-bond-type index $g=0,1,...,[N_{s1}/2D-1]$ labels 
the $N_{s1}/2D$ $T$ vectors 
${\vec{T}}_{d,l}^{\,g}$. For the model on the square
lattice it is defined in terms of the numbers $N_d$
and $N_{\bar{d}}$ given in Eq. (\ref{xd-xd}) and reads, 
\begin{eqnarray}
g & = & N_d + 2\vert N_{\bar{d}}\vert {N_{a_s}\over 4} 
\, ; \hspace{0.50cm} N_{\bar{d}} \leq 0 \, ,
\nonumber \\
& = & N_d + 2(N_{\bar{d}}-1){N_{a_s}\over 4} 
\, ; \hspace{0.50cm} N_{\bar{d}} > 0 \, .
\label{g-2D}
\end{eqnarray}
For 1D one has that $g=N_1=0,1,...,N_{s1}/2 -1$.

The values of the two-site-bond-type index $g$ are consecutive positive 
integers whose minimum value $g=0$ corresponds to
$N_1=N_2=0$ so that,
\begin{equation}
T_{d,l}^{\,0} = 0 \, .
\label{T0}
\end{equation}
For the model on the square lattice the maximum value 
$g=[N_{s1}/2D-1]$ refers to $N_d=N_{a_s}/4 -1$
and $N_{\bar{d}} =N_{a_s}/4$. 

Each pair of values (and value) of the Cartesian coordinates of 
the $T$ vector ${\vec{T}}_{d,l}^{\,g}=[T_{d,l,1}^{\,g},T_{d,l,2}^{\,g}]$ 
for the square lattice (and $\vec{T}_{1,l}^{\,g}=[T_{1,l,1}^{\,g}]$
for 1D) corresponds to exactly one of the values  
$g=0,1,...,[N_{s1}/2D-1]$ so that,
\begin{equation}
\sum_{N_d=0}^{N_{a_s}/4-1}
\sum_{N_{\bar{d}}=-N_{a_s}/4+1}^{N_{a_s}/4} \equiv 
\sum_{g=0}^{N_{s1}/4-1}  
\,  ; \hspace{0.50cm} D=2 
\, ; \hspace{0.50cm}
\sum_{N_1=0}^{N_{a_s}/4-1} \equiv  
\sum_{g=0}^{N_{s1}/2-1}  
\,  ; \hspace{0.50cm} D=1
\, .
\label{link-sum}
\end{equation}

Two-site bonds with the same $g$ value
and $d\neq d'$ and/or $l\neq l'$ are
{\it equivalent two-site bonds}. Those are of the same
type but belong to different families.
Furthermore, $T$ vectors
${\vec{T}}_{d,l}^{\,g}$ and ${\vec{T}}_{d',l'}^{\,g}$
with the same value of $g$ and $d\neq d'$ and/or 
$l\neq l'$ are related as follows,
\begin{equation}
{\vec{T}}_{d,l}^{\,g} = ll'\left[\delta_{d,d'} +
\delta_{{\bar{d}},d'}{\bf \sigma_x}\right]
{\vec{T}}_{d',l'}^{\,g} \, ; \hspace{0.50cm}
{\bf \sigma_x} =\left[
\begin{array}{cc}
0 & 1 \\
1 & 0 \nonumber
\end{array}\right] \, ,
\label{T-T'-g}
\end{equation}
where ${\bf \sigma_x}$ is the usual Pauli matrix. 

As described below in terms of suitable
operators, a $s1$ bond particle of real-space coordinate
$\vec{r}_j$ is a superposition of $N_{a_{s}}^D/2=N_{a_{s1}}^D=N_{s1}$
two-site bonds, each being associated with a 
link vector $\vec{r}_{d,l}^{\,g}$. For each site of the
spin effective lattice there is {\it exactly one}
other site of the same lattice such that the bond connecting
the two sites has center at $\vec{r}_{j}+{\vec{r}_{d,l}}^{\,0}$.
Therefore, any two-site bond of the same family involves two
sites of well-defined real-space coordinate 
$\vec{r}_{j}+{\vec{r}_{d,l}}^{\,0}-\vec{r}_{d,l}^{\,g}$ and 
$\vec{r}_{j}+{\vec{r}_{d,l}}^{\,0}+\vec{r}_{d,l}^{\,g}$,
which do not contribute together to any other two-site bond
of the same family. 

An important quantity is the distance between the two sites of
a two-site bond, which we call {\it two-site bond length}. It is independent of the real-space 
coordinate $\vec{r}=\vec{r}_{j}+{\vec{r}_{d,l}}^{\,0}$ of the two-site bond center 
and is fully determined by the link vector $\vec{r}_{d,l}^{\,g}$ and
thus depends only on the two-site bond type associated with the index $g$. For $D=2$ and $D=1$ it reads,
\begin{equation}
\xi_{g} \equiv \vert 2\vec{r}_{d,l}^{\,g}\vert  
= a_s\sqrt{(1+2N_d)^2 + (2N_{\bar{d}})^2} 
\, ; \hspace{0.5cm}
\xi_{g} \equiv \vert 2x_{1,l}^{\,g}\vert  
= a_s\,(1+2N_1) \, ,
\label{xi-L}
\end{equation}
respectively. Its minimum and maximum values are,
\begin{equation}
{\rm min}\,\xi_{g} = \xi_{0} = a_s 
\, ; \hspace{0.5cm}
{\rm max}\,\xi_{g} = \sqrt{2}\,a_s\,(N_{a_s}/2-1)
+ {\cal{O}} (1/N_a) \, ,
\label{max-min-xi-L}
\end{equation}
for the square lattice and 
${\rm min}\,\xi_{g} = \xi_{0} = a_s$
and ${\rm max}\,\xi_{g} = a_s\,(N_{a_s}/2-1)$
for 1D. 

For 1D, two-site bonds with different $g$ values have different length $\xi_{g}$. In turn, for
the square lattice there are two-site bonds of different
type and hence different $g$ values that have the same length $\xi_{g}$. Indeed,
analysis of the two-site-bond-length expression
of Eq. (\ref{xi-L}) reveals that for the latter lattice two-site bonds with different $g$
values and numbers 
$[N_d,N_{\bar{d}}]$ and $[N_d',N_{\bar{d}}']$, 
respectively, such that $N_d=N_d'$ and 
$N_{\bar{d}}=-N_{\bar{d}}'$ have the same length. 

The set of values of the numbers $N_1$ and $N_2$
given in Eq. (\ref{xd-xd}) for the model on the square lattice imply that the maximum value
of the two-site bond length is $\sqrt{D}\,a_s\,(N_{a_s}/2-1)$
rather than $\sqrt{D}\,a_s\,(N_{a_s}-1)$. Indeed and as
mentioned above, the two-site bonds contributing 
to a $s1$ bond particle of the $N^h_{s1}=0$ configuration state 
are independent. It is then required that each
bond involves two sites that participate simultaneously in 
exactly one two-site bond. Within the torus row and column periodic boundary
conditions for the spin effective square lattice 
such a requirement is fulfilled provided that 
the range of the numbers $N_1$ and $N_2$
is that given in Eq. (\ref{xd-xd}).  

For each family of two-site bonds associated with
a $s1$ bond particle of real-space coordinate $\vec{r}_{j}$
there is a {\it primary two-site bond}. It corresponds to $g=0$ and
thus connects two nearest-neighboring sites of the spin effective lattice,
one of them having the same real-space coordinate 
$\vec{r}_{j}=\vec{r}-{\vec{r}_{d,l}}^{\,0}$ as the $s1$ 
bond particle and corresponding local $s1$ fermion. For primary two-site bonds the link vector $\vec{r}_{d,l}^{\,g}$
reads $\vec{r}_{d,l}^{\,g}={\vec{r}_{d,l}}^{\,0}$ where the primary
link vector ${\vec{r}_{d,l}}^{\,0}$ is
given in Eq. (\ref{r-r0-T}). 

For the model on the square lattice
there are four primary two-site bonds, one per family.
Their link vectors ${\vec{r}_{d,l}}^{\,0}$ have
components such that $N_1=N_2=0$ in Eqs. (\ref{r-r0-T}) 
and (\ref{xd-xd}). Therefore, the primary two-site bonds have minimum
length $\xi_{{\vec{r}}_{d,l}^{\,0}}=a_s$. Alike the 
remaining two-site bonds of its family, the center of a primary 
two-site bond is located at $\vec{r}=\vec{r}_{j}+{\vec{r}_{d,l}}^{\,0}$.
For the square lattice there are two horizontal primary two-site bonds 
whose centers are located at $\vec{r}_{j}+\vec{r}_{1,l}^{\,0}$ with 
$l=\pm 1$ and two vertical primary two-site bonds whose centers are
located at $\vec{r}_{j}+\vec{r}_{2,l}^{\,0}$ with $l=\pm 1$.
In the case of the 1D lattice there are two 
primary two-site bonds whose centers are located at 
$\vec{r}_{j}+\vec{r}_{1,l}^{\,0}$ with $l=\pm 1$.

\subsection{Partitions and $g$-primary partitions}

The building blocks of a $N_{s1}^h=0$ configuration state
are singlet pairs of spinons on sites $\vec{r}_j^{\,-}$
and $\vec{r}_j^{\,+}$ of the spin effective lattice,
\begin{eqnarray}
\vert\vec{r}_j^{\,-},\vec{r}_j^{\,+}\rangle
& = & {1\over \sqrt{2}}\left(\vert\uparrow_{\vec{r}_j^{\,-}}
\downarrow_{\vec{r}_j^{\,+}}\rangle -
\vert\downarrow_{\vec{r}_j^{\,-}}
\uparrow_{\vec{r}_j^{\,+}}\rangle\right) \, ,
\nonumber \\
\vec{r}_j^{\,\mp} & = & \vec{r}_j + \vec{r}_{d,l}^{\,0}
\mp \vec{r}_{d,l}^{\,g} 
\, ; \hspace{0.5cm}
d = d (j) \, , \hspace{0.15cm}
l = l (j) \, , \hspace{0.15cm}
g = g (j) \, .
\label{state-+}
\end{eqnarray}
Here, as given above, the values of the integer indices $d$, $l$, and $g$ are
in the ranges $d=1,2$, $l=\pm 1$, and $g\in (0,N_{s1}/2D-1)$,
respectively. Such values are a function of the index $j=1,...,N_{a_{s1}}^D$,
where $N_{a_{s1}}^D=N_{a_{s}}^D/2=N_{s1}$, of the real-space coordinate
$\vec{r}_j$ of each $s1$ bond particle in the sub-lattice 1.
Indeed, the two sites of such pairs of sites are connected by two-site bonds 
and each bond is associated with exactly one $s1$ bond particle. 

Each connection involving $N_{a_{s}}^D/2 =N_{a_{s1}}^D=N_{s1}$ different bonds determines a {\it partition}. 
A partition is a $N_{a_{s}}^D$-spinon occupancy configuration where each site of
the spin effective lattice is linked to one site only
and all $N_{a_{s}}^D=2N_{s1}$ sites then correspond to $N_{a_{s}}^D/2 =N_{a_{s1}}^D=N_{s1}$
well-defined two-site bonds, each belonging to a different $s1$ bond 
particle. 

Each of the $N_{s1}$ $s1$ bond particles contributes with exactly one of its 
$N_{a_s}^D/2=N_{s1}$ two-site bonds to a partition. In a partition any site of the spin effective 
lattice participates in one bond only and there is
a single two-site bond attached to each site, which connects it to some
other site. And the latter site is attached to the former site only.
  
The $N_{s1}^h=0$ configuration state may be represented as,
\begin{equation}
\vert \phi \rangle =
\sum_{P} C_P
\prod_{j=1}^{N_{s1}}\vert\vec{r}_j^{\,-},\vec{r}_j^{\,+}\rangle 
\, ; \hspace{0.5cm}
C_P = \prod_{j=1}^{N_{s1}} e^{i\phi_{j,s1}}\,h_{g(j)}^* \, ,
\label{phi-+}
\end{equation}
where the coefficients $h_g$ are associated with the local $s1$ fermion bond
weights and appear in the  local $s1$ fermion operators 
defined below, $\phi_{j,s1}$ is given in Eq. (\ref{f-an-operators}) for the $\alpha\nu=s1$ branch,
the product of singlet states $\prod_{j=1}^{N_{s1}}\vert\vec{r}_j^{\,-},\vec{r}_j^{\,+}\rangle$ 
refers to a bond state associated with a given partition, and
the summation $\sum_{P}$ runs over all partitions.

A particular type of partition involves $N_{a_{s}}^D/2=N_{s1}$ 
identical two-site bonds. The indices $d$, $l$, and $g$ of identical two-site bonds 
have the same values but correspond to $s1$ bond
particles with different real-space coordinates $\vec{r}_j$.
Such a partition involves a set of $N_{a_{s}}^D/2=N_{s1}$ identical 
two-site bonds whose bonds connect different sites of the 
spin effective lattice, each site being linked to exactly one site.  
In this case the two real-space coordinates of the 
$N_{a_{s}}^D/2=N_{s1}$ pairs of sites are connected by the same 
real-space vector $2\vec{r}_{d,l}^{\,g}$, so that each two-site bond has the same length. 
\begin{figure}
\includegraphics[width=3.5cm,height=3.5cm]{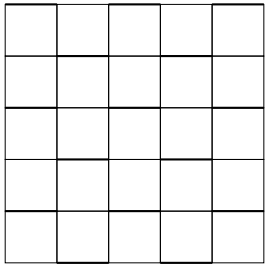}
\caption{Sub-domain of the spin effective lattice 
with a primary partition  of 
$d=1$ horizontal two-site bonds for the model on the square lattice. The primary two-site bonds are 
represented by the thick horizontal lines connecting 
two sites of the spin effective lattice. For a reference frame 
where the site located at the corner on the left-hand
side and lower limit of the squared sub-domain has 
Cartesian coordinates $(0,0)$ the family
indices read $d=1$, $l=+1$ if that site 
belongs to sub-lattice 2, whereas 
$d=1$, $l=-1$ if instead it belongs to 
sub-lattice 1.}
\end{figure}

Each of the $N_{a_s}^D/2=N_{s1}$ bonds of a partition 
involves two sites of real-space coordinates $\vec{r}_{j'}$ 
and $\vec{r}_{j'}+2\vec{r}_{d,l}^{\,g}$ that belong 
to different sub-lattices, where $j'=1,...,N_{a_{s1}}^D$. The relation to the notation
used above for the real-space coordinates of the
two sites of a bond is as follows,
\begin{equation}
\vec{r}_{j'} = \vec{r}_{j}+{\vec{r}_{d,l}}^{\,0}-\vec{r}_{d,l}^{\,g} 
\, ; \hspace{0.5cm}
\vec{r}_{j'}+2\vec{r}_{d,l}^{\,g} = 
\vec{r}_{j}+{\vec{r}_{d,l}}^{\,0}+\vec{r}_{d,l}^{\,g} \, ,
\hspace{0.25cm} j,j'=1,...,N_{a_{s1}}^D \, .
\label{corresp}
\end{equation}
Here both $\vec{r}_{j'}$ and $\vec{r}_{j}$ are real-space
coordinates of the sub-lattice 1 and thus of the $s1$ effective lattice. 
Except for a primary two-site bond, one has that $j\neq j'$. The site of 
real-space coordinate $\vec{r}_{j}$ 
is that of the corresponding local $s1$ fermion. It is the closest to the two-site bond center
at $\vec{r}_{j}+{\vec{r}_{d,l}}^{\,0}$. In turn,
$\vec{r}_{j'}$ is the real-space coordinate of one of
the two sites of the spin effective lattice involved
in the two-site bond. 
 
When a partition is a set of $N_{a_s}^D/2=N_{s1}$ identical primary
two-site bonds, all site pairs involve nearest-neighboring sites of the 
spin effective lattice. It is then called a {\it primary partition}. 
The family of a primary partition is labeled by the indices $d$
and $l$ of the corresponding identical two-site bonds. Figures 1 and 2
represent primary partitions
of $d=1$ horizontal and $d=2$ vertical two-site bonds, respectively, 
for a sub-domain of the spin effective lattice of the model on the 
square lattice.

A useful concept is that of a {\it $g$-primary partition}. It is defined 
as the superposition of the $2D=2,4$ primary partitions. It follows 
that a $g$-primary partition contains $D\,N_{a_{s}}^D=2D\,N_{s1}$ 
primary two-site bonds. In such a 
configuration each of the $N_{a_{s}}^D=2N_{s1}$ sites of the 
spin-effective lattice has $2D=2,4$ two-site bonds attached to it.
Figure 3 shows a sub-domain of the spin-effective 
lattice with the $g$-primary partition 
of the $N^h_{s1}=0$ configuration state
for the model on the square lattice. 

\subsection{The local $s1$ fermion and $s1$ bond-particle operators}

The spinon occupancy configurations considered
above are similar to those associated
with multi-spin wave functions of spin-singlet states
used by several authors \cite{Fazekas,Fradkin,Auerbach}.
Such wave functions are often constructed having as 
building blocks two-site and two-spin spin-singlet configurations
similar to that of Eq. (\ref{state-+}) except
that here the two corresponding spinons refer to sites singly
occupied by rotated electrons. Thus here they correspond to
$U/4t>0$ rather than only to $U/4t\gg 1$ for the previous approaches. In those
one also connects pairs of lattice sites with
bonds and each such a connection determines
a partition. However, here bonds involve sites of the
spin effective lattice whereas those of previous related
studies refer to the sites of the original lattice of
the corresponding quantum problems.

For a given partition one can define a valence bond state
\cite{Fazekas,Fradkin,Auerbach}
as a product of singlet states and represent
an arbitrary singlet by a superposition
of valence bond states of general form similar
to that of Eq. (\ref{phi-+}). That involves a sum
over all partitions of the lattice into set of pairs.
However, for general wave functions 
such a decomposition works in general very
badly. Indeed, valence-bond states are not
orthogonal and their basis is overcomplete.
Fortunately, here each
of the $N_{a_{s}}^D/2=N_{s1}$ two-site bonds of a partition belongs
to a different $s1$ bond particle so that
each of such particles contributes to a
partition with exactly
one bond. Such a restriction
eliminates the unwanted and unphysical
contributions and renders the bond states
free of the overcompleteness problem.

The $s1$ bond particle operators $g_{{\vec{r}}_{j},s1}$ and $g_{\vec{r}_{j},s1}^{\dag} = 
\left(g_{{\vec{r}}_{j},s1}\right)^{\dag}$ may be expressed in terms of superpositions of two-site bond operators.
We recall that according to Eq. (\ref{f-an-operators}) for the $\alpha\nu =s1$ branch, the operator $g^{\dag}_{{\vec{r}}_{j},s1}$
is related to the corresponding local $s1$ fermion operator by the transformation
$f^{\dag}_{{\vec{r}}_{j},s1} = e^{i\phi_{j,s1}}\,g^{\dag}_{{\vec{r}}_{j},s1}$. 
For the one- and two-electron subspace spanned by states whose numbers obey
the approximate selection rules of Eq. (\ref{srs-0-ss}) as $N_a^D\rightarrow\infty$, 
the operators $g_{{\vec{r}}_{j},s1}$ (and $g^{\dag}_{{\vec{r}}_{j},s1}$)
that annihilate (and create) a 
$s1$ bond particle at a site of the spin effective lattice
of real-space coordinate ${\vec{r}}_{j}$ have the following 
general form both for the $1D$ and square lattices,
\begin{eqnarray}
g_{\vec{r}_{j},s1} & = & \sum_{g=0}^{N_{s1}/2D-1} h_{g}\, a_{\vec{r}_{j},s1,g}  
\, ; \hspace{0.35cm} g_{\vec{r}_{j},s1}^{\dag} = 
\left(g_{{\vec{r}}_{j},s1}\right)^{\dag} \, ,
\nonumber \\
a_{\vec{r}_{j},s1,g} & = &
\sum_{d=1}^{D}\sum_{l=\pm1}
\, b_{\vec{r}_{j}+{\vec{r}_{d,l}}^{\,0},s1,d,l,g} 
\, ; \hspace{0.50cm} D=1,2 \, .
\label{g-s1+general}
\end{eqnarray}
The operators $a_{\vec{r}_{j},s1,g}^{\dag}$ and $a_{\vec{r}_{j},s1,g}$ appearing in these expressions
create and annihilate, respectively, a superposition of $2D=2,4$ two-site bonds of 
the same type and $b_{\vec{r},s1,d,l,g}^{\dag}$ and $b_{\vec{r},s1,d,l,g}$ are
two-site bond operators whose expression is given below. 
\begin{figure}
\includegraphics[angle=90,width=3.5cm,height=3.5cm]{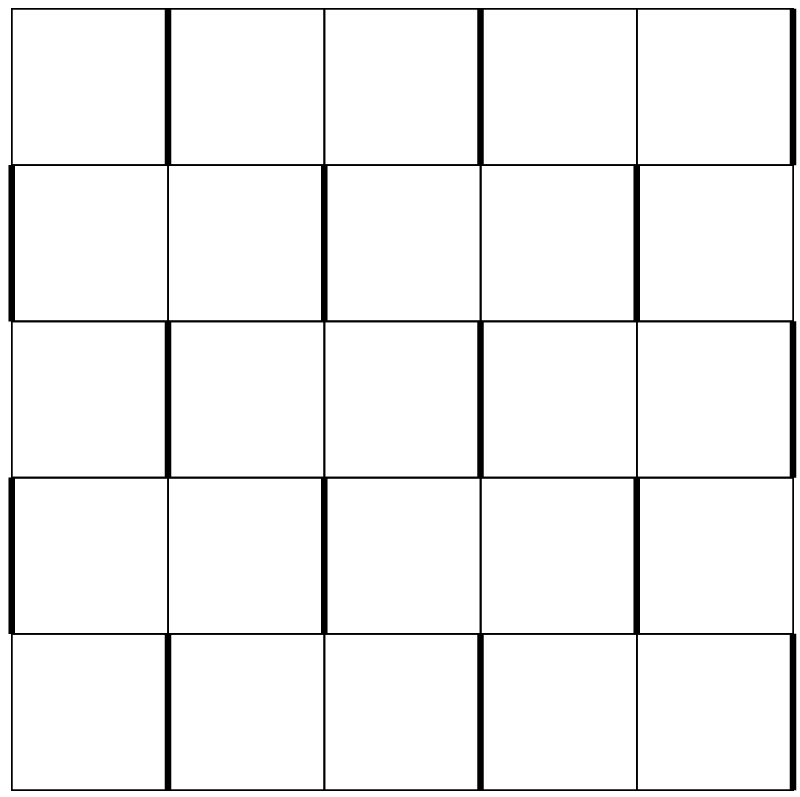}
\caption{Sub-domain of the spin effective lattice 
with a primary partition  of 
$d=2$ vertical two-site bonds for the model on the square lattice. The primary two-site bonds are 
represented by the thick vertical lines connecting 
two sites of the spin effective lattice. Alike in Fig. 1, 
the family indices read $d=2$, $l=+1$ if the site of
Cartesian coordinates $(0,0)$
belongs to sub-lattice 2, whereas 
$d=2$, $l=-1$ if instead it belongs to 
sub-lattice 1.}
\end{figure}

For the model on the square (and 1D) lattice the four (and two) primary two-site bonds
associated with the operators $a_{\vec{r}_{j},s1,0}^{\dag}$ and 
$a_{\vec{r}_{j},s1,0}$ are behind most of the spectral weight 
of a local $s1$ fermion and corresponding $s1$ bond particle of real-space coordinate 
$\vec{r}_{j}$. Consistently, the absolute value $\vert h_{g}\vert$
of the coefficients $h_{g}$ appearing in the expressions
of such operators given in Eq. (\ref{g-s1+general}) 
decreases for increasing two-site bond length $\xi_{g}$. 
These coefficients obey the normalization sum-rule,
\begin{equation}
\sum_{g=0}^{[N_{s1}/2D-1]} \vert h_{g}\vert^2 = {1\over 2D} 
\,  ; \hspace{0.50cm} D=1,2 \, .
\label{g-s1+sum-rule}
\end{equation}
The exact dependence of $\vert h_{g}\vert$ on the length $\xi_{g}$, value of $U/4t$, and hole concentration
$x$ remains for the Hubbard model an involved open problem. The suitable use
of this sum-rule and related symmetries leads though to useful information,
as discussed below. (The real-space coordinates ${\vec{r}}_{j}$ of the
local $s1$ fermion operators and corresponding $s1$ bond-particle operators of Eq. (\ref{g-s1+general}) have been chosen to inherently 
being those of the sub-lattice 1.)

The two-site bond operators $b_{\vec{r},s1,d,l,g}^{\dag}$ 
and $b_{\vec{r},s1,d,l,g}$
appearing in Eq. (\ref{g-s1+general}) are
associated with a well-defined bond connecting the
two sites of real-space coordinates $\vec{r}-\vec{r}_{d,l}^{\,g}$
and $\vec{r}+\vec{r}_{d,l}^{\,g}$, respectively.
Their expression can be obtained by 
considering the following related operator, 
\begin{eqnarray}
& & {(-1)^{d-1}\over\sqrt{2}}
[(1-{\tilde{n}}_{\vec{r}-\vec{r}_{d,l}^{\,g},\downarrow})
{\tilde{c}}^{\dag}_{\vec{r}-\vec{r}_{d,l}^{\,g},\uparrow}\,
{\tilde{c}}^{\dag}_{\vec{r}+\vec{r}_{d,l}^{\,g},\downarrow}\, 
(1-{\tilde{n}}_{\vec{r}+\vec{r}_{d,l}^{\,g},\uparrow})
\nonumber \\
& - & (1-{\tilde{n}}_{\vec{r}-\vec{r}_{d,l}^{\,g},\uparrow})\,
{\tilde{c}}^{\dag}_{\vec{r}-\vec{r}_{d,l}^{\,g},\downarrow}\,
{\tilde{c}}^{\dag}_{\vec{r}+\vec{r}_{d,l}^{\,g},\uparrow}\,
(1-{\tilde{n}}_{\vec{r}+\vec{r}_{d,l}^{\,g},\downarrow})]
\nonumber \\
& = & f_{\vec{r}-\vec{r}_{d,l}^{\,g},c}^{\dag}\,f_{\vec{r}+\vec{r}_{d,l}^{\,g},c}^{\dag} \,
b_{\vec{r},s1,d,l,g}^{\dag} \, .
\label{singlet-conf}
\end{eqnarray}
Its first expression provided here is in terms of the rotated-electron operators of Eq. (\ref{rotated-operators}).
The second expression is in terms of the $c$ fermion operators given in Eq. (\ref{fc+}) and the following two-site bond operator,
\begin{equation}
b_{\vec{r},s1,d,l,g}^{\dag} = 
{(-1)^{d-1}\over\sqrt{2}}\left(\left[{1\over 2}+s^{x_3}_{\vec{r}-\vec{r}_{d,l}^{\,g}}\right]
s^-_{\vec{r}+\vec{r}_{d,l}^{\,g}} - \left[{1\over 2}+s^{x_3}_{\vec{r}+\vec{r}_{d,l}^{\,g}}\right]
s^-_{\vec{r}-\vec{r}_{d,l}^{\,g}}\right) \, ,
\label{g-s-l}
\end{equation}
such that $b_{\vec{r},s1,d,l,g} = \left(b_{\vec{r},s1,d,l,g}^{\dag}\right)^{\dag}$.
Here the spinon operators are those given in Eq. (\ref{sir-pir}).
The second expression of Eq. (\ref{singlet-conf}) is obtained from the use of 
Eq. (\ref{c-up-c-down}). 
\begin{figure}
\includegraphics[width=3.75cm,height=4.50cm]{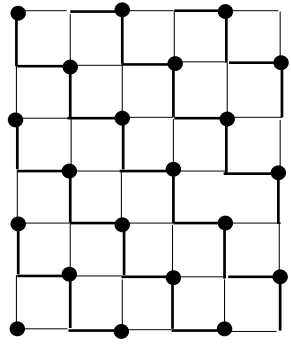}
\caption{Sub-domain of the spin-effective lattice 
representing the $g$-primary partition 
of a $N^h_{s1}=0$ configuration state
for the model on the square lattice. The sites belonging to the 
sub-lattice 1 are represented by 
filled circles. The horizontal (and vertical) thin 
and thick lines refer to $d=1$ (and $d=2$) $l=+1$ and 
$l=-1$ primary two-site bonds, respectively. In a $g$-primary partition of the 
model on the square lattice each 
site of the spin effective lattice has four two-site bonds attached to it.}
\end{figure}

On combining the local $s1$ fermion operator $f^{\dag}_{{\vec{r}}_{j},s1}$ expression
given in Eq. (\ref{f-an-operators}) for the $\alpha\nu=s1$ branch with Eqs. (\ref{g-s1+general})
and (\ref{g-s-l}) one arrives to the following equations for the operators
$f^{\dag}_{{\vec{r}}_{j},s1}$ and $f_{{\vec{r}}_{j},s1}$,
\begin{eqnarray}
f_{\vec{r}_{j},s1}^{\dag} & = & 
\left(f_{{\vec{r}}_{j},s1}\right)^{\dag} \, ; \hspace{0.35cm}
f_{\vec{r}_{j},s1} = e^{-i\phi_{j,s1}}\,\sum_{g=0}^{N_{s1}/2D-1} 
\sum_{d=1}^{D}\sum_{l=\pm1}
{(-1)^{d-1}h_{g}\over\sqrt{2}}
\nonumber \\
& \times &
\left(\left[{1\over 2}+s^{x_3}_{\vec{r}_{j}+{\vec{r}_{d,l}}^{\,0}-\vec{r}_{d,l}^{\,g}}\right]
s^-_{\vec{r}_{j}+{\vec{r}_{d,l}}^{\,0}+\vec{r}_{d,l}^{\,g}} - \left[{1\over 2}+s^{x_3}_{\vec{r}_{j}+{\vec{r}_{d,l}}^{\,0}+\vec{r}_{d,l}^{\,g}}\right]
s^-_{\vec{r}_{j}+{\vec{r}_{d,l}}^{\,0}-\vec{r}_{d,l}^{\,g}}\right) \, ,
\nonumber \\
\phi_{j,s1} & = & \sum_{j'\neq j}f^{\dag}_{{\vec{r}}_{j'},s1}
f_{{\vec{r}}_{j'},s1}\,\phi_{j',j,s1} \, ; \hspace{0.35cm}
\phi_{j',j,s1} = \arctan \left({{x_{j'}}_2-{x_{j}}_2\over {x_{j'}}_1-{x_{j}}_1}\right) \, .
\label{f-s1-operator}
\end{eqnarray}
The phase factor $(-1)^{d-1}$ that appears both here and in the
operator of Eq. (\ref{g-s-l}) is associated with the $d$-wave symmetry of the $s1$ fermion
spinon pairing of the model on the square lattice.
The introduction of such a phase-factor refers to a self-consistent
procedure. It follows from the $d$-wave symmetry of
the energy dispersion found in Ref. \cite{companion} for
the $s1$ fermions. Their energy-dispersion $d$-wave symmetry
arises naturally from symmetries beyond the form of
the operators introduced in Eqs. (\ref{singlet-conf}) and (\ref{g-s-l}).

According to the above discussions, the 
real-space coordinates $\vec{r}-\vec{r}_{d,l}^{\,g}$ and $\vec{r}+\vec{r}_{d,l}^{\,g}$ 
involved in the operators of Eqs. (\ref{singlet-conf}) and (\ref{g-s-l}) and
in the operators of Eq. (\ref{f-s1-operator}) for $\vec{r}=\vec{r}_{j}+{\vec{r}_{d,l}}^{\,0}$
correspond to two sites that belong to different sub-lattices of the 
spin effective lattice. Moreover, $\vec{r}=\vec{r}_{j}+{\vec{r}_{d,l}}^{\,0}$ 
is the two-site bond center and
the primary link vector ${\vec{r}_{d,l}}^{\,0}$ and link vector
$\vec{r}_{d,l}^{\,g}$ are given in Eqs. (\ref{r-r0-T}) and (\ref{xd-xd}), respectively.
In the configuration generated by the operator of Eq. (\ref{singlet-conf})
the two sites are singly occupied  by the rotated electrons 
associated with the operators appearing in the first expression 
of that equation. 

The $N^h_{s1}=0$ configuration state (\ref{phi-+}) can
be written in terms of $s1$ fermion operators $f^{\dag}_{{\vec{r}}_{j},s1}$
defined by Eq. (\ref{f-s1-operator}) as follows,
\begin{equation}
\vert \phi \rangle = \prod_{j=1}^{N_{s1}} f_{\vec{r}_{j},s1}^{\dag}\vert 0_{s};N_{a_{s}}^D\rangle \, .
\label{phi-+-s1}
\end{equation}
Here $\vert 0_{s};N_{a_{s}}^D\rangle$ is the spin $SU(2)$ vacuum on the right-hand side of Eq. (\ref{vacuum})
whose number $N_{a_{s}}^D=M^{de}_{s,+1/2}$ of deconfined $+1/2$ spinons equals
in the present case the number $M^{co}_s=2N_{s1}$ of confined spinons contained in the whole set
of $N_{s1}$ $s1$ fermions of the state $\vert \phi \rangle$ under consideration. Such a vacuum corresponds 
to a fully polarized spin-up spinon configuration. On imposing the equality of the general
configuration states given in Eqs. (\ref{phi-+}) and  (\ref{phi-+-s1}), respectively,
one defines which of the processes generated by application of the operators of 
Eqs. (\ref{g-s1+general})-(\ref{f-s1-operator}) onto the neutral states that span the subspaces 
of the one- and two-electron subspace are physical. Indeed the equality 
of the general configuration states given in such two equations
implies several restrictions onto the 
processes generated by the two-site bond operators (\ref{g-s-l}). Those
can be summarized in a few corresponding rules for exclusion of 
the unwanted spin configurations generated by unphysical processes. 

In order to introduce such rules, it is useful to briefly report and discuss a few
issues related to the local $s1$ fermion operators and corresponding $s1$ bond-particle
operators. Those follow from the algebra given in Eqs. 
(\ref{albegra-s-p-m})-(\ref{albegra-s-sz-com}) of Appendix D
for the basic spinon operators of the two-site bond
operators expressions provided in Eq. (\ref{g-s-l}).
The $s1$ bond-particle operators $g^{\dag}_{{\vec{r}}_{j},s1}$
and $g_{{\vec{r}}_{j},s1}$ of Eq. (\ref{g-s1+general}) and corresponding
local $s1$ fermion operators $f^{\dag}_{{\vec{r}}_{j},s1}$
and $f_{{\vec{r}}_{j},s1}$ of Eq. (\ref{f-s1-operator}) 
involve a sum of $N_{a_{s}}^D/2=N_{s1}$ two-site bond operators of general form
given in Eq. (\ref{g-s-l}) with $N_{a_{s}}^D/4D=N_{s1}/2D$ 
of such operators per family.
The number of unoccupied sites $N^h_{s 1}$ of 
Eq. (\ref{Nas1-Nhs1})
refers to a subspace with a fixed number $N_{s1}$ of $s1$
fermions. In turn, the creation and annihilation of one local $s1$ fermion by application
of the operators $f^{\dag}_{{\vec{r}}_{j},s1}$
and $f_{{\vec{r}}_{j},s1}$, respectively,
onto the ground state involves a superposition of
$N_{a_{s}}^D/2=N_{s1}$ elementary processes, which do not
conserve the number of these objects. Each 
such an elementary process is generated by 
an operator $b_{\vec{r},s1,d,l,g}^{\dag}$ and
$b_{\vec{r},s1,d,l,g}$, respectively, whose expression
is given in Eq. (\ref{g-s-l}). 

Within the LWS representation mostly used in this paper, application
of the rotated-electron operators of 
Eq. (\ref{singlet-conf}) onto two rotated-electron unoccupied sites
of the original lattice generates two virtual processes. The first process
involves creation of two $c$ fermions and two 
deconfined $+1/2$ spinons. The second process
refers to creation of a spin-singlet two-site and two-spinon
configuration upon annihilation of two deconfined $+1/2$ spinons.
Indeed, from analysis of the expression provided in Eq. (\ref{g-s-l}), 
one finds that application of the operator $b_{\vec{r},s1,d,l,g}^{\dag}$  
onto the sites of real-space coordinates 
$\vec{r}-\vec{r}_{d,l}^{\,g}$ and $\vec{r}+
\vec{r}_{d,l}^{\,g}$ gives zero except when
both these sites refer to a deconfined
$+1/2$ spinon. This is consistent with an "occupied site" 
of the $s1$ effective lattice corresponding to two sites of the spin effective lattice with real-space 
coordinates $\vec{r}-\vec{r}_{d,l}^{\,g}$ and 
$\vec{r}+\vec{r}_{d,l}^{\,g}$, respectively, which refer to 
deconfined $+1/2$ spinons in the initial configuration state. 

Consistently with the studies of Ref. \cite{companion}, for excitations involving transitions
between configuration states under which a $s1$ fermion moves around in the $s1$ effective lattice
through elementary processes that conserve the
numbers $N_{s1}$ and $N^h_{s1}$, a deconfined 
$+1/2$ spinon plays the role of a unoccupied site 
of such an effective lattice. In contrast, in 
elementary processes involving the creation of a $s1$ 
fermion the two annihilated deconfined $+1/2$
spinons give rise to a $s1$ effective lattice "occupied site" in the final occupancy configuration. 
\begin{figure}
\includegraphics[width=5.04cm,height=1.8cm]{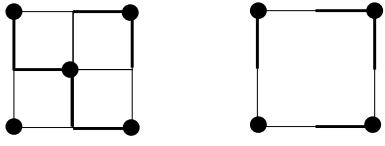}
\caption{Small sub-domain of two $g$-primary partitions
associated with $N^h_{s1}=0$ and $N^h_{s1}=2$ configuration states
for the model on the square lattice.
In the configuration of the figure on the left-hand side the 
filled circle at the middle of the sub-domain corresponds 
to the site whose real-space coordinate $\vec{r}_j$ is that of 
the $s1$ fermion. Annihilation of that 
object leads to the configuration on the
figure right-hand side. Note that for a primary $g$-basic 
partition this is equivalent to the suppression of 
the four two-site bonds attached to the above site of real-space coordinate 
$\vec{r}_j$.}
\end{figure}

According to the operator expression provided in Eq. (\ref{g-s-l}), 
upon acting onto the deconfined-spinon occupancies the operator 
$b_{\vec{r},s1,d,l,g}^{\dag}$ generates a superposition of two
configurations. For one of those the elementary
process generated by that operator flips the spin of the spinon at 
site $\vec{r}+\vec{r}_{d,l}^{\,g}$ and checks whether the spin 
of the spinon at site $\vec{r}-\vec{r}_{d,l}^{\,g}$ remains up. The 
elementary process generating the other configuration flips the spin
of the spinon at site $\vec{r}-\vec{r}_{d,l}^{\,g}$ and checks whether the 
spin of the spinon at site $\vec{r}+\vec{r}_{d,l}^{\,g}$ remains up. 
The relative phase factor $-1$ of the two configurations insures that the
$s1$ fermion created by the operator 
$f_{\vec{r}_j,s1}^{\dag}$ of Eq. (\ref{f-s1-operator}) is a
suitable superposition of spin-singlet configurations. 

Figure 4 shows a small sub-domain of two
$g$-primary partitions of $N^h_{s1}=0$
and $N^h_{s1}=2$ configuration states, respectively.
In the configuration on the figure left-hand
side the filled circle at the middle of the sub-domain 
corresponds to a site whose real-space coordinate 
$\vec{r}_j$ is that of a local $s1$ fermion. Application 
of the annihilation operator $f_{\vec{r}_{j},s1}$ of Eq. (\ref{f-s1-operator})  
onto that $g$-primary partition occurs
through the operator $a_{\vec{r}_{j},s1,0}$ 
also given in that equation. That leads to the configuration
on the figure right-hand side. Hence
annihilation of the local $s1$ fermion is 
for a $g$-primary partition equivalent
to the suppression of the four two-site bonds attached 
to the above site of real-space coordinate $\vec{r}_j$. 

A superficial analysis of the configurations shown
in Fig. 4 seems to indicate that there is one unoccupied 
site in the sub-domain of the final $N^h_{s1}=2$ configuration state. However, if
instead of the $g$-primary partition one
considers the corresponding four primary partitions, one finds that there are two
nearest-neighboring unoccupied sites. For the 
$d=1$ and $d=2$ primary partitions
these two sites belong to the same row and column, 
respectively.

Concerning the application of two-site bond operators and products of 
two-site bond operators onto spin configurations 
the restrictions arising from imposing that the representations
of the $N_{s1}^h=0$ configuration state given in 
Eqs. (\ref{phi-+}) and  (\ref{phi-+-s1}), respectively, 
are identical correspond to the following simple three rules whose fulfillment 
prevents the generation of unwanted and unphysical spin 
configurations:
\vspace{0.25cm}

{\it First rule} according to which application onto
a spin configuration of a two-site bond operator 
generates a physical spin configuration provided that 
in the initial spin configuration its 
sites of real-space coordinates $\vec{r}-\vec{r}_{d,l}^{\,g}$ and 
$\vec{r}+\vec{r}_{d,l}^{\,g}$, respectively, 
(i) refer to deconfined $+1/2$ spinons or (ii) are linked by a bond. 
\vspace{0.25cm}

{\it Second rule} states that in the processes generated by application onto a spin configuration of 
a product of a set of two-site bond operators the first rule applies to each
operator provided that (i) the two sites of each operator are not joined by
any other of the two-site bond operators of the set or (ii) both sites of some or all such operators
are the same as the two sites of one or several other operators. Otherwise,
application onto a spin configuration of the set of two-site bond operators
gives zero. Hence, each two-site bond operator cannot join a single site with
another operator of the set: It either joins none or both sites with one or several 
such operators. 
\vspace{0.25cm}

{\it Third rule} refers to when the two-site bond operators of a product of a set of such
operators correspond to the same $s1$ bond particle and the first and second rules are
obeyed. Then the processes generated by application onto a spin configuration of such a two-site 
bond operator product gives zero if in the initial spin configuration all involved
spin effective lattice sites refer to deconfined $+1/2$ spinons.
\vspace{0.25cm}

Such rules follow naturally from the definition of the subspace where 
the operators of Eqs. (\ref{g-s1+general})-(\ref{f-s1-operator}) act onto. The
main criterion is that such operators have been constructed to inherently 
generating a faithful representation provided that the corresponding
state (\ref{phi-+-s1}) represents the $N^h_{s1}=0$ configuration state 
and hence is identical to that given in Eq. (\ref{phi-+}) with the same 
value of $N_{s1}=N_{a_{s1}}^D$.
For instance, the second rule results from in all partitions of the 
summation on the right-hand side of Eq. (\ref{phi-+})
a site of the spin effective lattice being linked to exactly only one 
site. Indeed, in a given partition no two-site bonds
join the same site.  

\subsection{Consistency of the two-site bond weights to possible $x=0$ long-range and
$x>0$ short-range spin orders}

The exact dependence on the two-site bond length $\xi_{g}$ of the absolute value 
$\vert h_{g}\vert$ of the coefficients appearing in Eqs. (\ref{phi-+}) and (\ref{g-s1+sum-rule}) 
remains an open problem. However, if one assumes that $\vert h_{g}\vert$ 
has the following simple power-law dependence on that length,
\begin{equation}
\vert h_{g}\vert \approx  {C\over \xi_{g}^{\alpha_{s1}}} \, , 
\label{c-L-d}
\end{equation}
the two-site-bond-length expression (\ref{xi-L}) and
normalization condition (\ref{g-s1+sum-rule}) alone
imply $C^2$ be given by,
\begin{eqnarray}
C^2 & = & {a_s^{2\alpha_{s1}}\over 
4\sum_{N_d}
\sum_{N_{\bar{d}}}
[(1+2N_d)^2 + (2N_{\bar{d}})^2]^{-\alpha_{s1}}}
\,  ; \hspace{0.50cm} D=2 \, ,
\nonumber \\
C^2 & =  & {a_s^{2\alpha_{s1}}\over 
2\sum_{N_1}(1+2N_d)^{-2\alpha_{s1}}} \,  ; \hspace{0.50cm} D=1 \, .
\label{C-D-1-2}
\end{eqnarray}
Here the summations run over the range of $N_d$ and $N_{\bar{d}}$ 
(and $N_1$) values given in Eq. (\ref{link-sum}) for the model on the square (and 1D) 
lattice. An expression of the general form (\ref{c-L-d})
would imply that the dependence of $\vert h_{g}\vert$
on $U/4t$ and $x$ occurred only through that of the
exponent $\alpha_{s1} =  \alpha_{s1} (U/4t,x)$. 
\begin{figure}
\includegraphics[width=7cm,height=3.5cm]{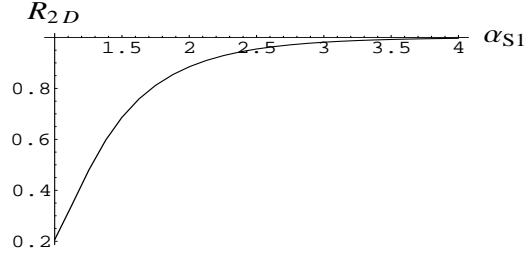}
\caption{The relative spectral weight $R_{2D}$ of the primary two-site bonds
of Eq. (\ref{rela-weight}) for the model on the
square lattice as a function of the exponent $\alpha_{s1}$
for the range $\alpha_{s1}\geq 1$. That
relative weight refers to
the simple power-law expressions given in 
Eqs. (\ref{c-L-d}) and (\ref{C-D-1-2}) for the coefficients
$\vert h_{g}\vert$ appearing in the 
summations of Eq. (\ref{g-s1+general}).
The physical range of the exponent $\alpha_{s1}$ corresponds
typically to $R_{2D}>0.9$ so that the primary
two-site bonds are behind most of the local $s1$ fermion
spectral weight. For $\alpha_{s1}\approx 2.5$ the
ratio $R_{2D}$ is larger than $0.9$ and approaches
quickly the unit upon further increasing $\alpha_{s1}$.}
\end{figure}
 
For the coefficients $\vert h_{g}\vert$ power-law expressions given in 
Eqs. (\ref{c-L-d}) and (\ref{C-D-1-2}), the ratio of the spectral weight 
of a primary two-site bond over the weight of the
corresponding two-site bond family,
\begin{equation}
R_{z} \equiv {\vert h_0\vert^2\over
\sum_{g=0}^{[N_{s1}/2D-1]} \vert h_{g}\vert^2} 
= 2D\,\vert h_0\vert^2
\,  , \hspace{0.50cm} z=1D,2D \, ,
\label{ratios-2-1}
\end{equation}
is given by,
\begin{eqnarray}
R_{2D} & = & {1\over \sum_{N_d}\sum_{N_{\bar{d}}}
[(1+2N_d)^2 + (2N_{\bar{d}})^2]^{-\alpha_{s1}}}
\,  ; \hspace{0.15cm} D=2 \, ,
\nonumber \\
R_{1D} & = & 
{1\over\sum_{N_1}(1+2N_d)^{-2\alpha_{s1}}}\,  ; \hspace{0.15cm} D=1 \, .
\label{rela-weight}
\end{eqnarray}
The ratios of Eq. (\ref{rela-weight}) refer to the  
$N_a^D\rightarrow\infty$ limit and are plotted in Figs. 5 and 6
for the model on the square and 1D lattice, respectively, as a function
of the exponent $\alpha_{s1}$ for the range $\alpha_{s1}\geq 1$.
For  $\alpha_{s1}$ slightly larger than
$5D/4$ where $D=1,2$ the ratios $R_{1D}$ ($D=1$) and $R_{2D}$ 
($D=2$) are larger than $0.9$. 

For $N_a^D\rightarrow\infty$ the exact coefficients $\vert h_{g}\vert$
are decreasing functions of the two-site bond length whose expressions most likely 
are not, at least for the whole two-site-bond-length range, of the simple power-law form given in Eqs. (\ref{c-L-d}) and 
(\ref{C-D-1-2}). However the exact sum-rules (\ref{g-s1+sum-rule}) 
together with the coefficients $\vert h_{g}\vert$ being decreasing
functions of the two-site bond length implies that at least for not a too small length the 
unknown exact coefficients $\vert h_{g}\vert$ fall off as in Eq. (\ref{c-L-d}) and
the corresponding ratios of Eq. (\ref{ratios-2-1}) have a behavior similar to that
shown in Figs. 5 and 6. Hence for the model on the square lattice one expects that 
$\alpha_{s1}>2.5$ so that the ratio $R_{2D}$ is larger than $0.9$ 
and the primary two-site bonds are behind most of the spectral weight of the
local $s1$ fermion, alike for the 1D lattice for $\alpha_{s1}>1.25$.

An important point is that at $x=0$, provided that $N_a^D\rightarrow\infty$  
one may have for the model on the square lattice a ground 
state with both long-range antiferromagnetic order and $s1$ fermion spinon pairing with $d$-wave symmetry. 
Indeed for values of the two-site bond length $\xi_{g}$ of Eq. (\ref{xi-L}) not too small the absolute value 
$\vert h_{g}\vert$ of the coefficients $h_{g}$ of the local $s1$ fermion
operators of Eq. (\ref{f-s1-operator}) is expected to fall off as $\vert h_{g}\vert \approx  C\,(\xi_{g})^{-\alpha_{s1}}$.
Provided that for the $x=0$, $\mu=0$, and $m=0$ absolute 
ground state the value of the exponent $\alpha_{s1}$ 
is approximately in the range $\alpha_{s1}\leq 5$, the corresponding
spin-singlet spinon $s1$ bond pairing of the 
$N_a^2/2=N_{s1}$ two-spinon $s1$ bond particles associated with such
operators can for $U/4t$ not too small have $d$-wave symmetry 
yet the corresponding spin occupancy configurations have 
long-range antiferromagnetic order \cite{Auerbach}. Note that at $x=0$ the spin effective lattice
equals the original lattice. The upper value $5$ is that obtained
from numerical results on spin-singlet two-spin bonds \cite{Fazekas,Fradkin,Auerbach}.

In turn, it is expected that for a well-defined domain of finite $x$ values the two-site bond lengths $\xi_{g}$ of Eq. (\ref{xi-L})
and corresponding absolute values $\vert h_{g}\vert$ of the coefficients $h_{g}$ of the local $s1$ fermion
operators of Eq. (\ref{f-s1-operator}) are such that the corresponding $s1$ fermion spinon pairing
refers to a short-range spin order rather than to a long-range antiferromagnetic order 
as that occurring at $x=0$ \cite{companion}.

\subsection{General $N^h_{s1}=0,1,2$ configuration states}

The one- and two-electron subspace as defined in this paper is spanned 
by states whose numbers obey the approximate selection rules of Eq. (\ref{srs-0-ss}). 
Such states have $s1$ fermion occupancies with none, one, and two $s1$ effective-lattice
unoccupied sites. There are several configuration states
with one or two unoccupied sites in that lattice. Those refer to different positions
of such sites associated with different local $s1$ fermion occupancy configurations.
We recall that the corresponding $s1$ fermion hole motion is independent of that of
the rotated electrons that singly occupy sites of the original lattice relative to its remaining
sites. The latter are instead described by the motion
of the $c$ fermions, whose occupancy configurations
correspond to the state representations of the hidden
global $U(1)$ symmetry. In turn, the configuration states
generated by the motion around in the $s1$ 
effective lattice of the $s1$ fermion holes associated with that lattice unoccupied sites refer to state
representations of the global spin $SU(2)$ symmetry.

The structure of the $N^h_{s1}=0$ configuration states is simpler than
that of the $N^h_{s1}=1,2$ configuration states. A key
property however simplifies the study of the latter states:
There is a one-to-one correspondence between the
occupancies of the $N_{s1}$ occupied sites of the $s1$ effective lattice 
of the partitions of a $N^h_{s1}=0$ configuration state 
and those of the occupied sites of the $s1$ effective lattice
of partitions of the $N^h_{s1}=1,2$ configuration states
with the same number $N_{s1}$ of local $s1$ fermions. The point
is that except for a change in the real-space coordinates 
of some of the sites, which accounts for the presence of one or two
unoccupied sites, the two-site bond configurations of the sites
of the spin effective lattice that are in one-to-one correspondence
with each other remain unaltered. Indeed under suitable boundary conditions accounting
for the presence of one or two unoccupied sites, the topology of the spinon occupancy configurations
associated with the $N_{s1}$ occupied sites of the $s1$ effective lattice 
remains unaltered.
\begin{figure}
\includegraphics[width=7cm,height=3.5cm]{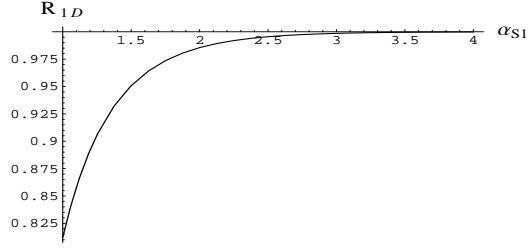}
\caption{The relative spectral weight $R_{1D}$ of the primary two-site bonds
of Eq. (\ref{rela-weight}) for the
1D model as a function of the exponent $\alpha_{s1}$
for the range $\alpha_{s1}\geq 1$. 
The ratio $R_{1D}$ approaches the unit upon increasing
the value of $\alpha_{s1}$ more quickly than
the ratio plotted in Fig. 5 for the model on the square lattice.
For $\alpha_{s1}=n_{s1}=1,2,3,...$ integer 
the relative weight reads $R_{1D}=1/[(1-2^{-2n_{s1}})\zeta (2n_{s1})]$
where $\zeta (x)$ is a Riemann's zeta function. For
instance, $R_{1D}=8/\pi^2\approx 0.811$ for 
$\alpha_{s1}=1$ and $R_{1D}=96/\pi^4\approx 0.986$ 
for $\alpha_{s1}=2$.}
\end{figure}

For simplicity in this paper we omit a detailed study of the $N^h_{s1}=1,2$ configuration states.
The main result is that the slightly changed $s1$ bond-particle operator expressions
that account for the presence of one or two $s1$ effective lattice unoccupied sites exactly
obey to the same algebra as the corresponding $s1$ bond-particle operators that act onto the 
$N^h_{s1}=0$ configuration state considered above. 

In Appendix D it is confirmed that for $U/4t>0$ and the subspaces where the $s1$ bond-particle operators of the 
one- and two-electron subspace are defined, upon acting onto the $s1$ effective lattice 
such operators anticommute on the same site of that lattice,
\begin{equation}
\{g^{\dag}_{{\vec{r}}_{j},s1},g_{{\vec{r}}_{j},s1}\} = 1 \, ;
\hspace{0.15cm}
\{g^{\dag}_{{\vec{r}}_{j},s1},g^{\dag}_{{\vec{r}}_{j},s1}\} =
\{g_{{\vec{r}}_{j},\alpha\nu},g_{{\vec{r}}_{j},s1}\}=0 \, ,
\label{g-local}
\end{equation}
and commute on different sites,
\begin{equation}
[g^{\dag}_{{\vec{r}}_{j},s1},g_{{\vec{r}}_{j'},s1}] =
[g^{\dag}_{{\vec{r}}_{j},s1},g^{\dag}_{{\vec{r}}_{j'},s1}]
= [g_{{\vec{r}}_{j},s1},g_{{\vec{r}}_{j'},1}] = 0 \, .
\label{g-non-local}
\end{equation}
Here $j\neq j'$. That algebra implies that upon acting onto the $s1$ effective lattice the $s1$ bond-particle 
operators are hard-core like. An important consequence of such a property is that one can perform an
extended Jordan-Wigner transformation $f^{\dag}_{{\vec{r}}_{j},s1} = e^{i\phi_{j,s1}}\,
g^{\dag}_{{\vec{r}}_{j},s1}$, whose phase operator $\phi_{j,s1}$ is given in Eq. (\ref{f-s1-operator}). 
It transforms the $s1$ bond particles into $s1$ fermions. As discussed in Section V-D, that for the model 
on the square lattice each $s1$ fermion has a flux tube of one flux quantum on average attached to it is 
consistent with the $s1$ fermion and $s1$ bond-particle wave functions obeying Fermi and Bose statistics, 
respectively.

The number of $N^h_{s1}=1$ configuration states is $N_{a_{s1}}^D = [N_{a_s}^D/2 + S_s] =N_{s1}+1$. Those
correspond to one-electron excited states. 
The $s1$ fermion occupancy configurations of Ref. \cite{companion} concerning $N^h_{s1}=1$ excited states generated 
by application onto $x\geq 0$ and $m=0$ ground states of one-electron operators can be expressed 
as suitable superpositions of such $N_{a_{s1}}^D = [N_{a_s}^D/2 + S_s] =N_{s1}+1$ configuration states.
On the other hand, the $s1$ fermion momentum occupancy of a $N^h_{s1}=2$ spin-triplet (and spin-singlet)
excited state is described by a suitable superposition of a set of $[N_{s1}+2]\,[N_{s1}+1]/2$ configuration 
states of spin $S_s=1$ with no $s2$ fermions, so that $N_{s2}=0$ (and of vanishing spin $S_s=0$ and 
a single $s2$ fermion so that $N_{s2}=1$.) Application of 
a local $s1$ fermion creation operator of real-space coordinate $\vec{r}_j$ onto such a ground state 
gives zero for most such configuration states. In turn, $N^h_{s1}=2$ configuration states whose two
unoccupied sites refer to the configuration plotted in Fig. 4 are transformed onto the $N^h_{s1}=0$ configuration 
state with one more $s1$ fermion than the initial state. This is so provided that the real-space coordinate of the 
corresponding central unoccupied site on the right-hand side of that figure coincides with that of the applied operator. 

Besides the unphysical processes excluded by the above three rules of $N^h_{s1}=0$ configuration states
and similar corresponding rules of $N^h_{s1}=1,2$ configuration states, one finds that at spinon number
$M_s=2S_c$ fixed value application of a local $s1$ 
fermion creation operator onto $N^h_{s1}=0,1$ configuration states gives always zero. In turn, application 
of a local $s1$ fermion annihilation operator whose real-space coordinate is that of an occupied site of the $s1$ 
effective lattice of a $N^h_{s1}=0,1,2$ configuration state transforms it into a state with two more unoccupied sites 
than the initial state.
        
\section{Concluding remarks}

This paper is a needed intermediate step between the extended global $SO(3)\times SO(3)\times U(1)$ symmetry 
recently found in Ref. \cite{bipartite} for the Hubbard model on any bipartite lattice and the results on the
square-lattice quantum liquid presented in Ref. \cite{companion}. The studies of that reference address such a symmetry physical 
consequences for the model on the square lattice described in terms of that quantum liquid of $c$ and $s1$ fermions. 
Specifically, here we introduce a general operator description valid for both the Hubbard model on the
1D and square lattice in terms of three types of elementary quantum objects whose occupancy configurations correspond
to state representations of the model new-found extended global $SO(3)\times SO(3)\times U(1)$ symmetry.

The description introduced here refers to the model full Hilbert space. Its starting point is a unitary transformation that generates
rotated electrons from the electrons. It is such that rotated-electron double and single occupancy are good quantum numbers for $U/4t>0$. The
complete set of states associated with that description are generated by occupancy configurations of 
spinless and $\eta$-spinless $c$ fermions, spin-$1/2$ spinons, and $\eta$-spin-$1/2$ spinons. Those refer to three degrees of
freedom of related rotated-electron occupancy configurations. Indeed the $c$ fermion and spinon occupancy configurations 
describe the hidden $U(1)$ symmetry and spin $SU(2)$ symmetry, respectively, degrees of freedom of the
rotated-electron singly occupied sites. In turn, the $c$ fermion hole and $\eta$-spinon occupancy configurations 
describe the hidden $U(1)$ symmetry and $\eta$-spin $SU(2)$ symmetry, respectively, degrees of freedom of the
rotated-electron doubly occupied and unoccupied sites. Such hidden $U(1)$ symmetry and two $SU(2)$ symmetries 
are those contained in the model extended global $SO(3)\times SO(3)\times U(1)=[SU(2)\times SU(2)\times U(1)]/Z_2^2$ symmetry.
The $c$, $\eta$-spin, and spin effective lattices and corresponding $c$, $\eta\nu$, and $s\nu$ momentum bands,
respectively, are concepts valid in the $N_a^D\rightarrow\infty$ limit. Such effective lattices have been constructed to inherently providing an 
approximate representation for the hidden $U(1)$ symmetry, spin $SU(2)$ symmetry, and $\eta$-spin $SU(2)$
symmetry degrees of freedom of the rotated-electron occupancy configurations that generate a complete set of 
momentum eigenstates. Such occupancy configurations generate representations of the 
$SO(3)\times SO(3)\times U(1)=[SU(2)\times SU(2)\times U(1)]/Z_2^2$ symmetry group.

The physical interest of the description introduced in this paper refers mostly to the model in the 
one- and two-electron subspace, as defined in Section V. However, the 
definition of that subspace and the confirmation of the conservation in it of a set of quantities of interest for the model 
on the square lattice physics require the use of general properties of the full-Hilbert-space description constructed in this paper. This fully
justifies the interest of our studies, which in spite of their extension and often technical character are a needed step 
for the introduction of the square-lattice quantum liquid further investigated in Ref. \cite{companion}. That simpler
quantum problem refers to the Hubbard model on the square lattice in the 
one- and two-electron subspace as defined in this paper. One of the main results of this paper is the introduction of such 
a quantum liquid.
       
Two accomplishments are the following. (i) The use of suitable rotated
electrons as defined in Section II allows the extension of spinon representations associated with singly
occupied sites to all $U/4t>0$ values rather than only for $U/4t\gg 1$, as for the usual schemes in terms of
electron singly occupancy. (ii) Such rotated electrons generate a description consistent with the extended new found
model global $SO(3)\times SO(3)\times U(1)=[SU(2)\times SU(2)\times U(1)]/Z_2^2$ symmetry, what reveals
that the corresponding rotated-electron occupancy configurations that generate the state representations
of that symmetry refer to three degrees of freedom rather than to two. The occurrence of only
two fully symmetrical and independent degrees of freedom associated with two $SU(2)$
symmetries was suggested from the previously assumed model global $SO(4)=[SU(2)\times SU(2)]/Z_2$ symmetry \cite{Zhang}.
For $U\neq 0$ the model contains indeed such a symmetry, yet it is not its full global symmetry.

The accomplishment (ii) has important consequences. For $x>0$ and excitation energy $\omega <2\mu$
it reveals the occurrence of two degrees of freedom for the $M_{\eta,-1/2}=0$ physics. Those are 
associated with the spinless $c$ fermion hidden $U(1)$ symmetry and spin-$1/2$ spinon $SU(2)$ symmetry,
respectively. This provides a new scenario for studies of the effects of doping on the physics
of the Hubbard model on the square lattice. Within our description, which emerges from the complete model global 
symmetry, doping creates $c$ fermion holes. In turn, the corresponding $x>0$ and $m=0$ ground states are 
spin-singlet states whose $s1$ fermion band remains full and thus all sites of the corresponding $s1$ effective 
lattice remain occupied, alike for the $x=0$ and $m=0$ ground state. As investigated elsewhere, the effects of 
doping can then be described in terms of $c$ - $s1$ fermion residual interactions. 

For small and intermediate values of $U/4t$ the explicit form of the electron - rotated-electron unitary operator $\hat{V}$ 
associated with the rotated-electron operators as defined in this paper remains an open problem. 
Fortunately, however, for a one- or two-electron operator ${\hat{O}}$ the terms of the general expression 
(\ref{OOr}) containing commutators involving the related operator $\hat{S}={\tilde{S}}$ generate nearly no spectral weight.
Hence one can reach a quite faithful representation of such a one- or two-electron operator ${\hat{O}}$ by replacing it by the corresponding
leading-order operator term ${\tilde{O}}$ of expression (\ref{OOr}) 
and writing it in terms of the $c$ and $s1$ fermion operators. 
(${\tilde{O}}$ is a one- or two-rotated-electron operator
whose expression in terms of rotated-electron creation and annihilation operators is the same
as that of ${\hat{O}}$ in terms of electron creation and annihilation operators.)
The physical reason why for one- and two-electron operators the operator terms of the general expression (\ref{OOr}) 
containing commutators involving the operator $\hat{S}={\tilde{S}}$ generate very little one- and two-electron spectral weight is that 
they are of higher order in terms $c$ and $s1$ fermion elementary processes. Indeed, the interactions of such objects
are residual and the $c$ and $s1$ fermion leading-order elementary processes are generated by the leading-order operator ${\tilde{O}}$
of expression (\ref{OOr}). This is confirmed by expressing it in terms of $c$ and $s1$ fermion operators \cite{companion}.

Although our operator description is compatible with and in part inspired in the exact solution of the 
1D model, it accounts for the basic differences between the physics of the Hubbard model on the 1D and square 
lattice, respectively. For instance, in 1D the occurrence of an infinite set of conservations laws associated with
the model integrability implies that the $c$ - $s1$ fermion residual interactions refer only to zero-momentum
forward-scattering. They merely give rise to phase shifts whose expressions may be extracted from the
BA solution. This allows the introduction of a pseudofermion dynamical theory, which provides finite-energy 
spectral and correlation function expressions involving phase shifts \cite{V,TTF}. Hence in 1D such interactions do not involve 
interchange of energy and momentum. In contrast, they do for the Hubbard model on the square lattice, yet they 
are much simpler than the corresponding electronic correlations. Indeed the quantum problem is non-perturbative
in terms of electron operators. It follows that in contrast to a 3D isotropic Fermi liquid \cite{Pines}, rewriting the 
square-lattice quantum-liquid theory in terms of the standard formalism of many-electron physics is in general an 
extremely complex problem. Fortunately, such a quantum liquid dramatically simplifies when expressed in terms 
of the $c$ fermion and $s1$ fermion operators \cite{companion}.

The problem is simplest at $x=0$ for spin excitations for which the $c$ band remains full and the effects of the $c$ - $s1$ fermion
interactions are frozen. The preliminary investigations of Ref. \cite{companion} on the physical consequences 
of the model on the square lattice new found global symmetry in actual materials in terms of the $c$ and $s1$ fermion description refer to 
$x=0$. The results of such investigations confirm that the description introduced in this paper is useful for
the further understanding of the role plaid by the electronic correlations in the spin spectrum of the parent compound 
La$_2$CuO$_4$ \cite{LCO-neutr-scatt}. Indeed, it is quantitatively
described in that paper by the corresponding spin spectrum of the
square-lattice quantum liquid at $U/4t\approx 1.525$ and $t\approx 295$ meV.

Elsewhere further investigations on more complex $x>0$ and $m=0$ 2D problems for which the
effects of doping are accounted for in terms of $c$ - $s1$ fermion residual inelastic interactions will
be fulfilled. Ou preliminary studies of such 2D problems seem to provide evidence that upon addition 
of a weak three-dimensional uniaxial anisotropy perturbation to the square-lattice quantum liquid
introduced in this paper, a short-range spin order associated with the $s1$ fermion spinon
pairing, which occurs for a range of finite $x$ values \cite{companion}, might coexist for low temperatures and 
a well-defined range of hole concentrations with a long-range superconducting order.

\begin{acknowledgments}
I thank discussions with M. A. N. Ara\'ujo, H. Q. Lin, P. D. Sacramento, and M. J. Sampaio, suggestions
on the figures by N. M. R. Peres, the support and hospitality of the Beijing Computational 
Science Research Center, and the support of the Portuguese FCT grant PTDC/FIS/64926/2006.
\end{acknowledgments}
\appendix

\section{Equivalence of the quantum numbers of the present 
description and those of the exact solution of the 1D Hubbard model}

The studies of Ref. \cite{1D} profit from the exact
solution of the 1D model whose connection to the present
description is discussed in this Appendix. Note though that no explicit relations to the rotated-electron operators 
as those given in Eqs. (\ref{fc+})-(\ref{c-up-c-down}) are derived in that reference.
Moreover, no relation to the $U(1)$ symmetry contained in the model 
$SO(3)\times SO(3)\times U(1)$ global symmetry is established.  
The eigenvalue of the generator (\ref{Or-ope}) of the 
$U(1)$ symmetry is one half the number of rotated-electron
singly occupied sites $2S_c$. Such an eigenvalue plays an
important role in the present description. It determines
the values of the number $N_c=2S_c$ of $c$ fermions, $M_s=2S_c$
of spinons, $N_c^h=[N_a-2S_c]$ of $c$ fermion holes,
and $M_{\eta} = [N_a-2S_c]$ of $\eta$-spinons. (For simplicity, in this
Appendix we omit the index $D=1$ from the site numbers $N_a^D$,
$N_{a_{\alpha}}^D$, and $N_{a_{\alpha\nu}}^D$ where $\alpha=\eta,s$
and $\nu=1,2,...,\infty$.)

\subsection{Relation to the quantum numbers of the exact solution
for $N_a\gg1$}

The studies of this paper establish that the spinons 
(and $\eta$-spinons) that are not invariant
under the electron rotated-electron transformation are
part of spin (and $\eta$-spin) neutral $2\nu$-spinon 
(and $2\nu$-$\eta$-spinon) composite $s\nu$ (and
$\eta\nu$) bond particles. Those have a binding (and an
anti-binding) character. 

Fortunately, within the $N_a\gg 1$ limit associated 
with our description the problem of the internal degrees of freedom
of such composite $\alpha\nu$ bond particles 
separates from that of their positions in the corresponding $\alpha\nu$ effective
lattice. Thus, the only general result needed for the goals 
of this Appendix is that the composite $\alpha\nu$ bond-particle operators 
and its $\alpha\nu$ effective lattice have been constructed to inherently such 
operators anticommuting on the same $\alpha\nu$ effective-lattice site,
\begin{equation}
\{g^{\dag}_{x_{j},\alpha\nu},g_{x_{j},\alpha\nu}\} = 1 \, ;
\hspace{0.35cm}
\{g^{\dag}_{x_{j},\alpha\nu},g^{\dag}_{x_{j},\alpha\nu}\} =
\{g_{x_{j},\alpha\nu},g_{x_{j},\alpha\nu}\}=0 \, ,
\label{g-local-an}
\end{equation}
and commuting on different sites,
\begin{equation}
[g^{\dag}_{x_{j},\alpha\nu},g_{x_{j'},\alpha\nu}] =
[g^{\dag}_{x_{j},\alpha\nu},g^{\dag}_{x_{j'},\alpha\nu}]
= [g_{x_{j},\alpha\nu},g_{x_{j'},\alpha\nu}] = 0 
\, ; \hspace{0.5cm} j\neq j' \, .
\label{g-non-local-an}
\end{equation}
Furthermore, such operators commute with the
$c$ fermion operators and operators corresponding to different 
$\alpha\nu$ branches also commute with each other.

In this Appendix we confirm that the general hard-core algebra
of Eqs. (\ref{g-local-an}) and (\ref{g-non-local-an}) combined with
the universal number expressions given in Section IV 
leads to discrete momentum values for the $c$ and $\alpha\nu$ fermions
that coincide with the quantum numbers of the exact 
solution. This holds for the whole LWS Hilbert subspace 
that such a solution refers to.

It follows from the algebra (\ref{g-local-an})-(\ref{g-non-local-an}) that one can perform an 
extended Jordan-Wigner transformation. It transforms the $\alpha\nu$ bond particles into $\alpha\nu$ fermions with 
operators $f^{\dag}_{x_{j},\alpha\nu}$. Alike in the general expressions
provided in Eq. (\ref{f-an-operators}), such operators are related to the corresponding 
bond-particle operators as,
\begin{equation}
f^{\dag}_{x_{j},\alpha\nu} = e^{i\phi_{j,\alpha\nu}}\,
g^{\dag}_{x_{j},\alpha\nu} 
\, ; \hspace{0.50cm} 
f_{x_{j},\alpha\nu} = e^{-i\phi_{j,\alpha\nu}}\,
g_{x_{j},\alpha\nu} \, .
\label{JW-f+-an}
\end{equation}
Here,
\begin{equation}
\phi_{j,\alpha\nu} = \sum_{j'\neq j}f^{\dag}_{x_{j'},\alpha\nu}
f_{x_{j'},\alpha\nu}\,\phi_{j',j,\alpha\nu} \, ,
\label{JW-phi-an}
\end{equation}
where $\phi_{j',j,\alpha\nu}$ is the phase given in Eq. (\ref{f-an-operators}).
However, for 1D the coordinates ${x_{j}}_2$ and ${x_{j'}}_2$ appearing in
that equation vanish. Hence $\vec{r}_j$ reduces to the
real-space coordinate $x_j\equiv {x_{j}}_1$ of the $\alpha\nu$ bond particle 
in its $\alpha\nu$ effective lattice. Therefore, 
for 1D the phase $\phi_{j',j,\alpha\nu}$ 
can for all $\alpha\nu$ branches have only the values
$\phi_{j',j,\alpha\nu}=0$ and
$\phi_{j',j,\alpha\nu}=\pi$. Indeed, the
relative angle between two sites of the
$\alpha\nu$ effective lattice in a 1D chain
can only be one of the two values. Then
the $\alpha\nu$ phase factor of Eq. (\ref{JW-phi-an}) is such that,
\begin{equation}
e^{ia_{\alpha\nu}{\partial\over\partial x}\phi_{\alpha\nu} (x)\vert_{x=x_j}}
= e^{i(\phi_{j+1,\alpha\nu}-\phi_{j,\alpha\nu})}  
= e^{i\pi\,f^{\dag}_{x_{j},\alpha\nu}\,f_{x_{j},\alpha\nu}} \, ,
\label{1D-rel}
\end{equation}
where $\phi_{\alpha\nu} (x_j) \equiv \phi_{j,\alpha\nu}$.

The $c$ fermion operators have the anticommuting relations given in Eq. (\ref{albegra-cf}).
For 1D those read,
\begin{equation}
\{f^{\dag}_{x_j,c}\, ,f_{x_{j'},c}\} = \delta_{j,j'} 
\, ; \hspace{0.50cm}
\{f_{x_j,c}^{\dag}\, ,f_{x_{j'},c}^{\dag}\} =
\{f_{x_j,c}\, ,f_{x_{j'},c}\} = 0 \, .
\label{albegra-cf-1D}
\end{equation}
Moreover, the $\alpha\nu$ fermion operators that emerge from
the Jordan-Wigner transformation associated with Eqs.
(\ref{JW-f+-an}) and (\ref{JW-phi-an}) have similar anticommuting 
relations given by,
\begin{equation}
\{f^{\dag}_{x_{j},\alpha\nu}\, ,f_{x_{j'},\alpha\nu}\} =
\delta_{j,j'} 
\, ; \hspace{0.50cm}
\{f^{\dag}_{x_{j},\alpha\nu}\, ,f^{\dag}_{x_{j'},\alpha\nu}\} =
\{f_{x_{j},\alpha\nu}\, ,f_{x_{j'},\alpha\nu}\}  = 0 \, .
\label{1D-anti-com-1D}
\end{equation}
In addition, the $c$ fermion operators commute with 
the $\alpha\nu$ fermion operators and $\alpha\nu$ 
and $\alpha'\nu'$ fermion operators such that
$\alpha\nu\neq \alpha'\nu'$ also commute.

For 1D the $c$ fermion operators of Eq. (\ref{fc+}) 
labeled by the discrete momentum values $q_j$ read,
\begin{equation}
f_{q_j,c}^{\dag} = 
{1\over{\sqrt{N_a}}}\sum_{j'=1}^{N_a}\,e^{+iq_j x_{j'}}\,
f_{x_{j'},c}^{\dag} 
\, ; \hspace{0.50cm}
f_{q_j,c} = {1\over{\sqrt{N_a}}}\sum_{j'=1}^{N_a}\,e^{-iq_j x_{j'}}\,
f_{x_{j'},c} 
\, ; \hspace{0.35cm}
j = 1,...,N_a \, ; \hspace{0.35cm} x_j = j a
\, ; \hspace{0.35cm} L = a N_{a} \, .
\label{fc-q-x-1D}
\end{equation}
The $\alpha\nu$ fermion operators of Eq. (\ref{f-an-operators})
are in 1D denoted by $f^{\dag}_{q_{j},\alpha\nu}$. They are
labeled by the discrete momentum values $q_j$
such that $j=1,...,N_{a_{\alpha\nu}}$. Those are the conjugate 
variables of the $\alpha\nu$ effective lattice real-space
coordinates $x_j$. For subspaces for which the ratio $N_{a_{\alpha\nu}}/N_a$ involving the number
$N_{a_{\alpha\nu}}$ of sites of the $\alpha\nu$ effective lattice
given in Eqs. (\ref{N*}) and (\ref{N-h-an}) is finite for $N_a\rightarrow\infty$
such operators are given by,
\begin{equation}
f_{q_j,\alpha\nu}^{\dag} = 
{1\over{\sqrt{N_{a_{\alpha\nu}}}}}\sum_{j'=1}^{N_{a_{\alpha\nu}}}\,e^{+iq_j x_{j'}}\,
f_{x_{j'},\alpha\nu}^{\dag} 
\, ; \hspace{0.5cm}
f_{q_j,\alpha\nu} = {1\over{\sqrt{N_{a_{\alpha\nu}}}}}
\sum_{j'=1}^{N_{a_{\alpha\nu}}}\,e^{-iq_j x_{j'}}\,
f_{x_{j'},\alpha\nu} 
\, ; \hspace{0.35cm}
j = 1,...,N_{a_{\alpha\nu}} \, ; \hspace{0.1cm} x_j = j a_{\alpha\nu} \, .
\label{fan-q-x-1D}
\end{equation}
Note that $L = a_{\alpha\nu} N_{a_{\alpha\nu}}$.

In 1D the phase factor $e^{i\phi_{j,\alpha\nu}}$ does not have any effect 
when operating before $f^{\dag}_{x_{j},\alpha\nu}$. It follows that in 1D
the expression of the Hamiltonian does not
involve the phase $\phi_{j,\alpha\nu}$. Moreover, expression
of the 1D ground-state normal-ordered Hamiltonian in terms of the $c$
and $\alpha\nu$ fermion operators reveals 
that such objects have zero-momentum forward-scattering only. 
This is consistent with the integrability of the model in 1D. In the present
$N_a\rightarrow\infty$ limit it is associated with the 
occurrence of an infinite number of conservations laws \cite{Martins}. 
For the 1D model the occurrence of such conservations laws is behind
the set of $\alpha\nu$ fermion numbers 
$\{N_{\alpha\nu}\}$ being good quantum numbers \cite{Prosen}.
This is in contrast to the model on the square lattice, for which
such numbers are not in general conserved, yet the number $M^{co}_{\alpha}=2\sum_{\nu}\nu\,N_{\alpha\nu}$ 
of confined $\eta$-spinons ($\alpha=\eta$) and spinons ($\alpha=s$) is.

The Jordan-Wigner transformations phases $\phi_{j,\alpha\nu}$
have direct effects on the boundary conditions. Those
determine the discrete momentum values $q_j$ of both the
$c$ and $\alpha\nu$ fermion operators of Eqs. (\ref{fc-q-x-1D}) and
(\ref{fan-q-x-1D}), respectively. In 1D the periodic 
boundary conditions of the original electron problem are 
ensured provided that one accounts for the effects 
of the Jordan-Wigner transformation on the boundary conditions  
of the $c$ fermions and $\alpha\nu$ fermions
upon moving one of such objects around the chain of length
$L$ once. 

As discussed in Section IV, for both the model on 
the 1D and square lattices the rotated-electron 
occupancies of the sites of the original lattice separate into
two degrees of freedom only. Those of the $2S_c$ sites
of the original lattice singly occupied by rotated electrons separate into
(i) $2S_c$ sites of the $c$ effective lattice occupied
by $c$ fermions and (ii) $2S_c$ sites of the spin effective
lattice occupied by spinons.
Those of the $[N_a-2S_c]$ sites
of the original lattice doubly occupied and unoccupied
by rotated electrons separate into
(i) $[N_a-2S_c]$ sites of the $c$ effective lattice unoccupied
by $c$ fermions and (ii) $[N_a-2S_c]$ sites of the $\eta$-spin effective
lattice occupied by $\eta$-spinons.

Consider for instance the $2\nu$ sites of the spin (and $\eta$-spin) 
effective lattice referring to the occupancy configuration
of one local $2\nu$-spinon composite $s\nu$ fermion
(and  $2\nu$-$\eta$-spinon composite $\eta\nu$ fermion).
Those correspond to the spin (and $\eta$-spin) degrees of
freedom of $2\nu$ sites of the original lattice. The corresponding
degrees of freedom associated with the global $U(1)$ symmetry found
in Ref. \cite{bipartite} are described  
by $2\nu$ sites of the $c$ effective lattice occupied
(and unoccupied) by $c$ fermions. 

An important point is that the deconfined spinons and 
deconfined $\eta$-spinons are not part of
the Jordan-Wigner transformations that map
the $\alpha\nu$ bond particles onto $\alpha\nu$
fermions. It follows that when one $c$ fermion 
moves around its effective lattice of length
$L$ it feels the effects of the Jordan-Wigner transformations 
through the sites of the spin and $\eta$-spin lattices
associated with those of the $s\nu$ and $\eta\nu$
effective lattices occupied by $s\nu$ and $\eta\nu$ fermions,
respectively. Indeed, we recall that the sites of 
the $c$ effective lattice on the one hand and those 
of the spin and $\eta$-spin effective lattices on the
other hand correspond to the different degrees of
freedom of rotated-electron occupancies of the same sites 
of the original lattice.

The $c$ fermions do not emerge from a Jordan-Wigner transformation.
Moreover, each $\alpha\nu$ fermion corresponds to a set of $2\nu$ sites of the original
lattice different from and independent of those of any other $\alpha'\nu'$
fermion. It then follows that upon moving
around its $\alpha\nu$ effective lattice of length $L$,
a $\alpha\nu$ fermion feels only the
Jordan-Wigner-transformation phases of its own 
lattice. Those are associated with both the $\alpha\nu$ fermions and 
$\alpha\nu$ fermion holes. Hence its discrete momentum 
values obey the following periodic or anti-periodic boundary conditions,
\begin{equation}
e^{iq_j\,L} = \prod_{j=1}^{N_{a_{\alpha\nu}}}\left\{
\left[e^{i(\phi_{j+1,\alpha\nu}-\phi_{j,\alpha\nu})}\right]^{\dagger}
e^{i(\phi_{j+1,\alpha\nu}-\phi_{j,\alpha\nu})}\right\} =
e^{i\pi [N_{a_{\alpha\nu}}-1]} = -e^{i\pi N_{a_{\alpha\nu}}} \, .
\label{pbcanp}
\end{equation}
Here the phase factor reads $1$ and $-1$ for $[N_{a_{\alpha\nu}}-1]$
even and odd, respectively. The term $-1$ in $[N_{a_{\alpha\nu}}-1]$ can be
understood as referring to the site occupied by the $\alpha\nu$ fermion moving
around its effective lattice and must be excluded. For the $\alpha\nu$
fermions the unoccupied sites of their $\alpha\nu$ effective lattice
exist in their own right. Indeed, note that according to Eq. (\ref{JW-f+-an}) 
both the creation and annihilation operators of such objects involve 
the Jordan-Wigner-transformation phase $\phi_{j,\alpha\nu}$. 
As a result, such a phase affects both the $\alpha\nu$ fermions and 
$\alpha\nu$ fermion holes. That justifies why the phase factor
$e^{i\pi [N_{a_{\alpha\nu}}-1]}$ of
Eq. (\ref{pbcanp}) involves all the $N_{a_{\alpha\nu}}=[N_{\alpha\nu}+
N_{\alpha\nu}^h]$ sites of the $\alpha\nu$ effective lattice. The only exception is
that occupied by the moving $\alpha\nu$ fermion. Hence it involves both
the $[N_{\alpha\nu}-1]$ sites occupied by the remaining fermions of
the same $\alpha\nu$ branch and the corresponding $N_{\alpha\nu}^h$ 
$\alpha\nu$ fermion holes.

In contrast, the $c$ fermions are only affected by the
sites occupied by $\alpha\nu$ fermions. Indeed, only the 
sets of $2\nu$ sites of the spin (and $\eta$-spin) effective lattice
associated with each occupied site of the $s\nu$ (and $\eta\nu$)  
$\nu=1,2,3,...$ effective lattices and the sites of the spin (and $\eta$-spin) effective 
lattice occupied by deconfined spinons (and deconfined 
$\eta$-spinons) correspond to sites of the original lattice whose 
degrees of freedom associated with the $c$ fermion $U(1)$ 
symmetry are described by the occupancy configurations
of the $c$ effective lattice. However, the  
deconfined spinons (and deconfined 
$\eta$-spinons) do not undergo any 
Jordan-Wigner transformation. Thus due to the Jordan-Wigner-transformation
phase $\phi_{j',\alpha\nu}$ of each of the $N_{\alpha\nu}$ 
$\alpha\nu$ fermions at sites $j'=1,...,N_{a_{\alpha\nu}}$ 
of their $\alpha\nu$ effective lattice the 
$c$ fermion discrete momentum values are determined by the 
following periodic or anti-periodic boundary condition,
\begin{equation}
e^{iq_j\,L} = 
\prod_{\alpha\nu}\prod_{j'=1}^{N_{\alpha\nu}}e^{i(\phi_{j'+1,\alpha\nu}-\phi_{j',\alpha\nu})} 
= e^{i\pi\sum_{\alpha\nu}N_{\alpha\nu}} \, .
\label{pbccp}
\end{equation}
Again, the phase factor on the right-hand side
of Eq. (\ref{pbccp}) reads $1$ and $-1$ for $\sum_{\alpha\nu}N_{\alpha\nu}$
even and odd, respectively. 

The above results imply that the discrete momentum values $q_j$ of both
$c$ and $\alpha\nu$ fermions have the usual 
momentum spacing $q_{j+1}-q_j=2\pi/L$ and read,
\begin{equation}
q_j = {2\pi\over L}\,I^{\alpha\nu}_j \, ; \hspace{0.35cm} j=1,...,N_{a_{\alpha\nu}} 
\, ; \hspace{0.5cm}
q_j = {2\pi\over L}\,I^c_j \, ; \hspace{0.35cm} j=1,...,N_a \, .
\label{q-j-f-repr}
\end{equation}
However, following the boundary conditions (\ref{pbcanp}) [and (\ref{pbccp})] the numbers 
$I^{\alpha\nu}_j$ (and $I^{c}_j$) where $j=1,2,...,N_{a_{\alpha\nu}}$ (and $j=1,2,...,N_a$)
appearing in this equation are not always integers. They are 
integers and half-odd integers for $[N_{a_{\alpha\nu}}-1]$
(and $\sum_{\alpha\nu}N_{\alpha\nu}$ ) even and odd, respectively. Furthermore,
as a result of the periodic or anti-periodic character of such boundary conditions
these numbers obey the inequality $\vert I^{\alpha\nu}_j\vert\leq [N_{a_{\alpha\nu}}-1]/2$ for both 
$[N_{a_{\alpha\nu}}-1]$ odd and even (and the inequality $\vert I^{c}_j\vert\leq [N_a-1]/2$ 
for $\sum_{\alpha\nu}N_{\alpha\nu}$ even
and  $-[N_a-2]/2\leq I^{c}_j \leq N_a/2$ for $\sum_{\alpha\nu}N_{\alpha\nu}$ odd).

For the one- and two-electron subspace as defined in
Section V one can separate the numbers $I^{s1}_j$ 
and $I^{c}_j$ of Eq. (\ref{q-j-f-repr}) in two terms corresponding to an integer
number and a small deviation as $I^{s1}_j\equiv [{\cal{N}}^{s1}_j + {q_{s1}^0\over 2\pi}{L\over N_{s1}}]$
and $I^{c}_j\equiv [{\cal{N}}^c_j + {q_{c}^0\over 2\pi}{L\over N_c}]$, respectively. The corresponding
$c$ and $s1$ fermion discrete momentum values then read,
\begin{eqnarray}
q_j & = & {2\pi\over L}\,{\cal{N}}^{c}_j + q_{c}^0/N_{c} 
\, ; \hspace{0.50cm}
{\cal{N}}^{c}_j= j - {N_{a}\over 2} = 0,\pm 1, \pm 2, ... 
\, ; \hspace{0.35cm} j=1,...,N_{a} \, .
\nonumber \\
q_j & = & {2\pi\over L}\,{\cal{N}}^{s1}_j + q_{s1}^0/N_{s1} 
\, ; \hspace{0.50cm}
{\cal{N}}^{s1}_j= j - {N_{a_{s1}}\over 2} = 0,\pm 1, \pm 2, ... 
\, ; \hspace{0.35cm} j=1,...,N_{a_{s1}} \, ,
\label{q-j-f-Q-c-0}
\end{eqnarray}
Here $q_{c}^0$ (and $q_{s1}^0$) is given either by $q_{c}^0=0$ or $q_{c}^0=\pi [N_c/L]$ 
(and $q_{s1}^0=0$ or $q_{s1}^0= \pi [N_{s1}/L]$) for all $j=1,...,N_a$ (and $j=1,...,N_{a_{s1}}$)
discrete momentum values of the $c$ (and $s1$) band whose momentum occupancy describes a given state.
The $s1$ effective lattice length $L=N_{a_{s1}}\,a_{s1}$ where
$a_{s1}=L/N_{a_{s1}}=[N_a/N_{a_{s1}}]\,a$ is the $s1$ effective lattice spacing.

Importantly, the $c$ fermion and $\alpha\nu$ fermion
discrete momentum values obtained from our $N_a\gg1$ operational 
description of the quantum problem correspond to the BA quantum 
numbers of the exact solution. Indeed, the discrete momentum 
values of Eq. (\ref{q-j-f-repr}) can be expressed as, 
\begin{equation}
q_j = {2\pi\over L}\,I_j \, ; \hspace{0.35cm} j=1,...,N_a 
\, ; \hspace{0.5cm}
q_j = {2\pi\over L}\,J^{'\nu}_j \, ; \hspace{0.35cm} j=1,...,N_{a_{\eta\nu}} 
\, ; \hspace{0.5cm}
q_j = {2\pi\over L}\,J^{\nu}_j \, ; \hspace{0.35cm} j=1,...,N_{a_{s\nu}} \, .
\label{q-j}
\end{equation}
Here $I_j\equiv I^c_j$, $J^{'\nu}_j\equiv I^{\eta\nu}_j$, and 
$J^{\nu}_j\equiv I^{s\nu}_j$ are the exact-solution
integers or half integers quantum numbers involved 
in Eqs. (2.12a)-(2.12c) of Ref. \cite{Takahashi} and defined 
in the unnumbered equations provided below these equations.
(In the notation of that reference, $\nu =n$ 
and $j=\alpha$ in $J^{'\nu}_j$ and $J^{\nu}_j$). 
Moreover, the numbers on the right-hand side of the two 
inequalities given just above
Eq. (2.13a) of that reference correspond to 
$N_{a_{\eta\nu}}/2$ and $N_{a_{s\nu}}/2$, respectively. It follows
that these inequalities read $\vert J^{'\nu}_j\vert<N_{a_{\eta\nu}}/2$
and $\vert J^{\nu}_j\vert<N_{a_{s\nu}}/2$. This
is fully consistent with the above inequality
$\vert I^{\alpha\nu}_j\vert\leq [N_{a_{\alpha\nu}}-1]/2$
where $\alpha=\eta,s$ and $N_{a_{\alpha\nu}}$ is 
given in Eq. (\ref{N*}). A careful comparison of the notations and definitions
used in Ref. \cite{Takahashi} and here confirms that
there is also full consistency between the even or odd character
of the integer numbers $[N_{a_{\eta\nu}}-1]$, $[N_{a_{s\nu}}-1]$,
and $\sum_{\alpha\nu}N_{\alpha\nu}$ considered here and those
that determine the integer of half-integer character
of the quantum numbers $J^{'\nu}_j\equiv I^{\eta\nu}_j$, 
$J^{\nu}_j\equiv I^{s\nu}_j$, and $I_j\equiv I^c_j$,
respectively, in that reference.

We emphasize that for the $\alpha\nu$ fermions the discrete momentum values
$q_j$ of Eq. (\ref{q-j}) are the eigenvalues of the translation generator in the
presence of the fictitious magnetic field of Eq. (\ref{A-j-s1-3D}). For
1D it reads ${\vec{B}}_{\alpha\nu} (x_j) = 
\sum_{j'\neq j} n_{x_{j'},\alpha\nu}\,\delta (x_{j'}-x_{j})\,{\vec{e}}_{x_3}$.
Hence the corresponding exact-solution quantum numbers are
the eigenvalues of such a translator operator in units of $2\pi/L$. 
 
For 1D the numbers $\{N_{\alpha\nu}\}$ of $\alpha\nu$ fermions are conserved.
Thus that the discrete momentum values $q_j$ of the $c$ and $\alpha\nu$ fermions are
good quantum numbers is consistent with the momentum operator commuting with the
unitary operator $\hat{V}^{\dag}$ as defined in this paper. That operator generates exact
$U/4t>0$ energy and momentum eigenstates
$\vert \Psi_{LWS;U/4t}\rangle ={\hat{V}}^{\dag}\,\vert \Psi_{LWS;\infty}\rangle$ 
of general form given in Eq. (\ref{LWS-full-el}). Indeed, in 1D 
$\vert \Psi_{LWS;U/4t}\rangle=\vert \Phi_{LWS;U/4t}\rangle$. Such states have $U/4t$-dependent energy
eigenvalues and $U/4t$-independent momentum eigenvalues.

The use of the exact solution of the 1D problem confirms that the momentum eigenvalues 
have the general form given in Eq. (\ref{P-1-2-el-ss}). For 1D they may be written as,
\begin{equation}
P =\sum_{j=1}^{N_a} q_j\, N_c (q_j)
+ \sum_{\nu=1}^{\infty}\sum_{j=1}^{N_{a_{s\nu}}}
q_{j}\, N_{s\nu} (q_{j}) 
+ \sum_{\nu=1}^{\infty}\sum_{j=1}^{N_{a_{\eta\nu}}}
[\pi -q_{j}]\, N_{\eta\nu} (q_{j}) 
+\pi\,M_{\eta,-1/2} \, .
\label{P}
\end{equation}
Here $M_{\eta,-1/2}$ is the total number of $\eta$-spin-projection $-1/2$ $\eta$-spinons
and the distributions $N_c (q_j)$ and 
$N_{\alpha\nu} (q_{j'})$ are the eigenvalues of
the operators $\hat{N}_{c}(q_{j}) = f^{\dag}_{q_{j},c}\,f_{q_{j},c}$
and $\hat{N}_{\alpha\nu}(q_{j}) = 
f^{\dag}_{q_{j},\alpha\nu}\,f_{q_{j},\alpha\nu}$, respectively.
Those have values $1$ and $0$ for occupied and unoccupied 
momentum values, respectively. One may obtain the expression
(\ref{P}) from analysis of the 1D problem for $U/4t\gg 1$ without the use
of BA. The starting point of such a procedure is that for $U/t\rightarrow\infty$ the 
electrons that singly occupy sites do not feel the 
on-site repulsion. Consistently, expression (\ref{P})
is that also provided by the exact solution after the
rapidities are replaced by the quantum numbers
$I_j\equiv I^c_j$, $J^{'\nu}_j\equiv I^{\eta\nu}_j$, and 
$J^{\nu}_j\equiv I^{s\nu}_j$ related to the discrete
momentum values $q_j$ by Eq. (\ref{q-j}).

That the physical momentum (\ref{P}) is for $U/4t>0$ additive 
in the $c$ and $\alpha\nu$ fermion discrete momentum values 
results from the latter being good quantum numbers whose occupancy 
configurations generate the energy eigenstates. 
It follows from the direct relation to the thermodynamic Bethe
ansatz equations of Ref. \cite{Takahashi} that the $c$
fermions obtained here from the rotated electrons
through Eq. (\ref{fc+}) and the $\alpha\nu$ fermions 
that emerge from the Jordan-Wigner transformations
of Eq. (\ref{JW-f+-an}) are for the 1D model the
$c$ pseudoparticles and $\alpha\nu$ pseudoparticles,
respectively, associated in Ref. \cite{1D} with the 
BA quantum numbers. The momentum quantum numbers
of Eq. (\ref{q-j}) are precisely those given in Eq. (A.1)
of Ref. \cite{1D} and the numbers $N_{a_{\alpha\nu}}$
of Eq. (\ref{N*}) equal the numbers $N_{\alpha\nu}^*$ 
defined by its Eqs. (B.6), (B.7), and (B.11) with the index $c$ replaced by
$\eta$. Furthermore, the $\eta$-spinons
and spinons considered here are for 1D the holons
and spinons of that reference, respectively. Also the
deconfined $\eta$-spinons and deconfined spinons
are for 1D the Yang holons
and HL spinons, respectively, of Ref. \cite{1D}.

\subsection{Relation to the algebraic formulation of the exact solution}

We just confirmed that for 1D the discrete momentum values of
the $c$ fermion operators $f^{\dag}_{q_{j},c}$,  
and those of the $\alpha\nu$ fermion operators $f^{\dag}_{q_{j},\alpha\nu}$
equal the quantum numbers of the exact BA solution. Such a result
was obtained in the $N_a\gg1$ limit that the description 
considered in this paper and in Ref. \cite{companion} 
refers to.  Within our description the $c$ fermions emerge from 
the electron - rotated-electron unitary transformation, as given
in Eq. (\ref{fc+}). In turn, the $\alpha\nu$ fermions emerge from that 
transformation and an extended Jordan-Wigner 
transformation, as given in Eqs. (\ref{f-an-operators}) and (\ref{JW-f+-an}).
However, such a connection corresponds to the quantum
numbers only. The relation of the $c$ and $\alpha\nu$
fermion operators to the exact solution of the 1D model
remains an open problem. 

The relation of the building blocks of our description to the
original electrons is uniquely defined yet corresponds to
a complex problem. Such building blocks are the
$c$ fermions, $\eta$-spinons, and spinons. For 
$U/4t\gg 1$ the rotated electrons
become electrons and the $c$ fermion creation operator 
$f_{\vec{r}_j,c}^{\dag}$ becomes the quasicharge annihilation 
operator $\hat{c}_r$ of Ref. \cite{Ostlund-06}.
Therefore, in that limit the $c$ fermions are the ``holes"
of the quasicharge particles of that reference. In turn,
the spinons and $\eta$-spinons are associated with the
local spin and pseudospin operators, respectively, of
the same reference. The transformation considered in 
Ref. \cite{Ostlund-06} does not introduce Hilbert-space 
constraints. It follows that suitable occupancy 
configurations of the objects associated with the local 
quasicharge, spin, and pseudospin operators considered
in that reference exist that generate a complete set 
of states. However, in 1D only in the limit $U/4t\gg 1$ 
suitable occupancy configurations of such basic objects 
generate exact energy eigenstates. 

The point is that rotated electrons as defined in 
this paper are related to electrons by a unitary 
transformation. And such a transformation is such 
that for $U/4t>0$ rotated-electron occupancy configurations 
of the same form as those that generate energy eigenstates for
$U/4t\gg 1$ in terms of electron operators
do generate energy eigenstates for finite values of $U/4t$.
The $c$ fermion, $\eta$-spinon, and spinon operators
are related to the rotated-electron operators as the
quasicharge, spin, and pseudospin operators of
Ref. \cite{Ostlund-06} are related to electron operators.

Importantly, note that the validity of the $c$ fermion,
spinon, and $\eta$-spinon operational description
constructed in this paper and in Ref. \cite{companion}
for the Hubbard model on a square and 1D lattices
is for the 1D problem independent of its relation to the
exact solution. For the LWS subspace that such a solution refers
to the validity of our operational description
follows from the transformations behind it
not introducing Hilbert-space constraints.
Such transformations correspond to explicit 
operator expressions in terms of rotated-electron 
operators. For the $c$ fermion operators it is given in Eq. (\ref{fc+}). 
For the spinon and $\eta$-spinon operators they are provided
in Eqs. (\ref{sir-pir})-(\ref{rotated-quasi-spin}). And
the rotated-electron operators are related to
the original electron operators by the unitary
transformation ${\tilde{c}}_{\vec{r}_j,\sigma}^{\dag} =
{\hat{V}}^{\dag}\,c_{\vec{r}_j,\sigma}^{\dag}\,{\hat{V}}$.
Its unitary operator ${\hat{V}}^{\dag}$ is within our description uniquely defined. For $U/4t>0$ it 
has been constructed to inherently generating a complete set of energy
eigenstates of the general form 
$\vert \Psi_{LWS;U/4t}\rangle ={\hat{V}}^{\dag}\,\vert \Psi_{LWS;\infty}\rangle$. Here
$\{\vert \Psi_{LWS;\infty}\rangle\}$ is a complete
set of suitably chosen $U/4t\gg 1$ energy eigenstates.
For 1D such states are such that $\vert \Psi_{LWS;U/4t}\rangle =\vert \Phi_{LWS;U/4t}\rangle$
where $\vert \Phi_{LWS;U/4t}\rangle ={\hat{V}}^{\dag}\,\vert \Phi_{LWS;\infty}\rangle$ 
are states of general form given in Eq. (\ref{LWS-full-el}).

In order to clarify the relation of the $c$ and $\alpha\nu$
fermion operators to the exact solution of the 1D model
rather than the so called coordinate 
BA \cite{Lieb,Takahashi} it is convenient to consider 
the solution of the problem by an algebraic operator formulation. 
Within it the HWSs or LWSs of the $\eta$-spin and spin algebras are built up in 
terms of linear combination of products of several types of
creation fields acting onto the hole or electronic vacuum, 
respectively \cite{Martins,ISM}. The 1D model energy eigenstates 
that are HWSs or LWSs of these algebras are often 
called {\it Bethe states}. Here we briefly discuss how in 1D
and for $N_a\gg1$ the $c$ and $\alpha\nu$ fermion
operators emerge from the creation fields of the 
algebraic formulation of the Bethe states. 

That the general description introduced in this paper for the Hubbard
model on the square and 1D lattices is consistent 
with the exact solution of the 1D problem at the operator level as
well, confirms its validity for 1D. This is the only 
motivation and aim of this Appendix. However, 
since that refers to a side problem of that studied in
this paper, in the following discussion we skip 
most technical details that are unnecessary for its general 
goals. Nevertheless provided that our analysis
of the problem is complemented with the detailed information
provided in Refs. \cite{Takahashi,Martins}, the resulting
message clarifies the main issues under consideration.
 
The algebraic formulation of the Bethe states refers to the 
transfer matrix of the classical coupled spin model, which is the "covering" 
1D Hubbard model \cite{CM}. Indeed, within the inverse scattering method 
\cite{Martins} the central object to be 
diagonalized is the quantum transfer matrix rather than the underlying 
1D Hubbard model. The transfer-matrix eigenvalues provide the spectrum 
of a set of $[N_a-1]$ conserved charges. 
The creation and annihilation fields are labeled by the BA rapidities 
$\lambda$. Those may be generally complex and are not the ultimate quantum 
numbers of the model. Many quantities are functions of 
such rapidities. For instance, the weights $a(\lambda)$ and $b(\lambda)$ 
considered in the derivation of Ref. \cite{Martins}
satisfy the free-fermion condition $a(\lambda)^2 +b(\lambda)^2 =1$. 
(A possible and often used parametrization is
$a(\lambda)=\cos (\lambda)$ and $b(\lambda) =\sin (\lambda)$.)
The reparametrization $\tilde{\lambda}=[a(\lambda)/b(\lambda)]\,e^{2h(\lambda)}-
[b(\lambda)/a(\lambda)]\,e^{-2h(\lambda)}-U/2$
where the constraint $h(\lambda)$ is defined by the relation
$\sinh [2h(\lambda)]=[U/2]\,a(\lambda)b(\lambda)$,
plays an important role in the derivation in the context of the quantum
inverse scattering method of the non-trivial Boltzmann
weights of the isotropic six-vertex model given in Eq. (33) of Ref.
\cite{Martins}.

The diagonalization of the charge degrees of freedom involves a transfer 
matrix of the form provided in Eq. (21) of that reference. Its off-diagonal 
entries are some of the above mentioned creation and annihilation fields. 
The commutation relations of such important operators play a major role 
in the theory. They are given in Eqs. (25), (40)-(42), (B.1)-(B.3), (B.7)-(B.11), 
and (B.19)-(B.22) of the same reference. 
The solution of the spin degrees of freedom involves the 
diagonalization of the auxiliary transfer matrix associated with the monodromy 
matrix provided in Eq. (95) of Ref. \cite{Martins}. Again, the off-diagonal entries 
of that matrix play the role of creation and annihilation operators. Their 
commutation relations are given in Eq. (98) of that reference. 
The latter commutation relations correspond to the usual
Faddeev-Zamolodchikov algebra associated with the traditional
ABCD form of the elements of the monodromy matrix. In turn, the above 
relations associated with the charge monodromy matrix refer to a 
different algebra. The corresponding form of that matrix was
called ABCDF by the authors of Ref. \cite{Martins}.

As discussed in Ref. \cite{bipartite},
the main reason why the solution of the problem by the algebraic
inverse scattering method \cite{Martins} was achieved only thirty years
after that of the coordinate BA \cite{Lieb,Takahashi} is
that it was expected that the charge and spin monodromy 
matrices had the same traditional
ABCD form \cite{ISM}. This would be consistent with the occurrence
of a spin $SU(2)$ symmetry and a charge (and
$\eta$-spin) $SU(2)$ symmetry known long ago \cite{Heilmann},
associated with a global $SO(4)=[SU(2)\times SU(2)]/Z_2$ 
symmetry \cite{Zhang}.
Fortunately, the studies of Ref. \cite{Martins} used an
appropriate representation of the charge and spin monodromy 
matrices. Its structure is able to distinguish creation and
annihilation fields as well as possible {\it hidden symmetries},
as discussed by the authors of that reference.

A hidden symmetry beyond $SO(4)$ was indeed identified  
recently. It is the global $U(1)$ symmetry found 
in Ref. \cite{bipartite}. The studies of that reference reveal 
that for $U/4t>0$ the model charge and spin degrees of 
freedom are associated with $U(2)=SU(2)\times U(1)$ and
$SU(2)$ symmetries, rather than with two $SU(2)$ symmetries,
respectively. The occurrence of such charge $U(2)=SU(2)\times U(1)$
symmetry and spin $SU(2)$ symmetry is fully consistent
with the different ABCDF and ABCD forms of the charge and spin monodromy 
matrices of Eqs. (21) and (95) of Ref. \cite{Martins}, respectively.
Indeed, the former matrix is larger than the latter. It involves 
more fields than expected from the global
$SO(4)=[SU(2)\times SU(2)]/Z_2$ symmetry alone. This is
consistent with the global $SO(3)\times SO(3)\times U(1)=[SO(4)\times U(1)]/Z_2$
symmetry of the model on the 1D and any other bipartite lattice \cite{bipartite}.

Our general description of the model on a square and 1D lattices 
takes such an extended global symmetry into
account. Furthermore, for 1D it was also implicitly taken into
account by the appropriate representation of the charge
and spin monodromy matrices used in Ref. \cite{Martins}. Such
a consistency is a necessary condition for the $c$ and
$\alpha\nu$ fermion operators emerging from the
fields of the monodromy matrices upon diagonalization
of the 1D problem for $N_a\gg1$.

Initially the expressions obtained by the algebraic
inverse scattering method for the Bethe states include both
wanted terms and several types of unwanted terms. The latter terms are 
eliminated by imposing suitable restrictions to the rapidities. Such constrains 
lead to the BA equations. Ultimately they lead to the real integer
and half-integer quantum numbers. In units of
$2\pi/L$ those are the discrete momentum values that label the $c$ and
$\alpha\nu$ fermion operators. 

After solving the BA 
for both the charge and spin degrees of freedom, one reaches the charge
rapidities $\lambda_j$ with $[N_a -2S_{\eta}]$ values 
such that  $j=1,...,[N_a -2S_{\eta}]$. In addition, one reaches
the spin rapidities ${\tilde{\lambda}}_j$ with $[N_a/2-S_{\eta}-S_s]$ values 
such that  $j=1,...,[N_a/2-S_{\eta}-S_s]$. Here $S_{\eta}=S_{\eta}^{x_3}$ and
$S_{s}=S_{s}^{x_3}$ for a HWS and $S_{\eta}=-S_{\eta}^{x_3}$ and
$S_{s}=-S_{s}^{x_3}$ for a LWS of both the $\eta$-spin and spin algebras.
However, the rapidities can be extended to non-Bethe tower states provided
that the number of their values are expressed in terms of
$S_{\eta}$ and $S_{s}$ rather than of $S_{\eta}^{x_3}$ and
$S_{s}^{x_3}$, respectively. Equations of the same form as
those obtained by the coordinate Bethe anstaz are reached by
the algebraic operator formulation. This is so provided that one 
introduces the charge momentum rapidities $k_j$ and spin rapidities 
${\bar{\lambda}}_j$ given by \cite{Martins}
$z_{-}(\lambda_j)=[a(\lambda_j)/b(\lambda_j)]\,e^{2h(\lambda_j)}=
e^{ik_j a}$ where $j=1,...,[N_a -2S_{\eta}]$ and  
${\bar{\lambda}}_j = - i {\tilde{\lambda}}_j/2$ where
$j=1,...,[N_a/2-S_{\eta}-S_s]/2$. 

Before discussing the structure of the rapidities for $N_a\gg1$, 
we confirm that the numbers $[N_a -2S_{\eta}]$
and $[N_a/2-S_{\eta}-S_s]$ of discrete charge momentum and spin rapidity 
values, respectively, provided by the BA solution are closely
related to the occupancy configurations of the rotated electrons that
are not invariant under the electron - rotated-electron unitary
transformation. Indeed, such numbers can be rewritten as,
\begin{equation}
[N_a -2S_{\eta}] = [2S_c + M^{co}_{\eta}] \, ;
\hspace{0.50cm} [N_a/2-S_{\eta}-S_s] = [M^{co}_{\eta} + M^{co}_{s}]/2 \, .
\label{NN-CC}
\end{equation}
Here the numbers $M^{co}_{\eta}$ and $M^{co}_{s}$ are those of Eq. (\ref{M-L-Sum}).

$2S_c$ is the number of elementary rotated-electron
charges associated with the singly
occupied sites. Moreover, $M^{co}_{\eta}=2\sum_{\nu}\nu\,N_{\eta\nu}$
is the number of such charges associated with the rotated-electron
doubly occupied sites whose occupancy configurations are
not invariant under the above unitary transformation.
Furthermore, $[M^{co}_{\eta} + M^{co}_{s}]/2=\sum_{\alpha\nu}\nu\,N_{\alpha\nu}$ 
is the number $M^{co}_{\eta}/2=\sum_{\nu}\nu\,N_{\eta\nu}$ of down spins
(and up spins) of such rotated-electron doubly occupied sites  
plus the number $M^{co}_{s}/2=\sum_{\nu}\nu\,N_{s\nu}$ of down spins 
(and up spins) of the rotated-electron singly occupied sites
whose occupancy configurations are
not invariant under that transformation.
 
The structure of the charge momentum and spin rapidities simplifies in the
$N_a\gg1$ limit \cite{Takahashi}. Then the $[N_a -2S_{\eta}] = 
[2S_c + M^{co}_{\eta}]$ charge momentum rapidity values separate 
into two classes. Those correspond to $2S_c$ and $M^{co}_{\eta}$ of 
these values, respectively. The set of charge momentum 
rapidity values $k_j$ such that $j=1,...,2S_c$ are real and are
related to the integer or half-integer quantum numbers
$I^c_j$ of Eq. (\ref{q-j-f-repr}). Those are such that the numbers $q_j = [2\pi/L]\,I^c_j$ are
the $2S_c$ discrete momentum values occupied by the $c$
fermions, out of a total number $N_a$ of
such values. They corresponds to the $2S_c$ elementary 
charges of the rotated electrons that singly occupy sites.

Moreover, for $N_a\gg1$ the $[N_a/2-S_{\eta}-S_s] = [M^{co}_{\eta} + M^{co}_{s}]/2$
spin rapidity values separate into two classes. Those correspond to 
$M^{co}_{\eta}/2$ and $M^{co}_{s}/2$ of these values, respectively. The point
is that the $M^{co}_{\eta}/2$ down spins and $M^{co}_{\eta}/2$ up spins
of the rotated electrons that doubly occupy sites and
whose occupancy
configurations are not invariant under the electron -
rotated-electron unitary transformation combine with 
the $M^{co}_{\eta}$ elementary charges left over by the
above separation of the $[N_a -2S_{\eta}] = 
[2S_c + M^{co}_{\eta}]$ charge momentum rapidity values.
Therefore, $M^{co}_{\eta}/2$ spin rapidity values out of 
$[M^{co}_{\eta} + M^{co}_{s}]/2$ combine with $M^{co}_{\eta}$ 
charge momentum rapidity values out of
$[2S_c + M^{co}_{\eta}]$. This leads to $M^{co}_{\eta}/2$ new
rapidity values associated with the $\eta$-spin
singlet configurations. Those describe the
$\eta$-spin degrees of freedom of the 
rotated electrons that doubly occupy sites and
whose occupancy configurations are not invariant under the electron -
rotated-electron unitary transformation.

In turn, the $M^{co}_{s}/2$ spin rapidity values left over by
the separation of the $[N_a/2-S_{\eta}-S_s] = 
[M^{co}_{\eta} + M^{co}_{s}]/2$ spin rapidity values describe the 
spin-singlet configurations associated with the 
rotated electrons that singly occupy sites and
whose occupancy
configurations are not invariant under the electron -
rotated-electron unitary transformation. The charge
degrees of freedom of these rotated electrons are
described by the above set of charge momentum 
rapidity values $k_j$ such that $j=1,...,2S_c$.

According to the $N_a\gg1$ results of
Ref. \cite{Takahashi}, the $M^{co}_{\eta}/2$ $\eta$-spin rapidity values
(and $M^{co}_{s}/2$ spin rapidity values) further separate into
$\eta\nu$ rapidities of length $\nu=1,2,...$
with $N_{\eta\nu}$ values (and $s\nu$ rapidities of length $\nu=1,2,...$ with 
$N_{s\nu}$ values). Obviously, the sum-rules
$M^{co}_{\eta}/2=\sum_{\nu}\nu\,N_{\eta\nu}$ and 
$M^{co}_{s}/2=\sum_{\nu}\nu\,N_{s\nu}$ are obeyed. The
results of that reference reveal
that $s\nu$ rapidities of length $\nu=1$ are real and
the imaginary part of the remaining branches of
$\alpha\nu$ rapidities where $\alpha =\eta,s$ has a simple
form. 

Therefore, for $N_a\gg1$ the original BA equations
lead to a system of $[1+M^{co}_{\eta}/2+M^{co}_s/2]$ (infinite in the limit
$N_a\rightarrow\infty$) coupled thermodynamic equations.
Their solution gives the charge momentum 
rapidity values $k_j$ such that $j=1,...,2S_c$ and 
the values $\Lambda_{j,\alpha\nu}$ of the real part
of the $\alpha\nu$ rapidities such that $j=1,...,N_{\alpha\nu}$.
Those are given as a function of the occupancies of the real integer
or half-integer quantum numbers $I^{c}_j$ 
and $I^{\alpha\nu}_j$ of Eq. (\ref{q-j-f-repr}), respectively.
Alike for the $c$ fermions, the quantum
numbers $I^{\alpha\nu}_j$ such that $q_j = [2\pi/L]\,I^{\alpha\nu}_j$ are
the $N_{\alpha\nu}$ discrete momentum values occupied by $\alpha\nu$
fermions, out of a total number $N_{a_{\alpha\nu}}$ of
such values. (Note that here we used a notation suitable
to the extended global symmetry of the problem and
that in Ref. \cite{Takahashi} such quantum numbers are 
denoted as in Eq. (\ref{q-j}) with $\nu$ replaced by $n$
and the values $\Lambda_{j,\eta\nu}$ and $\Lambda_{j,s\nu}$ 
of the real part of the rapidities by 
$\Lambda^{'n}_{\alpha}$ and $\Lambda^{n}_{\alpha}$,
respectively, where $\alpha$ plays the role of $j$.)

Summarizing the above discussion, in order to reach the $c$ 
and $\alpha\nu$ fermion operators 
the algebraic operator formulation of the 
diagonalization of the quantum problem
starts by building up the Bethe states in terms of linear 
combination of products of the above mentioned
several types of creation fields 
acting onto a suitable vacuum. The diagonalization of the charge 
and spin degrees of freedom involves the transfer 
matrices given in Eqs. (21) and (95) of Ref. \cite{Martins}, 
respectively. Their off-diagonal entries are some of these
creation and annihilation fields. The different form of such
matrices is consistent with the model global $SO(3)\times SO(3)\times U(1)$
symmetry. It is taken into account as well by our general description 
both for the model on the 1D and
square lattices used in studies of the square-lattice 
quantum liquid of $c$ and $s1$ fermions. 
The creation and annihilation fields 
obey the very involved commutation relations given in 
Eqs. (25), (40)-(42), (98), (B.1)-(B.3), (B.7)-(B.11), and
(B.19)-(B.22) of Ref. \cite{Martins} and are labeled by rapidities. 
Those may be generally complex and are not the ultimate quantum numbers.

However, such creation and annihilation fields and their involved
algebra generate expressions for the Bethe states that include both
wanted terms and several types of unwanted and unphysical terms.
Indeed, they act onto an extended and partially unphysical Hilbert
space, larger than that of the model.
The unwanted and unphysical terms are eliminated by imposing suitable restrictions to 
the rapidities that change the nature of the fields. For
$N_a\gg1$ they are replaced by the $c$ and $\alpha\nu$ fermion operators
labeled by the real integer and half-integer quantum numbers
of the diagonalized model. Hence the BA equations 
obtained by imposing suitable restrictions to the rapidities
describe the relation between the rapidities and the
ultimate quantum numbers associated with the
$c$ and $\alpha\nu$ fermion operators.

In addition to emerging from the elimination of the unwanted and 
unphysical terms of the Bethe states generated by the initial 
creation and annihilation fields, the $c$ and $\alpha\nu$ fermion operator
algebra refers to well-defined subspaces. Those are spanned by energy eigenstates
whose number of $\eta$-spinons, spinons, and $c$ fermions is
fixed. It is given by $N_{a_{\eta}}=[N_a-2S_c]$, $N_{a_{s}}=2S_c$,
and $N_c=2S_c$, respectively. Hence the number $2S_c$ of 
rotated-electron singly occupied sites and the numbers $N_{a_{\eta}}$
and $N_{a_{s}}$ of sites of the $\eta$-spin and spin lattices,
respectively, are fixed. 

As discussed in this paper, for both the model on 1D lattice considered in this Appendix and
the model on the square lattice the $S_c>0$ vacuum of the
LWS subspace $\vert 0_{\eta s}\rangle$ given in Eq. (\ref{vacuum}) is 
invariant under the electron - rotated-electron unitary transformation.
For $D=1$ the $N_{a_{s}}$ deconfined $+1/2$ spinons 
of such a vacuum are the spins of $N_{a_{s}}$ 
spin-up electrons, the $N_{a_{\eta}}$ deconfined $+1/2$ $\eta$-spinons refer 
to the $N_{a_{\eta}}$ sites unoccupied by electrons, and
the $N_c$ $c$ fermions describe the charge degrees of freedom
of such electrons of the fully polarized state.
 
The corresponding LWSs that span the above subspace
refer to rotated electrons rather than to electrons. They
have the general form given in Eq. (\ref{LWS-full-el}).
In it the set of numbers $\{N_{\alpha\nu}\}$ obey the
sum-rule associated with the expressions provided in Eq.
(\ref{M-L-Sum}), the $\eta$-spin ($\alpha =\eta$) and
spin ($\alpha =s$) can have values $S_{\alpha}=0,..., N_{a_{\alpha}}/2$
such that $S_{\alpha} =[N_{a_{\alpha}}/2 -M^{co}_{\alpha}/2]$,
and the set of numbers $\{N_{a_{\alpha\nu}}\}$ of discrete 
momentum values of each $\alpha\nu$ band are well
defined and given by Eq. (\ref{N*}) and (\ref{N-h-an}).

As mentioned previously, each subspace with fixed values of $S_c$ and hence
also with fixed values of $N_{a_{\eta}}=[N_a-2S_c]$ and $N_{a_{s}}=2S_c$,
which the vacuum of Eq. (\ref{vacuum}) refers to,
can be divided into smaller subspaces. Those have fixed values for $S_c$, $S_{\eta}$, 
and $S_s$ and hence also fixed values for $M^{co}_{\eta}$ and $M^{co}_s$.
Furthermore, the latter subspaces can be further divided into even smaller 
subspaces with fixed values for the set of numbers $\{N_{\eta\nu}\}$ and 
$\{N_{s\nu}\}$, which obey the sum-rule of Eq. (\ref{M-L-Sum}).   

In contrast to the initial creation and annihilation fields,
the $c$ and $\alpha\nu$ fermion operators generate the Bethe
states free from unwanted and unphysical terms,
are labeled by the quantum 
numbers of the diagonalized model, and their algebra does
not refer to the full LWS or HWS subspace but instead
to subspaces spanned by well-defined types of
Bethe states. Hence there is no contradiction whatsoever between the 
charge ABCDF algebra \cite{Martins} and spin ABCD traditional Faddeev-Zamolodchikov 
algebra \cite{ISM} associated with involved commutation relations of the initial fields
and the anticommutation relations of the $c$ and $\alpha\nu$ fermion operators
provided in Eqs. (\ref{albegra-cf-1D}) and (\ref{1D-anti-com-1D}).
Indeed, the initial creation and annihilation fields act onto an
extended and partially unphysical Hilbert space. In turn, the
$c$ and $\alpha\nu$ fermion operators act onto well-defined 
subspaces of the model physical Hilbert space.

Upon acting onto such subspaces, the operators $f^{\dag}_{q_{j},c}$
of the $c$ fermions and $f^{\dag}_{q_{j},\alpha\nu}$ of the
$\alpha\nu$ fermions have the expected simple anticommutation
relations associated with those provided in Eqs. (\ref{albegra-cf-1D}) 
and (\ref{1D-anti-com-1D}) for the corresponding operators 
$f^{\dag}_{x_{j},c}$ and $f^{\dag}_{x_{j},\alpha\nu}$, respectively.
Moreover, $c$ fermion operators commute with the $\alpha\nu$
fermion operators and $\alpha\nu$ and $\alpha'\nu'$ fermion
operators belonging to different $\alpha\nu\neq\alpha'\nu'$ 
branches also commute with each other.

\section{Full information about the quantum problem when
defined in the LWS subspace}

In this Appendix we confirm that full information about the quantum problem
can be achieved by defining it in the LWS subspace spanned
by the energy eigenstates that are both LWSs of the $\eta$-spin
and spin algebras. By that we mean that all ${\cal{N}}$-electron 
operator matrix elements between energy eigenstates such
that at least one of them is a non-LWS can be evaluated exactly 
in terms of a  corresponding quantum problem involving 
another well-defined ${\cal{N}}$-electron operator acting 
onto the LWS subspace. Here ${\cal{N}}=1,2,...$ refers
to one, two, or any other finite number of electrons. Hence the expression
of the general ${\cal{N}}$-electron operator under consideration  
reduces to an elementary creation or annihilation electronic
operator (${\cal{N}}=1$) or involves the product of two or more
such elementary operators (${\cal{N}}\geq 2$). 

In addition, we show here that the use of the model
global symmetry provides a simple relation between
the energy of any non-LWS and the corresponding
LWS. It follows that all contributions to the physical quantities
from non-LWSs can be evaluated by considering 
a related problem defined in the LWS subspace. 
The use of such a symmetry 
reveals that for hole concentration
$x>0$ and spin density $m>0$ the ground state
is always a LWS of both the $\eta$-spin and
spin algebras. For simplicity here we consider matrix elements
between the ground state and a non-LWS. Those
appear in Lehmann representations of  
zero-temperature spectral functions
and correlation functions. Similar results can be
obtained for matrix elements between any excited 
energy eigenstates. 
 
Expressions (\ref{fc+})-(\ref{rotated-quasi-spin})
for the $c$ fermion, $\eta$-spinon, and spinon operators refer to the
LWS subspace. A similar representation can be used for instance for
the HWS subspace. It is spanned by all energy eigenstates 
with $S_{\alpha}=S^{x_3}_{\alpha}$ where $\alpha =\eta ,s$. 
(There are also two mixed subspaces such that $S_{\eta}=\pm S^{x_3}_{\eta}$
and $S_{s}=\mp S^{x_3}_{s}$.) The HWS representation is suitable 
for canonical ensembles referring to electronic densities larger than 
one and negative spin densities. In that case the ground states are 
HWSs of both the $\eta$-spin and spin algebras.

We start by confirming that there is a well-defined ${\cal{N}}$-electron operator
${\hat{\Theta}}_{{\cal{N}}}$ such that any matrix element 
$\langle f\vert\,{\hat{O}}_{{\cal{N}}}\vert\psi_{GS} \rangle$ of
a ${\cal{N}}$-electron operator ${\hat{O}}_{{\cal{N}}}$ 
can be written as $\langle f\vert\,{\hat{O}}_{{\cal{N}}}\vert\psi_{GS} \rangle =
\langle f.LHS\vert{\hat{\Theta}}_{{\cal{N}}} \vert\psi_{GS} \rangle$.
Here $\vert f\rangle$ is a non-LWS of the $\eta$-spin algebra and/or spin algebra,
$\vert\psi_{GS} \rangle$ is the ground state of the Hubbard
model on the 1D or square lattice, and $\vert f.LWS\rangle$ is the LWS 
that corresponds to the state $\vert f\rangle$. Within the two $SU(2)$ algebras the
latter state can be expressed as,
\begin{equation}
\vert f\rangle = \prod_{\alpha
=\eta,s}\frac{1}{\sqrt{{\cal{C}}_{\alpha}}}({\hat{S}}^{\dag}_{\alpha})^{M^{de}_{\alpha,-1/2}}\vert f.LWH\rangle \, . 
\label{IRREG}
\end{equation}
In this expression,
\begin{equation}
{\cal{C}}_{\alpha} = \delta_{M^{de}_{\alpha,-1/2},0} +
\prod_{l=1}^{M^{de}_{\alpha,-1/2}}l\,[\,M^{de}_{\alpha}+1-l\,] \, , 
\label{Calpha}
\end{equation}
is a normalization constant and $M^{de}_{\alpha,-1/2}\leq M^{de}_{\alpha}=2S_{\alpha}$.
The $\eta$-spin flip ($\alpha =\eta$) and spin flip ($\alpha =s$)
operators ${\hat{S}}^{\dag}_{\alpha}$ are the off-diagonal generators of the
corresponding $SU(2)$ algebras provided in Eq. (\ref{Scs}).
These operators remain invariant under the electron - rotated-electron unitary
transformation. Thus as given in that equation they have the same expression 
in terms of electron and rotated-electron creation and annihilation operators. 

As confirmed in Section IV,
for a hole concentration $x\geq 0$ and spin density $m\geq 0$
the ground state is a LWS of both the $\eta$-spin and spin $SU(2)$ algebras. 
Thus it has the following property,
\begin{equation}
{\hat{S}}_{\alpha}\,\vert\psi_{GS} \rangle = 0 \, ; \hspace{0.5cm} \alpha = \eta,s \, .
\label{GS=0}
\end{equation}
The operator ${\hat{\Theta}}_{{\cal{N}}}$ is then such that,
\begin{eqnarray}
\langle f\vert\,{\hat{O}}_{{\cal{N}}}\vert\psi_{GS} \rangle & = & \langle f.LWS\vert 
\prod_{\alpha =\eta,s}{1\over
\sqrt{{\cal{C}}_{\alpha}}}({\hat{S}}_{\alpha})^{M^{de}_{\alpha,-1/2}}\,
{\hat{O}}_{{\cal{N}}} \vert\psi_{GS} \rangle 
\nonumber \\
& = & \langle f.LHS\vert
{\hat{\Theta}}_{{\cal{N}}} \vert\psi_{GS} \rangle  \, . 
\label{SYMME}
\end{eqnarray}

By the suitable use of Eq. (\ref{GS=0}), it is straightforward to show that 
the operator ${\hat{\Theta}}_{{\cal{N}}}$ is given by the following commutator,
\begin{equation}
{\hat{\Theta}}_{{\cal{N}}} = \Bigl[\prod_{\alpha =\eta,s}{1\over
\sqrt{{\cal{C}}_{\alpha}}}\Bigr]\,\Bigl[\prod_{\alpha
=\eta,s}({\hat{S}}_{\alpha})^{M^{de}_{\alpha,-1/2}},
{\hat{O}}_{{\cal{N}}}\Bigr] \, , 
\label{Th}
\end{equation}
for $M^{de}_{\eta,-1/2}>0$ and/or $M^{de}_{s,-1/2}>0$ and by,
\begin{equation}
{\hat{\Theta}}_{{\cal{N}}} = {\hat{O}}_{{\cal{N}}} \, ,  
\label{OL-0}
\end{equation}
for $M^{de}_{\eta,-1/2}=M^{de}_{s,-1/2}= 0$. If the commutator on
the right-hand side of Eq. (\ref{Th}) vanishes then the matrix element
$\langle f\vert\,{\hat{O}}_{{\cal{N}}}\vert\psi_{GS} \rangle$ under
consideration also vanishes.

We denote by $E_f$ the energy eigenvalue of the non-LWS $\vert f\rangle$ and 
by $E_{f.LWS}$ that of the corresponding LWS $\vert f.LWS\rangle$. To find the 
relation between $E_f$ and $E_{f.LWS}$ one adds chemical-potential and 
magnetic-field operator terms to the Hamiltonian (\ref{H}), what lowers its 
symmetry. Such operator terms commute with the Hamiltonian (\ref{H}). 
Thus the rotated-electron occupancy configurations 
of all energy eigenstates correspond to state representations of its 
global symmetry for all densities. Moreover, the use of such commutation relations
reveals that the energy eigenvalues $E_f$ and $E_{f.LWS}$ are related as,
\begin{equation}
E_f = E_{f.LWS} + \sum_{\alpha =\eta,s}\epsilon_{\alpha,-1/2}\,M^{de}_{\alpha,-1/2} \, .
\label{energies}
\end{equation}
Here $\epsilon_{\eta,-1/2} =2\mu$ and $\epsilon_{s,-1/2} = 2\mu_B\,H$ are the
deconfined $-1/2$ $\eta$-spinon energy of Eq. (\ref{energy-eta}) for $x>0$ and the 
deconfined $-1/2$ spinon energy of Eq. (\ref{energy-s}) for $m>0$, respectively.

It follows from Eq. (\ref{energies}) that $E_f\geq E_{f.LWS}$. 
For a hole concentration $x>0$ and spin density $m>0$ 
such an inequality can be shown to be consistent with 
the ground state being always a LWS of both the $\eta$-spin and spin algebras.
On the other hand, the model global symmetry requires that $E_f = E_{f.LWS}$ for the
half-filing and zero-magnetization absolute ground state. Such a requirement
is fulfilled: For half filling one has that $\mu\in (-\mu^0,\mu^0)$. 
The value $\mu=0$ corresponds to the middle of the Mott-Hubbard gap.
In turn, the magnetic field $H$ vanishes for zero magnetization. The
absolute ground state corresponds to $S_{\eta}=S_s=0$ and $S_c =N_a^D/2$.
Hence it is both a LWS and a HWS of the $\eta$-spin and spin algebras.

Within a Lehmann representation the ${\cal{N}}$-electron spectral functions
are expressed as a sum of terms, one for each excited energy eigenstate. In
spectral-function terms associated with non-LWSs one can then replace the 
matrix element $\langle f\vert\,{\hat{O}}_{{\cal{N}}}\vert\psi_{GS} \rangle$ by 
$\langle f.LHS\vert{\hat{\Theta}}_{{\cal{N}}} \vert\psi_{GS} \rangle$. This holds
provided that the excited-state energy $E_f$ is expressed as in Eq. (\ref{energies}).
Then the original ${\cal{N}}$-electron spectral function can be written as a 
sum of spectral functions of well-defined ${\cal{N}}$-electron operators 
of the form given in Eqs. (\ref{Th}) or (\ref{OL-0}), all acting onto the LWS subspace.
 
These results confirm that full information about the present quantum problem
can be obtained by defining it in the LWS subspace. 
 
\section{Additional information on the quantum liquid of $c$ and $s1$ fermions}

In the following we provide further information about why the square-lattice quantum liquid corresponding to the
Hubbard model on the square lattice in the one- and two-electron subspace
may be described only by $c$ and $s1$ fermions on their $c$ and $s1$
effective lattices, respectively. Some of our results apply to the 1D model as well.
For simplicity, in this Appendix we limit our considerations to $x>0$ initial ground states and their excited states
of energy below $2\mu$. However, similar results hold for the model in the
one- and two-electron subspace spanned by the $x=0$, $\mu=0$, and $m=0$
ground state and its excited states whose number value ranges are those 
provided in Eq. (\ref{srs-0-ss}). 

In addition, below we address the issue concerning the states of Eq. (\ref{non-LWS}) that belong to the one- and 
two-electron subspace being both energy and momentum eigenstates and provide their specific form in that subspace. 

\subsection{Effects of the objects other than the $c$ and $s1$ fermions}

According to the number value ranges of Eqs. (\ref{srs-0-ss}) and (\ref{deltaNcs1}), the
one- and two-electron subspace is for hole concentrations $x>0$ spanned by an initial $m=0$ 
ground state plus its excited states of energy $\omega <2\mu$ having either none $N_{s2}=0$ or one $N_{s2}=1$
spin-neutral four-spinon $s2$ fermion. $N_{s2}=1$ spin-singlet excited
states belonging to that subspace have no deconfined spinons. One then finds 
from the use of Eq. (\ref{N-h-an}) for the $\alpha\nu =s2$ branch
that $N^h_{s2}=0$ for the latter states, so that they have no holes in the $s2$ momentum band 
and thus $N_{a_{s2}}^D=1$. This means that for such states the $s2$ fermion occupies a $s2$ 
band with a single vanishing momentum value. 
Since such a $s2$ fermion meets the criterion of Eq. (\ref{invariant-V}) of invariance 
under the electron - rotated-electron unitary transformation, the only explicit effect of its creation  
is onto the numbers of occupied and unoccupied sites of 
the $s1$ effective lattice and corresponding numbers of 
$s1$ band $s1$ fermions and $s1$ fermion holes. Specifically, according to the
expressions provided in Eq. (\ref{Nas1-Nhs1}), 
the deviations $\delta S_c =\delta S_s =0$ and $\delta N_{s2}=1$ generated by a state
transition involving creation of one $s2$ fermion lead to deviations in the number
of $s1$ fermions and $s1$ fermion holes given by 
$\delta N_{s1} = -2\delta N_{s2}=-2$ and
$\delta N^h_{s1} = 2\delta N_{s2}=2$, respectively.
 
Moreover, the ranges of Eqs. (\ref{srs-0-ss}) and (\ref{deltaNcs1}) confirm that such $N_{s2}=1$ excited states 
have zero spin, $S_s=0$. According to Eq. (\ref{Nas1-Nhs1}),
the number of holes in the $s1$ band is $N^h_{s 1} = 2N_{s2} =2$
for such states, in contrast to $N^h_{s 1}=0$ for the initial ground state.
In turn, the number $N_{a_{s 1}}^D$ of sites of the $s1$ effective lattice
remains unaltered. Following the annihilation of two $s1$ fermions and
creation of one $s2$ fermion, two unoccupied sites appear in 
the $s1$ effective lattice. As a result, two holes emerge in the $s1$ band as well. The emergence of these 
unoccupied sites and holes involves two virtual processes where 
(i) two $s1$ fermions are annihilated and
four deconfined spinons are created and (ii)
the latter deconfined spinons are annihilated
and the $s2$ fermion is created. 

Hence the only explicit net effect of the creation of a single vanishing-energy and zero-momentum
$s2$ fermion is the annihilation of two $s1$ fermions and corresponding emergence
of two holes in the $s1$ band and two unoccupied sites in the $s1$ effective lattice. Therefore, 
in the case of the one- and two-electron subspace at excitation energy $\omega <2\mu$ one can
ignore that object in the theory provided that the corresponding changes in the
$s1$ band and $s1$ effective lattice occupancies are accounted for.
Within neutral $s1$ fermion particle-hole processes of transitions between
two excited states with a single $s2$ fermion, two of the four spinons of such 
an object are used in the motion of $s1$ fermions around in the $s1$ effective 
lattice. Indeed, such two spinons play the role of unoccupied sites of that lattice, consistently with the expression
$N^h_{s 1} = 2N_{s2}$ obtained from Eq. (\ref{Nas1-Nhs1}) for $S_s=0$.

Spin-singlet excitations of the model on the square lattice generated by application onto a $x> 0$ and $m=0$
initial ground state of the operator $f^{\dag}_{0,s2}\,f_{{\vec{q}},s1}\,f_{{\vec{q}}\,',s1}$ where ${\vec{q}}$ and
${\vec{q}}\,'$ are the momenta of the two emerging $s1$ fermion holes are
neutral states that conserve $S_c$, $S_s$, and $N_{a_{s1}}^2$. The implicit role of the $s2$ fermion creation operator
$f^{\dag}_{0,s2}$ is exactly canceling the contributions of the
annihilation of the two $s1$ fermions of momenta ${\vec{q}}$ and ${\vec{q}}\,'$   
to the commutator $[\hat{q}_{s1\,x_1},\hat{q}_{s1\,x_2}]$ of the square-lattice model $s1$ translation generators
in the presence of the fictitious magnetic field ${\vec{B}}_{s1}$ of Eq. (\ref{A-j-s1-3D}).
This ensures that the overall excitation is neutral. Since the $s2$ fermion has vanishing
energy and momentum and the $s1$ band and its number $N_{a_{s1}}^2$ of
discrete momentum values remain unaltered, one can effectively consider that the generator
of such an excitation is $f_{{\vec{q}},s1}\,f_{{\vec{q}}\,',s1}$ and omit the $s2$
fermion creation operator. Its only role is ensuring that the overall excitation is neutral
and the two components of the square-lattice model $s1$ fermion microscopic momenta can be specified. It follows
that for the one- and two-electron subspace considered here the operators $f_{{\vec{q}},s1}\,f_{{\vec{q}}\,',s1}$,
$f^{\dag}_{{\vec{q}}\,',s1}\,f^{\dag}_{{\vec{q}},s1}$, $f^{\dag}_{{\vec{q}},s1}\,f_{{\vec{q}}\,',s1}$,
and $f_{{\vec{q}},s1}\,f^{\dag}_{{\vec{q}}\,',s1}$ generate neutral excitations.

Also the $M^{de}_{s} =2S_s$ deconfined spinons play the role of unoccupied sites of
the $s1$ effective lattice. Again, this is consistent with the expression
$N^h_{s 1} = M^{de}_{s} =2S_s=1,2$ obtained from Eq. (\ref{Nas1-Nhs1}) for $N_{s2}=0$ .
As given in Eq. (\ref{srs-0-ss}), for $x>0$ and excitation energy $\omega <2\mu$
the one- and two-electron subspace $M^{de}_{s}$ allowed
values are ${\cal{N}}$-electron-operator dependent and given by $M^{de}_{s} =0,...,{\cal{N}}$ where ${\cal{N}}=1,2$
for $x>0$. For $M^{de}_{s}=2S_s=1,2$ one has that $N_{s2} =0$. 
Now in contrast to creation of a single $s2$ fermion, a deviation $\delta 2S_s=1,2$ 
generated by a transition from the ground state to such $2S_s=1,2$ 
excited states may lead to deviations in the numbers
of occupied and unoccupied sites of the $s1$
effective lattice and corresponding $s1$ fermion and $s1$
fermion holes that do not obey the
usual equality $\delta N_{s1}= -\delta N^h_{s1}$.
Indeed, in the present case $2\delta S_c=\pm 1$ for 
$\delta N^h_{s 1} = 2\delta S_s=1$ and $2\delta S_c=0,\pm 2$
for $\delta N^h_{s 1} = 2\delta S_s=2$. Hence 
according to the expressions provided in Eq. (\ref{Nas1-Nhs1}),
such deviations lead to deviations in the numbers
of occupied and unoccupied sites of the $s1$
effective lattice and corresponding numbers of $s1$ fermions and
$s1$ fermion holes. Those read
$\delta N_{s1} = [\delta S_c-\delta S_s]$ and
$\delta N^h_{s1} = \delta 2S_s$, respectively.
It follows that the total number of sites and thus of
discrete momentum values of the $s1$ band may  
change under such transitions. This leads to an additional deviation
$\delta N_{a_{s1}}^D = [\delta S_c +\delta S_s]$. For one-electron excited states one has that
$\delta N^h_{s 1} = 2\delta S_s=1$ and $2\delta S_c=\pm 1$. As a result,
$\delta N_{s1} = \pm 1/2-1/2=-1,0$ and $\delta N_{a_{s1}}^D = \pm 1/2 +1/2=0,-1$.
In turn, for $N_{s2}=0$ two-electron excited states one has $\delta N^h_{s 1} = 2\delta S_s=2$
and $2\delta S_c=0,\pm 2$. Thus $\delta N_{s1} = -1, (\pm 1-1)=-2,-1,0$
and $\delta N_{a_{s1}}^D = 1,(\pm 1+1)=0,1,2$. 

For the $s1$ fermion operators $f^{\dag}_{{\vec{q}},s1}$ and $f_{{\vec{q}},s1}$,
excitations that involve changes $\delta N_{a_{s1}}^D = [\delta S_c +\delta S_s]$
in the number of sites and discrete momentum values of the $s1$ effective
lattice and $s1$ band, respectively, correspond to transitions between different quantum problems. Indeed,
such operators act onto subspaces spanned by neutral states, which conserve 
$S_c$, $S_s$, and $N_{a_{s1}}^D$. In turn, the generator of a non-neutral excitation
is in general the product of two operators. The first operator adds sites to or removes sites from
the $s1$ effective lattice or adds discrete momentum values to or removes discrete momentum values from
the corresponding $s1$ momentum band. Such small changes account for the above deviations $\delta N_{a_{s1}}^D = [\delta S_c +\delta S_s]$.
The second operator is a $s1$ fermion operator or a product of such operators appropriate to
the excited-state subspace.

Also the vanishing momentum and energy $M^{de}_{\eta,+1/2}=x\,N_a^D$ deconfined $+1/2$
$\eta$-spinons (or $M^{de}_{\eta,-1/2}=x\,N_a^D$ deconfined $-1/2$ $\eta$-spinons if one
uses a HWS representation suitable to $x<0$)
are invariant under the electron - rotated-electron unitary 
transformation. Their creation or annihilation may be accounted for 
by small suitable changes in occupancies of the $c$ effective lattice and $c$ 
momentum band. For $x>0$, excitation energy below $2\mu$, and the one- and two-electron subspace 
considered here such deconfined $+1/2$ $\eta$-spinons correspond to a single
occupancy configuration associated with the $\eta$-spin vacuum $\vert 0_{\eta};N_{a_{\eta}}^D\rangle$
of Eq. (\ref{vacuum}). In turn, the degrees of freedom
of the rotated-electron occupancies of such $x\,N_a^D$ sites of the original 
lattice associated with the hidden $U(1)$ symmetry
refer to the unoccupied sites of the $c$ effective lattice of Eq. (\ref{Nac-Nhc}) and
corresponding $c$ band holes. Hence the number $M^{de}_{\eta,+1/2}=x\,N_a^D$ 
of deconfined $+1/2$ $\eta$-spinons equals that of $N^h_c=x\,N_a^D$
unoccupied sites of the $c$ effective lattice and corresponding $c$ band holes.
This confirms that for $x>0$ and excitation energy $\omega <2\mu$ the deviations $\delta M^{de}_{\eta,+1/2}=(\delta x)\,N_a^D$
originated from creation and annihilation of 
deconfined $+1/2$ $\eta$-spinons have within the one- and two-electron subspace
no effects on the physics other than the corresponding 
deviation $\delta N^h_c=(\delta x)\,N_a^D$
in the number of unoccupied sites of the $c$ effective lattice and $c$ band holes. 

Concerning the effects of the presence of a single $\eta 1$ fermion or
of one or two deconfined $-1/2$ or $+1/2$ $\eta$-spinons in the excited states 
of the $x=0$, $\mu=0$, and $m=0$ ground state, similar results to those reported
above concerning the effects of a single $s2$ fermion or
one or two deconfined $-1/2$ or $+1/2$ spinons apply. 
Again, except for a fixed uniquely defined finite-energy gap for each type of excitation,
which vanishes for spin excitations \cite{companion},
the half-filling Hubbard model on the square lattice in the one- and two-electron
subspace as defined in Section V can be described by a quantum liquid of
$c$ and $s1$ fermions.

\subsection{The $c$ and $s1$ fermion momentum values and the energy eigenstates}
                 
Here we provide the specific form that the momentum eigenstates of Eq. (\ref{non-LWS})
have in the one- and two-electron subspace for $x>0$ and excitation energy $\omega <2\mu$. 
Such states refer to a complete set of states in the full Hilbert space. 
Corresponding simple expressions apply to the excited states of the $x=0$, $\mu=0$, and $m=0$ ground state
that belong to the one- and two-electron subspace as defined in Section V.

In general the states of Eq. (\ref{non-LWS}) are not energy eigenstates of the model on the square lattice. 
Fortunately, in the one- and two-electron subspace such momentum energy eigenstates
are energy eigenstates. This confirms the usefulness of both the general description introduced 
in this paper and the square-lattice quantum liquid that refers to the Hubbard model in that subspace.
      
The $s1$ band discrete momentum values ${\vec{q}}_j$ where $j=1,...,N_{a_{s1}}^D$ 
are the conjugate of the real-space coordinates ${\vec{r}}_j$ of the $s1$ effective lattice
for which also $j=1,...,N_{a_{s1}}^D$. The same applies to the $c$ band discrete momentum values 
${\vec{q}}_j$ and the $c$ effective lattice real-space coordinates ${\vec{r}}_j$ 
where in both cases $j=1,...,N_{a}^D$. (The latter lattice is identical to the original lattice.)
As discussed in Section IV-B, the $c$ translation generators ${\hat{{\vec{q}}}}_c$
commute with both the Hamiltonian and momentum operator for the
whole Hilbert space. This is why the $c$ band discrete momentum values 
are good quantum numbers. In turn, the $s1$ translation generators ${\hat{{\vec{q}}}}_{s1}$ in the 
presence of the fictitious magnetic field ${\vec{B}}_{s1}$ of Eq. (\ref{A-j-s1-3D})
do not commute in general
with the Hamiltonian of the Hubbard model on the square lattice. 

The one- and two-electron subspace as defined in Section V is a subspace of type (A)
considered in Section IV-F. Hence, the $U/4t\rightarrow\infty$ local spin $SU(2)$ gauge symmetry  
implies that the $s1$ band momenta are good quantum numbers 
for the Hubbard model on the square lattice. That is so provided that for the model in the one- and two-electron subspace 
the numbers $N_{s1}^h$ and $N_{s1}$ are conserved. Since in that subspace those are indeed good quantum numbers, the unitarity of the operator $\hat{V}$ then 
implies that in such a subspace the $s1$ band momentum values are conserved for $U/4t>0$ as well.
Thus the Hamiltonian of the Hubbard model on the square lattice in the one- and two-electron subspace 
commutes with the $s1$ translation generators $\hat{q}_{s1\,x_1}$ and $\hat{q}_{s1\,x_2}$ 
in the presence of the corresponding fictitious magnetic field ${\vec{B}}_{s1}$. Indeed
in the neutral subspaces of the one- and two-electron subspace the $s1$ translation generators ${\hat{{\vec{q}}}}_{s1}$ 
can be constructed to inherently commuting with both the Hamiltonian and momentum operator of the Hubbard
model on the square lattice. Note however that this requires a suitable definition of the $s1$ fermions in terms of all spin-effective
lattice two-site bonds centered at the local $s1$ fermion spatial coordinate that in the $U/4t\rightarrow\infty$ limit
profits from the model local spin $SU(2)$ gauge symmetry. 

Since in contrast to the $c$ fermions, the $s1$ fermions have internal structure,
how is the $s1$ fermion momentum ${\vec{q}}$ related to the two underlying spinons?
Deconfined spinons carry no momentum and are invariant under the electron - rotated-electron unitary 
transformation. On the other hand, within the LWS
representation of the spin $SU(2)$ algebra, for the states that span the one- and two-electron
subspace for which $N_{s1}^h$ may have the values $N_{s1}^h=0,1,2$, the spin-down
spinon of the spin-singlet two-spinon $s1$ fermion of
momentum ${\vec{q}}$ carries momentum ${\vec{q}}$ and its 
spin-up spinon carries momentum $-{\vec{q}}$.
In turn, within the HWS representation of that algebra, 
its spin-down spinon carries momentum $-{\vec{q}}$ and its spin-up spinon carries momentum ${\vec{q}}$.
Within the present LWS representation, the spin-singlet two-spinon
$s1$ fermions of momenta (i) ${\vec{q}}$ and (ii) $-{\vec{q}}$ involve (i) a spin-down
spinon of momentum ${\vec{q}}$ and a spin-up spinon of momentum $-{\vec{q}}$
and (ii) a spin-down spinon of momentum $-{\vec{q}}$ and a spin-up spinon of momentum ${\vec{q}}$,
respectively.  We emphasize that for $N_{s1}^h=0$ states, one has that ${{\vec{q}}}_{s1}=0$ in the
momentum spectrum ${\vec{P}} = {{\vec{q}}}_c + {{\vec{q}}}_{s1}$,
so that the total $s1$ band momentum vanishes. As discussed elsewhere, the momenta
${\vec{q}}$ or ${\vec{q}}$ and ${\vec{q}}\,'$ of the $s1$ band holes of $N_{s1}^h=1,2$ states 
can alternatively be associated with that of the annihilated $s1$ fermion or corresponding
spinons of the broken spinon pair.  

The set of neutral energy eigenstates 
that for $x>0$ and excitation energy $\omega <2\mu$ span the subspaces of the one- and two-electron subspace  
are particular cases of the general momentum eigenstates $\vert \Phi_{U/4t}\rangle$ of Eq. (\ref{non-LWS}). The use
of the general expression of such states given in that equation leads to the following
corresponding general form for such energy eigenstates $\vert \Psi_{U/4t}\rangle =\vert \Phi_{U/4t}\rangle$,
\begin{equation}
\vert \Psi_{U/4t}\rangle = \frac{1}{\sqrt{{\cal{C}}_{s}}}({\hat{S}}^{\dag}_{s})^{M^{de}_{s,-1/2}}\vert \Psi_{LWS;U/4t}\rangle 
\, ; \hspace{0.5cm} {\cal{C}}_{s} = \delta_{M^{de}_{s,-1/2},0} +
\prod_{l=1}^{M^{de}_{s,-1/2}}l\,[\,M^{de}_{s}+1-l\,] = 1, 2, 4 \, .
\label{non-LWS-c-s1}
\end{equation}
The LWS appearing in this equation refers to a particular form of the general
LWS of Eq. (\ref{LWS-full-el}) and reads,
\begin{equation}
\vert \Psi_{LWS;U/4t}\rangle = [\vert 0_{\eta};N_{a_{\eta}}^D\rangle]
[\prod_{{\vec{q}}\,'}f^{\dag}_{{\vec{q}}\,',s1}\vert 0_{s1};N_{a_{s}^D}\rangle]
[\prod_{{\vec{q}}}f^{\dag}_{{\vec{q}},c}\vert GS_c;0\rangle] \, .
\label{LWS-full-el-c-s1}
\end{equation}
Here $f^{\dag}_{{\vec{q}}\,',s1} ={\hat{V}}^{\dag}\,{\mathcal{F}}^{\dag}_{{\vec{q}}\,',s1}\,{\hat{V}}$,
$f^{\dag}_{{\vec{q}},c} ={\hat{V}}^{\dag}\,{\mathcal{F}}^{\dag}_{{\vec{q}},c}\,{\hat{V}}$, and
${\mathcal{F}}^{\dag}_{{\vec{q}}\,',s1}$ and ${\mathcal{F}}^{\dag}_{{\vec{q}},c}$ are the
creation operators of a $U/4t\rightarrow\infty$ $s1$ fermion of momentum ${\vec{q}}\,'$ and $c$ fermion of
momentum ${\vec{q}}$, respectively. Moreover, $\vert 0_{\eta};N_{a_{\eta}}^D\rangle$ is
the $\eta$-spin $SU(2)$ vacuum associated with $N_{a_{\eta}}^D$ deconfined $+1/2$ $\eta$-spinons,
$\vert 0_{s};N_{a_{s}}^D\rangle$ is the spin $SU(2)$ vacuum associated with $N_{a_{s}}^D$ 
deconfined $+1/2$ spinons, and $\vert GS_c;0\rangle$ is the $c$ fermion $U(1)$
vacuum. Such three vacua are invariant under the electron - rotated-electron unitary
transformation, refer to the model global
$[SU(2)\times SU(2)\times U(1)]/Z_2^2 =SO(3)\times SO(3)\times U(1)$ symmetry, 
and appear in the theory vacuum of Eq. (\ref{vacuum}). 
(In that equation, $\vert GS_c;2S_c\rangle=\prod_{{\vec{q}}}f^{\dag}_{{\vec{q}},c}\vert GS_c;0\rangle$.)

\section{Hard-core character of the $s1$ bond-particle operators}

The goal of this Appendix is to confirm the validity of the relations 
provided in Eqs. (\ref{g-local}) and (\ref{g-non-local}).
All the results given in the following apply both to the model on the square
and 1D lattices. The 1D expressions are readily obtained if one considers
in the general expressions given below only $d=1$ contributions and terms,
together with the choice $D=1$ in the $D$-dependent quantities. Here the 
bond index $d$ is that appearing in the $\sum_{d=1}^{D}$ summation 
of the operators $a_{\vec{r}_{j},s1,g}^{\dag}$ and $a_{\vec{r}_{j},s1,g}$ 
expression provided in Eq. (\ref{g-s1+general}).

In order to probe the relations of Eqs. (\ref{g-local}) and (\ref{g-non-local}), we consider 
without any loss of generality that the initial state is a $N^h_{s1}=4$ configuration state.
Indeed, application of a $s1$ bond-particle creation 
operator onto $N^h_{s1}=0,1$ configuration states gives 
zero. Application of two $s1$ bond-particle creation 
operators onto $N^h_{s1}=2$ configuration states gives 
zero as well. Fortunately, the discussions about $N^h_{s1}=0$ configuration
states of Section VI can be generalized to $N^h_{s1}=4$ configuration states. 
One finds that application onto some of such configuration states of two two-site
bond operators may not give zero. Furthermore, it turns out that for such $N^h_{s1}=4$
configuration states the presence of the four unoccupied sites in the $s1$ effective lattice
does not affect the spinon occupancy configurations of the $N_{s1}$ $s1$
bond particles so that their operator expressions are the same as for 
the $N^h_{s1}=0$ configuration state studied in Section VI.

We start by combining the expressions given in Eq. (\ref{sir-pir}) with the algebraic relations
provided in Eqs. (\ref{albegra-s-h})-(\ref{albegra-q-com}), what readily leads to the
following usual algebra for the spinon operators defined in Eq. (\ref{sir-pir}),
\begin{equation}
\{s^{+}_{\vec{r}_j},s^{-}_{\vec{r}_j}\} = 1 \, ,
\hspace{0.5cm}
\{s^{\pm}_{\vec{r}_j},s^{\pm}_{\vec{r}_j}\} = 0 \, ,
\label{albegra-s-p-m}
\end{equation}
\begin{equation}
[s^{+}_{\vec{r}_j},s^{-}_{\vec{r}_{j'}}] =
[s^{\pm}_{\vec{r}_j},s^{\pm}_{\vec{r}_{j'}}]=0 \, ,
\label{albegra-s-com}
\end{equation}
for $j\neq j'$ and,
\begin{equation}
[s^{x_3}_{\vec{r}_j},s^{\pm}_{\vec{r}_{j'}}] =
\pm\delta_{j,j'}s^{\pm}_{\vec{r}_{j}} \, .
\label{albegra-s-sz-com}
\end{equation}
It follows that the spinon operators $s^{\pm}_{\vec{r}_j}$ anticommute 
on the same site and commute on different sites. Consistently 
with the rotated-electron singly-occupied site projector 
$n_{\vec{r}_j,c}$ appearing in the expression of the spinon 
operators $s^{\pm}_{\vec{r}_j}$ and $s^{x_3}_{\vec{r}_j}$
provided in Eq. (\ref{sir-pir}) and the discussion of Section III on
the justification of the validity of the concept of a spin effective lattice
in the $N_a^D\rightarrow\infty$ limit, the real-space coordinates 
$\vec{r}_j$ of such operators can in that limit be identified
with those of the latter lattice. The corresponding operator index values $j=1,...,N_{a_s}^D$
are thus those of that effective lattice.

For the $N^h_{s1}=4$ configuration states under consideration
the operators $g^{\dag}_{{\vec{r}}_{j},s1}$ (and $g_{{\vec{r}}_{j},s1}$) 
that create (and annihilate) a $s1$ bond particle at a site of 
the spin effective lattice of real-space coordinate ${\vec{r}}_{j}$ have 
the general form given in Eq. (\ref{g-s1+general}) both for the 
Hubbard model on the 1D and square lattices. The absolute value $\vert h_{g}\vert$
of the coefficients $h_{g}$ appearing in that equation decreases for increasing two-site bond length 
$\xi_{g}$ and obeys the normalization sum-rule
(\ref{g-s1+sum-rule}). Hence the expression of 
the $s1$ bond-particle operator $g_{\vec{r}_{j},s1}^{\dag} $ involves
the operators $a_{\vec{r}_{j},s1,g}^{\dag}$ and $a_{\vec{r}_{j},s1,g}$ 
of Eq. (\ref{g-s1+general}), which 
create and annihilate, respectively, a superposition of $2D=2,4$ two-site bonds of 
the same type. The expression of such two-site bond operators 
$b_{\vec{r},s1,d,l,g}^{\dag}$ and $b_{\vec{r},s1,d,l,g}$ is given in Eq. (\ref{g-s-l}).

In order to confirm the validity of Eqs. (\ref{g-local}) and (\ref{g-non-local}),
we use Eqs. (\ref{g-s1+general})-(\ref{g-s-l}) to rewrite the anti-commutation 
relations of Eq. (\ref{g-local}) in terms of anti-commutators of two-site bond 
operators as follows,
\begin{equation}
\{g^{\dag}_{{\vec{r}}_{j},s1},g_{{\vec{r}}_{j},s1}\} = 
\sum_{d,l,g}\sum_{d',l',g'} h^*_g\,h_{g'}
\{b_{{\vec{r}}_j +\vec{r}_{d,l}^{\,0},s1,d,l,g}^{\dag},
b_{{\vec{r}}_j +\vec{r}_{d',l'}^{\,0},s1,d',l',g'}\} \, ,
\label{g-local-AP+-}
\end{equation}
\begin{equation}
\{g^{\dag}_{{\vec{r}}_{j},s1},g^{\dag}_{{\vec{r}}_{j},s1}\} = 
\sum_{d,l,g}\sum_{d',l',g'} h^*_g\,h^*_{g'}
\{b_{{\vec{r}}_j +\vec{r}_{d,l}^{\,0},s1,d,l,g}^{\dag},
b_{{\vec{r}}_j +\vec{r}_{d',l'}^{\,0},s1,d',l',g'}^{\dag}\} \, ,
\label{g-local-AP++}
\end{equation}
\begin{equation}
\{g_{{\vec{r}}_{j},s1},g_{{\vec{r}}_{j},s1}\} = 
\sum_{d,l,g}\sum_{d',l',g'} h_g\,h_{g'}
\{b_{{\vec{r}}_j +\vec{r}_{d,l}^{\,0},s1,d,l,g},
b_{{\vec{r}}_j +\vec{r}_{d',l'}^{\,0},s1,d',l',g'}\} \, .
\label{g-local-AP--}
\end{equation}
For simplicity here we have used the abbreviated summation notation,
\begin{equation}
\sum_{d,l,g} \equiv \sum_{d=1}^{D}\sum_{l=\pm 1}\sum_{g=0}^{N_{s1}/2D-1} \, .
\label{sum}
\end{equation}
Moreover, on using of the same equations, the commutation relations of 
Eq. (\ref{g-non-local}) can be expressed in terms of the commutators of two-site 
bond operators. That leads to,
\begin{equation}
[g^{\dag}_{{\vec{r}}_{j},s1},g_{{\vec{r}}_{j'},s1}] = 
\sum_{d,l,g}\sum_{d',l',g'} h^*_g\,h_{g'}
[b_{{\vec{r}}_j +\vec{r}_{d,l}^{\,0},s1,d,l,g}^{\dag},
b_{{\vec{r}}_{j'} +\vec{r}_{d',l'}^{\,0},s1,d',l',g'}] \, ,
\label{g-non-local-AP+-}
\end{equation}
\begin{equation}
[g^{\dag}_{{\vec{r}}_{j},s1},g^{\dag}_{{\vec{r}}_{j'},s1}] = 
\sum_{d,l,g}\sum_{d',l',g'} h^*_g\,h^*_{g'}
[b_{{\vec{r}}_j +\vec{r}_{d,l}^{\,0},s1,d,l,g}^{\dag},
b_{{\vec{r}}_{j'} +\vec{r}_{d',l'}^{\,0},s1,d',l',g'}^{\dag}] \, ,
\label{g-non-local-AP++}
\end{equation}
\begin{equation}
[g_{{\vec{r}}_{j},s1},g_{{\vec{r}}_{j'},s1}] = 
\sum_{d,l,g}\sum_{d',l',g'} h_g\,h_{g'}
[b_{{\vec{r}}_j +\vec{r}_{d,l}^{\,0},s1,d,l,g},
b_{{\vec{r}}_{j'} +\vec{r}_{d',l'}^{\,0},s1,d',l',g'}] \, ,
\label{g-non-local-AP--}
\end{equation}
where $j\neq j'$.

According to the studies of Section VI-C, three rules follow 
from the definition of the subspace where the operators of Eqs. 
(\ref{g-s1+general})-(\ref{g-s-l}) act onto.
The evaluation of the anti-commutators and commutators
of the two-site bond operators on the right-hand side
of Eqs. (\ref{g-local-AP+-})-(\ref{g-local-AP--}) and
(\ref{g-non-local-AP+-})-(\ref{g-non-local-AP--}),
respectively, relies on both such rules and the algebra 
given in Eqs. (\ref{albegra-s-p-m})-(\ref{albegra-s-sz-com})
of the spinon operators
$s^{\pm}_{\vec{r}_j}$ and $s^{x_3}_{\vec{r}_j}$ of Eq. (\ref{sir-pir}),
which are the building blocks of the two-site
bond operators of Eq. (\ref{g-s-l}).
Fortunately, according to Eqs. (\ref{albegra-s-p-m})-(\ref{albegra-s-sz-com})
the spinon operators $s^{\pm}_{\vec{r}_j}$ obey the
usual algebra: They anticommute 
on the same site of the spin effective lattice
and commute on different sites. 

The two-site bond operators of Eq. (\ref{g-s-l}) can be rewritten as,
\begin{equation}
b_{\vec{r}_1,\vec{r}_2}^{\dag} = 
{(-1)^{d-1}\over\sqrt{2}}\left(\left[{1\over 2}+s^{x_3}_{\vec{r}_1}\right]
s^-_{\vec{r}_2} - \left[{1\over 2}+s^{x_3}_{\vec{r}_2}\right]
s^-_{\vec{r}_1}\right) \, ,
\label{g-s-l-simple}
\end{equation}
and $b_{\vec{r}_1,\vec{r}_2} = \left(b_{\vec{r}_1,\vec{r}_2}^{\dag}\right)^{\dag}$
where recalling that the real-space coordinate of their two-site bond
center reads $\vec{r} = \vec{r}_j+\vec{r}_{d,l}^{\,0}$ the
real-space coordinates $\vec{r}_1$ and $\vec{r}_2$ are
given by,
\begin{equation}
\vec{r}_1 = \vec{r}_j+\vec{r}_{d,l}^{\,0}-\vec{r}_{d,l}^{\,g} \, ;
\hspace{0.35cm}
\vec{r}_2 = \vec{r}_{j'}+\vec{r}_{d',l'}^{\,0}+\vec{r}_{d',l'}^{\,g'} \, .
\label{variables}
\end{equation}

The evaluation of the anti-commutators and commutators
of the two-site bond operators on the right-hand side
of Eqs. (\ref{g-local-AP+-})-(\ref{g-local-AP--}) and
(\ref{g-non-local-AP+-})-(\ref{g-non-local-AP--}),
respectively, then relies on straightforward manipulations based on Eqs. 
(\ref{albegra-s-p-m})-(\ref{albegra-s-sz-com}) and the use of the 
three rules given in Section VI-C. The latter exclude unphysical processes.
This is equivalent defining the subspace that the operator algebra under consideration refers to. For the two general
anti-commutators needed to evaluate the two-site bond operators 
on the right-hand side of Eq. (\ref{g-local-AP+-}) we find the
following expressions,
\begin{eqnarray}
\{b_{\vec{r}_1,\vec{r}_2}^{\dag},b_{\vec{r}_1,\vec{r}_2}\} & = &
\sum_{i=1,2}\{{1\over 2}\left({1\over 2}+s^{x_3}_{\vec{r}_i}\right)^2 -
s^+_{\vec{r}_i}s^-_{\vec{r}_{\bar{i}}}\left({1\over 2}+s^{x_3}_{\vec{r}_i}\right)
\left({1\over 2}+s^{x_3}_{\vec{r}_{\bar{i}}}\right) 
\nonumber \\
& + & {1\over 2}s^+_{\vec{r}_i}s^-_{\vec{r}_{\bar{i}}}
\left[\left({1\over 2}+s^{x_3}_{\vec{r}_i}\right) -
\left({1\over 2}+s^{x_3}_{\vec{r}_{\bar{i}}}\right)\right]\} \, ,
\label{anti-com-b+b-on}
\end{eqnarray}
\begin{eqnarray}
\{b_{\vec{r}_1,\vec{r}_2}^{\dag},b_{\vec{r}_{1'},\vec{r}_{2'}}\} & = & (-1)^{d+d'}
\sum_{i=1,2}\{s^+_{\vec{r}_{i'}}s^-_{\vec{r}_{i}}\left({1\over 2}+s^{x_3}_{\vec{r}_{\bar{i}'}}\right)
\left({1\over 2}+s^{x_3}_{\vec{r}_{\bar{i}}}\right)
\nonumber \\
& - & s^+_{\vec{r}_{i'}}s^-_{\vec{r}_{\bar{i}}}\left({1\over 2}+s^{x_3}_{\vec{r}_{\bar{i}'}}\right)
\left({1\over 2}+s^{x_3}_{\vec{r}_{i}}\right)\} \, ,
\label{anti-com-b+b-off}
\end{eqnarray}
where $\bar{1}=2$, $\bar{2}=1$, $\vec{r}_1\neq \vec{r}_{1'}, \vec{r}_{2'}$, and
$\vec{r}_2\neq \vec{r}_{1'}, \vec{r}_{2'}$. Indeed, according to the second rule 
of Section VI-C only general operators $b_{\vec{r}_1,\vec{r}_2}^{\dag}b_{\vec{r}_{1'},\vec{r}_{2'}}$
whose two-site bond operators $b_{\vec{r}_1,\vec{r}_2}^{\dag}$ and $b_{\vec{r}_{1'},\vec{r}_{2'}}$ 
do not join sites or join both sites of the spin effective
lattice lead to wanted and physical spin configurations. 

Moreover, in the initial configuration onto which the four-site operator pairs of two-site bond
operators appearing in Eq. (\ref{anti-com-b+b-on}) act one has according to the first and second rules 
that the two sites of real-space coordinates $\vec{r}_1$
and $\vec{r}_2$ either (i) refer to two deconfined $+1/2$ spinons or
(ii) are linked by a bond. In turn, the
anti-commutator of Eq. (\ref{anti-com-b+b-off}) has the same
form as those on the right-hand side of
Eq. (\ref{g-local-AP+-}). Hence one has that $j=j'$ yet $\vec{r}_{d,l}^{\,g}\neq \vec{r}_{d',l'}^{\,g'}$
in Eq. (\ref{variables})  so that the restrictions imposed by the third rule 
are fulfilled. Since we find that
$b_{\vec{r}_1,\vec{r}_2}^{\dag}b_{\vec{r}_{1'},\vec{r}_{2'}}=
b_{\vec{r}_{1'},\vec{r}_{2'}}b_{\vec{r}_1,\vec{r}_2}^{\dag}$ 
and each of such operators is given by one half the
operator on the right-hand side of Eq. (\ref{anti-com-b+b-off}),
when the two sites of real-space coordinates $\vec{r}_1$ and $\vec{r}_2$ 
(and $\vec{r}_{1'}$ and $\vec{r}_{2'}$) are linked by a bond (and
refer to two deconfined $+1/2$ spinons)
one must consider both initial configurations for which the sites
of real-space coordinates $\vec{r}_{1'}$ and $\vec{r}_{2'}$
(and $\vec{r}_1$ and $\vec{r}_2$) (i) are linked by a bond
and (ii) refer to two deconfined $+1/2$ spinons.
In turn, when the two sites of real-space coordinates 
$\vec{r}_1$ and $\vec{r}_2$ (and $\vec{r}_{1'}$ and $\vec{r}_{2'}$) 
refer to two deconfined $+1/2$ spinons 
(and are linked by a bond) one must consider only initial 
configurations for which the sites
of real-space coordinates $\vec{r}_{1'}$ and $\vec{r}_{2'}$
(and $\vec{r}_1$ and $\vec{r}_2$) are linked by a bond
(and refer to two deconfined $+1/2$ spinons).

It then follows from analysis of the operator expression on
the right-hand side of Eq. (\ref{anti-com-b+b-on}) that
when in the initial configuration the two sites $\vec{r}_1$ and 
$\vec{r}_2$ refer to two deconfined $+1/2$ spinons
the operator term $\sum_{i=1,2}[1/2](1/2+s^{x_3}_{\vec{r}_i})^2$
transforms that configuration onto itself whereas the
remaining operator terms give zero. In turn, when in the initial 
configuration the two sites $\vec{r}_1$ and $\vec{r}_2$
are linked and correspond to a bond configuration, the operator terms
$\sum_{i=1,2}[1/2]\{(1/2+s^{x_3}_{\vec{r}_i})^2
-s^+_{\vec{r}_i}s^-_{\vec{r}_{\bar{i}}}(1/2+s^{x_3}_{\vec{r}_{\bar{i}}})\}$
transform that configuration onto itself whereas the
remaining operator terms give zero. 
On the other hand, when acting onto the above initial configurations 
the operator on the right-hand side of Eq.
(\ref{anti-com-b+b-off}) gives always zero so that when acting
onto the subspace that the operators of Eqs. (\ref{g-s1+general})-(\ref{g-s-l})
refer to the anti-commutators provided in Eqs. (\ref{anti-com-b+b-on})
and (\ref{anti-com-b+b-off}) simplify and are given by,
\begin{equation}
\{b_{\vec{r}_1,\vec{r}_2}^{\dag},b_{\vec{r}_1,\vec{r}_2}\} = 1 \, ;
\hspace{0.50cm}
\{b_{\vec{r}_1,\vec{r}_2}^{\dag},b_{\vec{r}_{1'},\vec{r}_{2'}}\} = 0 \, ,
\label{anti-com-b+b-on-off}
\end{equation}
where $\vec{r}_1\neq \vec{r}_{1'}, \vec{r}_{2'}$ and
$\vec{r}_2\neq \vec{r}_{1'}, \vec{r}_{2'}$.

Next concerning the two general anti-commutators needed to evaluate the 
two-site bond operators on the right-hand side of Eq. (\ref{g-local-AP++}) 
we find the following expressions,
\begin{equation}
\{b_{\vec{r}_1,\vec{r}_2}^{\dag},b_{\vec{r}_1,\vec{r}_2}^{\dag}\} =
-2\left({1\over 2}+s^{x_3}_{\vec{r}_1}\right)\left({1\over 2}+s^{x_3}_{\vec{r}_2}\right)
s^-_{\vec{r}_1}s^-_{\vec{r}_2} -
\sum_{i=1,2}\left({1\over 2}+s^{x_3}_{\vec{r}_i}\right)s^-_{\vec{r}_1}s^-_{\vec{r}_2} \, ,
\label{anti-com-b+b+on}
\end{equation}
\begin{eqnarray}
\{b_{\vec{r}_1,\vec{r}_2}^{\dag},b_{\vec{r}_{1'},\vec{r}_{2'}}^{\dag}\} & = & (-1)^{d+d'}
\sum_{i=1,2}\{\left({1\over 2}+s^{x_3}_{\vec{r}_{i}}\right)
\left({1\over 2}+s^{x_3}_{\vec{r}_{i'}}\right)s^-_{\vec{r}_{\bar{i}}}s^-_{\vec{r}_{\bar{i}'}} 
\nonumber \\
& - & \left({1\over 2}+s^{x_3}_{\vec{r}_{i}}\right)
\left({1\over 2}+s^{x_3}_{\vec{r}_{\bar{i}'}}\right)s^-_{\vec{r}_{\bar{i}}}s^-_{\vec{r}_{i'}}\} \, ,
\label{anti-com-b+b+off}
\end{eqnarray}
where as above $\bar{1}=2$, $\bar{2}=1$, $\vec{r}_1\neq \vec{r}_{1'}, \vec{r}_{2'}$, and
$\vec{r}_2\neq \vec{r}_{1'}, \vec{r}_{2'}$. 

It follows from analysis of the operator on
the right-hand side of Eq. (\ref{anti-com-b+b+on}) that,
according to the first and second rules reported in Section VI-C,
when in the initial spin configuration both the sites of
real-space coordinates $\vec{r}_1$ and 
$\vec{r}_2$ (i) refer to two deconfined $+1/2$ spinons
and (ii) are linked by a bond, application of that operator onto
such a configuration gives zero. In turn the
anti-commutator of Eq. (\ref{anti-com-b+b+off}) is of the
form of those on the right-hand side of
Eq. (\ref{g-local-AP++}) so that in Eq. (\ref{variables}) one
has that $j=j'$ yet $\vec{r}_{d,l}^{\,g}\neq \vec{r}_{d',l'}^{\,g'}$
and thus the third rule of Section VI-C applies. Since we find that
$b_{\vec{r}_1,\vec{r}_2}^{\dag}b_{\vec{r}_{1'},\vec{r}_{2'}}^{\dag}=
b_{\vec{r}_{1'},\vec{r}_{2'}}^{\dag}b_{\vec{r}_1,\vec{r}_2}^{\dag}$ 
and each of such operators is given by one half the
operator on the right-hand side of Eq. (\ref{anti-com-b+b+off}), concerning
the latter operator when the sites of real-space 
coordinates $\vec{r}_1$ and $\vec{r}_2$ (and $\vec{r}_{1'}$ and $\vec{r}_{2'}$)
are linked by a bond one must consider both initial configurations for which the 
sites of real-space coordinates $\vec{r}_{1'}$ and $\vec{r}_{2'}$
(and $\vec{r}_1$ and $\vec{r}_2$) (i) are linked by a bond
and (ii) refer to two deconfined $+1/2$ spinons.
In turn, when the two sites of real-space coordinates 
$\vec{r}_1$ and $\vec{r}_2$ (and $\vec{r}_{1'}$ and $\vec{r}_{2'}$) 
refer to two deconfined $+1/2$ spinons 
one must consider only initial configurations for which the sites
of real-space coordinates $\vec{r}_{1'}$ and $\vec{r}_{2'}$
(and $\vec{r}_1$ and $\vec{r}_2$) are linked by a bond \cite{companion}.
It follows then from analysis of the operator on the right-hand side of Eq. (\ref{anti-com-b+b+off}) 
that application of it onto any of such spin configurations gives zero.

A similar analysis for the two general anti-commutators needed to evaluate the 
two-site bond operators on the right-hand side of Eq. (\ref{g-local-AP--}) leads to, 
\begin{equation}
\{b_{\vec{r}_1,\vec{r}_2},b_{\vec{r}_1,\vec{r}_2}\} =
-2s^+_{\vec{r}_1}s^+_{\vec{r}_2}\left({1\over 2}+s^{x_3}_{\vec{r}_1}\right)\left({1\over 2}+s^{x_3}_{\vec{r}_2}\right) -
\sum_{i=1,2}s^+_{\vec{r}_1}s^+_{\vec{r}_2}\left({1\over 2}+s^{x_3}_{\vec{r}_i}\right) \, ,
\label{anti-com-b-b-on}
\end{equation}
\begin{eqnarray}
\{b_{\vec{r}_1,\vec{r}_2},b_{\vec{r}_{1'},\vec{r}_{2'}}\} & = & (-1)^{d+d'}
\sum_{i=1,2}\{s^+_{\vec{r}_{\bar{i}}}s^+_{\vec{r}_{\bar{i}'}}\left({1\over 2}+s^{x_3}_{\vec{r}_{i}}\right)
\left({1\over 2}+s^{x_3}_{\vec{r}_{i'}}\right) 
\nonumber \\
& - & s^+_{\vec{r}_{\bar{i}}}s^+_{\vec{r}_{i'}}\left({1\over 2}+s^{x_3}_{\vec{r}_{i}}\right)
\left({1\over 2}+s^{x_3}_{\vec{r}_{\bar{i}'}}\right)\} \, ,
\label{anti-com-b-b-off}
\end{eqnarray}
where as above $\bar{1}=2$, $\bar{2}=1$, $\vec{r}_1\neq \vec{r}_{1'}, \vec{r}_{2'}$, and
$\vec{r}_2\neq \vec{r}_{1'}, \vec{r}_{2'}$. 

Again analysis of the operator on
the right-hand side of Eq. (\ref{anti-com-b-b-on}) reveals that
according to the first and second rules of Section VI-C
when in the initial spin configuration both the sites of
real-space coordinates $\vec{r}_1$ and 
$\vec{r}_2$ (i) refer to two deconfined $+1/2$ spinons
and (ii) are linked by a bond, application of that operator onto
such a configuration gives zero. On the other hand, the
anti-commutator of Eq. (\ref{anti-com-b-b-off}) is of the
form of those on the right-hand side of
Eq. (\ref{g-local-AP++}) so that in Eq. (\ref{variables}) one
has that $j=j'$ yet $\vec{r}_{d,l}^{\,g}\neq \vec{r}_{d',l'}^{\,g'}$
and then the third rule applies. Since we find that
$b_{\vec{r}_1,\vec{r}_2}b_{\vec{r}_{1'},\vec{r}_{2'}}=
b_{\vec{r}_{1'},\vec{r}_{2'}}b_{\vec{r}_1,\vec{r}_2}$ 
and each of such operators is given by one half the
operator on the right-hand side of Eq. (\ref{anti-com-b+b+off}), concerning
the latter operator when the sites of real-space 
coordinates $\vec{r}_1$ and $\vec{r}_2$ (and $\vec{r}_{1'}$ and $\vec{r}_{2'}$)
refer to two deconfined $+1/2$ spinons one must 
consider both initial configurations for which the 
sites of real-space coordinates $\vec{r}_{1'}$ and $\vec{r}_{2'}$
(and $\vec{r}_1$ and $\vec{r}_2$) (i) are linked by a bond
and (ii) refer to two deconfined $+1/2$ spinons.
In turn, when the two sites of real-space coordinates 
$\vec{r}_1$ and $\vec{r}_2$ (and $\vec{r}_{1'}$ and $\vec{r}_{2'}$) 
are linked by a bond one must consider only initial configurations for which the sites
of real-space coordinates $\vec{r}_{1'}$ and $\vec{r}_{2'}$
(and $\vec{r}_1$ and $\vec{r}_2$) refer to two 
deconfined $+1/2$ spinons \cite{companion}. 
Analysis of the operator on the right-hand side of Eq. (\ref{anti-com-b+b+off}) 
then reveals that application of it onto any of such spin configurations gives zero.

It then follows from the above results that when acting
onto the subspace that the operators of Eqs. (\ref{g-s1+general})-(\ref{g-s-l})
refer to the anti-commutators provided in Eqs. (\ref{anti-com-b+b+on})
and (\ref{anti-com-b+b+off}) and Eqs. (\ref{anti-com-b-b-on})
and (\ref{anti-com-b-b-off}) simplify and read,
\begin{equation}
\{b_{\vec{r}_1,\vec{r}_2}^{\dag},b_{\vec{r}_1,\vec{r}_2}^{\dag}\} =
\{b_{\vec{r}_1,\vec{r}_2}^{\dag},b_{\vec{r}_{1'},\vec{r}_{2'}}^{\dag}\} = 0 
\, ; \hspace{0.5cm}
\{b_{\vec{r}_1,\vec{r}_2},b_{\vec{r}_1,\vec{r}_2}\} =
\{b_{\vec{r}_1,\vec{r}_2},b_{\vec{r}_{1'},\vec{r}_{2'}}\} = 0 \, ,
\label{anti-com-b+-b+-on-off}
\end{equation}
where $\vec{r}_1\neq \vec{r}_{1'}, \vec{r}_{2'}$ and
$\vec{r}_2\neq \vec{r}_{1'}, \vec{r}_{2'}$.

The use in Eq. (\ref{g-local-AP+-}) of the anti-commutators of Eq. (\ref{anti-com-b+b-on-off}) with 
the two-site bond operators related to those of Eq. (\ref{g-s-l}) by 
the expressions provided in Eqs. (\ref{g-s-l-simple}) and (\ref{variables}) leads to,
\begin{equation}
\{g^{\dag}_{{\vec{r}}_{j},s1},g_{{\vec{r}}_{j},s1}\} =
\sum_{d,l,g} \vert h_g\vert^2 = 2D\sum_g \vert h_g\vert^2 =1 \, ,
\label{g-local-AP+-final}
\end{equation}
which is the first relation of Eq. (\ref{g-local}). To perform the
summation of Eq. (\ref{g-local-AP+-final}) the sum-rule (\ref{g-s1+sum-rule}) 
is used. Furthermore, the use of the anti-commutators of Eq. (\ref{anti-com-b+-b+-on-off})
in Eqs. (\ref{g-local-AP++}) and (\ref{g-local-AP--}) leads to the remaining
relations of Eq. (\ref{g-local}).

The evaluation of the commutators of Eq. (\ref{g-non-local}) by the use of the 
expressions given in Eqs. (\ref{g-non-local-AP+-})-(\ref{g-non-local-AP--})
is much simpler. First it is simplified by the property that two-site bonds belonging
to $s1$ bond-particle operators with different real-space coordinates
are always different. Second the evaluation of such commutators further relies 
on straightforward manipulations based on Eqs. 
(\ref{albegra-s-p-m})-(\ref{albegra-s-sz-com}), which lead directly to,
\begin{equation}
[b_{\vec{r}_1,\vec{r}_2}^{\dag},b_{\vec{r}_{1'},\vec{r}_{2'}}] =
[b_{\vec{r}_1,\vec{r}_2}^{\dag},b_{\vec{r}_{1'},\vec{r}_{2'}}^{\dag}] =
[b_{\vec{r}_1,\vec{r}_2},b_{\vec{r}_{1'},\vec{r}_{2'}}] = 0 \, ,
\label{com-all}
\end{equation}
for $\vec{r}_1\neq \vec{r}_{1'}, \vec{r}_{2'}$ and
$\vec{r}_2\neq \vec{r}_{1'}, \vec{r}_{2'}$.

Finally, the use in Eqs. (\ref{g-non-local-AP+-})-(\ref{g-non-local-AP--}) of the commutators of Eq. (\ref{com-all}) with 
the two-site bond operators related to those of Eq. (\ref{g-s-l}) by the expressions provided in Eqs. (\ref{g-s-l-simple}) 
and (\ref{variables}) leads readily to the commutation relations provided in Eq. (\ref{g-non-local}).
                                                                      

\end{document}